\documentclass[%
prd,preprint,
nofootinbib,preprintnumbers,
superscriptaddress,tightenlines
 reprint,
]{revtex4-2}
\usepackage[utf8]{inputenc}
\usepackage[T1]{fontenc}
\usepackage{custom}
\usepackage{graphicx}
\usepackage{subcaption}
\usepackage{dcolumn}
\usepackage{bm}

\graphicspath{{images/}}

\begin{document}

\preprint{MIT-CTP/5417}

\title{Freezing Out Fluctuations in Hydro+ Near the QCD Critical Point}

\author{Maneesha Pradeep}
\affiliation{Department of Physics, University of Illinois at Chicago, Chicago, IL 60607, USA}
\author{Krishna Rajagopal}%
\affiliation{Center for Theoretical Physics, Massachusetts Institute of Technology,
Cambridge, MA 02139, USA}


\author{Mikhail Stephanov}
\affiliation{Department of Physics, University of Illinois at Chicago, Chicago, IL 60607, USA}
\author{Yi Yin}
\affiliation{Quark Matter Research Center, Institute of Modern Physics, Chinese Academy of Sciences, Lanzhou, Gansu, 073000, China}
\date{April 2, 2022}

\begin{abstract}
We introduce a freeze-out procedure to convert the critical fluctuations in a droplet of quark-gluon plasma (QGP) that has, as it expanded and cooled, passed close to a posited critical point on the phase diagram into cumulants of hadron multiplicities that can subsequently be measured.
The procedure connects the out-of-equilibrium critical fluctuations described in concert with the hydrodynamic evolution of the droplet of QGP by extended hydrodynamics, known as Hydro+, with the subsequent kinetic description in terms of observable hadrons.
We introduce a critical scalar isoscalar field sigma whose fluctuations cause
correlations between observed hadrons due to the couplings of the
sigma field to the hadrons via their masses.
We match the QGP fluctuations obtained by solving the Hydro+ equations
describing the evolution of critical fluctuations before freeze-out 
to the correlations of the sigma field. In turn, these are imprinted onto
correlations and fluctuations in the multiplicity of hadrons, most importantly protons, after freeze-out via 
the generalization of the familiar half-century-old Cooper-Frye freeze-out prescription which we introduce.
The proposed framework allows us to study the effects of critical slowing down 
and the consequent deviation of the observable predictions from equilibrium expectations
quantitatively.
We also quantify the suppression of cumulants due to conservation of baryon number. We demonstrate the procedure in practice by freezing out a Hydro+ simulation in an azimuthally symmetric and boost invariant background that includes radial flow discussed in  Ref.~\cite{Rajagopal:2019xwg}.

\end{abstract}

\maketitle

\tableofcontents

\section{Introduction}
\label{sec:introduction}

Understanding the physics of strongly interacting matter in extreme
conditions and mapping the phase diagram of QCD has been a major goal
of theoretical and experimental efforts from high-energy heavy-ion
collisions to neutron star mergers \cite{Aprahamian:2015qub, Busza:2018rrf, Bzdak:2019pkr}. One possible central feature of the
phase diagram of quantum chromodynamics (QCD) --- a QCD critical point 
at the baryon chemical potential $\mu_B$ above which the crossover via which cooling
quark-gluon plasma (QGP) becomes ordinary hadronic matter becomes a first order phase 
transition --- still remains a theoretical
conjecture. The challenge of its discovery is being taken up by the
current Beam Energy Scan (BES) program at the Relativistic Heavy Ion Collider (RHIC) at Brookhaven National Laboratory as well as at planned facilities worldwide. 
The
intriguing hints observed in the first phase of the BES~\cite{Aprahamian:2015qub,STAR:2021iop}, in particular the deviations of certain measures of fluctuations from their non-critical baseline, deviations that vary non-monotonically as a function of $\sqrt{s}$, motivate the ongoing
experimental efforts in its second, higher statistics, phase (BES II).  BES II data has been taken by the STAR collaboration over the course of 2019-2021 in AuAu collisions at a sequence of collision energies, which corresponds to a scan in $\mu_B$~\cite{Stephanov:1998dy,Stephanov:1999zu}.  The data is currently being analyzed and we look forward with considerable anticipation to learning much from these measurements.

On the theory side, there have been many studies of the observable consequences of critical fluctuations in heavy ion collisions that 
produce a droplet of QGP that cool close to a critical point upon making the (greatly simplifying, but unrealistic) assumption that these fluctuations stay in equilibrium~\cite{Stephanov:1998dy,Stephanov:1999zu,Hatta:2003wn,Stephanov:2008qz,Asakawa:2009aj,Athanasiou:2010kw,Stephanov:2011pb,Ling:2015yau,Brewer:2018abr,Bzdak:2019pkr}.
A part of the essence of critical fluctuations is that 
since their correlation length, $\xi$, grows near the critical point, the
typical timescale for their evolution grows also -- this is referred to as critical slowing down.  This means that in the rapidly cooling droplets of QGP produced in heavy ion collisions critical fluctuations cannot be described by their equilibrium values slaved to hydrodynamic fields.
Fortunately, much progress has also been achieved in describing
the {\em non-equilibrium} evolution of hydrodynamic
fluctuations \cite{Berdnikov:1999ph,  Mukherjee:2015swa, Mukherjee:2016kyu,Stephanov:2017ghc, Akamatsu:2016llw, Akamatsu:2018vjr, Bluhm:2019yfb, Akamatsu:2017rdu, An:2019osr, Martinez:2018wia,  Martinez:2019bsn, An:2019csj, An:2020vri, Bluhm:2020mpc, Shen:2017bsr, Rajagopal:2019xwg, Du:2020bxp, An:2021wof, Kapusta:2011gt,Murase:2013tma, Young:2014pka, Murase:2016rhl,Singh:2018dpk,Nahrgang:2017oqp,Nahrgang:2018afz, Nahrgang:2020yxm,Sakai:2020pjw}, whose distinctive behavior at long wavelengths is  governed
by the universality of critical behavior and thus can serve as a signature of the
critical point. The challenge which still needs to be addressed is
establishing a connection between the hydrodynamic fluid --
including its critical fluctuations -- and the observed particle yields -- and
their fluctuations. This work is aimed at closing this gap.

Traditionally one thinks of hydrodynamics as a deterministic theory of
fluid evolution. The hydrodynamic variables are the local fluid
velocity $u(x)$ as well as 
{\em average} densities
of conserved quantities like the energy density $\varepsilon$ and the baryon number density $n$, or, equivalently, the corresponding conjugate
variables such as the temperature $T$ and chemical potentials (like
the chemical potential for baryon number, $\mu_B$) characterizing the local
equilibrium conditions. These variables evolve deterministically according to hydrodynamic equations. Heavy ion collision experiments, of
course, do not measure these hydrodynamic or thermodynamic quantities
directly. The conversion to experimentally observable
hadron multiplicities occurs at freezeout -- a point in the evolution of the
expanding cooling droplet of matter where the density becomes low enough that the kinetic
description in terms of a hadron gas becomes applicable and the scattering
rates for processes that modify the particle species (i.e., chemical) composition is
negligible. At that point one can convert the hydrodynamic variables
into particle yields and momentum distributions, i.e.~spectra. 
The well-known procedure known as
Cooper-Frye freezeout~\cite{Cooper:1974mv} maps the local fluid
velocity $u(x)$ and 
hydrodynamic fields such as  $T(x)$ and
$\mu_B(x)$ on the freezeout surface (a hypersurface in spacetime) 
to a simplified hadronic description in terms of kinetic variables of an expanding ideal resonance gas of hadrons, namely a
gas of particles with momenta distributed
according to boosted Fermi-Dirac or Bose-Einstein distributions. The interactions
are encoded in resonances, which later decay, modifying the
distributions ultimately measured by experiment. The average densities
of conserved quantities such as energy or baryon number are guaranteed
to match as long as the hadron resonance gas provides a good
description of the equation of state, including the relations between $T$ and
$\mu_B$ and $\varepsilon$ and $n$ in that regime.

The Cooper-Frye freezeout procedure~\cite{Cooper:1974mv}
has been successfully employed in the description of high energy
heavy-ion collision data for more than four decades to describe average particle yields and spectra.
The procedure ensures that the event-averaged baryon number and energy-momentum
densities are matched between the hydrodynamic and kinetic theory descriptions.
At sufficiently high collision energies $\sqrt{s}$, the data from many experiments are in
reasonable agreement with this description across a broad kinematic regime.
The Cooper-Frye framework, however, does not describe fluctuations in either the hydrodynamic fluid or the kinetic theory particles.
Our goal is to
extend the Cooper-Frye procedure to the description of critical
fluctuations.  Such a description is crucial in the special case of heavy ion collisions
that freeze out in the vicinity of a critical point. In this case, fluctuations are both enhanced
and of considerable interest, since it is via detecting critical fluctuations that we hope
to discern the presence of a critical point~\cite{Stephanov:1998dy,Stephanov:1999zu}.
These fluctuations are due to thermal noise and their
magnitude, or more importantly, their correlations are a sensitive
signature of the proximity of a thermodynamic singularity, such as the
critical point. Obviously, we cannot expect to match these
critical fluctuations using a {\em free} gas of hadrons. The effect of the
critical correlations can be captured by a critical scalar field, which we
call $\sigma$ --- a collective mode which becomes ``soft'', long-range
correlated and slow, at the critical point, justifying its treatment as
a collective field. One can then match the singular part of fluctuations of
hydrodynamic variables by the fluctuation of the field $\sigma$
which, via its coupling to the observed particles, causes their masses to fluctuate at the time of freezeout and consequently yields observable fluctuations in particle multiplicities.

In this paper we show how to implement such a procedure and
demonstrate its application in a model of the hydrodynamic evolution
near the critical point already studied in Ref.~\cite{Rajagopal:2019xwg}.

Our paper is organized as follows.  In the remainder of this
  Introduction, in Section~\ref{introduction2} we review some
  foundational aspects of critical fluctuations in equilibrium and of
  critical slowing down. We also introduce the Hydro+ equations that describe the dynamics of out-of-equilibrium fluctuations near a critical point. In Section~\ref{CooperFrye} we review the standard Cooper-Frye freezeout procedure that neglects fluctuations.
With this groundwork in place, in Section~\ref{presc1}, we derive and explain
our freezeout procedure that extends the Cooper-Frye approach so as to match the critical fluctuations just before freezeout, as described by Hydro+, to observable fluctuations in particle multiplicities just after.
In Section~\ref{Sectbjk} we apply our freezeout procedure to the Bjorken scenario: a fluid that is undergoing boost-invariant longitudinal expansion, meaning that it is cooling, but that is translation-invariant and at rest in the transverse directions.  In this simplified setting, we are able to push much of the calculation through analytically and in so doing gain intuition and elucidate general features
that arise again in the next Section.  In Section~\ref{RRWY} we illustrate the use of the freezeout procedure that we have introduced and fully exercise its salient features 
by obtaining the two-point correlations of particle multiplicities
from the more realistic Hydro+ simulation of Ref.~\cite{Rajagopal:2019xwg} in which the fluid is boost invariant and azimuthally symmetric but  is 
finite in transverse extent and thus exhibits radial flow.
We conclude in Section~\ref{SummaryOutlook} with a summary of the main qualitative lessons
that we draw from our results of Sections~\ref{Sectbjk} and \ref{RRWY} as well as a look ahead 
at important next steps.

\subsection{Equilibrium fluctuations,  critical slowing down, and evolution equations for out-of-equilibrium fluctuations near a critical point}
\label{introduction2}

Thermodynamic equilibrium systems near a critical point are
characterized by a certain singular behavior of fluctuations. Due to the
divergence of the correlation length $\xi$  the overall
magnitude of fluctuations diverges in the thermodynamic limit
$V\gg\xi^3$, where $V$ is the system volume, as $\xi\to\infty$.  The
universality of critical behavior means that the leading singular behavior
of the magnitude of fluctuations with $\xi$ is insensitive to 
microscopic details of the physical system
for different systems in the same universality
class. The QCD critical point falls in the static universality class of the 3D
Ising model \cite{Berges:1998rc, Halasz:1998qr, Guida:1996ep,
  Zinn-Justin:1989rgp}, with a single scalar field becoming soft and slow at the critical 
  point.\footnote{In a world with massless up and down quarks and, consequently, three massless pion fields, there would be four soft fields at the critical point, which would then be in the $O(4)$ universality class. See Refs.~\cite{Rajagopal:1992qz,Berges:1998rc,Halasz:1998qr,Rajagopal:2000wf} for further discussion.}
  In QCD, the critical field is a linear combination of scalar operators such as the chiral condensate $\langle \bar q q \rangle$ and the baryon number density.
This means, in particular, that,
in equilibrium near the critical point, the $k^{th}$ cumulants of the
baryon number density, which are related to the the derivatives of pressure $P$
with respect to baryon chemical potential $\mu_B$ at fixed temperature
$T$, diverge as certain powers of $\xi$~\cite{Stephanov:2008qz}:
\begin{eqnarray}
\label{eq-cumulants}
\left<\delta n_B^{k}\right>_{\text{eq}}= \left(\frac{T}{V}\right)^{k-1}\,\frac{\partial^{k}P(T,\mu_B)}{\partial \mu_B^k}\sim \xi^{\frac{k(5-\eta)}{2}-3}\,,
\end{eqnarray}
where $\eta\approx 0.04$ is the well-known Ising critical exponent
\cite{Guida:1996ep, Zinn-Justin:1989rgp}. The smallness of this
exponent makes it negligible in practical applications we will be
concerned with here. In this paper we shall only be concerned with the
variance of particle multiplicities, which is to say we shall only need Eq.~(\ref{eq-cumulants}) with
$k=2$.

The singular behavior of fluctuations predicted by
Eq.~(\ref{eq-cumulants}) points the way toward finding signatures of the presence of a critical
point in the phase diagram of QCD in heavy-ion collision experiments.
As the location of the freeze-out point
on the QCD phase diagram, which moves in response to experimentally
varying a parameter such as center of mass collision energy $\sqrt s$,
approaches and then passes the location of the QCD critical point, the
magnitude and higher cumulants of fluctuations should show a
characteristic non-monotonic dependence. There are, however, two essential and quite nontrivial
steps which must be taken in order to connect the elegant
scaling equation~(\ref{eq-cumulants}) to experimental
observables.

First, due to the fact that the droplet of matter produced in a heavy-ion
collision expands and cools rapidly  it is far from being a static thermodynamic system, meaning that
non-equilibrium effects on fluctuations must be
considered ~\cite{Berdnikov:1999ph,Mukherjee:2015swa}; 
as the temperature of the fluid drops 
the fluctuations in the fluid do not
have time to develop in the way that they would in equilibrium.
Furthermore, critical slowing down is an essential feature of physics near a critical point, meaning that the time needed for fluctuations to grow becomes longer the closer one gets to a critical point.
This means that 
non-equilibrium effects on fluctuations become even more important near a critical
point than elsewhere. The description of such non-equilibrium effects is provided by
an extension of hydrodynamics known as Hydro+~\cite{Stephanov:2017ghc} that we shall review briefly in this Section.

The second necessary advance  arises 
due to the fact that experimental measurements do not access
hydrodynamic variables, such as the densities of conserved quantities arising in Eq.(\ref{eq-cumulants}),
directly. Therefore, a connection needs to be made between such
quantities, and their fluctuations, and the experimentally observable
multiplicities of protons, pions, etc. In this paper we propose how
this can be done via introducing a scalar field $\sigma$ in the kinetic theory description of the observable hadrons whose fluctuations can be determined via matching to the critical fluctuations of hydrodynamic quantities, and illustrate it using simplified, but realistic, examples.

Hydro+ extends hydrodynamics by considering the evolution of
fluctuations of hydrodynamic variables towards their local
thermodynamic equilibrium distribution. These fluctuations are
characterized by correlation functions. In this paper, both for simplicity and as a necessary first step,
we shall focus on the {\em magnitude\/} of fluctuations and defer the discussion
of non-Gaussianity measures to future work. The magnitude of the
fluctuations is characterized by a two-point correlation
function. Near the critical point the slowest hydrodynamic mode is the
entropy per baryon $\hs\equiv s/n$ and its fluctuations relax on the
(parametrically) longest times scale \cite{Stephanov:2017ghc}. Thus the non-equilibrium
dynamics is most important in this mode.  
Furthermore, near the critical point, we can imagine mapping the QCD
energy density $\varepsilon$ and baryon number density $n$ to the
3D Ising entropy density $S$ and Ising magnetization density $M$
or, equivalently, mapping the QCD phase diagram variables $T$ and
$\mu_B$ to the 3D Ising variables $r$ (reduced Ising model temperature) and
  $h$ (magnetic field) as set up
explicitly in Refs.~\cite{Parotto:2018pwx,Pradeep:2019ccv}, and then
determining which combination of $\varepsilon$ and $n$ corresponds to the most singular Ising model fluctuations.  
In practice, though, any
combination of $S$ and~$M$ has the same leading singular behavior in powers of
the correlation length $\xi$ as long as it involves fluctuations of
$M$, and a similar statement applies to any combination of
$\varepsilon$ and~$n$ as long as it involves fluctuations of the entropy per baryon $\hat{s}\equiv s/n$
~\cite{Stephanov:2017ghc,Akamatsu:2018vjr}. It therefore suffices
to use the fluctuations of $\hat{s}$, which
also happens to be the slowest hydrodynamic mode. We shall match these fluctuations to the leading
singular contribution to the fluctuations of the $\sigma$ field that we shall introduce in Section \ref{presc1}.

In the local rest frame of the fluid, the equal-time correlation function of $\hat s$ can be expressed in terms of its
Wigner transform:
	 \begin{eqnarray}\label{eq:phiQ=<ss>}
\phi_\mathbf{Q}(x) \equiv \int_{\Delta \bm x} \left<\delta \hat
           s\left(x_+\right)\delta \hat s\left(x_-\right)\right>\,
           e^{i\mathbf{Q}\cdot{\Delta \bm x}}\,.
\end{eqnarray}
Here $x=(x_++x_-)/2$ and  $\Delta x=x_+-x_-=
(0,\Delta\bm x)$ in the local rest frame of the fluid at the point $x$. The relaxation of this
quantity to its local equilibrium value $\bar{\phi}_{\mathbf{Q}}$ is
governed by the equation \cite{Stephanov:2017ghc}:
\begin{eqnarray}
	\label{phiev1}
	u(x)\cdot \partial \phi_\mathbf{Q}(x)=-\Gamma(\bm Q)\, \left(\phi_{\mathbf{Q}}(x)-\bar{\phi}_{\mathbf{Q}}(x)\right)\,.
	\end{eqnarray}		  
        For the purposes of this paper, the equilibrium $\bar{\phi}_\mathbf{Q}$ can be adequately approximated by the Ornstein-Zernike
        ansatz~\footnote{While the value of $\bar\phi_{\bm Q}$ at
          $\bm Q=0$ in Eq.~(\ref{phiev1}) is determined by thermodynamics, the dependence on
          $\bm Q$ in this expression is an often used approximation which takes into
          account the nonzero correlation length. A more sophisticated
          form can be found in Ref.~\cite{Stephanov:2017ghc}.}
\begin{eqnarray}
	\label{phieq1}
	\bar{\phi}_\mathbf{Q}\approx\frac{{ c_p/n^2}}{1+(Q\xi)^2}\,,
	\end{eqnarray}
where $Q\equiv|\bm Q|$ and $c_p$ and $n$ are the heat capacity at constant pressure and the baryon number density.

The $Q$-dependent relaxation rate,
$\Gamma$, controls how slowly $\phi_{\mathbf{Q}}(x)$ evolves toward its equilibrium value $\bar{\phi}_{\mathbf{Q}}$
via Eq.~(\ref{phiev1}).
The leading critical behavior of $\Gamma$ 
depends on the dynamic universality class. For the QCD
critical point~\cite{Son:2004iv}, it is the one of Model H (liquid-gas
critical point) in
Halperin and Hohenberg's classification~\cite{Hohenberg:1977ym}, where the linear 
combination of $\delta \varepsilon$ 
and $\delta n$ given by $\delta \hat s = (\delta\varepsilon-(w/n)
\delta n)/(Tn)$ is the slow, diffusive, scalar  mode nonlinearly coupled to
diffusive (transverse) momentum modes. At
the same level of approximation as in Eq.~(\ref{phieq1}), the leading
critical behavior of the relaxation rate in model H is given by~\cite{KAWASAKI19701}: 
	\begin{eqnarray}
	\label{GammaH}
\Gamma(\bm Q)=\frac{2\dip\xi_0}{\xi^{3}} K(Q\xi),
\end{eqnarray}
where $\xi_0$ is a typical value of the correlation length well away from the critical point, $D_0$ is a constant with the dimensions of length that we shall take as a free parameter,  and
\begin{equation}
\label{Kawasaki1}
    K(x)\equiv \frac{3}{4}\left[1+x^2+(x^3-x^{-1})\arctan x\right]\,.
\end{equation}
As we shall demonstrate, the most important property of the critical
mode relaxation rate given by Eqs.~(\ref{GammaH}-\ref{Kawasaki1}) is
that it vanishes as $Q\to0$:
\begin{equation}
  \label{eq:Gamma0}
  \Gamma(\bm Q) = \frac{2D_0\xi_0}{\xi} Q^2 + \mathcal O( Q^4)\,.
\end{equation}
This reflects the fact that $\phi_{\bm Q}$ measures the
fluctuation of hydrodynamic variables, which are conserved. The
relaxation rate of the 2-point correlator of fluctuations is twice the
rate of the relaxation of the corresponding mode, whose relaxation is
also diffusive with a diffusion coefficient given by 
\begin{equation}
\label{DD0}
D=D_0\frac{\xi_0}{\xi}
\end{equation}
which vanishes at the critical point, where $\xi\to\infty$. We can
think of the parameter $D_0$ which we introduced as the diffusion
constant at some reference point well away from the critical point
.
A crude estimate for  $\dip$ could be obtained by using~\cite{KAWASAKI19701,Hohenberg:1977ym}
	\begin{eqnarray}
	\dip\approx \frac{T}{6\pi\eta\xi_0}\approx \frac{2T}{3 s(T)\xi_0}\,,
	\end{eqnarray}
	 where $\eta$ and $s$ are the shear viscosity and entropy density, respectively, and where
	 we have taken $\eta\approx s/(4\pi)$. Taking $s(T)=\tilde{s} T^3$ with $\tilde{s}\approx 6$ as is reasonable around $T=T_c$~\cite{Borsanyi:2013bia,HotQCD:2014kol} and choosing $T_c=160$~MeV and $\xi_0=0.5$~fm as we shall throughout, we estimate a critical contribution of 
	 $\dip\approx 0.3$~fm.
Assuming that the non-critical contribution to $\dip$ is not too large, 
we expect $\dip>0.3$~fm but not $\dip\gg 0.3$~fm.  To bracket the uncertainty in this estimate, we shall 
typically illustrate our results by plotting the results obtained from calculations employing $\dip=0.25$~fm and
$\dip=1$~fm.

To elucidate and emphasize the importance of the conservation laws in the dynamics
of fluctuations and, consequently, in the experimental signatures of the critical point
we shall compare and contrast results obtained using the model H
universality class with those which one would have obtained using
model A universality class. In the model A universality class, the critical
order parameter is {\it not} a conserved quantity and the relaxation rate of
the fluctuations does not vanish as $\bm Q\to 0$. To the same level of
approximation as we have been using so far we can utilize the following
ansatz for the relaxation rate in model A:
	\begin{eqnarray}
	\label{GammaA}
          \Gamma(\bm Q)=\frac{\Gamma_0\xi_0^2}{\xi^{2}}\,
          \left(1+(Q\xi)^2\right)\,,
          \qquad \mbox {(model A),}
\end{eqnarray}
where $\Gamma_0$ is a constant with the dimensions of rate ($1/$time) which we can think
of as the relaxation rate at a point well away from the critical point where the correlation length is
$\xi_0$.

\begin{figure}[t]
  \centering
  \includegraphics[width=0.7\textwidth]{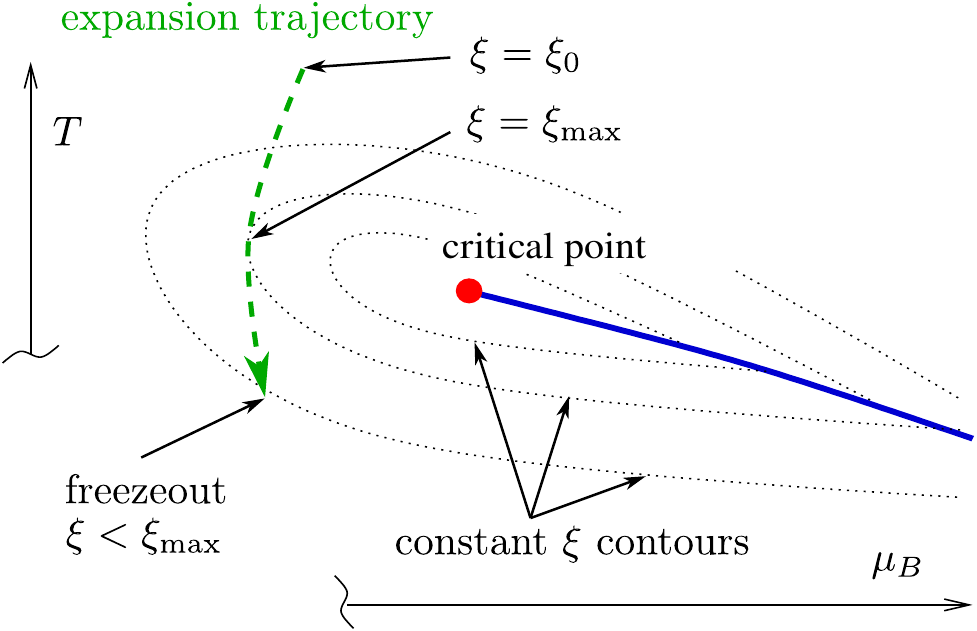}
  \caption{Schematic view of a trajectory followed by an expanding cooling droplet of matter produced in a heavy ion collision on the
    QCD phase diagram in the vicinity of the critical point.}
  \label{fig:trajectory}
\end{figure}

As the fireball expands, its temperature $T$ drops and the point
characterizing the thermodynamic state of the system on the QCD phase
diagram moves past the critical point. The correlation length $\xi$
reaches the maximum value $\xi_{\rm max}$ (see
Fig.~\ref{fig:trajectory}) which depends on how close the trajectory
is to the critical point. The parameter $\xi_{\rm max}$ is controlled
experimentally by varying the collision energy $\sqrt s$~\cite{Stephanov:1998dy,Stephanov:1999zu} since collisions with lower $\sqrt{s}$ produce droplets of matter containing a greater excess in the number of quarks over the number of antiquarks, meaning a higher baryon chemical potential $\mu_B$.  Lowering the collision energy in steps, as in the Beam Energy Scan program at RHIC, moves the entire expansion trajectory in Fig.~\ref{fig:trajectory}, including the freezeout point, rightward in steps.

The fluctuation evolution equation (\ref{phiev1}) depends on the correlation length
$\xi$ via the dependence of $\bar\phi_{\bm Q}$ and $\Gamma(\bm Q)$ on
$\xi$. In a realistic hydrodynamic simulation, $\xi$ will be determined
upon solving the hydrodynamic equations with a given equation of state.
Since our purpose in this paper is to describe how to freeze out
critical fluctuations in hydrodynamics and translate them into
observables based on particle multiplicity fluctuations, we shall
instead, for simplicity, choose a plausible parametrization of $\xi$
along the expansion trajectory in terms of $T$.

\begin{figure}[t]
\begin{center}
\includegraphics[scale=0.7]{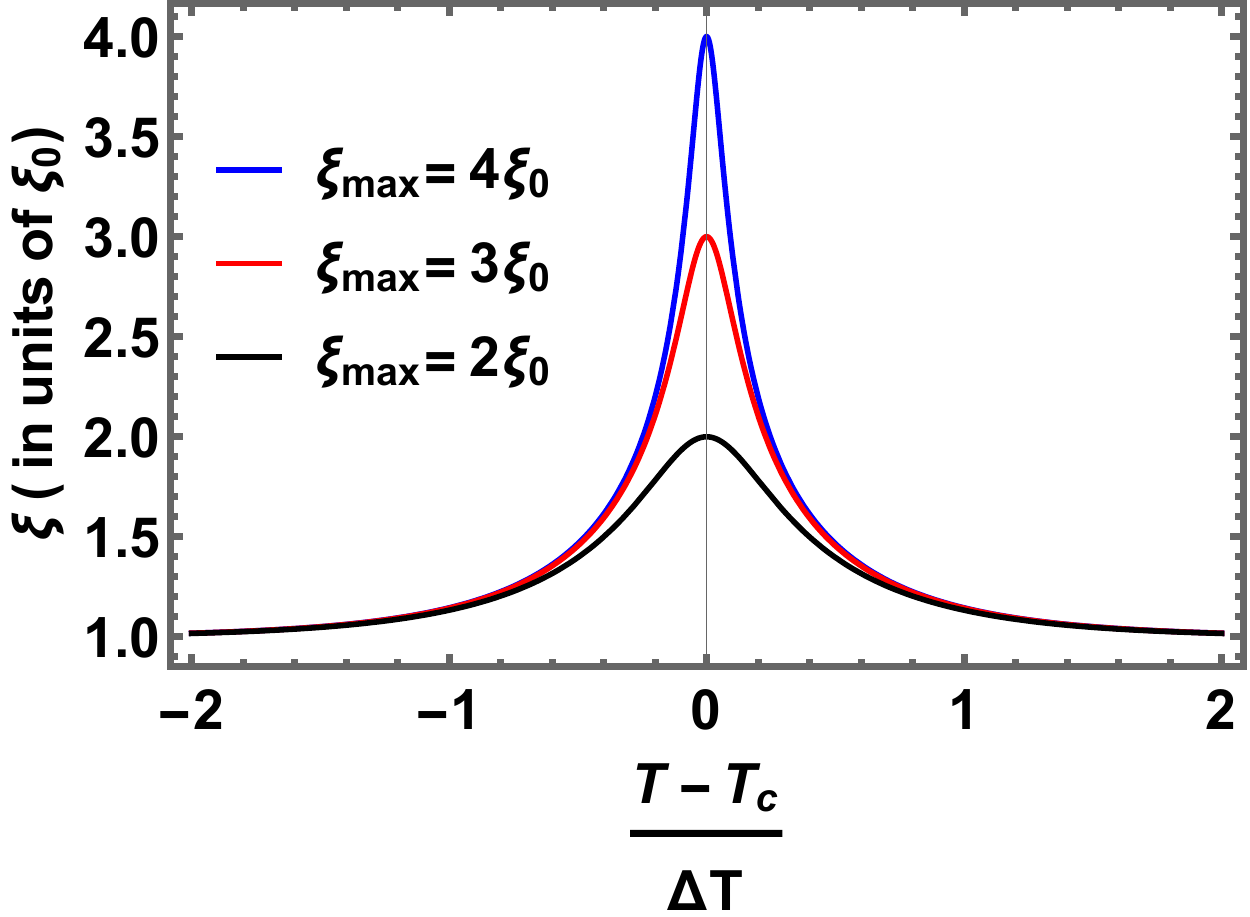}
\end{center}
\caption{The dependence of the correlation length $\xi$ on temperature
for different trajectories of the fireball expansion (i.e., different
$\xi_{\rm max}$).}
\label{fig:xi}
\end{figure}

Specifically, we shall adopt the parametrization of the correlation
length along the trajectory of the expanding fireball on the QCD phase
diagram in terms of temperature used previously in Ref.~\cite{Rajagopal:2019xwg}:
\begin{eqnarray}\label{xiparam1}
\left(\frac{\xi}{\xi_0}\right)^{-4}=\tanh^{2}\left(\frac{T-T_c}{\Delta T}\right)\left[1-\left(\frac{\xi_{\text{max}}}{\xi_0}\right)^{-4}\right]+\left(\frac{\xi_{\text{max}}}{\xi_0}\right)^{-4}\,,
\end{eqnarray}
with $\Delta T=T_c/5$. We shall not attempt to refine this parametrization in this
work. Alternate parametrizations for the correlation length are
discussed, e.g., in Refs.~\cite{Athanasiou:2010kw, Du:2020bxp}. The ansatz
in Eq.~(\ref{xiparam1}) reflects the main features of $\xi(T)$ relevant
for this work --- the correlation length reaches a maximum value
$\xi_{\rm max}$ at a certain temperature $T_c$ (close to the crossover
temperature) and then decreases as the system continues to cool
on its way to freezeout --- as shown in Fig.~\ref{fig:xi} which is how we imagine $\xi$ varying along a trajectory like that illustrated by the green dashed line in Fig.~\ref{fig:trajectory}.
In our explicit calculations, we shall choose $T_c=160$~MeV and
$\xi_0=0.5$~fm.

\subsection{Cooper-Frye freeze-out}
\label{CooperFrye}

Hydrodynamics describes hot and dense QCD matter created in heavy-ion
collisions in terms of densities of conserved quantities such as
energy or baryon charge, or the corresponding thermodynamic variables
such as temperature $T$ or baryochemical potential $\mu_B$, as well as the local 
fluid velocity. Heavy ion collision
experiments, on the other hand, measure multiplicities and momentum
distributions of particles which emerge from the expanding and cooling
droplet of fluid as it breaks up into hadrons. These multiplicities and distributions are
well described by a procedure which we shall summarize below known as Cooper-Frye freezeout
\cite{Cooper:1974mv} which starts from the output of a hydrodynamic simulation.

    In the traditional Cooper-Frye procedure, the macroscopic evolution of the conserved charges and fluid velocity field obtained from a hydrodynamic calculation are converted into a microscopic description in terms of particles in a Hadron Resonance Gas model. The freeze-out hypersurface where this switching is done is determined based on some thermodynamic condition for eg., fixed temperature or energy density.  
The averages of the conserved densities are equated to those of a hadron resonance gas of particles via the Cooper-Frye formula.  Let $dS_\mu$ be the differential element pointing along the normal vector to the freeze-out surface. The mean multiplicity of particle species $A$ ($\<N_A\>$) according to the Cooper-Frye formula is given by,
	\begin{equation}
	\label{NA1}
	\<N_A\>=d_A\int dS_\mu\, \int Dp_A \,p^{\mu}\,\<f_A(x,p)\>
	\end{equation}
	Here, $d_A$ is the degeneracy of particle species $A$ and $Dp_A$ is the Lorentz invariant measure:
	\begin{equation}
	\label{Dp1}
	Dp_A=2\frac{d^4p}{(2\pi)^{3}}\delta \left(p^{2}-m_A^2\right)
       {  \theta(p_0)}\,.
	\end{equation}
	$\<f_A\>$ is the momentum-dependent particle distribution function which is either taken to be Fermi-Dirac or Bose-Einstein based on the spin and statistics of hadron species $A$. For simplicity, throughout this paper we shall ignore the spin and statistics and consider $\<f_A\>$ to be the Boltzmann distribution
        \begin{equation}
	\<f_A(x,p)\> 
        = \exp\left(\frac{-p\cdot u(x)+\mu_A(x)}{T(x)} \right)\,,
	\label{fABoltz}
	\end{equation}
	where $T(x)$, $\mu_A(x)$ and $u(x)$ are the temperature, the chemical
        potential of species $A$, and the local fluid velocity at a point $x$ on the freeze-out hypersurface. For mesons, $\mu_A=0$,
        while for baryons/antibaryons $\mu_A = \pm\mu_B$,
        respectively. In addition to ignoring the modification of the
        distribution function due to spin and statistics, we also ignore
        further modifications to $\left<f_A\right>$ due to viscous
        effects \cite{https://doi.org/10.1002/cpa.3160020403,ANDERSON1974466, Pratt:2010jt,McNelis:2019auj,McNelis:2021acu,
          Denicol:2018wdp} in this preliminary study.

 In Section~\ref{presc1} we shall turn to describing our extension of the Cooper-Frye procedure
that will enable us to translate the output of a Hydro+ simulation, with traditional hydrodynamic variables as well as fluctuations described by $\phi_{\bm{Q}}(x)$, into particles in a way that faithfully turns the critical fluctuations in the fluid into fluctuations and correlations of the hadrons. We are pursuing this goal within what is often referred to as a deterministic framework for describing the fluctuations: Hydro+ adds new deterministic equations of motion to the equations of hydrodynamics, equations that describe the evolution of quantities that characterize the fluctuations starting 
with $\phi_{\bm{Q}}(x)$ that describes their two-point correlation function. Fluctuations can also be described stochastically, where one evolves an ensemble of configurations each with its own realization of the fluctuations~\cite{Kapusta:2011gt, Murase:2013tma, Young:2014pka, Murase:2016rhl,Singh:2018dpk,Nahrgang:2017oqp,Nahrgang:2018afz, Nahrgang:2020yxm,Sakai:2020pjw}.  It would be natural in a stochastic description to analyze freezeout via extending the Cooper-Frye procedure in a manner that follows an analogous logic to that we shall employ here, but we leave this to future work.

\section{Cooper-Frye freeze-out for critical fluctuations}
\label{presc1}

The Cooper-Frye freezeout procedure described in the previous section
converted hydrodynamic variables into (event) mean multiplicities. In
order to describe the signatures of the QCD critical point we need to
be able to describe also the fluctuations and correlations of these
multiplicities. We shall now describe a freezeout procedure which
extends Cooper-Frye freezeout by connecting the fluctuations of
hydrodynamic variables to the fluctuations of the particle
multiplicities. We shall focus on the fluctuations of the slowest, and
thus most out of equilibrium mode -- $\hs$. In Hydro+ the two point
function of this mode is given by (its Fourier transform)
$\phi_\mathbf{Q}$. Our goal is to connect it to the two-point correlation
function of the multiplicity fluctuation $\delta f$
       \begin{eqnarray}
        \label{Bolzmod}
        f_A(x,p)=\left<f_A(x,p)\right>+\delta f_A(x,p)
        \end{eqnarray}
where $\left<f_A(x,p)\right>$ is given by Eq.~(\ref{fABoltz}).  We
shall use the model of critical correlations which incorporates
critical fluctuations in the hadronic description via the interaction
of the hadrons with a critical $\sigma$ field. Such a description of
critical fluctuations in a hadron gas has been
used in equilibrium \cite{Stephanov:1999zu,Stephanov:2008qz,Athanasiou:2010kw,Stephanov:2011pb,Jiang:2015hri,Ling:2015yau,Bluhm:2016byc,Brewer:2018abr,Wu:2018twy,Szymanski:2019yho} as well as with some out-of-equilibrium effects included~\cite{Stephanov:2009ra}.
In this approach the interaction with the $\sigma$ field modifies the
masses of the hadrons, to linear order in $\sigma$, as follows:
\begin{equation}
	\label{ms1}
	\delta m_A=g_A\sigma \,,
	\end{equation}
	where $\langle \sigma \rangle = 0$.
	The proportionality constant $g_A$ plays the role of the
        coupling constant between the hadron species $A$ and the
        $\sigma$ field. The critical contribution to the fluctuations
        of $f_A$ is due to the dependence of the averaged particle
        distribution function $\langle f_A\rangle$ on the mass, and is given by
	\begin{equation}
	\label{fAs1}
	(\delta f_A(x,p))_\sigma=
        g_A\frac{\partial \left<f_A(x,p)\right>}{\partial m_A}\sigma(x)\,,
	\end{equation}
	where $\<f_A\>$ is the Boltzmann distribution in
        Eq.~(\ref{fABoltz}). 
As a result, fluctuations of the $\sigma$ field translate into
fluctuations and correlations between particles, as in
\begin{multline}
  \label{eq:dfdf}
  \langle\delta f_{A_1}(x_1,p_1)\delta f_{A_2}(x_2,p_2)\rangle
  = \langle f_{A_1}(x_1,p_1)\rangle\delta_{1,2}
  \\
  + g_{A_1}g_{A_2}\frac{\partial \left<f_{A_1}(x_1,p_1)\right>}{\partial m_{A_1}}
  \frac{\partial \left<f_{A_2}(x_2,p_2)\right>}{\partial m_{A_2}}
    \langle\sigma(x_1)\sigma(x_2)\rangle\,.
\end{multline}
Since hydrodynamic variables, such as baryon density, are expressed in
terms of momentum space integrals of the particle distribution functions,
the correlation functions of hydrodynamic variables are proportional
to the correlation functions of the $\sigma$ field. This reproduces the
essential property of fluctuations at the critical point --
the critical (most singular at the critical point) contribution of all
correlation functions are proportional to the correlator of a single
critical scalar field.

Our main focus is on the correlation functions of $\hs$. Universality
of critical behavior dictates that in equilibrium this critical
contribution to this correlator should be also proportional to the
correlator of $\sigma$. In this paper we shall also assume that this
remains true out of equilibrium. This allows us to connect the
correlations of $\hs$ at the end of the Hydro+ evolution to the
fluctuations of $\sigma$ in the kinetic description at freezeout and consequently to observable fluctuations and correlations of hadron multiplicities.

The equilibrium fluctuations of the critical field are dictated by the
universality of critical behavior and are controlled by the
probability functional $P=\exp\{-\Omega/T\}$, where $\Omega[\sigma]$ the
effective action (or free energy) which can be written for small
fluctuations at long wavelengths in an expansion in powers of the
field $\sigma$ and its gradients
 around its equilibrium value, $\langle\sigma\rangle=0$, as follows:
        \begin{eqnarray}
        \label{Omegapot1}
        \Omega(\sigma)=\int d^{3}x\,
          \left[\frac{(\nabla\sigma)^2}{2}+\frac{m_\s^{2}}{2}\s^2+
          \frac{\lambda_3}{3}\s^{3}+\frac{\lambda_4}{4}\s^4+
          \dots\right]\,.
        \end{eqnarray}
The equilibrium two-point correlator can be then found from Eq.~(\ref{Omegapot1})
and is given by
\begin{eqnarray}
	\label{sseq1uv}
    \<\s(\bm x_+)\s(\bm x_-)\>\approx\frac{T}{4\pi \left|\Delta \bm{
  x}\right|} e^{{-\left|\Delta{\bm x}\right|}/{\xi}} 
\end{eqnarray}
where $\Delta\bm x = \bm x_+-\bm x_-$ and $\xi\equiv 1/m_\sigma$ is the correlation length of the $\sigma$-field fluctuations. 
As we shall only be interested in the two-point correlator in this work and as we are neglecting the (small) nonzero value of the critical exponent $\eta$, we will be able to neglect the terms of order $\sigma^3$ and higher in the expansion (\ref{Omegapot1}). The
Fourier/Wigner transform of the two-point correlator is then given by
\begin{eqnarray}
	\label{sseq1uvFT}
\chi_{\bm Q}\equiv
	\int_{\Delta\bm x}e^{-i\bm Q\cdot\bm\Delta \bm x}\<\s(\bm  x_+)\s(\bm x_-)\>
  \approx\frac{T\xi^2}{1+(Q\xi)^2}\,.
\end{eqnarray}
In the approximate equalities in Eqs.~(\ref{sseq1uv})
and~(\ref{sseq1uvFT}) we ignore loop corrections, which are known to
be small in the 3D Ising universality class in which $\eta$ is small. 

We choose the units of length in Eq.~(\ref{Omegapot1}) so that the value of $\xi$ introduced in Eqs.(\ref{Omegapot1}) and (\ref{sseq1uv})  matches the value of the correlation length of the thermodynamic fluctuations introduced above, in Section~\ref{introduction2}.
The universality of
the critical behavior then dictates that the relationship
between the two-point correlators of the fluctuating soft mode in the hydrodynamic description of the physics at the freezeout point and the fluctuating $\sigma$-field in the kinetic theory description of the physics at the same point takes the simple form
\begin{equation}
  \label{eq:<s.s>=<sigma.sigma>}
  \<\delta\hs(\bm x_+)\delta\hs(\bm x_-)\>=Z\<\s(\bm x_+)\s(\bm x_-)\>\,.
\end{equation}
Equivalently, the Wigner transforms are related via the same proportionality constant $Z$:
\begin{equation}
  \label{eq:phiQ=<sigma.sigma>}
  \bar\phi_{\bm Q} = Z \chi_{\bm Q}\,.
\end{equation}
Using Eqs.~(\ref{phieq1}) and~(\ref{sseq1uvFT}) we find
\begin{equation}
  \label{eq:Z=}
  Z\approx \frac{c_p}{Tn^2\xi^2}\,.
\end{equation}
Note that, while both $c_p$ and $\xi^2$ diverge at the critical point,
their ratio is finite in the approximation we are using.\footnote{Our
  approximation sets the critical exponent to its mean-field value $\eta = 0$, which is a good approximation to make for a critical point in the 3D Ising universality class where $\eta\sim 0.04$. If one uses a more
  sophisticated, non-mean-field equation of state as in, e.g.,
  Ref.~\cite{Parotto:2018pwx}, and/or  more sophisticated form of
  $\bar\phi_{\bm Q}$ and $\chi_{\bm Q}$ as in Ref.\cite{Stephanov:2017ghc}, the value of the normalization constant will
  nevertheless be determined by the 
  matching equation~(\ref{eq:phiQ=<sigma.sigma>}), which is more general than the approximation in which we have derived it.}

We shall apply the relationship in Eq.~(\ref{eq:<s.s>=<sigma.sigma>}),
or equivalently in Eq.~(\ref{eq:phiQ=<sigma.sigma>}), to express the fluctuations of
$\sigma$ at freezeout also when these fluctuations are out of
equilibrium. Although not strictly justifiable by the universality of
critical phenomena in equilibrium, it does allow us to match critical fluctuations
at the kinetic and hydrodynamic stage in a way which preserves the
information about important non-equilibrium effects, including the effects of
conservation laws.

We shall thus determine the correlation functions of $\sigma$ at freezeout as follows:
\bes
	\label{s1}
	\begin{eqnarray}
	\label{s11}
	\left<\s(x)\right>&\equiv& 0
	\\ \label{s21}
	\<\s(x_+)\s(x_-)\>&=&Z^{-1}\,\<\delta\hat{s}(x_+)\delta\hat{s}(x_-)\>
	\end{eqnarray}
\ees
where $Z$ is a normalization constant which can be obtained by matching the fluctuations
obtained in the kinetic description to fluctuations (i.e.,
susceptibilities) obtained from the QCD equation of state 
using Eq.~(\ref{eq:phiQ=<sigma.sigma>}). Since in this paper our focus is entirely on
developing and exploring the implementation of the freezeout prescription that we introduce to describe
fluctuations, we shall take the constants $Z$ in Eq.~(\ref{s21}) and $g_A$ in Eqs.~(\ref{ms1})-(\ref{eq:dfdf})
as given
and postpone their determination by matching a particular QCD EoS to
future work. We also note that we shall find ways to express our results that are
independent of those unknown parameters. Note that there is a subtlety in defining
Eq.~(\ref{s21}) relating to the choice of frame in which $x_+$ and $x_-$ are at equal time; we shall discuss this in Section~\ref{sec:toward}.

Due to Eq.~(\ref{s11}), the mean number of particles is unmodified by
critical fluctuations and is given by Eq.~(\ref{NA1}).
Integrating the spatial correlations given by Eq.~(\ref{eq:dfdf}) over the full freeze-out
hypersurface and using Eq.~(\ref{s21}), we can express the leading critical
contribution to the correlator of particle multiplicities $N_A$ and $N_B$ as:
	\begin{equation}
	\label{NABsigma}
	\<\delta N_A\delta N_B\>_\sigma=\,\int dS_{\mu,+}\, \int dS_{\nu,-}\, \, J_A^{\mu}({x_+})\,J_B^{\nu}(x_-)\,Z^{-1}\left<\delta\hs(x_+)\delta\hs(x_-)\right>
	\end{equation}
	
	\begin{equation}
	\label{JA1}
	J_A^{\alpha}(x_\pm)\equiv g_Ad_A\,\int Dp_{A} \,p^{\alpha}\,\frac{\partial\<f_A(x_\pm,p)\> }{\partial m_A}
   \end{equation}
   with $d_A$ the spin (and/or isospin) degeneracy of the particle species $A$.
   The subscript $\sigma$ in $\left<\delta N_A\delta N_B\right>_\sigma$ is there to
remind us that this is the contribution due to critical
fluctuations.  The expressions (\ref{NABsigma}) and (\ref{JA1}) constitute the central result whose consequences we shall explore over the course of the rest of this paper by making them explicit in simplified settings.  Note that the field $\sigma$ has now done its job and has now disappeared; in (\ref{NABsigma}) and (\ref{JA1}) we have a relationship between the critical fluctuations of hydrodynamic variable on the right-hand side of (\ref{NABsigma}) and the correlator of observable particle multiplicities on the left-hand side.

Although we shall not go beyond two-point correlators in our explicit explorations to come, we note with future work in mind that
a straightforward generalization of Eq. (\ref{NABsigma}) yields the form
\begin{eqnarray}
\left<\delta N_A^{k}\right>^c_{\sigma}=\int dS_{\mu_{1}} \dots \int dS_{\mu_{k}}\, J_A^{\mu_{1}}(x_1)\dots J_A^{\mu_k}(x_k)\, Z^{-k/2}\, \left<\hat{s}(x_1)\dots\hat{s}(x_k)\right>_c \label{kthcumulant}
\end{eqnarray}
for the critical contribution to the $k^{th}$ cumulant of the multiplicity of particle species $A$.
We have extended the Cooper-Frye procedure in a way that will allow us to see how the critical, i.e.~most singular, contribution  the two-point correlations of $\hat{s}$ 
translates into the variance of observed particle multiplicities. 
We leave continuing onward to higher-point correlations and non-Gaussian cumulants for future work.

Finally, we note that the total variance of the particle multiplicity has an additional non-critical contribution which is usually taken as Poissonian:
\begin{eqnarray}\label{delNatot1}
   \left<\delta N_A^2\right>=\left<N_A\right>+\left<\delta N_A^2\right>_\sigma\,.
   \end{eqnarray}
There can certainly be corrections to the non-critical contributions that we represent here by the Poisson distribution. These may arise from global charge conservation \cite{Bzdak:2012an,Vovchenko:2020gne} or initial fluctuations \cite{Luzum:2013yya}, for example, but we shall not discuss them in this work. This work is intended only as a prescription for freezing out the fluctuations near the critical point that encode information about the leading singularity.

\subsection{Toward explicit evaluation}
\label{sec:toward}

The extended Cooper-Frye procedure that we have derived above involves expressions with a certain formality to them.  Noting that sometimes the devil turns out to be found in the details, we shall now begin to take the steps need to turn these expressions into tools that can be used in explicit calculations.

The Wigner transform $\phi_{\bm Q}$, as defined in
Eq.~(\ref{eq:phiQ=<ss>}), involves integration over the hyperplane
orthogonal to the 4-vector $u(x)$. That is, the points $x_+$ and $x_-$ are
equal-time points in the rest frame of the fluid at point $x$, or
$\Delta x\cdot u(x)=0$ 
However, in general, the freezeout surface over
which the integration in Eq.~(\ref{NABsigma}) is to be performed does not
necessarily have the property that the points $x_+$ and $x_-$ are
simultaneous in the rest frame of the fluid at the point $x$. While, for example, this
property holds for boost-invariant Bjorken flow, it does not hold for a
flow with radial component, such as the one considered in
Ref.~\cite{Rajagopal:2019xwg} that we shall analyze in Section~\ref{RRWY}. 
In order to translate $\phi_{\bm Q}$ into
the correlator $\langle\delta\hs(x_+)\delta\hs(x_-)\rangle$ in such a case one
needs to be able to evolve this correlator not only in time $u\cdot x$
(using Eq.~(\ref{phiev1})) but also in time {\em difference}
$u\cdot\Delta x$. We shall show below that because this evolution is
slow (and especially slow at the critical point) one can neglect the
effect of such evolution and can therefore nevertheless express the unequal-time correlator of interest in
terms of $\phi_{\bm Q}$.

Let us consider a small region of the freezeout surface around a point $x$ that lies on the surface
and let us assume that the surface is not perpendicular to the 4-vector $u(x)$. This means
that freezeout does not happen simultaneously at all points in this
region. Let us denote the velocity of the frame in which this patch of the freezeout surface
{\rm is} an equal-time surface by $\beta$. ($\beta=0$ for Bjorken flow). If
the typical range of the correlator is of order $\ell_*$, then the
typical value of the time difference
$u(x)\cdot\Delta x\sim \beta \ell_*$. The typical scale $\ell_*$ can
be determined by the condition that the relaxation rate
$\Gamma(Q_*)\sim D Q_*^2$ of the corresponding modes $Q_*=1/\ell_*$ is
of order the expansion rate $1/\tau$, where $D$ was introduced in Eq.~(\ref{DD0}). That is, $\ell_*\sim \sqrt{D\tau}$.

The evolution of the correlator $\langle\delta\hs(x_+)\delta\hs(x_-)\rangle$ as a function of
the time separation $u\cdot\Delta x$ occurs with the same rate, also of order $\Gamma(Q_*)$.
As a result, the correction to the correlator is of
order $\Gamma(Q_*)\, u\cdot\Delta x\sim \beta\sqrt{D/\tau}$. This
quantity is small already because $\tau$ is a macroscopic scale, while
$D$ is microscopic, i.e., $\tau\gg D$. Furthermore, near the
critical point, $D$ itself is vanishing: as seen in Eq.~(\ref{DD0}), it is smaller than the
microscopic scale by another factor of $\xi_0/\xi$.

\newcommand\dxp{{\Delta x_\perp}}

\begin{figure}[t]
  \centering
  \includegraphics[width=.5\textwidth]{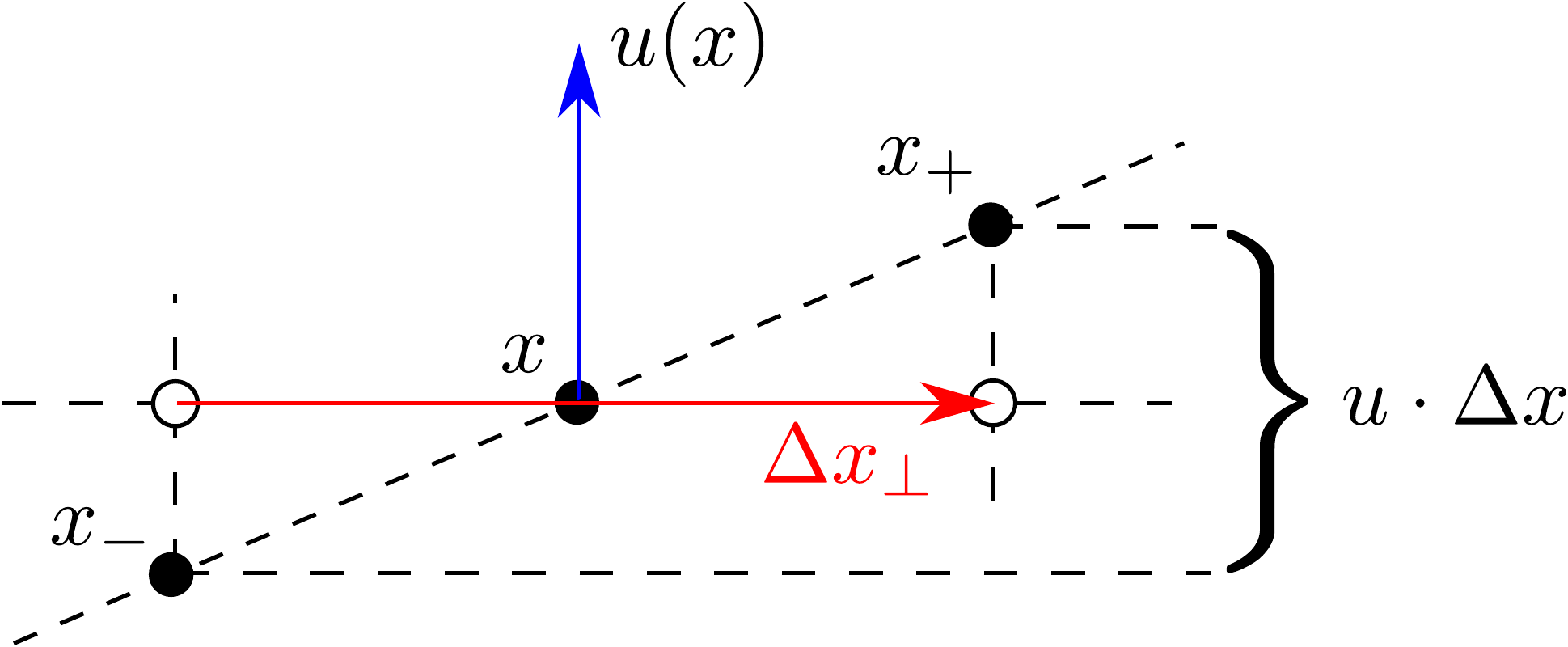}
  \caption{Geometric representation of Eq.~(\ref{delxproj1}). $x_+$ and $x_-$ are on the freezeout surface; $x$ is the midpoint between them. The four-vector $\Delta x_\perp$ (red) is perpendicular to the fluid four-velocity at the point $x$, $u(x)$, meaning that in the local fluid rest frame it is a four-vector with no time-component.}
\label{fig:delta-x}
\end{figure}

More formally, let us define a projection of the separation four-vector
$\Delta x$ onto the plane perpendicular to $u(x)$:
\begin{equation}
  \label{delxproj1}
\dxp \equiv \Delta x- \left[u(x)\cdot \Delta x\right]\, u(x) \,.
\end{equation}
Then we can express the correlator
$\langle\delta\hs(x_+)\delta\hs(x_-)\rangle$ in
Eq.~(\ref{NABsigma}) in terms of $\phi_{\bm Q}$ defined in
Eq.~(\ref{eq:phiQ=<ss>}), obtaining
\begin{equation}
  \label{eq:<ss>=phiQ}
  \langle\delta\hs(x_+)\delta\hs(x_-)\rangle
  = \tilde\phi(x;\Delta \bm x_\perp) +  \mathcal O(\sqrt{D/\tau})\,,
\end{equation}
where the three-component vector $\Delta\bm x_\perp$ is the projection of
the four-vector $\Delta x$ onto the equal-time hyperplane orthogonal to the vector
$u(x)$ as defined in Eq.~(\ref{delxproj1}) and illustrated in Fig.~\ref{fig:delta-x}, and $\tilde\phi(x;\Delta
\bm x_\perp)$ is the inverse Fourier (Wigner) transform of $\phi_{\bm
  Q}(x)$:
\begin{equation}
  \label{eq:phitilde}
  \tilde\phi(x;\Delta \bm x_\perp)
  \equiv \int \frac{d^3 Q}{(2\pi)^3} e^{i{\bm Q}\cdot {\Delta \bm x_\perp}}
  \phi_{\bm Q}(x)\,.
\end{equation}
Eq.~(\ref{eq:<ss>=phiQ}) formalizes the qualitative argument from the preceding paragraph.

We shall be comparing our results with what one would obtain upon assuming
the fluctuations are in equilibrium. Up to corrections suppressed by
ratios of microscopic (e.g., correlation length $\xi$) to macroscopic
scales (such as hydrodynamic gradient scales, e.g., $\tau$) we can
replace the correlation function in Eq.~(\ref{sseq1uv}) with a delta
function:
\begin{eqnarray}
	\label{ss=delta}
  Z^{-1}\left<\delta\hs(\bm x_+)\delta\hs(\bm x_-)\right>
  =	\<\s(\bm x_+)\s(\bm x_-)\>=T\xi^2\delta^{(3)}(\Delta \bm x)\,.
\end{eqnarray}
Substituting into Eq.~(\ref{NABsigma}) we find
\begin{equation}
  \label{eq:NAvariance-eq}
	\<\delta N_A\delta N_B\>_\sigma^{\rm eqbm}=\,\int dS_{\mu}(x)\, \, J_A^{\mu}({x})\,\hat{n}_{\nu}(x)\,J_B^{\nu}(x)\,T(x)\xi^2(x)\,.
\end{equation}
where $\hat{n}(x)$ is the unit vector along the normal on the freeze-out hypersurface at $x$.
This expression straightforwardly generalizes existing results for
equilibrium fluctuations, see for example Ref.~\cite{Athanasiou:2010kw}, to locally equilibrated fluctuations in a (more realistic)
inhomogeneous fireball. 
We shall make comparisons between our full results and the equilibrium
fluctuation predictions (\ref{eq:NAvariance-eq})
in order to highlight the importance of 
non-equilibrium effects, especially the effects due to
conservation laws.

\subsection{Ratios of observables}

We shall calculate the contribution of critical fluctuations to the variance of
the particle multiplicity of species $A$  (we shall consider protons, $A=p$, and pions, $A=\pi$) in a specified finite rapidity and transverse momentum
acceptance window. To eliminate the dependence on the volume (i.e.,
the transverse size) of the droplet of QGP we shall introduce the intensive ratio  
\begin{eqnarray}\label{omegaAdef1}
  \omega_A \equiv \frac{\<\delta N_A^2\>_{\sigma}}{ \<N_A\>}\,,
\end{eqnarray}
which was referred to as $\omega_{A,\sigma}$ in Ref.~\cite{Athanasiou:2010kw}.
We note that $\omega_A$ depends on the choice of acceptance window. (See, e.g.,
Ref.~\cite{Ling:2015yau}.) 
  Since this acceptance dependence is not the main focus
of the present study, while criticality and non-equilibrium effects are, we
shall often illustrate our results by plotting the ratio
\begin{eqnarray}\label{omegaAdef2}
  \tilde{\omega}_A=\frac{\omega_A}{\omega_A^{\text{nc}}}\,,
\end{eqnarray}
where $\omega_A^{\text{nc}}$ is the $\omega_A$ calculated upon assuming
freezeout well away from the critical point, i.e., upon setting
$\xi_{\text{max}}=\xi_0$. We have checked (for a few sets of parameters) that the acceptance dependence
of the numerator and denominator in Eq.~(\ref{omegaAdef2}) is similar
and, thus, largely cancels.
  In contrast, the numerator $\omega_A$ is strongly
enhanced by critical fluctuations (for example, in equilibrium $\tilde{\omega}_A^{\text{eqbm}}=\xi^2/\xi_0^2$)  and is sensitive
to the non-equilibrium effects of critical slowing down, while
the denominator $\omega_A^{\rm nc}$ is, by construction, not affected by
critical fluctuations. Although $\omega_A$ defined in Eq.~(\ref{omegaAdef2}) depends on the unknown parameters $g_A$ and $Z$ via the ratio $g_A/\sqrt{Z}$, all dependence on these unknowns cancels in 
the ratio of ratios defined in Eq.~(\ref{omegaAdef2}), within the approximations that we shall make.
This is a second significant benefit, and we suggest employing $\tilde{\omega}_A$ (and its generalizations to higher cumulants) in the comparison between future theoretical calculations and experimental measurements. 

In subsequent Sections, we shall compute $\tilde{\omega}_A$ in two different model hydrodynamic backgrounds. 
In Section~\ref{Sectbjk}, we study an analytically solvable scenario with longitudinal boost invariance and no dependence on transverse coordinates, which is to say Hydro+ in Bjorken flow.  Then, in Section~\ref{RRWY} we shall freeze out the numerical simulation of Hydro+ with azimuthal symmetry, radial expansion, and longitudinal boost invariance from Ref.~\cite{Rajagopal:2019xwg}.  Before we conclude this Section, though, we need to establish some notation with which to describe two points on the freeze-out hypersurface and the separation between them.

\subsection{Establishing notation for azimuthally symmetric boost invariant freezeout}
\label{azsbifo}

To specify the shape of an azimuthally symmetric boost-invariant freezeout surface, it is convenient to use Bjorken coordinates defined in terms of the Cartesian
coordinates $(\,t,\, x_1,\,x_2,\, x_3)$ in the lab frame via
\begin{equation}
  \label{eq:bj-coords}
  t = \tau\cosh\eta,\quad x_1 = r\cos\varphi,\quad x_2 = r\sin\varphi,\quad x_3=\tau\sinh\eta\,.
\end{equation}
The mutually orthogonal set of unit vectors corresponding to each of
the Bjorken coordinates can be expressed in terms of the Cartesian
coordinates as
\begin{eqnarray}\label{hattau}
  \hat{\tau}&=&(\,\cosh \eta,\, 0,\,0,\,\sinh\eta\,)\\ \label{hateta}
  \hat{\eta}&=&(\,\sinh \eta,\, 0,\,0,\,\cosh\eta\,)\\ \label{hatr}
  \hat{r}&=&(\,0, \cos\varphi,\sin\varphi,0\,)\\ \label{hatvarphi}
  \hat{\varphi}&=&(\,0,\,
                         -\sin\varphi,\,\cos\varphi,\,0\,)\,.
\end{eqnarray}
The radial profile of a boost-invariant and azimuthally-symmetric
freezeout surface can then be expressed in a parametric form using an
arbitrary parameter $\alpha$ as in Ref.~\cite{Floerchinger:2013hza}
\begin{equation}
  \label{eq:tauralpha}
  \tau = \tau_f(\alpha),\quad r=r_f(\alpha)\,,
\end{equation}
so that the point on the freeze-out hypersurface corresponding to parameters
$\alpha,\eta,\varphi$ is given by:
\begin{eqnarray}\label{xdelxsym1a}
	x(\alpha,\eta,\varphi)&=&\tau_f(\alpha)\hat{\tau}(\eta)+r_f(\alpha) \hat{r}(\varphi)\,.
\end{eqnarray}
Then, the volume vector normal to the freeze-out hypersurface can be
written as $d^3S=n\,d\alpha \tau d\eta r d\varphi$ where the vector $n$ is given by:
\begin{equation}
  \label{nazsymm_boostinv}
  n(\alpha,\eta,\varphi) = \frac{\partial x}{\partial \alpha} \wedge 
\frac{\partial x}{\tau\partial \eta} \wedge 
\frac{\partial x}{r \partial \varphi}
= r_f'(\alpha)\hat\tau(\eta)
  + \tau_f'(\alpha)\hat r(\varphi)\,.
\end{equation}
The flow four-velocity on the freezeout surface is given by
\begin{equation}\label{uazsymm_boostinv}
  u(\alpha,\eta,\varphi) = u^{r}(\alpha)\hat{r}(\phi)+u^{\tau}(\alpha)\hat{\tau}(\eta)
\end{equation}
in the coordinates with which we are working.

In defining the two-point correlation function we shall need to specify two points on the freeze-out hypersurface. Let $x_\pm\equiv x(\alpha_\pm, \, \eta_\pm,\varphi_\pm)$ be any two such points on with
$x\equiv (x_++x_-)/2$ being their midpoint and $\Delta x\equiv x_+-x_-$ being the separation vector between
them. Let us denote similarly
$\tau=(\tau_++\tau_-)/2, \, r=(r_++r_-)/2, \, \eta=(\eta_++\eta_-)/2,
\, \varphi=(\varphi_++\varphi_-)/2$ and
$\Delta \tau=\tau_+-\tau_-, \,\Delta r=r_+-r_-, \,\Delta
\eta=\eta_+-\eta_-, \,\Delta \varphi=\varphi_+-\varphi_- $. Then
\bes
\label{xdelxsym1}
\begin{eqnarray} \label{xdelxsym1b}
	x&=&\tau \cosh\frac{\Delta \eta}{2}\hat{\tau}+r \cos \frac{\Delta \varphi}{2} \hat{r}+\frac{\Delta\tau}{2}\sinh\frac{\Delta\eta}{2}\hat{\eta}+\frac{\Delta r}{2}\sin\frac{\Delta\varphi}{2}\hat{\varphi}\\ \label{xdelxsym1c}
	\Delta x
	&=&2\tau \sinh\frac{\Delta \eta}{2}\hat{\eta}+2r \sin\frac{\Delta\varphi}{2} \hat{\varphi}+\Delta r \cos\frac{\Delta \varphi}{2}\hat{r}+
	\Delta \tau \cosh\frac{\Delta\eta}{2}\hat{\tau}\\ \label{xdelxsym1d}
	u\cdot\Delta x &=& u^{\tau}(x)\Delta \tau \cosh\frac{\Delta\eta}{2}-u^{r}(x) \Delta r \cos\frac{\Delta \varphi}{2}\\  \nonumber
		{\Delta{ x}_\perp}&\equiv&2\tau \sinh\frac{\Delta
                                      \eta}{2}\hat{\eta}+2r
                                      \sin\frac{\Delta\varphi}{2}
                                      \hat{\varphi}\\
&& \qquad  +\left[-\Delta \tau u^{r}(x)\cosh\frac{\Delta\eta}{2}+\Delta r u^{\tau}(x)\cos\frac{\Delta\varphi}{2}\right]\, \left(u^{r}(x) \hat{\tau}+u^{\tau}(x)\hat{r}\right)\,,\label{tildedeltax1e}
\end{eqnarray}
\ees
where $\Delta{x}_\perp$ was defined in Eq.~(\ref{delxproj1}).
While the points $x_+$ and $x_-$ are on the freezeout surface by construction, the
midpoint $x$, in general, is not. The displacement between the midpoint and the freezeout surface is, however, small
when the typical range of the correlator is much shorter than the
typical curvature radius of the freezeout surface. We can use an argument
similar to the one preceding Eq.~(\ref{delxproj1}) to simplify the
calculation by neglecting the
difference between the correlator at the actual midpoint
and the correlator at the point on the freezeout surface given by
\begin{eqnarray}\label{midpointapprox1}
  x\equiv \tau_f(\alpha)\hat{\tau}(\eta)+ r_f(\alpha) \hat{r}(\varphi)
\end{eqnarray}
where $\alpha \equiv (\alpha_++\alpha_-)/2$. Henceforth, we shall use
$x$ to denote the on-hypersurface approximation
(\ref{midpointapprox1}) to the actual midpoint.  Again, neglecting the
effect of the curvature
and linearizing in $\Delta\eta$, $\Delta\varphi$ and $\Delta\alpha=\alpha_+-\alpha_-$, the projection of the separation vector	${\Delta {x}}_\perp$ from Eq.~(\ref{tildedeltax1e}) onto the
hyperplane normal to the four-vector $u$ can be approximated as
\begin{eqnarray}\label{tildex1}
 \Delta{x}_\perp&\approx& n\cdot u\, \Delta \alpha\, \hat\alpha_\perp
 +\tau_f\,\Delta\eta\, \hat{\eta}\,
+r_f\Delta\varphi \hat{\varphi}\,,
\end{eqnarray}
where $u$ is the 4-velocity of the fluid at the point $x$, the vector $n$ is
defined in Eq.~(\ref{nazsymm_boostinv}), and we have introduced a spacelike
unit vector 
\begin{equation}
  \label{eq:rho}
  \hat\alpha_\perp \equiv u^r\hat\tau+u^\tau\hat r\,.
\end{equation}
The vectors $\hat\alpha_\perp$, $\hat\eta$ and $\hat\varphi$ form a basis in the
hyperplane perpendicular to the four-vector $u$ given by Eq.~(\ref{uazsymm_boostinv}) (the equal-time hyperplane in the
rest frame of the fluid at the point $x$.)
 
 With all of this notation established, we can now take a step toward making the expression Eq.~(\ref{NABsigma}) for the squared variance of the multiplicity of species $A$
that we have derived above as our central result more explicit, writing it as
\begin{multline} 
  \left<(\delta N_A)^2\right>_\sigma
  = \int d\alpha_+\, \tau_+d\eta_+\, r_+d\varphi_+\,
       \int d\alpha_-\, \tau_-d\eta_-\, r_- d\varphi_-\,
       (n\cdot J_A)_+\,  (n\cdot J_A)_-\,
       Z^{-1}\tilde{\phi}(x,\mathbf{\Delta{x}_\perp})\,, \label{delNA2RRWYagain3}
\end{multline}
where $\mathbf{\Delta{x}}_\perp$ is a three-vector whose components in the
$\hat\alpha_\perp$, $\hat\eta$, $\varphi$ basis are given by
Eq.~(\ref{tildex1})  and $(n\cdot J_A)_\pm\equiv n(x_\pm)\cdot
J_A(x_\pm)$, where $n$  is given by Eq.~(\ref{nazsymm_boostinv}).

The integral in Eq.~(\ref{JA1}) expressed in Bjorken coordinates
takes the form
\begin{equation}
  \label{eq:J-Bjorken}
 J_A(x_\pm) = \frac{d_A\,m_{A}}{T}
  \int_{y_{\text{min}}}^{y_{\text{max}}}
  \frac{dy}{2\pi}\,\int_{0}^{2\pi}
  d\phi\,\int_{p_{T,\text{min}}}^{p_{T,\text{max}}}
  \frac{p_{T}dp_{T}}{(2\pi)^2}\,
 {\<f_A(x_\pm,p)\>}\,
  \frac{p}{p\cdot u(x_\pm)}\,,
\end{equation}
where we used Eq.~(\ref{fABoltz}). The Cartesian coordinates in the
lab frame of the particle four-momentum are given by
\begin{eqnarray}\label{p1}
   p=(m_T\cosh y, p_T\cos \phi,p_T\sin \phi, m_T\sinh y)
\end{eqnarray}
in terms of the particle rapidity $y$ and transverse mass
$m_T\equiv\sqrt{p_T^2+m^2}$.


\section{Freezing out a Hydro+ simulation with Bjorken flow}
\label{Sectbjk}

In this Section, we apply our approach to the well-known Bjorken
scenario: a hot fluid that is undergoing idealized boost-invariant longitudinal
expansion, so that it cools as a function of proper time, but that is translation-invariant and at rest in the transverse directions~\cite{Bjorken:1982qr}. We shall obtain some results
in this simplified scenario in analytic form, thus
allowing us to elucidate general features that we shall observe again
in a more realistic scenario with transverse expansion in the next
Section.

\subsection{Evolution of $\phi_{\boldmath Q}$}

The Bjorken scenario implies that all thermodynamics quantities such
as the temperature, $T$, or the energy density, $\varepsilon$, or net baryon number density, $n$, as
well as quantities describing the fluctuations of these conserved densities depend only on the
Bjorken proper time $\tau\equiv \sqrt{t^2-z^2}$, and are independent of 
the longitudinal spacetime-rapidity, $\eta$,
as well as of the transverse coordinates. Thus, the hydrodynamic equations reduce to
{\em ordinary} differential equations for functions of $\tau$ which can be solved easily for a
given equation of state. Throughout this work, in the Bjorken scenario of this Section and in the semi-realistic scenario of the next Section, we shall employ the simplified equation of state from  Ref.~\cite{Rajagopal:2019xwg} that we summarize briefly in Appendix~\ref{appendeos}. Throughout this work, we shall furthermore assume that the
dynamical back-reaction of the fluctuations on the equation of state and on the hydrodynamic solution is
negligible. This assumption has been tested quantitatively in different model calculations~\cite{Rajagopal:2019xwg,Du:2020bxp} and is
a good approximation: the effects of such back-reaction are typically at the sub-percent level. 
The hydrodynamic evolution
sets in at $\tau=\tau_I$ where the temperature $T(\tau_I)=T_I$ and it
continues until freeze-out at $\tau=\tau_f$ where the temperature
$T(\tau_f)=T_f$.  In the Bjorken scenario where there is no radial flow, Eq.~(\ref{uazsymm_boostinv}) becomes the statement that
the flow velocity unit-four-vector is $u=\hat\tau$ in Bjorken coordinates. The evolution equation (\ref{phiev1}) for the
fluctuation measure $\phi_{\bm Q}$ then takes the form of an ordinary
differential equation:
\begin{equation}\label{Bjkphiev}
	\partial_\tau \phi_{\bm Q}=-\Gamma(\bm Q)
        \,\left(\phi_{\bm Q}-\bar{\phi}_{\bm Q}\right)\,, 
\end{equation}
where $\Gamma(\bm Q)$ depends on $\tau$ through $\xi(\tau)$ and is specified via Eqs.~(\ref{GammaH},\ref{Kawasaki1},\ref{xiparam1}). Since our focus throughout is on the effects caused by fluctuations near a critical point, for simplicity we
shall assume that the initial fluctuations are in equilibrium, i.e.,
\begin{eqnarray}
	\label{phiqinit}
	\phi_{\bm Q}(\tau_i)=\bar{\phi}_{\bm Q}\Big|_{T=T_i}\,.
\end{eqnarray}
This assumption could certainly be improved in future, but for our present purposes any choice in which the initial fluctuations are small compared to those that develop later will suffice.
Since, in the Bjorken scenario, the temperature depends only on
$\tau$, the unit four-vector normal to the isothermal freeze-out hypersurface
$T(\tau_f)=T_f$ at a spacetime point $x$ is given by
$\hat n(x)=\hat{\tau}(x)=u(x)$.

\begin{figure}[t]
	\begin{subfigure}{0.49\textwidth}
	\includegraphics[width=\textwidth]{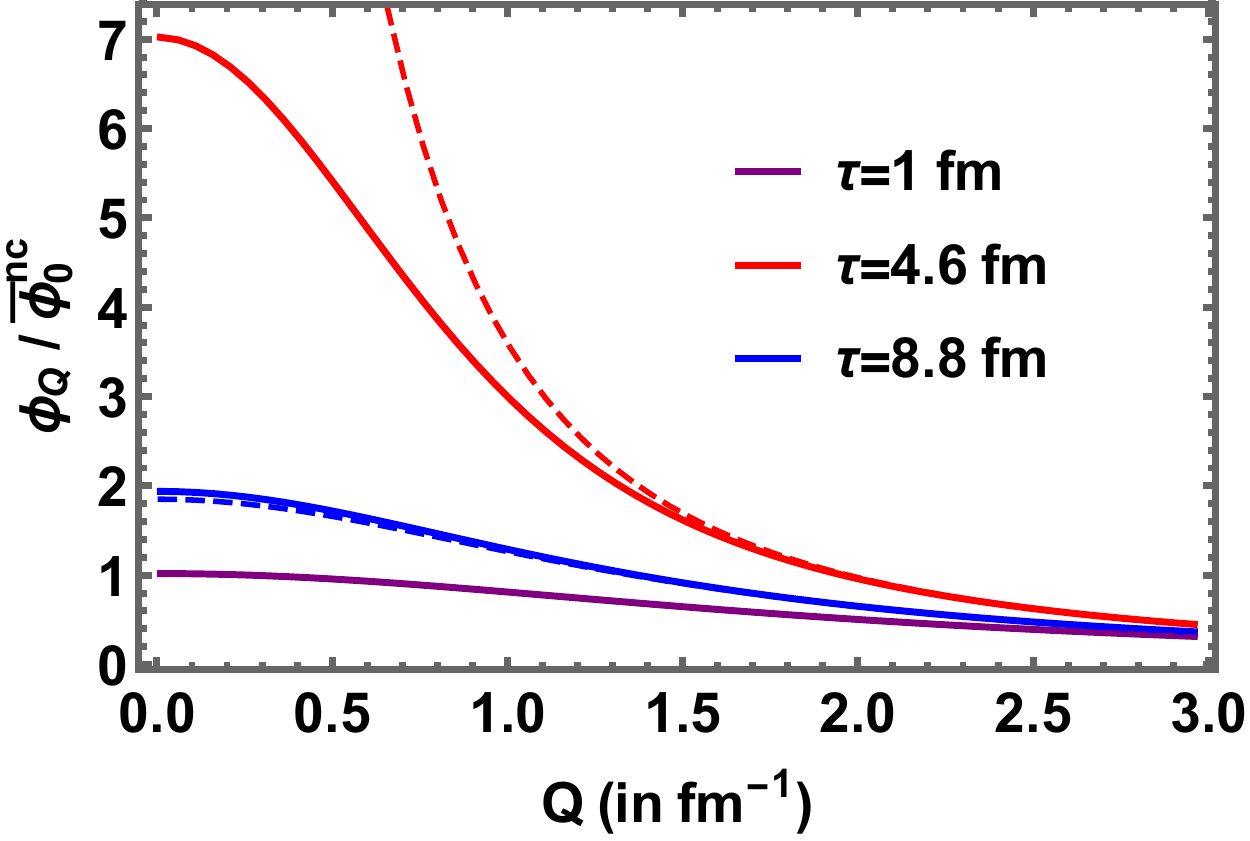}
	\subcaption{Model A}
	\end{subfigure}
	\begin{subfigure}{0.49\textwidth}
	\includegraphics[width=\textwidth]{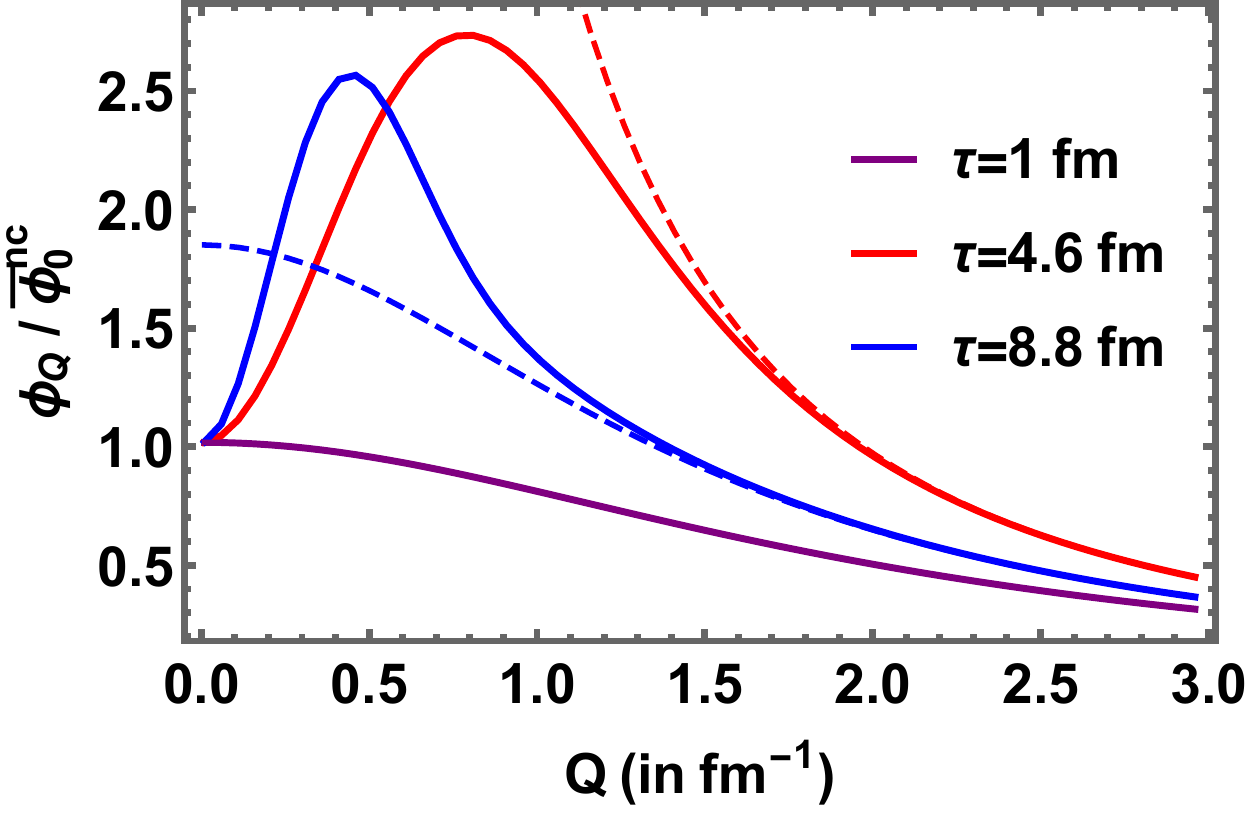}
	\subcaption{Model H}
	\end{subfigure}
	\caption{Evolution of $\phi_{\bm Q}$ as a function of  Bjorken time $\tau$,
          using model A and model H dynamics, corresponding to the
          relaxation rates given by Eqs.~(\ref{GammaA})
          and~(\ref{GammaH}), respectively. We have taken
          $\Gamma_0=1\, \text{fm}^{-1}$, $\dip=1\, \text{fm}$ and $\xi_{\text{max}}=3$~fm in both panels. The three solid curves in each figure
          correspond to different times $\tau$ as the boost-invariant, spatially homogeneous, Bjorken fluid is
          expanding and cooling in the vicinity of a critical
          point. The temperatures are given by $T=235$, 160 and 140
          MeV, for times $\tau=1$, 4.6 and 8.8~fm, respectively.  The
          dashed curves represent the equilibrium values
          $\bar\phi_{\bm Q}$ for the corresponding
          temperatures (times).  We have initialized the hydrodynamic solution and the fluctuations at $\tau_i=1$~fm: at that time $\phi_{\bm Q}=\bar\phi_{\bm Q}$ at $T_i=235$~MeV.
          The dashed curves are highest at $\tau=4.6$~fm because that is when
         the evolution trajectory was closest to the critical point; the fluctuations would be largest at that time if they were in equilibrium. We see that in Model H the fluctuations (in our full, out-of-equilibrium, calculation) remain considerably enhanced at $\tau=8.8$~fm over a range of nonzero values of ${\bm Q}$. It is evident from the right plot that
          $\phi_{\bm Q}$ does not evolve at $\bm Q=0$ in Model H. This
          is a consequence of conservation laws. 
          In both plots, at all times shown,
          $\phi_{\bm Q}$ and $\bar{\phi}_{\bm Q}$ are both normalized by
          their non-critical value (their value at a location far enough away from the critical
          point that $\xi=\xi_0$) at $\bm Q=0$ in equilibrium, i.e.,
          $\bar{\phi}^{\text{nc}}_{\bm 0}\,=\,Z\,T\,\xi_0^{2}$.}
	\label{phicharbjkfigA}
      \end{figure}

In Fig.~\ref{phicharbjkfigA} we plot $\phi_{\bm Q}$ obtained by solving Eq.~(\ref{Bjkphiev})  numerically at three values of $\tau$. In
order to highlight the significance of the conservation laws, we
compare and contrast the results obtained with two choices of the
relaxation rate: model A where $\Gamma({\bm Q})$ is 
as given in Eq.~(\ref{GammaA}) and model H (the
universality class of the QCD critical point) where $\Gamma({\bm Q})$ is as given by
Eq.~(\ref{GammaH}). The most important feature of the model H
evolution of $\phi_{\bm Q}$ is the ``stickiness'' of the
$\bm Q=0$ mode: $\phi_{\bm 0}$ remains ``stuck'' at its initial value, whatever that value is.
If the time evolution later takes the fluid along a trajectory that passes 
near a critical point, $\phi_{\bm 0}$ remains stuck in model H whereas in model A it evolves
with time, ``trying'' 
to follow the dynamics that would have been obtained in equilibrium.
The stickiness of $\phi_{\bm 0}$  is, obviously, a consequence of the
conservation laws in hydrodynamics, since the $Q=0$ mode corresponds to
the fluctuations of conserved quantities (volume
integrals of hydrodynamic variables).
This important feature is absent in model A
result which describes the evolution of fluctuations of quantities which are not conserved.

	 \begin{figure}[t]
	    \centering
	    \begin{subfigure}{0.49\textwidth}
	    \includegraphics[width=\textwidth]{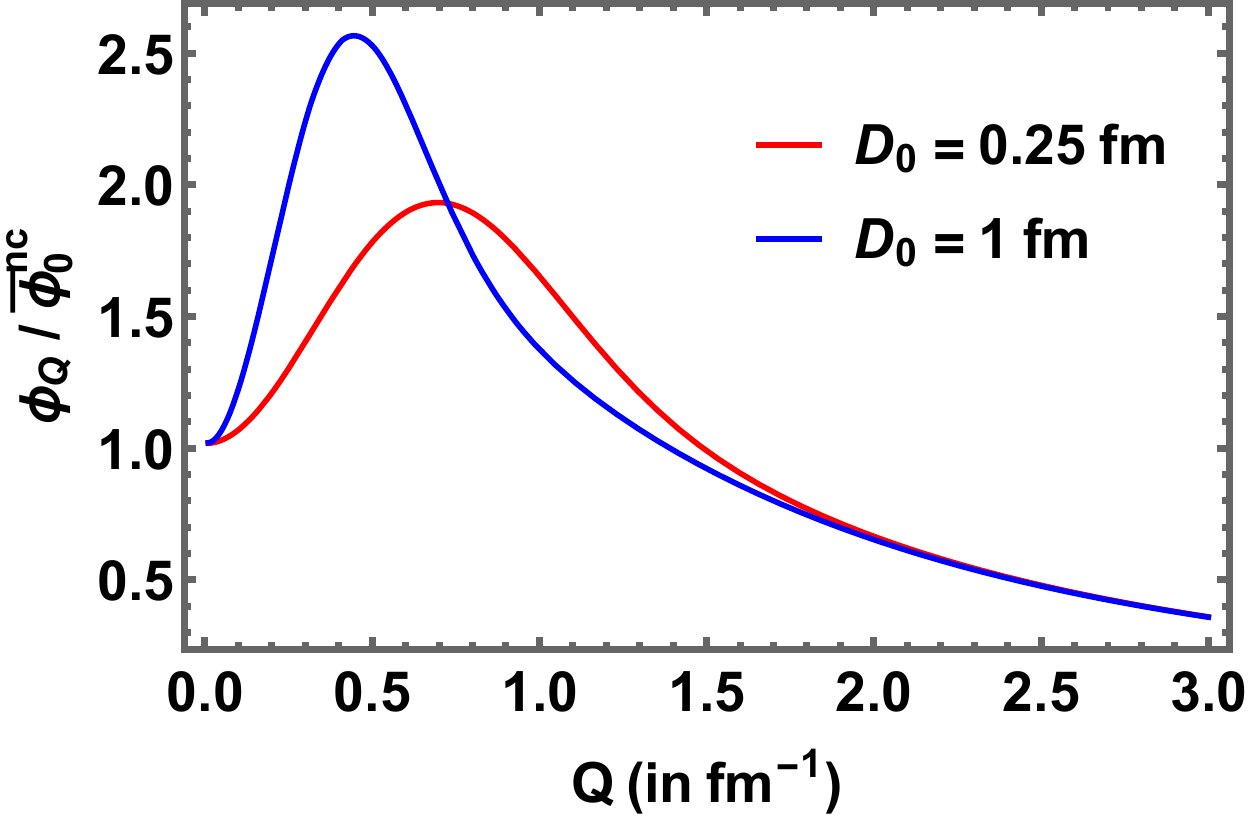}
	    \caption{$\phi$ at freeze-out}
	    \label{phiQdelnaGHa}
	    \end{subfigure}
	    \begin{subfigure}{0.49\textwidth}
	    \includegraphics[width=\textwidth]{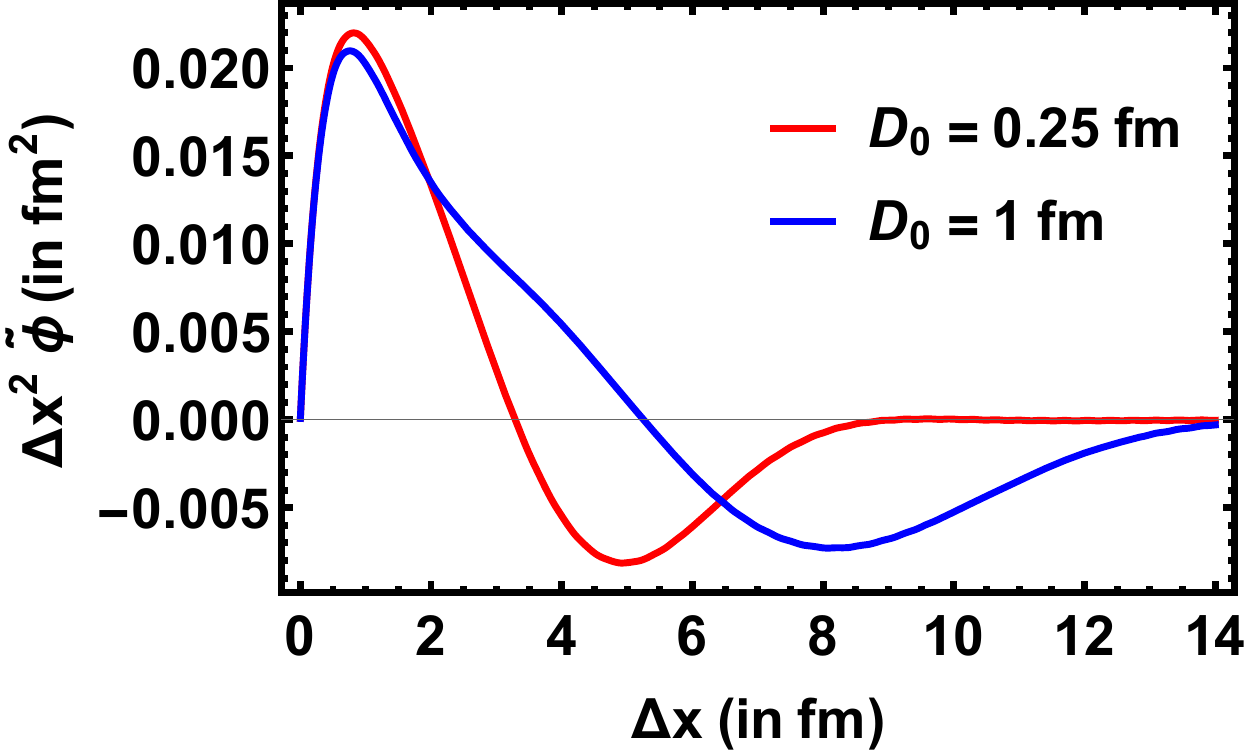}
	    \caption{$\tilde{\phi}$ at freeze-out}
	    \label{FTphiQdelnaGHa}
	    \end{subfigure}
	   	\caption{Normalized $\phi_{\bm Q}$ (a) and its inverse
              Fourier transform $\tilde\phi$ (b) at freezeout
              $T_f=140\, \text{MeV}$  after evolution according to Model H dynamics with two values of $\dip$.  In the text, we explain the dependence of the shapes of the curves in both panels on $D_0$, and the consequences of the conservation laws on the shapes of these curves.
            }
	    \label{phiQdelnaGH}
	\end{figure}

      In Fig.~\ref{phiQdelnaGH} we show the effect of  varying
      the parameter $D_0$ in the Model H relaxation rate (\ref{GammaH}) on the 
      fluctuation measure $\phi_{\bm Q}$ and its inverse Fourier tranform $\tilde\phi(\Delta{\bm x_\perp})$ defined in Eq.~(\ref{eq:phitilde}) at
      freezeout, after Hydro+ evolution from $\tau=\tau_i$ to $\tau=\tau_f$.
      The large $Q$, and correspondingly, small $\Delta x$ (defining $\Delta x \equiv |\Delta {\bm x_\perp}|$)
      behavior of the fluctuations is not affected at all, since the
      fluctuations at short length scales equilibrate quickly. The
      characteristic $Q$ where the peak of $\phi_{\bm Q}$ is situated shifts
      to smaller values of $Q$ with increasing $D_0$ because stronger diffusion
      tends to homogenize the system, including fluctuations. This can
      also be seen in spatial correlator $\tilde\phi$ becoming longer
      ranged. In addition, stronger diffusion (larger $D_0$) enhances the effect of
      the critical point on the fluctuations since the system 
      is able to  
      equilibrates more quickly towards
      the large equilibrium fluctuation values as it passes the
      critical point on its way to freezeout. This effect results in a
      more pronounced (higher) peak in $\phi_Q$ and, correspondingly,
      in an enhancement of $\tilde\phi$ at corresponding intermediate
      values of $\Delta x$ (of order 3 fm or so).

      The conservation
      laws keep $\phi_{\bm Q}$ stuck at ${\bm Q}=0$, which corresponds to keeping the integral of $\Delta x^2\, \tilde\phi$ constant. This means that their effect in panel (b) of Fig.~\ref{phiQdelnaGH} is that if there is a (large, critical) correlation at small $\Delta x$ they produce
      a corresponding compensating anticorrelation at
      longer $\Delta x$.  
      One can also show generally that a peak in $\phi_{\bm Q}$ away
      from $Q=0$
      corresponds to the anticorrelation (i.e., negative tail) in its
      Fourier transform $\tilde\phi(\Delta x_\perp)$. Indeed, if there
      exist a value of $\bm Q$ at which $\phi_{\bm Q}>\phi_{\bm 0}$, then
\begin{eqnarray}\label{eq:0>Q}
\int \, d^{3}{\Delta\bm  x} \,  \tilde{\phi} ({\Delta\bm  x})\,<\,\int\, d^{3}{\Delta\bm  x} \,  \cos ({\bm Q}\cdot{\Delta\bm  x}) \,\tilde{\phi} ({ \Delta\bm x}),
\end{eqnarray}
where we used the fact that $\tilde\phi(\Delta\bm x)$ is an even
function. Since $|\cos(\bm Q\cdot\Delta\bm x)|\leq1$, the
inequality~(\ref{eq:0>Q}) cannot be satisfied if
$\tilde\phi(\Delta \bm x )$ is always positive.

\subsection{Multiplicity fluctuations and their rapidity correlator}
      
        Upon substituting the solution to Eq.~(\ref{Bjkphiev}), or rather its
        inverse Fourier transform $\tilde\phi$, into
        Eq.~(\ref{delNA2RRWYagain3}), we can now calculate the square
        variance of multiplicity fluctuations
        $\left<\delta N_A^2\right>$.   In the simplified setup that we are employing in this Section, we can go one step farther and
       exploit Bjorken
        boost invariance to compute
        explicit results for the rapidity correlator defined as
	\begin{eqnarray}
	\label{cdely0}
	C_A(y_+,y_-)&=&\left\<\frac{dN_A}{dy}\right\>^{-1}\left<\delta\frac{ dN_A}{dy_+}\delta\frac{ dN_A}{dy_-}\right>_\sigma\,.
	\end{eqnarray}
	(In the next Section where we employ a more realistic hydrodynamic solution, we shall only compute $\left<\delta N_A^2\right>$, not $C_A$.)
	The correlator $C_A$ measures the correlations between
          the multiplicity of particle species $A$ at rapidities $y_+$ and
          $y_-$ and can be determined similarly to
          Eq.~(\ref{delNA2RRWYagain3}) in terms of $\phi_{\bm Q}$ or
          its inverse Fourier transform $\tilde\phi(\Delta \bm x_\perp)$.  For the
          idealized Bjorken scenario, some of the integrals in
          Eq.~(\ref{delNA2RRWYagain3}) (e.g., over transverse
          coordinates) can be taken analytically. In order to make
          even further analytical progress we shall consider the case
          of particles with mass much bigger than the temperature,
          $m_A\gg T$.  This is an adequate approximation for
          protons and will allow us to perform an additional integral
          analytically in that case. We shall {\it not} use this approximation in the
          next Section, where we shall anyway be doing the analogous integrals numerically,
          but it will be helpful here to make the result
          and its general features more explicit.  As described in detail in Appendix~\ref{append1}, upon doing the integrals we obtain
 \begin{eqnarray}
	\nonumber
	\left<\delta\frac{d N_A}{dy_+}\delta\frac{d N_A}{dy_-}\right>_\sigma&\approx& \frac{g_A^2\, d_A^2}{8\pi^{7/2}Z}\,m_A^{7/2}\,T_f^{1/2} \,  A_\perp\,\tau_f^2\,\int \frac{d\eta}{\cosh^{5/2}\eta}\,e^{-\frac{2m_A \cosh \eta}{T_f}}  \\\label{NAdydyvariance2bjk2}
	& &\times \,\int \frac{dQ_\eta}{2\pi} \, e^{iQ_\eta\tau_f\Delta y}\, e^{-\frac{Q_\eta^2\tau_f^2 T_f}{m_A\cosh \eta}}\,\phi_{\mathbf{Q_\parallel}}(\tau_f)
 \end{eqnarray}
which is Eq.~(\ref{delNadyplusminus_app}), where ${\mathbf{Q_\parallel}} \equiv Q_\eta\hat \eta$, $\Delta y\equiv y_+-y_-$, and $A_\perp$ is the transverse area.

We see that in the simplified setup of this Section in which the fluid is translation invariant in the transverse directions, the modes that contribute in Eq.~(\ref{NAdydyvariance2bjk2}) are those whose ${\mathbf Q}$ is directed along the $\hat \eta$ direction. Also,
the effect of the last Gaussian factor in the $Q_\eta$ integral in Eq.~(\ref{NAdydyvariance2bjk2}) is to
limit the range of that integral to values of order
\begin{equation}
Q_\eta \lesssim  \left(\tau_f\sqrt{\frac{T_f}{m_A\cosh \eta}}\,\right)^{-1}\,.\label{eq:Qtauf}
\end{equation}
The fact that the characteristic wavenumber $Q$ of the fluctuations
responsible for the correlations at freezeout is not zero (despite
considering a volume of fluid that is infinite in extent in rapidity $\eta$ in this idealized scenario) is
ultimately due to the fact that in the laboratory frame the fireball is not spatially homogeneous: the
fluid velocity varies over a longitudinal distance of
order $\tau_f$ due to the longitudinal expansion. One can check that
taking $\tau_f\to\infty$ results in only $Q=0$ contributing. However,
the characteristic $Q_\eta$ is not just $1/\tau_f$, but rather depends
on the mass of the particle. This is due to the thermal smearing, or
``blurring'', which translates spatial Bjorken rapidity $\eta$ into
kinematic particle rapidity $y$~\cite{Ling:2015yau,Ohnishi:2016bdf}. As we 
are assuming that $m_A\gg T$, the factor
$\sqrt{T_f/(m_A\cosh \eta)}$ in Eq.~(\ref{eq:Qtauf}) is the typical
thermal rapidity of the particles at temperature $T_f/\cosh\eta$,
which can be understood as the freezeout temperature ``red-shifted''
by longitudinal expansion.

 	 \begin{figure}[t]
	    \centering
	    \includegraphics[width=0.5\textwidth]{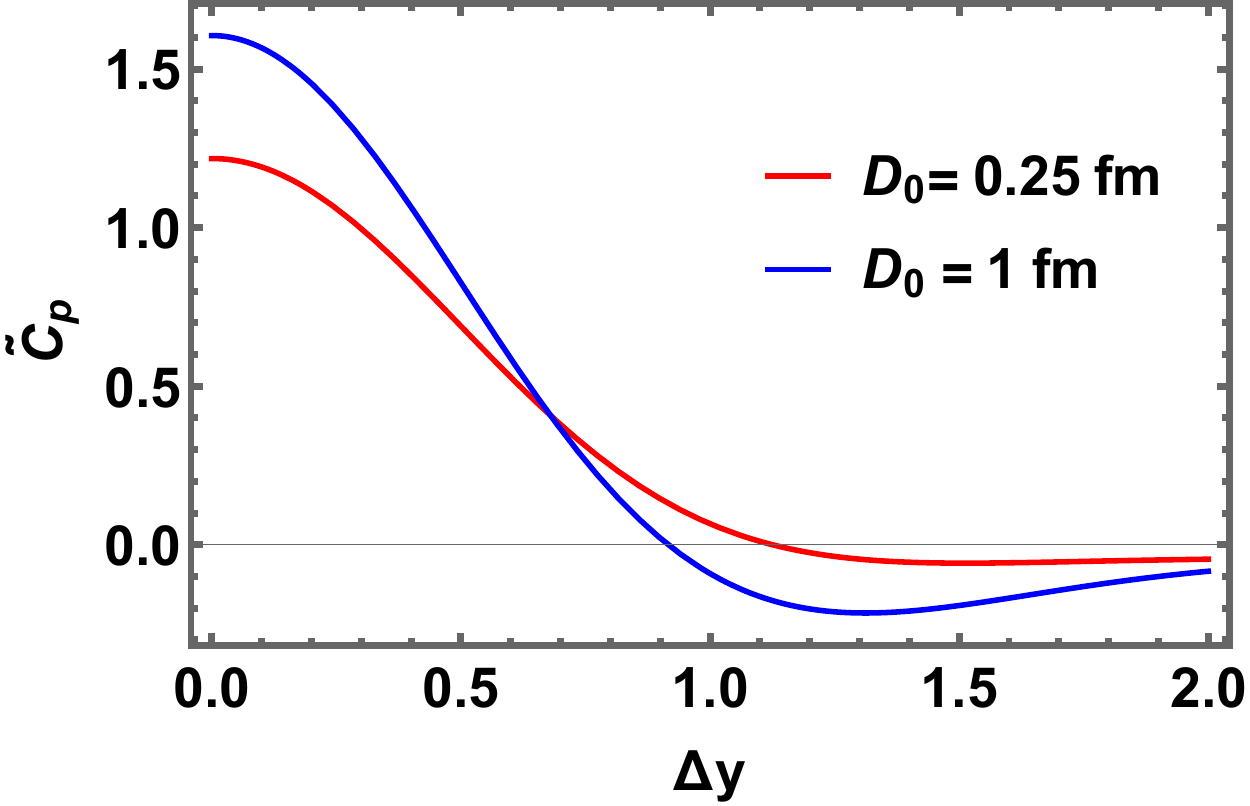}
	   \label{phiQdelnaGHb}
	    \caption{
              Normalized  proton multiplicity correlator $\tilde{C}(\Delta y)$ for protons from Eq.~(\ref{cdely1}) as a function of the rapidity gap $\Delta y$ in the
              Bjorken scenario for two choices of the diffusion
              parameter~$D_0$. }
	    \label{CDeltay}
          \end{figure}

The final piece that we need in order to compute $C_A$ is the denominator in Eq.~(\ref{cdely0}). By explicit calculation starting from Eq.~(\ref{NA1}), in the Bjorken scenario in which we are working
$\<dN_A/dy\>$  is given by:
 \begin{eqnarray}\label{Bjkmean}
\left\<\frac{dN_A}{dy}\right\>=d_A\,A_\bot\,\tau_f\,	(2\pi)^{-2}\int_{m_A}^{\infty} m_T^2 dm_T\int d\eta \, e^{-\frac{m_T\cosh\eta}{T_f}}\cosh\eta\,.
	\end{eqnarray}	

Substituting Eqs.~(\ref{NAdydyvariance2bjk2}) and (\ref{Bjkmean}) into Eq.~(\ref{cdely0}), we can evaluate $C_A$ which, because of boost invariance, is a function of $\Delta y$ only.  
In
 Fig.~\ref{CDeltay}, we plot our results for $C_A$, normalized by the its non-critical
 value at $\Delta y =0$,
 $C_A^{\text{nc}}(0)$, which we estimate by substituting
 $\phi_{\bm Q}=Z\,T\,(Q^2+\xi_0^{-2})^{-1}$ into
 Eq.~(\ref{NAdydyvariance2bjk2}),  where $\xi_0$ is, as before, the correlation
 length away from critical point defined in Eq.~(\ref{xiparam1}). That is, we
 define the ratio that we have plotted in Fig.~\ref{CDeltay}  as
\begin{eqnarray}
	\label{cdely1}
	\tilde{C}_A(\Delta y)&\equiv &\frac{C_A(\Delta y)}{C_A^{\text{nc}}(0)}\,.
	\end{eqnarray}
Since $\xi_{\text{max}}>\xi_0$, critical fluctuations make the ratio 
$\tilde C_A(0)$  larger than unity.

In Fig.~\ref{CDeltay} we also illustrate the dependence of the
rapidity correlations on the value of the
diffusion parameter $D_0$. Stronger diffusion (larger $D_0$) enhances the
effects of the critical point in $\phi_{\bm Q}$, as we saw in
Fig.~\ref{phiQdelnaGH}. This enhancement is reflected in the
corresponding particle rapidity correlations, as seen in
Fig.~\ref{CDeltay} at small $\Delta y$. Due to conservation laws,
anticorrelations at large $\Delta y$ are also enhanced. 
(For any value of $\dip$, the consequence of conservation is that the integral under the curve $\tilde{C}_A$ over all separations $\Delta y$ is independent of $\dip$ and is determined by the initial fluctuations.	This means that when we increase $D_0$ and see $\tilde C_A(\Delta y)$ growing at small $\Delta y$, it must also become more negative at large $\Delta y$.)
However, unlike the direct effect of
diffusion on the {\em range} of the spatial correlator $\tilde\phi$ in Fig.~\ref{FTphiQdelnaGHa},
the effect  on the {\em range} of $C(\Delta y)$ in Fig.~\ref{CDeltay} is minor. This
is due to the fact that this range is mostly determined by the effect of
thermal rapidity smearing or
``blurring''~\cite{Ling:2015yau}.

Now, the variance
of the particle multiplicity $\<\delta N_A^2\>_\sigma$ that in the general case takes the form (\ref{delNA2RRWYagain3})
can in this Bjorken scenario be obtained from the
rapidity correlator $C(\Delta y)$ 
by
integration over the rapidity window
$y_\pm\in(-y_{\text{max}},y_{\text{max}})$, i.e.,
  \begin{eqnarray}
    \frac{\left<\delta N_A^2\right>_\sigma}{\left<N_A\right>}
    =\int_{-2y_{\text{max}}}^{2y_{\text{max}}}
    \,d\Delta y\,
    \left(1-\frac{\left|\Delta y\right|}{2y_{\text{max}}}\right)\,C(\Delta y)\,.
  \end{eqnarray}
Upon using our expression for $C(\Delta y)$ from Eqs.~(\ref{cdely0}), (\ref{NAdydyvariance2bjk2}), and (\ref{Bjkmean}) 
and using the fact that boost invariance implies that $\langle N_A \rangle = 2 y_{\text{max}} \langle d N_A/dy\rangle$, 
we now obtain the result
  \begin{eqnarray} \nonumber \<\delta N_A^2\>_\sigma&\approx&
  \frac{1}{2}\,Z^{-1}\,m_A^{7/2}\,T_f^{1/2} \, \pi^{-7/2}\,g_A^2\,
  d_A^2\, A_\perp\,\tau_f^2\,\int d\eta\, \text{sech}^{5/2}
  \,\eta\,e^{-\frac{2m_A \cosh \eta}{T_f}}
  \,\times\\ \label{NAvariance2bjk3} & &\int \frac{dQ_\eta}{2\pi} \,
  \frac{\sin^2 \left(\tau_f\, Q_\eta\, y_{\text{max}}\right)}{\tau_f^2
    Q_\eta^2}\, e^{-\frac{Q_\eta^2\tau_f^2 T}{m_A\cosh
      \eta}}\,\phi_{\mathbf{Q_\parallel}}(\tau_f)
\end{eqnarray}
The $\Delta y$ dependence of $C(\Delta y)$ translates into the
rapidity acceptance window dependence of  $\<\delta
N_A^2\>_\sigma$, which has been discussed in the literature, e.g., in
Ref.~\cite{Ling:2015yau}, and will not be discussed here.

		\begin{figure}[t]
\begin{center}
\begin{subfigure}{0.49\textwidth}
  \includegraphics[scale=0.55]{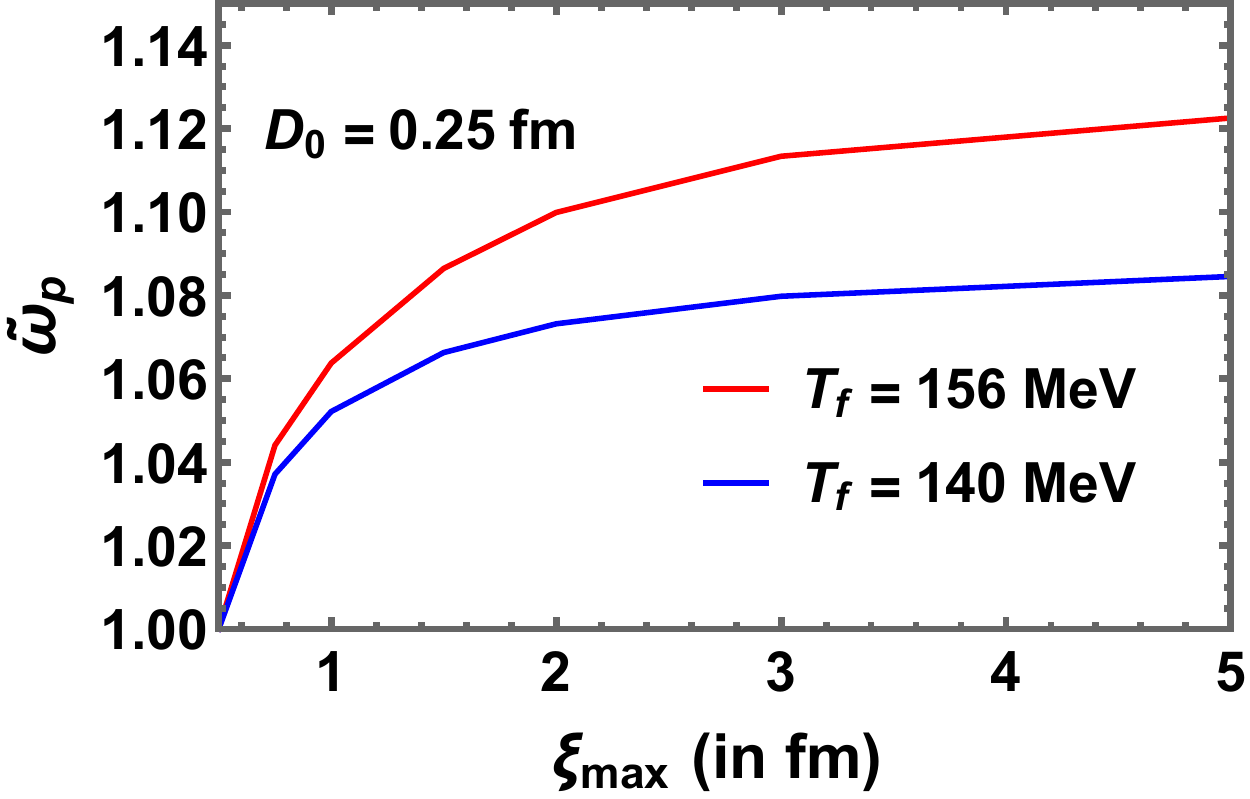}
  \caption{}
\end{subfigure}
\begin{subfigure}{0.49\textwidth}
\includegraphics[scale=0.55]{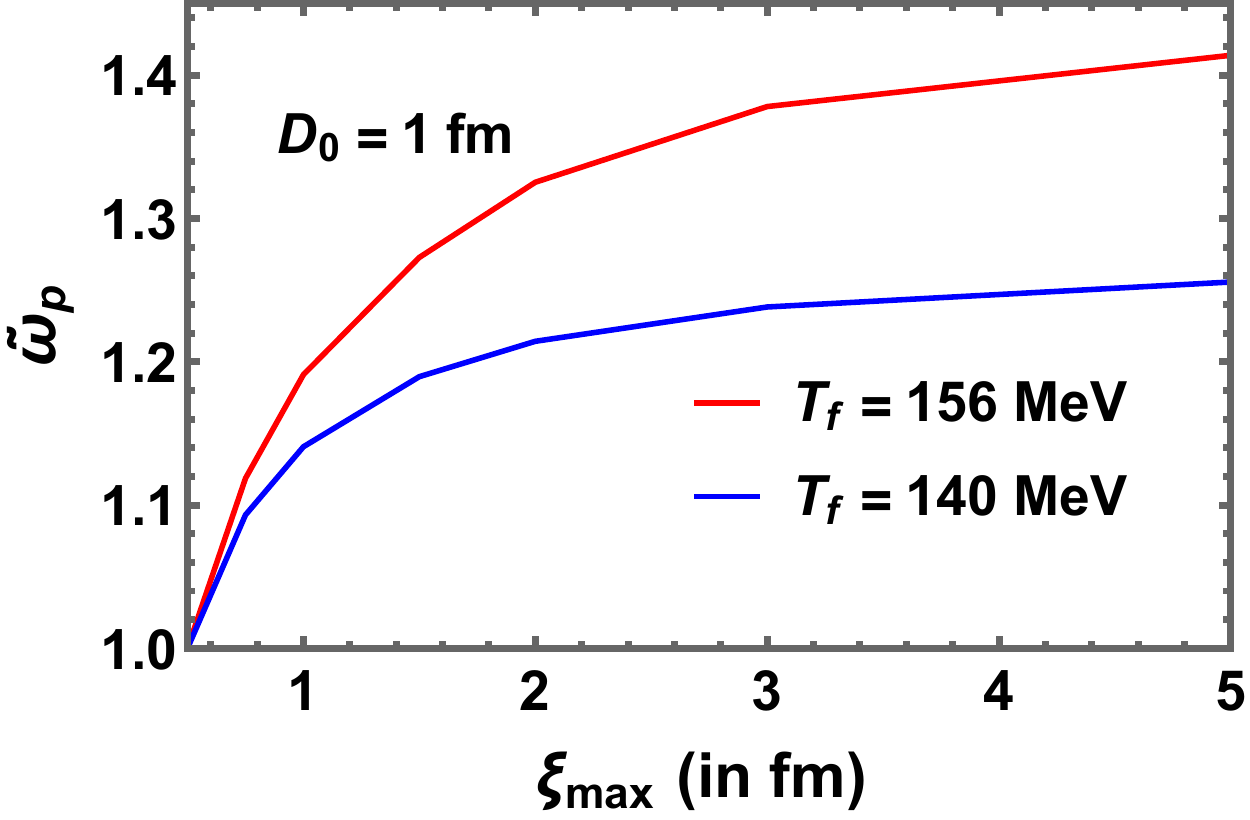}
  \caption{}
\end{subfigure}
\begin{subfigure}{0.49\textwidth}
\includegraphics[scale=0.55]{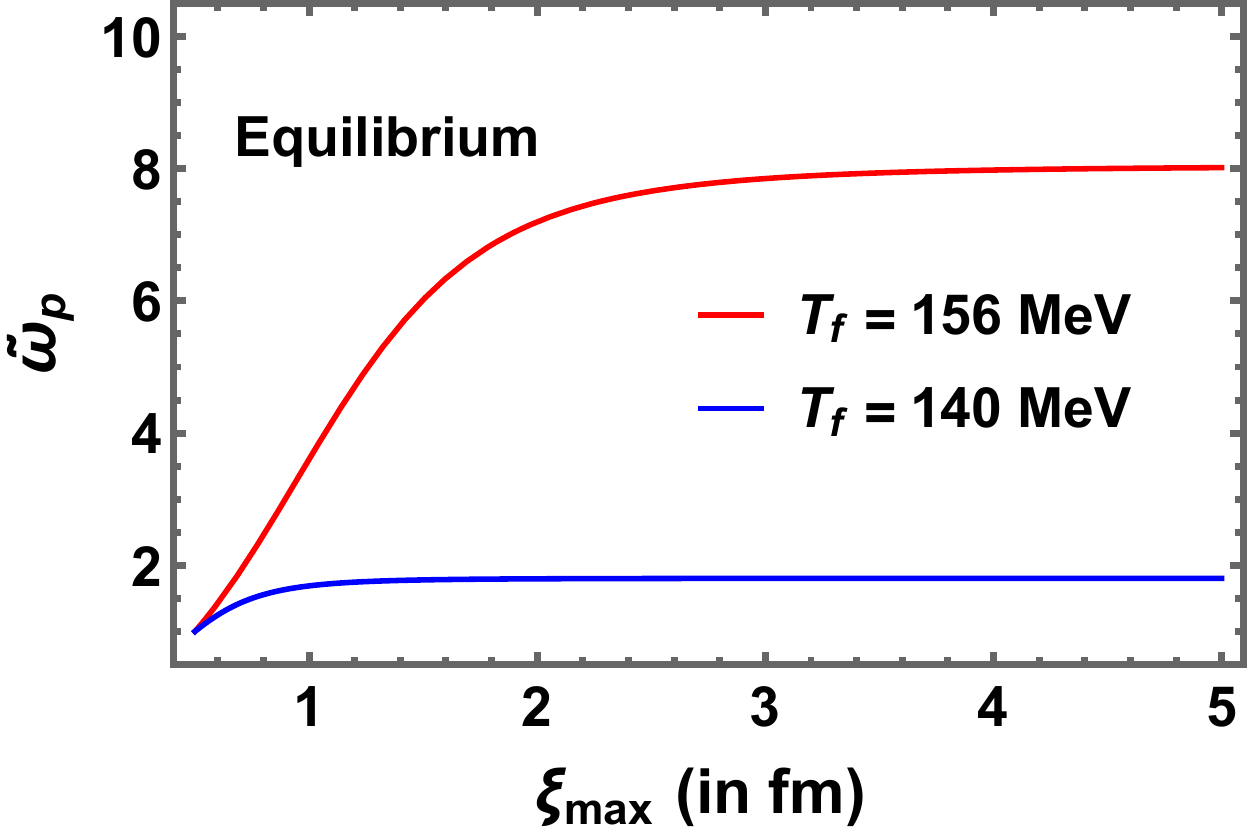}
  \caption{}
\end{subfigure}
\end{center}
\caption{The normalized fluctuation measure for protons,
Eq.~(\ref{omegaAdef2}), 
  as a
  function of $\xi_{\rm max}$, the maximum value of the equilibrium correlation length achieved along the system trajectory. Panels (a) and (b)
  correspond to different diffusion strengths, quantified by $D_0$,
  while red and blue curves correspond to different freezeout
  temperatures. Panel (c) shows the result that would have been
  obtained under the assumption that fluctuations are in equilibrium at
  freezeout. }
\label{fhs-protonsHBjk}
\end{figure}

\subsection{Dependence of fluctuations on dynamics and on proximity of freezeout to the critical point}

In this paper we focus on the magnitude of fluctuations and their
dependence on the proximity of freezeout to the critical point as well
as on the diffusion parameter $D_0$. The proximity of the freezeout to
the critical point is controlled by two major factors. One is the
proximity of the trajectory to the critical point, which in our
analysis is quantified by $\xi_{\rm max}$ -- the maximum equilibrium
correlation length along the expansion trajectory on the phase
diagram. The larger the value $\xi_{\rm max} $, the closer the system has
passed to the critical point on its way to freezeout. In
Fig.~\ref{fhs-protonsHBjk}, we plot the normalized critical
contribution to the squared variance of the proton multiplicity from
Eq.~(\ref{omegaAdef2}) on $\xi_{\rm max}$. One can see the
effect of the critical point on the fluctuations at freezeout increases
as the trajectory approaches the critical point (as $\xi_{\rm max}$
increases).

The other factor
controlling the proximity of the critical point to the freezeout is
the freezeout temperature $T_f$. The higher the freezeout temperature is
(while still below critical temperature), i.e., the earlier the freezeout occurs and the
closer the freezeout is to the critical point. Correspondingly, the
fluctuations at freezeout increase with $\xi_{\rm max}$ as well as
with $T_f$ in Fig.~\ref{fhs-protonsHBjk}.

The results in Fig.~\ref{fhs-protonsHBjk} also demonstrate that the magnitude of the
critical point signatures crucially depends on dynamics.  As we
already discussed in Figs.~\ref{phiQdelnaGH} and~\ref{CDeltay},
stronger diffusion (larger $D_0$) leads to larger effects of the
critical point. We can see this in Fig.~\ref{fhs-protonsHBjk} by
comparing the plots for two different values of $D_0$. 
In addition, as
a result of the conservation laws the magnitude of fluctuations is
significantly smaller than the equilibrium expectation at freezeout,
as can be seen by comparing to panel (c) in
Fig.~\ref{fhs-protonsHBjk}.
It is also apparent from this comparison
that, while the equilibrium expectation in panel (c) depends very strongly on the
freezeout temperature (the higher the temperature the closer is the
freezeout to the critical point, since $T_c>T_f$), the dynamical
predictions in panels (a) and (b)  of Fig.~\ref{fhs-protonsHBjk} are much less sensitive to the freezeout temperature. This
can be understood as a ``memory'' effect:
the fluctuations at freeze-out encode some information about 
fluctuations at earlier times because they do not have time to equilibrate, an effect which is enhanced by critical slowing down. This has the consequence that even though freeze-out happens 
at a temperature below that of the critical point, effects of critical fluctuations persist until freeze-out
and yield signatures in observables.
The magnitude of such memory effects depends on the value of $D_0$. We see that the results of our out-of-equilibrium calculation at freeze-out illustrated in panels (a) and (b) of Fig.~\ref{fhs-protonsHBjk} arise via an interplay between the suppression of fluctuations relative to their magnitude in equilibrium due to conservation and the enhancement of fluctuations arising from memory effects.

In panel (c) of Fig.~\ref{fhs-protonsHBjk}, the equilibrium magnitude of
fluctuations saturates as $\xi_{\text{max}}$ increases and the trajectory followed by the cooling
plasma gets closer to the critical point.
This happens
because the freezeout occurs at a
temperature $T_f$ below the critical point, where $\xi(T_f)<\xi_{\rm
  max}$, 
 and as the trajectory approaches the critical point, $\xi_{\rm max}$ diverges while $\xi(T_f)$ 
  goes to some (large) constant value
  which  is independent of $\xi_{\rm max}$ as $\xi_{\rm max}\to\infty$. 
  The
saturation is less pronounced in panels (a) and (b) because of 
dynamical memory effects: the blue curve ``remembers where it was'' at earlier times.

	To summarize some central results of this Section: (i) in the Bjorken scenario considered here there is a clear suppression in the normalized fluctuation measure $\tilde{\omega}_A$ at freezeout for the values of $\dip$ considered, because the slow modes are fluctuations of a conserved quantity; (ii) in Model H dynamics, $\tilde{\omega}_A$  is less sensitive to the freeze-out temperature than would be the case if the fluctuations were in equilibrium throughout. In the next Section, we shall investigate how radial flow modifies these and other observations.


\section{Freezing out a semi-realistic numerical hydro+ simulation}
\label{RRWY}

In this Section, we demonstrate the use of the freezeout procedure
introduced in Section.~\ref{presc1} and employed in a Bjorken scenario in Section~\ref{Sectbjk} 
to obtain the two-point
correlations of particle multiplicities from the Hydro+ simulation
that was introduced and analyzed, but not frozen out, 
in Ref.~\cite{Rajagopal:2019xwg}. As in the previous Section, the
system under consideration is boost invariant and has azimuthal
symmetry. Unlike in the previous Section, the system we consider here is
finite in transverse extent and thus exhibits radial flow.  We give a brief description of
the details of simulation here. For more details the reader may refer to
Ref.~\cite{Rajagopal:2019xwg}.

The evolution of the energy density, $\varepsilon(r,\tau)$ and the fluid four-velocities
$u(r,\tau)$ in our simulation is determined using the standard MIS second order
hydrodynamic equations as implemented in the publicly available
VH$1 + 1$ hydrodynamic code~\cite{Baier:2006gy, Baier:2006um,
  Romatschke:2007jx}.  The equation of state $p(\varepsilon)$ used in the
simulation was introduced in Ref.~\cite{Rajagopal:2019xwg} and was already employed in
Section~\ref{Sectbjk} and, for
convenience, is reviewed in Appendix~\ref{appendeos}. 
We set the shear viscosity to entropy density ratio to $\eta/s=0.08$ throughout, 
and solve the equations numerically using a spatial (radial) lattice with 1024 points spaced by $0.0123$~fm and a time step of $0.005$~fm.
In this Section, we initialize the
hydrodynamic simulation at $\tau_i=1\, \text{fm}$, with an initial
central temperature of 330 MeV, following Ref.~\cite{Baier:2006gy}. We set the
radial flow $v_r$ and the viscous part of the stress-energy tensor $\Pi_{\mu\nu}$
  to zero initially  at
$\tau=\tau_i$. 
We employ the standard Glauber model radial profile 
 corresponding to a central
Au-Au collision at $\sqrt{s}=200$ GeV for $\varepsilon(r)$ at
$\tau=\tau_i$, again following Ref.~\cite{Baier:2006gy}.

As in Ref.~\cite{Rajagopal:2019xwg}, in our Hydro+ simulation the hydrodynamic densities 
$\varepsilon(r,\tau)$ and $\Pi_{\mu\nu}(r,\tau)$ and the four-velocities
$u(r,\tau)$  provide the
background for the evolution of the fluctuations described by $\phi_{\bm Q}$ according to
Eq.~(\ref{phiev1}). Again following Ref.~\cite{Rajagopal:2019xwg},
we choose to initialize the fluctuations $\phi_{\bm Q}$ at
$\tau=\tau_i$ by setting them to the corresponding equilibrium values
determined by the local temperature at this initial time, i.e.
\begin{eqnarray}\label{initalphiQ}
\phi_{\bm Q}(r,\tau_i)=\bar{\phi}_{\bm Q}\Big|_{T(r,\tau_i)}\,.
\end{eqnarray}
For the interested reader, the limitations of the various assumptions made in setting 
up this Hydro+ simulation, as well as 
possible future improvements to it, are detailed in
Ref.~\cite{Rajagopal:2019xwg}.

We calculate the evolution of $\phi_{\bm Q}$ using the same code as in  Ref.~\cite{Rajagopal:2019xwg}, with two important changes. For simplicity, in their pioneering calculation the authors of Ref.~\cite{Rajagopal:2019xwg} 
chose to simulate the evolution of $\phi_{\bm Q}$ according to 
Model A relaxation dynamics, as if the fluctuations were those of a quantity that is not conserved. 
In this work, as in Section~\ref{Sectbjk}
we employ Model H dynamics which takes into account conservation
laws. This gives us an opportunity to highlight the effects of conservation laws on the 
dynamics of $\phi_{\bm Q}$ and on the resulting particle multiplicity fluctuations by comparing the results of this Section to those obtained by repeating the calculations of this Section using Model A evolution.  We perform this comparison in  
Appendix~\ref{RRWYModelA}.  The second change that we make relative to Ref.~\cite{Rajagopal:2019xwg} is that here we shall neglect the back-reaction of the fluctuations on the hydrodynamic densities.  The modifications to the bulk dynamics of the hydrodynamic fluid, in particular its entropy density $s(r,\tau)$, introduced by the presence of the fluctuations was computed in Ref.~\cite{Rajagopal:2019xwg,Du:2020bxp} and in fact the fluctuations and the hydrodynamic densities were computed self-consistently. However, these authors showed that including back reaction self-consistently introduces fractional changes to $\varepsilon(r,\tau)$ and $v_r(r,\tau)$ that are small, rarely comparable to $1\%$ and typically much smaller.  For this reason we neglect these effects.

In the remainder of this Section, we demonstrate the implementation of the freezeout prescription introduced
in Section~\ref{presc1} to freeze out the Hydro+ simulation with the
background described above for some reasonable values of $\dip$.
After describing and illustrating the evolution of $\phi_{\rm Q}$ in Subsection~\ref{Subsect_phi} and 
 the fluctuations on the freeze-out surface in Subsection~\ref{Subsect_freezeoutsurfacefluctuations}, in Subsection~\ref{freeze2ptRRWY1} we describe the resulting fluctuations in particle multiplicities.

\subsection{Evolution of $\phi_{\mathbf Q}$}
\label{Subsect_phi}

\begin{figure}[t]
\begin{center}
\includegraphics[scale=0.75]{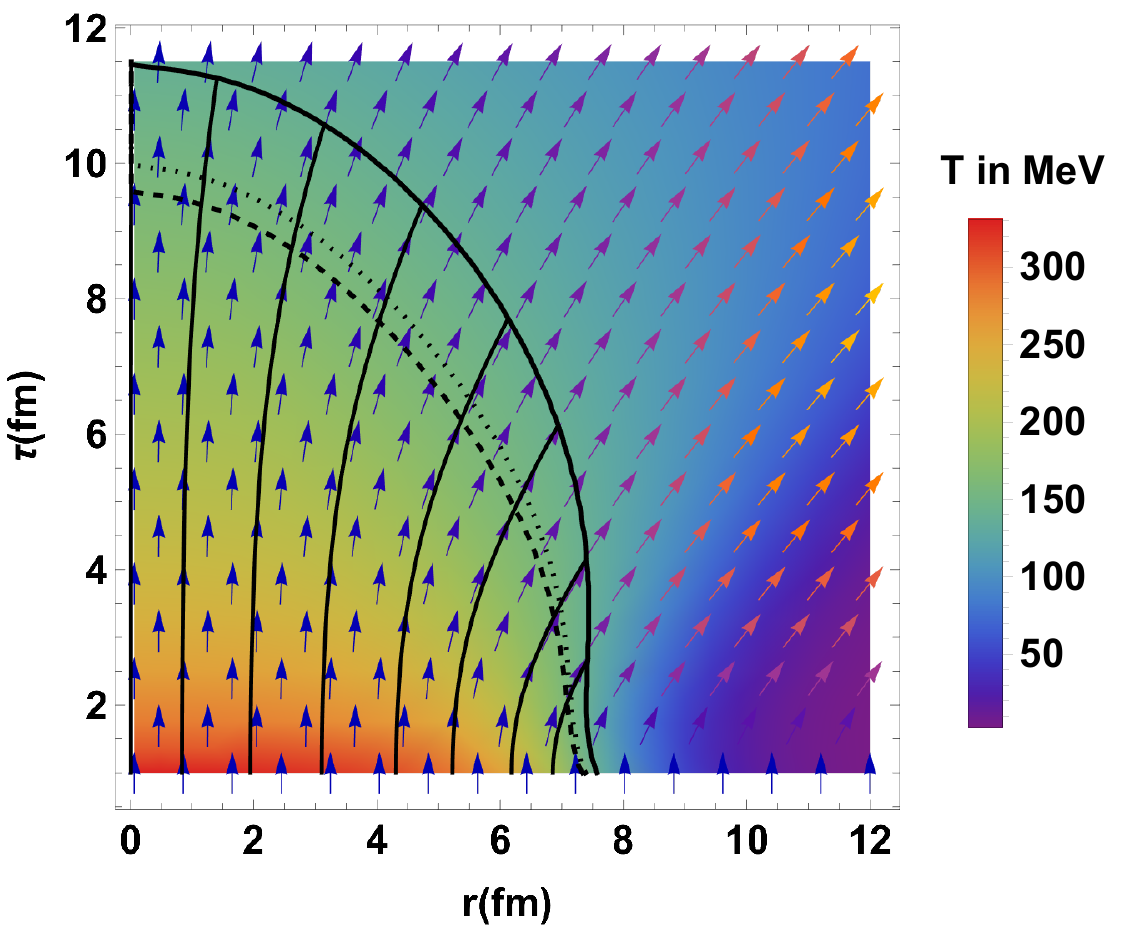}
\end{center}
\caption{The space-time dependence of the temperature (represented by color) 
and flow velocity in the
  hydrodynamic simulation of the expanding cooling droplet of quark-gluon plasma. 
    The magnitude of the radial flow
  at each space-time point is indicated by the tilt of the arrows. The dashed, dotted and
  solid black curves are the isothermal curves at $T=160\, \text{MeV}$,
  $156\, \text{MeV}$ and $140\, \text{MeV}$, respectively. Examples of 
   fluid cell trajectories, or hydrodynamic flow lines, are illustrated by
  solid black lines tangential to local flow vectors. 
}
\label{temperaturedensity}
\end{figure}

In this Subsection, we present and discuss the space-time dependence
of the fluctuation measure $\phi_{\bm Q}$ as it evolves according to
the relaxation equation given by Eq.~(\ref{phiev1}) with the Model H
relaxation rate given by Eq.~(\ref{GammaH}). The radial dependence of
the flow and temperature profiles makes $\phi_{\bm Q}$ dependent on
the radial variable $r$ in addition to the Bjorken time $\tau$.  Several
representative characteristic curves, or flow lines, determined by the
flow velocity $u$, are shown in Fig.~\ref{temperaturedensity}.

	\begin{figure}[t]
\begin{center}
\begin{subfigure}{0.45\textwidth}
\includegraphics[scale=0.57]{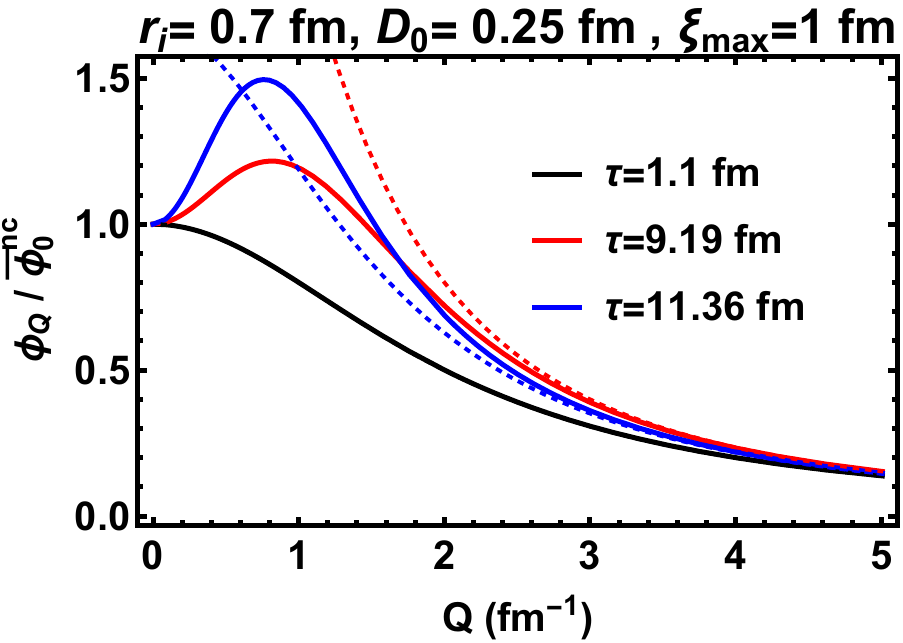}
\end{subfigure}
\begin{subfigure}{0.45\textwidth}
\includegraphics[scale=0.57]{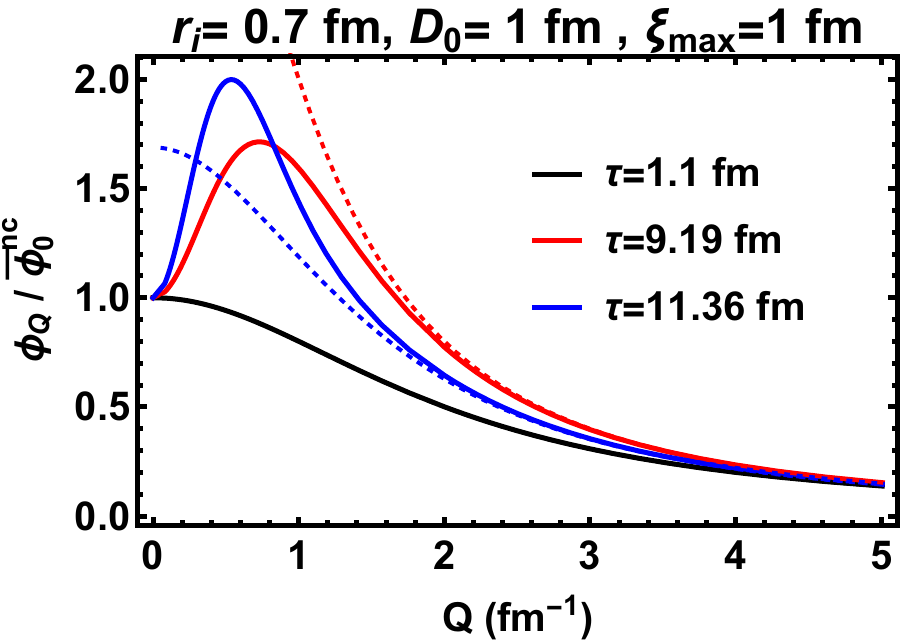}
\end{subfigure}
\begin{subfigure}{0.45\textwidth}
\includegraphics[scale=0.57]{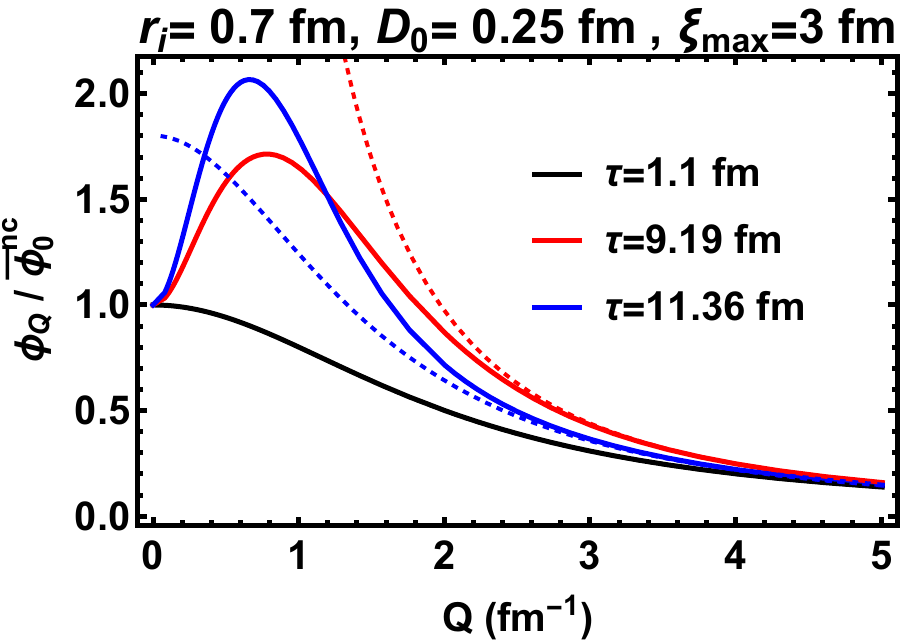}
\end{subfigure}
\begin{subfigure}{0.45\textwidth}
\includegraphics[scale=0.57]{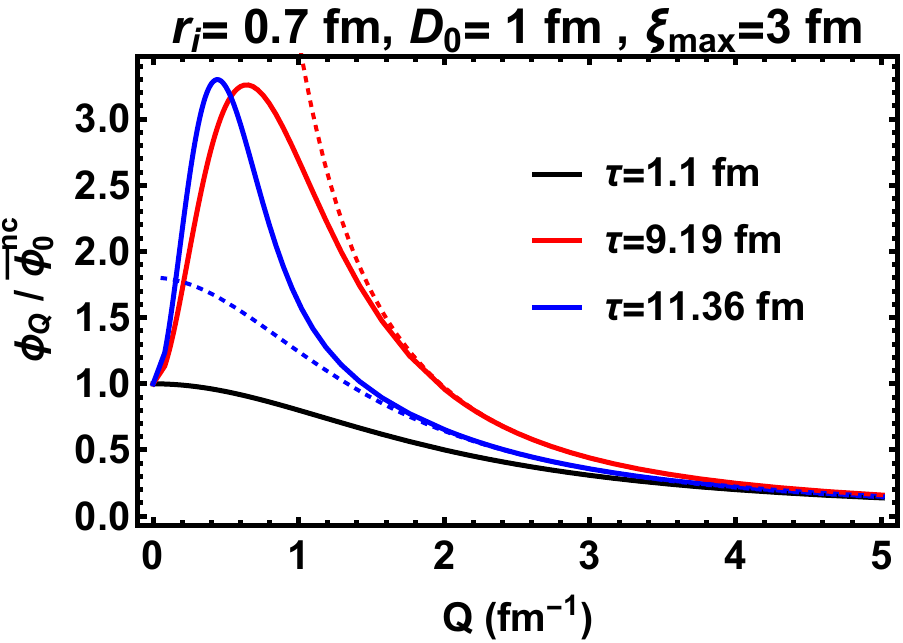}
\end{subfigure}
\begin{subfigure}{0.45\textwidth}
\includegraphics[scale=0.57]{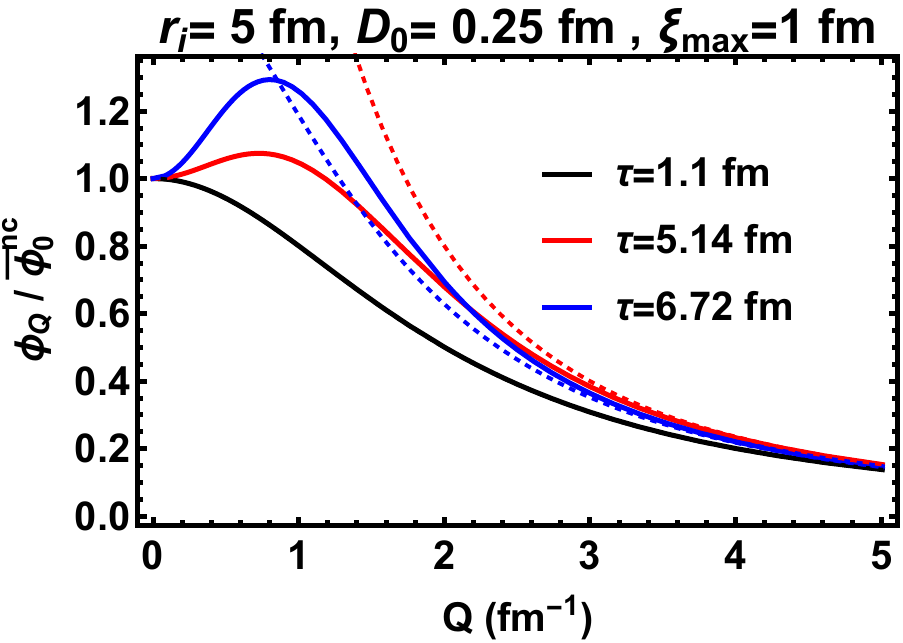}
\end{subfigure}
\begin{subfigure}{0.45\textwidth}
\includegraphics[scale=0.57]{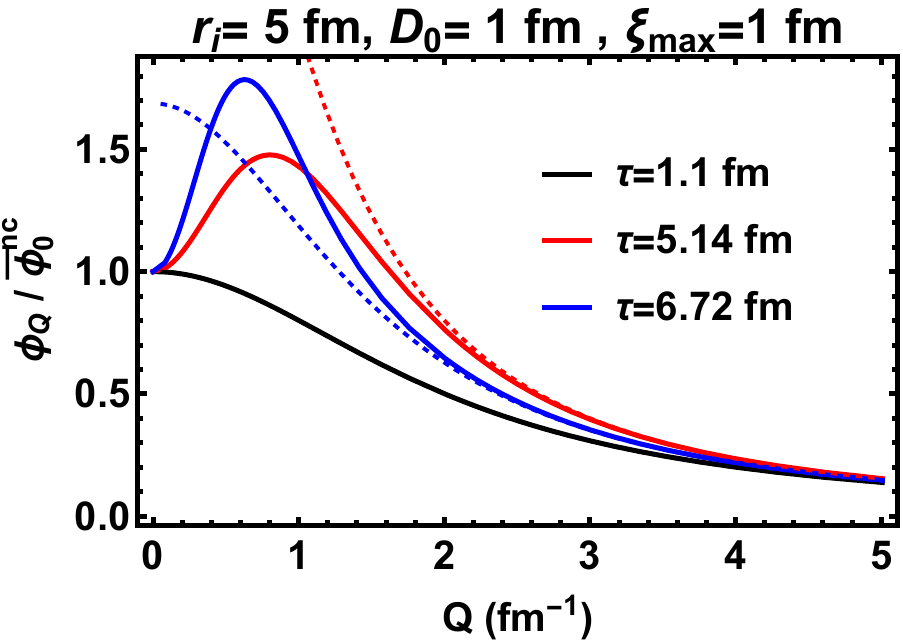}
\end{subfigure}
\begin{subfigure}{0.45\textwidth}
\includegraphics[scale=0.57]{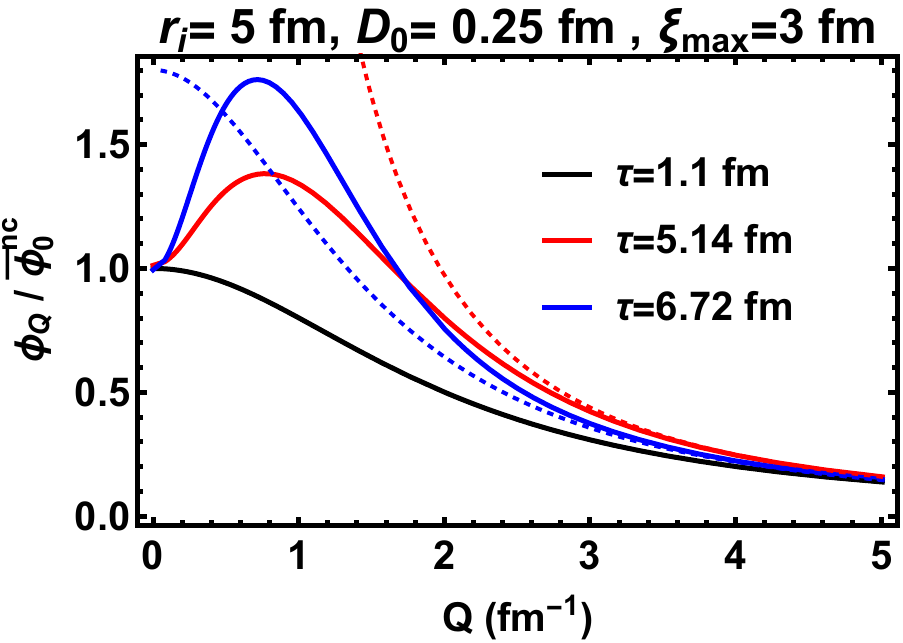}
\end{subfigure}
\begin{subfigure}{0.45\textwidth}
\includegraphics[scale=0.57]{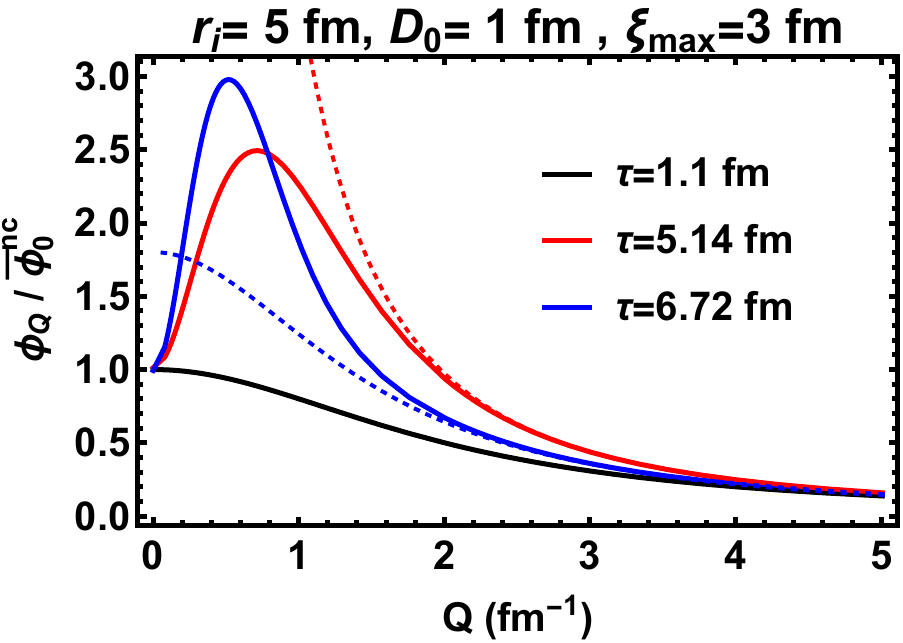}
\end{subfigure}
\end{center}
\caption{Hydro+ fluctuation measure $\phi_{\bm Q}$ along two hydrodynamic flow lines passing through $r=r_i$ at initial time $\tau=\tau_i$,
  with $r_i=0.7\, \text{fm}$ (top four panels) and $5\, \text{fm}$ (bottom
  four panels). The four plots in the left (right) column are for 
  $\dip = 0.25\, \text{fm}$ ($\dip = 1\, \text{fm}$), with
  $\xi_{\text{max}}=1\, \text{fm}$ and
  $\xi_{\text{max}}=3\, \text{fm}$ in alternating rows.  
  The solid and dashed curves
   are, respectively, the $\phi_{\bm Q}$ and $\bar{\phi}_{\bm Q}$
  (normalized to their values at $Q=0$ away from the critical point,
  where $\xi=\xi_0$) at three times $\tau$ indicated in the plot legends; the choice of $\tau$'s is explained in the text.
}
\label{phi-1ch}
\end{figure}

In Fig.~\ref{phi-1ch}, we plot our results for the fluctuation measure $\phi_{\bm Q}$ 
in the hydrodynamic background illustrated in Fig.~\ref{temperaturedensity}
at three different times $\tau$
along two hydrodynamic flow lines, one close to the center of
the fireball ($r(\tau_i)=0.7$ fm) and one further out ($r(\tau_i)=5$ fm). 
We display results from simulations
performed with $\dip=0.25\, \text{fm}$ (slower diffusion) and
$\dip=1\, \text{fm}$ (faster diffusion) and
$\xi_{\text{max}}=1\, \text{fm}$ (trajectory further away from the critical
point) and $\xi_{\text{max}}=3\, \text{fm}$ (trajectory closer to the
critical point). 
In all the panels, at $\tau=1$~fm (black curve) $\phi_{\rm Q}$ is given by its equilibrium value.
In the upper (lower) four panels, the red curves at $\tau=9.19$~fm ($\tau=5.14$~fm)
are drawn at the time when when the fluid cell moving along the flow line 
that started at $r_i=0.7$~fm ($r_i=5$~fm) has cooled to the temperature $T=T_c=160$~MeV
and the blue curves at $\tau=11.36$~fm ($\tau=6.72$~fm) 
are drawn at the time when these fluid cells have cooled further to $T=140$~MeV.
Increasing $\xi_{\text{max}}$, i.e.,
bringing the evolution trajectory closer to the critical point, causes
the magnitude of equilibrium fluctuations to increase. However the
relaxation to the equilibrium value becomes slower since its rate
$\Gamma(Q)\propto DQ^2$ and $D=\dip\xi_0/\xi$ is proportional to $1/\xi$. 
We find that the former
effect ``wins'' in the sense that $\phi_{\bm Q}$ at the time the fluid cell trajectory 
cools to $T=140$~MeV (i.e. the blue curves in Fig.~\ref{phi-1ch}),
which is well after the fluid cell trajectory passes the point where $\xi=\xi_{\rm max}$, see Fig.~\ref{fig:trajectory},  increases with
$\xi_{\text{max}}$, at least in the range of the parameters we have
considered. 

We see by comparing the left and right columns
of Fig.~\ref{phi-1ch} that increasing the diffusion parameter $\dip$, which 
increases the relaxation rate,
has the consequence that $\phi_{{\bm Q}}$ is closer to its instantaneous
equilibrium form $\bar{\phi}_{\bm Q}$ during the course of the evolution. The
value of $\phi_{\bm Q}$ at $\bm Q=0$, however, remains invariant
during the evolution due to the conservation laws inherent in model H:
$\Gamma(\bm Q=0)=0$.

The $Q$-dependence of $\phi_{\bm Q}$ is shaped by two competing
effects. As a given hydrodynamic cell, represented by a point on the
phase diagram (see Fig.~\ref{fig:trajectory}) moving along the expansion
trajectory, approaches the critical point, the ``desired'' equilibrium
values of $\bar\phi_{\bm Q}$, to which $\phi_{\bm Q}$ is forced to
relax by Eq.~(\ref{phiev1}), increases across all values of $\bm
Q$. However, while at larger $Q$, the relaxation is fast enough to
effectively equilibrate $\phi_{\bm Q}$ to these larger equilibrium
values, at lower $Q$ conservation laws slow down the evolution, making
the $\phi_{\bm Q}$ values lag behind $\bar\phi_{\bm Q}$ more
significantly. This produces a peak in $\phi_{\bm Q}$ at a
characteristic value of $Q$ denoted by $Q^{\text{peak}}$ in
Ref.~\cite{Du:2020bxp} which moves to lower values of $Q$ as $\dip$ is
increased.
These features are evident in
Fig.~\ref{phi-1ch} across the range of parameters we have considered. It is
also instructive to compare and contrast Fig.~\ref{phi-1ch} with the results
that would be obtained if the fluctuations followed model A dynamics
where the relaxation rate of low-$Q$ modes is not
suppressed and, consequently, $Q^{\rm peak}=0$.
We perform this comparison in Appendix~\ref{RRWYModelA}; see
Fig.~\ref{phi-ch2A} from that Appendix which is to be compared with Fig.~\ref{phi-1ch}.

\begin{figure}[t]
\begin{center}
\begin{subfigure}{0.32\textwidth}
\includegraphics[scale=0.55]{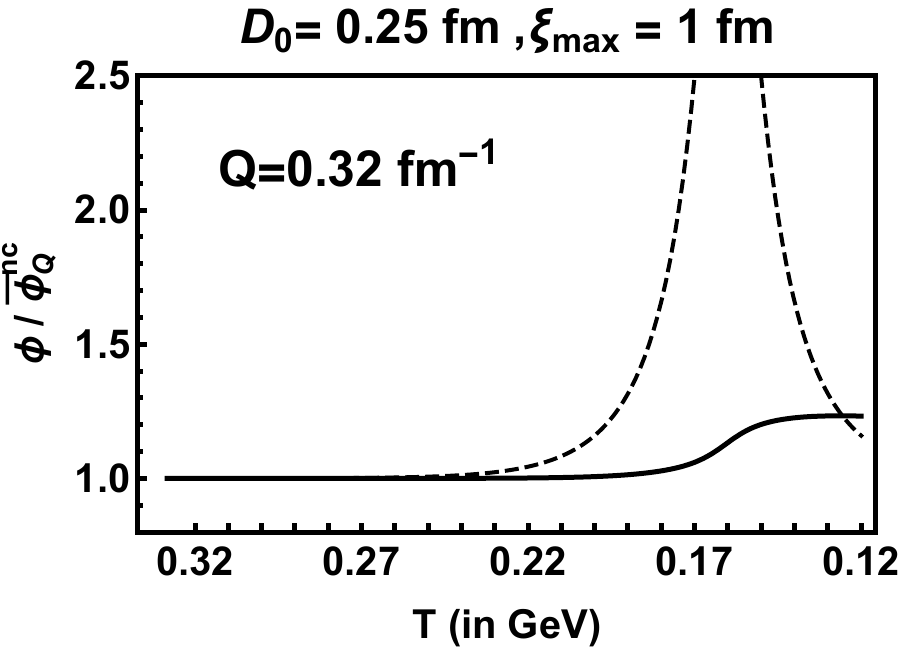}
\end{subfigure}
\begin{subfigure}{0.32\textwidth}
\includegraphics[scale=0.55]{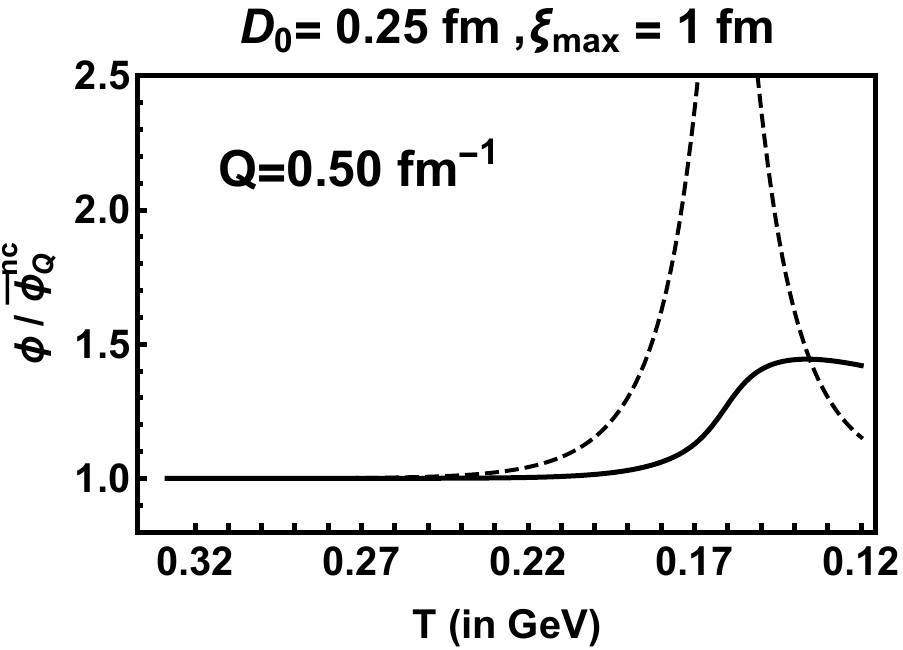}
\end{subfigure}
\begin{subfigure}{0.32\textwidth}
\includegraphics[scale=0.55]{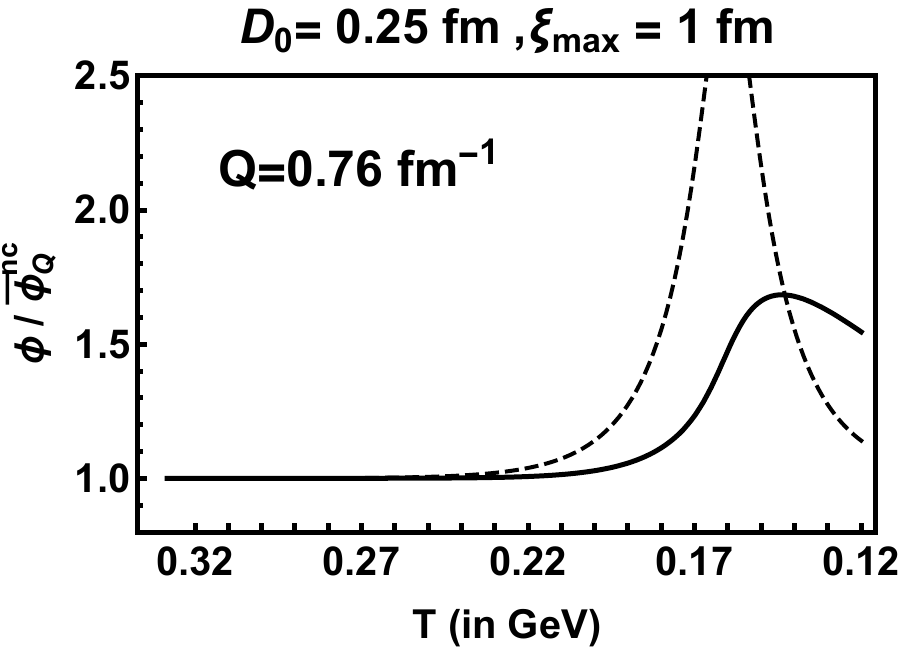}
\end{subfigure}
\begin{subfigure}{0.32\textwidth}
\includegraphics[scale=0.55]{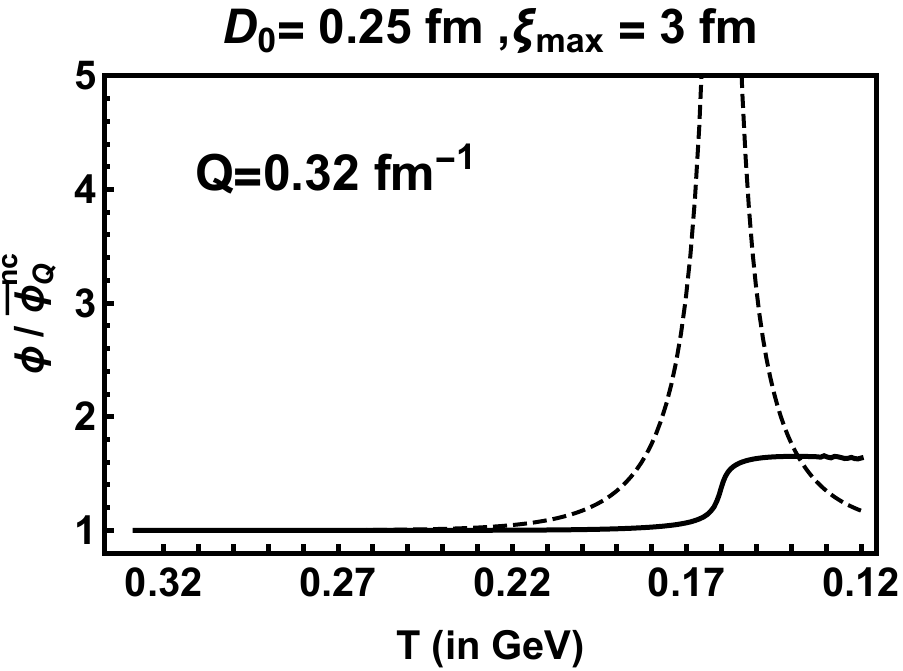}
\end{subfigure}
\begin{subfigure}{0.32\textwidth}
\includegraphics[scale=0.55]{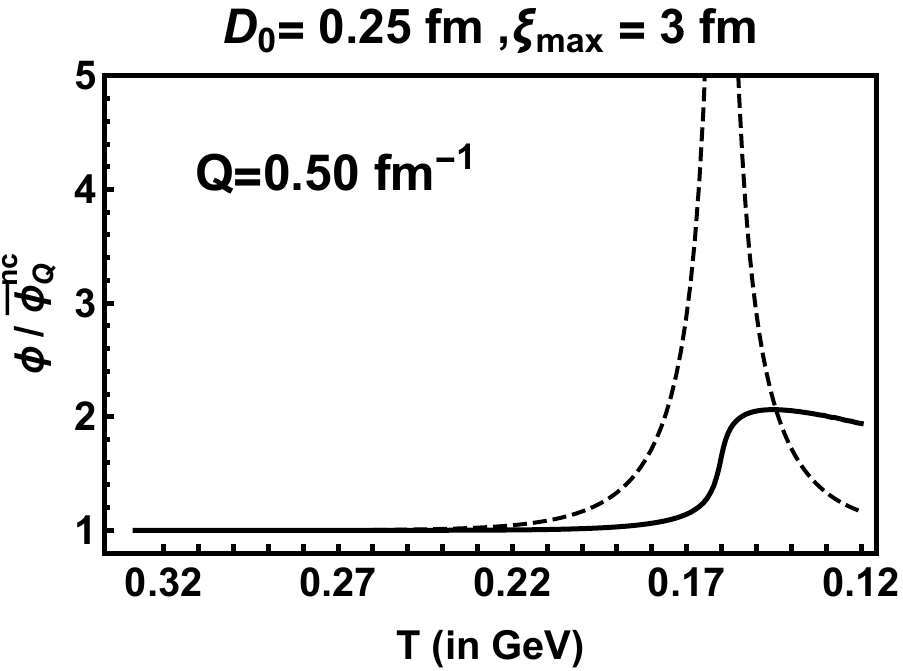}
\end{subfigure}
\begin{subfigure}{0.32\textwidth}
\includegraphics[scale=0.55]{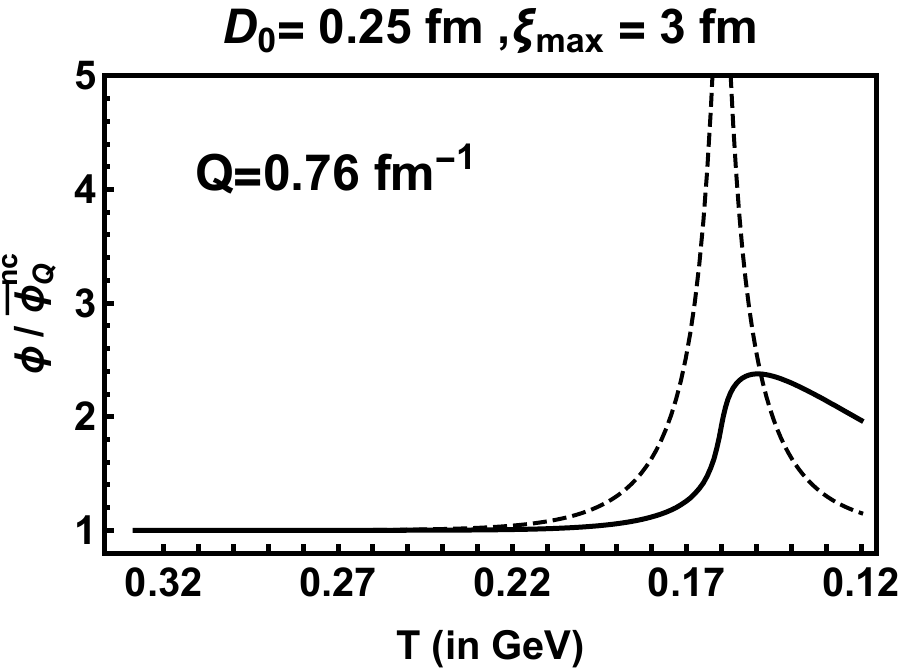}
\end{subfigure}
\begin{subfigure}{0.32\textwidth}
\includegraphics[scale=0.55]{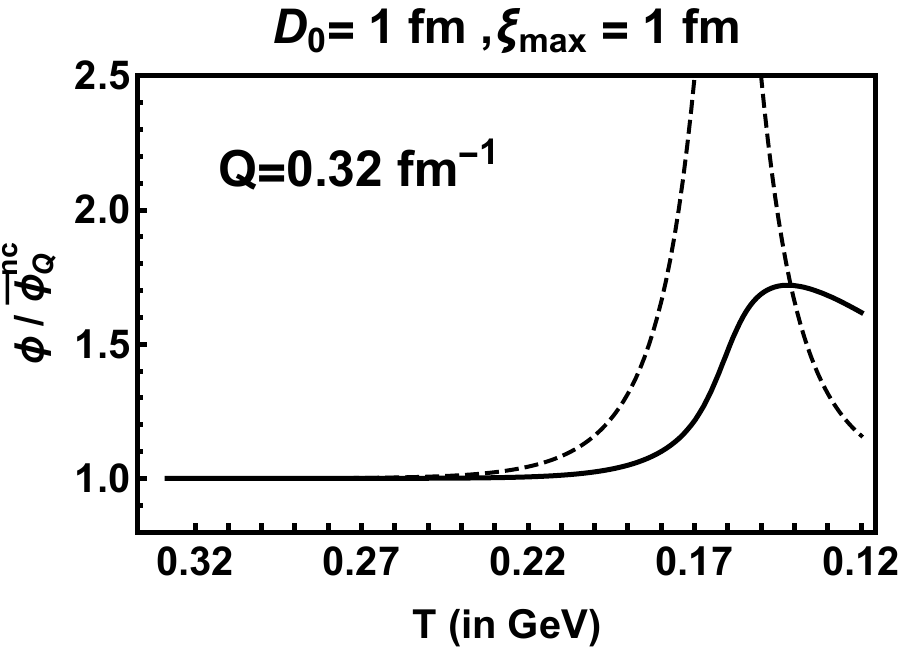}
\end{subfigure}
\begin{subfigure}{0.32\textwidth}
\includegraphics[scale=0.55]{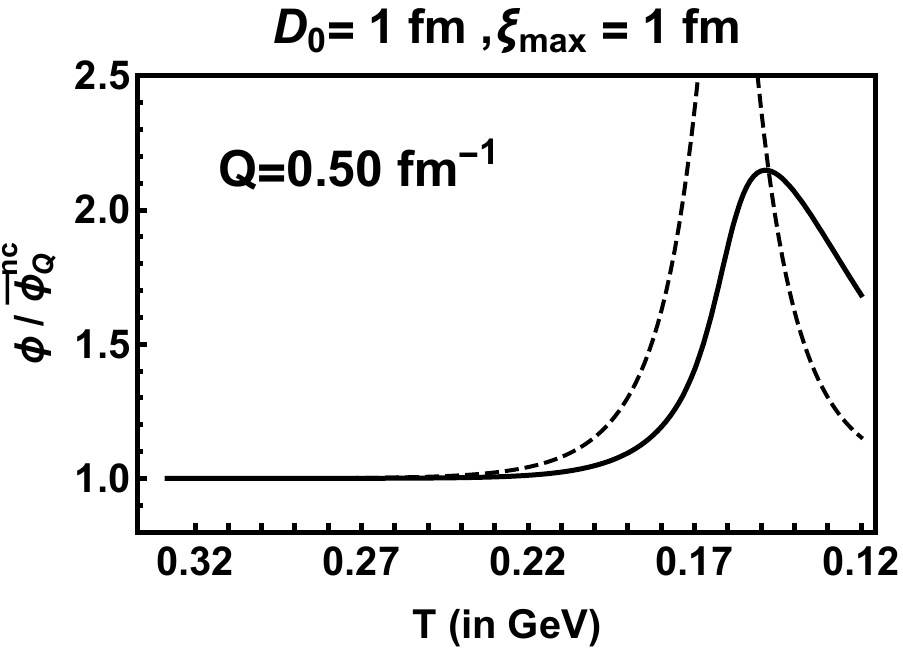}
\end{subfigure}
\begin{subfigure}{0.32\textwidth}
\includegraphics[scale=0.55]{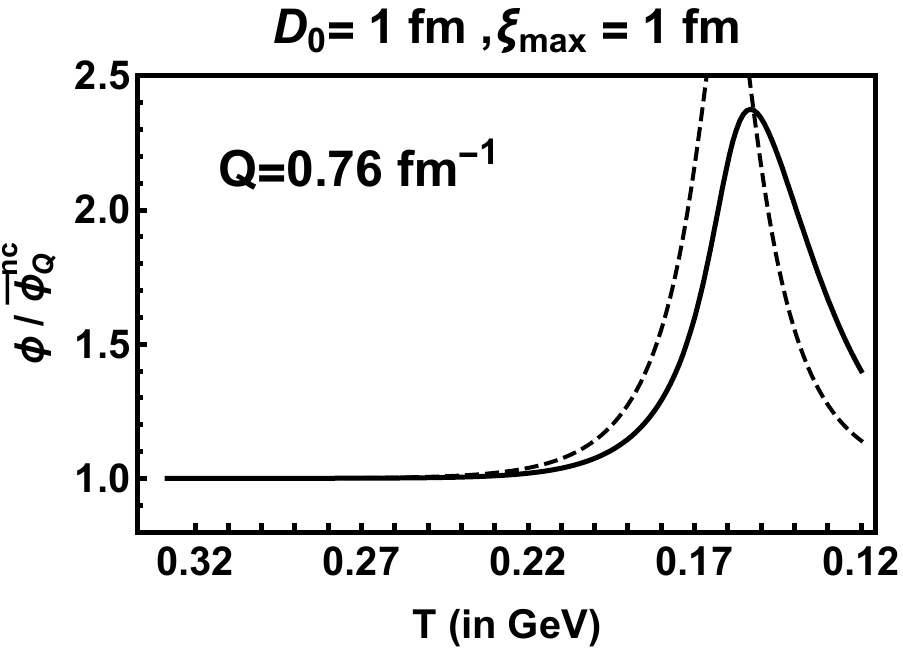}
\end{subfigure}
\begin{subfigure}{0.32\textwidth}
\includegraphics[scale=0.55]{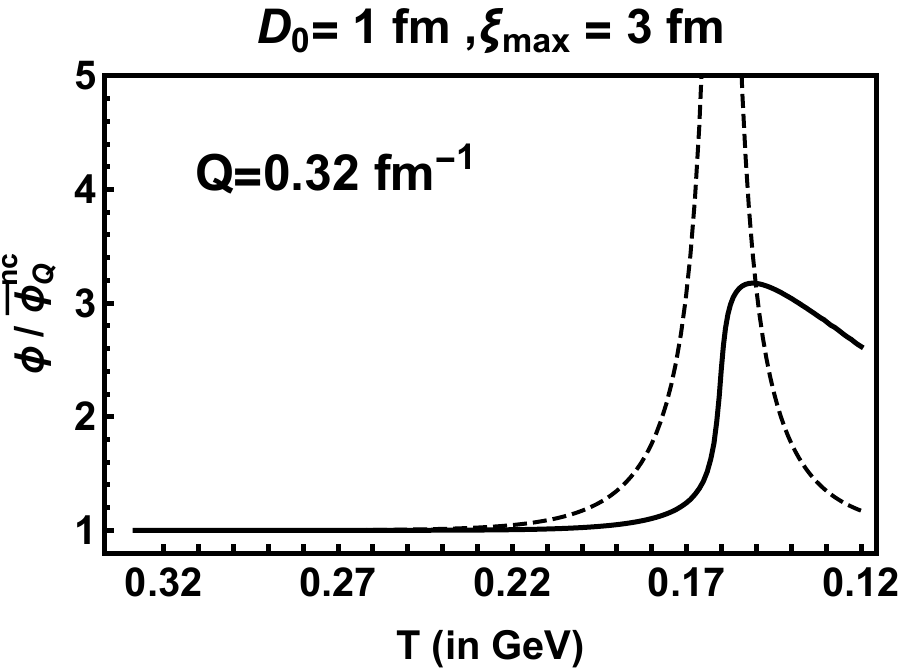}
\end{subfigure}
\begin{subfigure}{0.32\textwidth}
\includegraphics[scale=0.55]{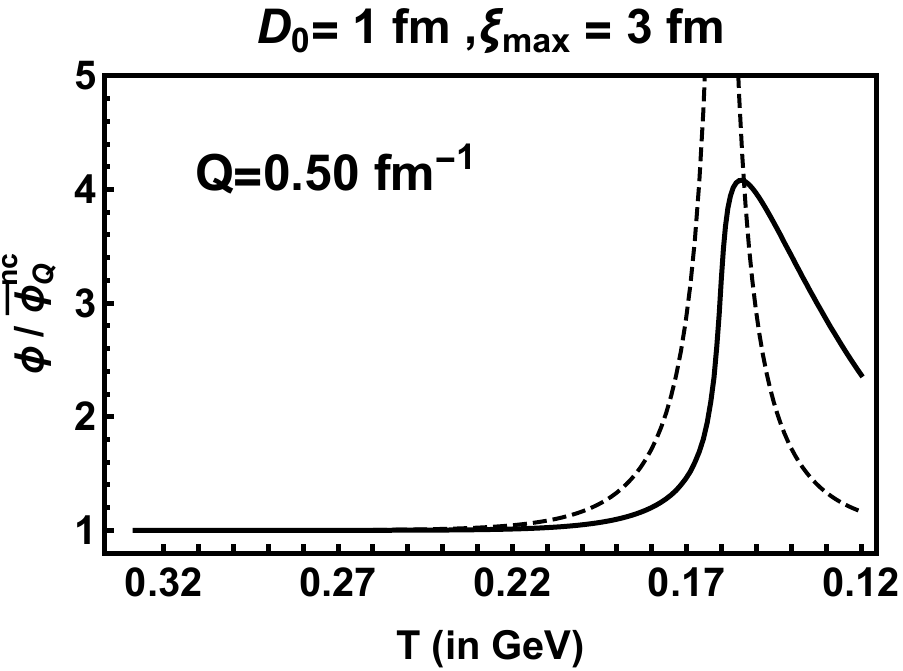}
\end{subfigure}
\begin{subfigure}{0.32\textwidth}
\includegraphics[scale=0.55]{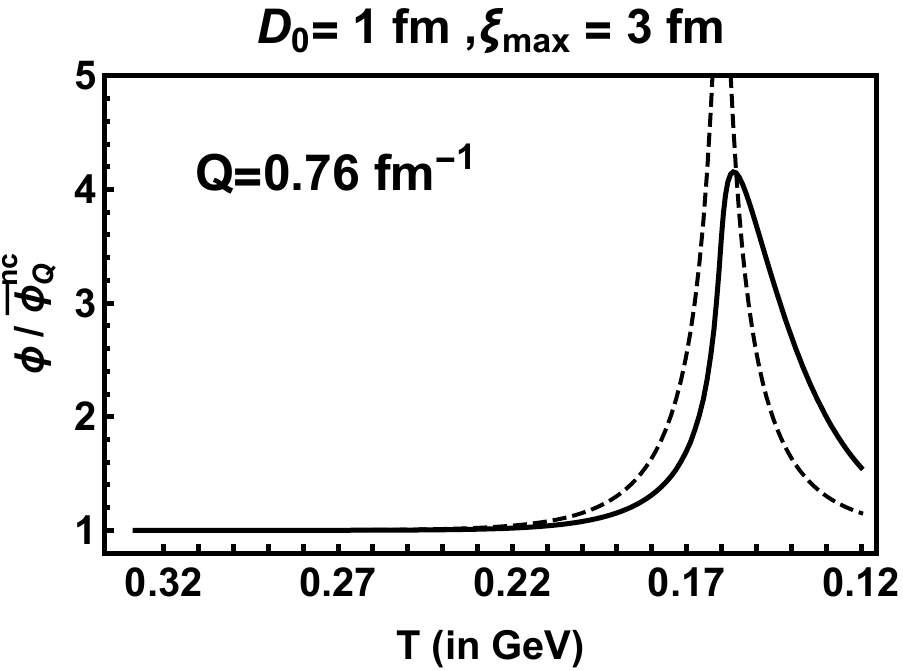}
\end{subfigure}
\end{center}
\caption{The values of $\phi_{\bm Q}$ (suitably normalized) for three
  representative values of $Q$ (same for each column), and for values
  $D_0$ (same in top and bottom six panels)
  and $\xi_{\rm max}$ (same in alternating rows) as in Fig.~\ref{phi-1ch}. The values of
  $\phi_{\bm Q}$ are taken along a fluid cell trajectory and
  plotted as a function of temperature, which is a monotonous function
  of time $\tau$ along the trajectory. The trajectory chosen for these
  plots begins at $r_i=r(\tau_i)=1.8\, \text{fm}$. The dashed and solid curves
  represent the equilibrium $\bar{\phi}_{\bm Q}$ and non-equilibrium
  $\phi_{\bm Q}$, respectively.}
\label{phi-ch2}
\end{figure}
    
In the simpler Bjorken scenario of Section~\ref{Sectbjk}, we described ``memory effects''
and looked at their dependence on $Q$ and the diffusion parameter $D_0$.
We can do the same here, for the $\phi_{\rm Q}$ obtained in this more realistic $r$-dependent
calculation, by displaying our results as in
Fig.~\ref{phi-ch2}, where suitably
normalized $\phi_{\bm Q}$ and $\bar{\phi}_{\bm Q}$ are plotted as a
function of the local temperature along a fluid cell trajectory for
three different values of $Q$. In accordance with
Eq.~(\ref{Bjkphiev}), the value of $\phi_{\bm Q}$ increases when
$\phi_{\bm Q}<\bar\phi_{\bm Q}$ and decreases when $\phi_{\bm
  Q}>\bar\phi_{\bm Q}$, as $\phi_{\bm Q}$ ``tries'' to relax toward the rising, and later
falling, equilibrium
value $\bar\phi_{\bm Q}$, as the
critical point is approached and later passed. 
For larger values of $\dip$ (as in the bottom half of Fig.~\ref{phi-ch2}) the rate of relaxation 
is greater, meaning that $\phi_{\bm Q}$ rises more rapidly, and therefore higher, in the critical region.
Although it also drops more rapidly as the temperature drops further, overall a larger $\dip$ yields larger fluctuations, at least within the reasonable range of values of $\dip$ that we explore.
For small $Q$ (see the left column in Fig.~\ref{phi-ch2}), the
value of $\phi_{\bm Q}$ grows very slowly, and reaches values much
lower that the equilibrium $\bar\phi_{\bm Q}$ before it starts
decreasing. However, for low $Q$, the rate at which $\phi_{\rm Q}$ decreases after 
the critical point has been passed
is also slow, and as a
result significant memory of the fluctuation magnitude near the
critical point (albeit itself smaller than equilibrium magnitude) is
retained at freezeout. This dynamics is qualitatively similar to the dynamics
first described in Ref.~\cite{Berdnikov:1999ph} in a very simplified
model of the out-of-equilibrium evolution of critical fluctuations 
with no spatial- or $Q$-dependence.

\subsection{Fluctuations on the freezeout surface}
\label{Subsect_freezeoutsurfacefluctuations}

As in the Bjorken scenario discussed in Section~\ref{Sectbjk}, we
consider two isothermal freeze-out scenarios with $T_f=140\, \text{MeV}$
and $T_f=156\, \text{MeV}$. The main difference here 
relative to Section~\ref{Sectbjk} is that the temperature is now not only a function of $\tau$
but also of the radial coordinate $r$. An isothermal surface,
therefore, is not simply $\tau={\rm const}$ for all $r$, as in the
previous section. The surface $T(\tau,r)=T_f$ can be
parametrized according to the discussion in Section~\ref{azsbifo}
and we use the notations and approximations discussed in that
section. For simplicity, we choose the parameter $\alpha$ introduced
in Eq.~(\ref{eq:tauralpha}) according to $\alpha=r$.

The magnitude of fluctuations at $T=156\, \text{MeV}$ 
in equilibrium is several times higher than that at $T_f=140\, \text{MeV}$, since
$T=156\, \text{MeV}$ is closer to the critical temperature $T_c=160$
MeV.  However, 
the time that the system spends in the critical region
before freezing out is shorter for $T_f=156\, \text{MeV}$ than for
$T_f=140\, \text{MeV}$. 
By comparing these two freeze-out scenarios, we can
understand the sensitivity of out-of-equilibrium fluctuations to the
proximity of the freeze-out temperature to the critical point.

\begin{figure}[t]
\begin{center}
\begin{subfigure}{0.49\textwidth}
\includegraphics[scale=0.8]{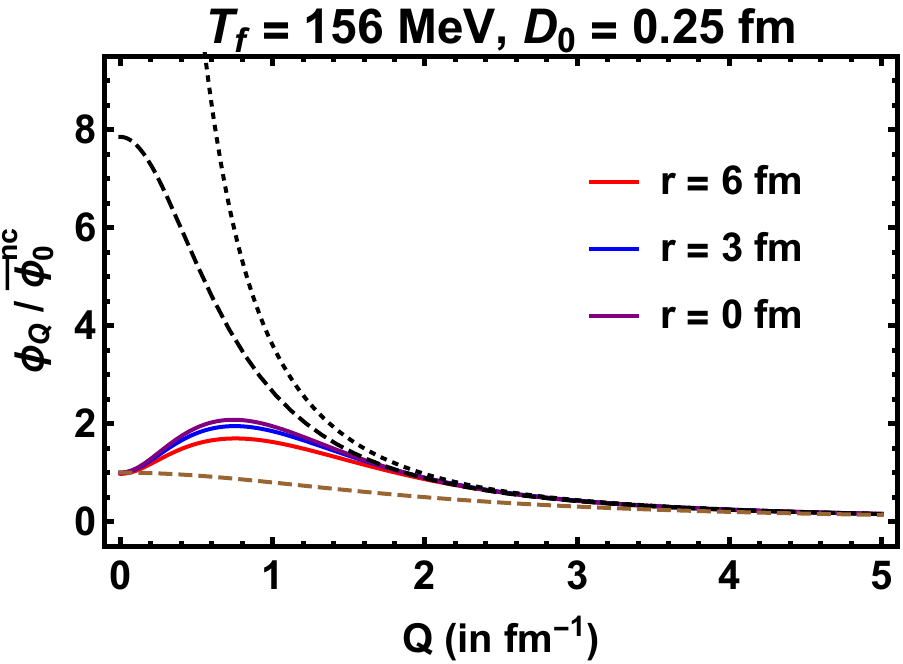}
\end{subfigure}
\begin{subfigure}{0.49\textwidth}
\includegraphics[scale=0.8]{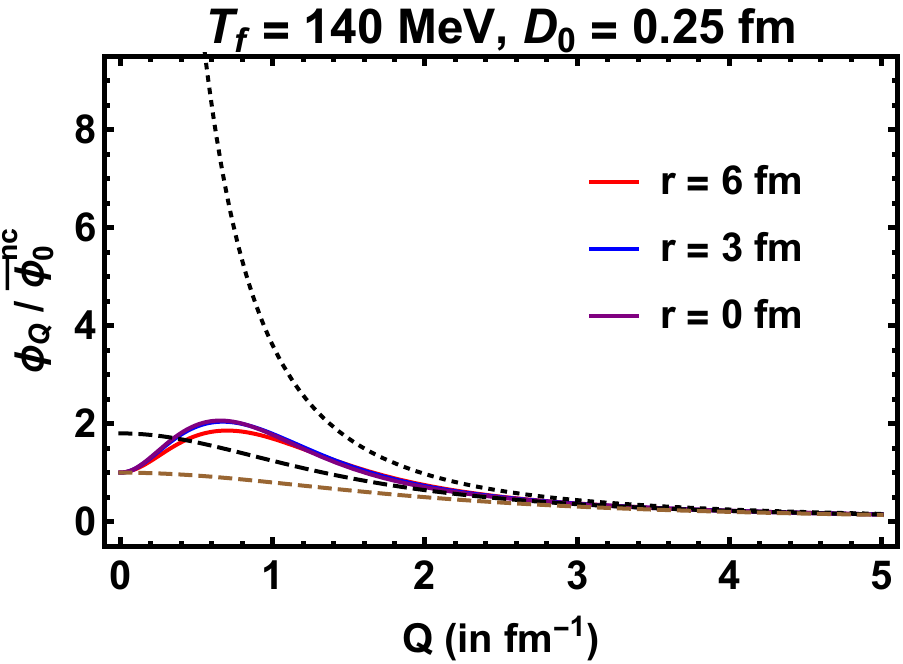}
\end{subfigure}
\begin{subfigure}{0.49\textwidth}
\includegraphics[scale=0.8]{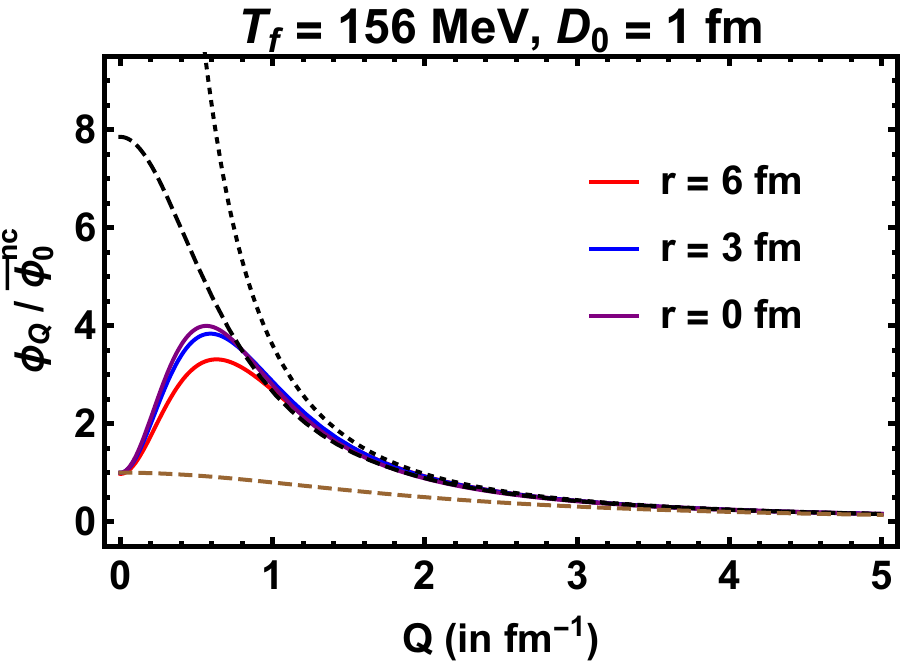}
\end{subfigure}
\begin{subfigure}{0.49\textwidth}
\includegraphics[scale=0.8]{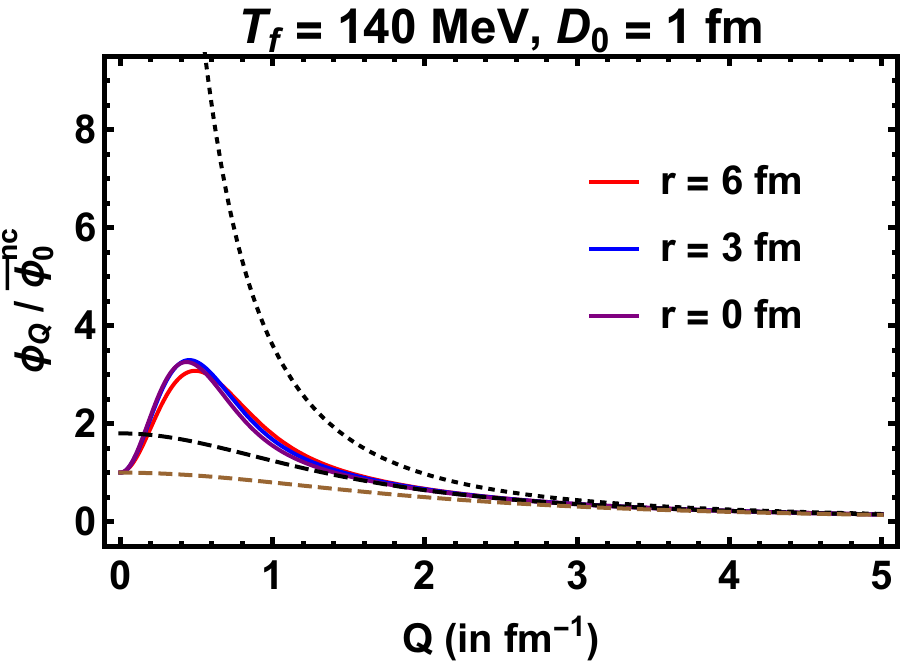}
\end{subfigure}
\end{center}
\caption{The Hydro+ variable $\phi_{\bm Q}$ (normalized to its value
  at $Q=0$ away from the critical point, where $\xi=\xi_0$) at
  freezeout evolved with two different diffusion parameters
  $D_0=0.25$~fm (upper panels) and $1$~fm (lower panels) and $\xi_{\text{max}}=3\, \text{fm}$. 
 The left (right) panels show results for evolution until the decreasing temperature has reached 
  a higher (lower) freeze-out temperature.
    The
  blue, red and purple curves show the values of $\phi_{\bm Q}$ at
  different points on the freezeout hypersurface, characterized by the radial
  coordinate $r$. The black dashed and dotted
  curves are the equilibrium curves at $T=T_f$ and
  $T=T_c$ respectively. The dashed brown curve is
  the (non-critical) equilibrium curves corresponding to
  $\xi=\xi_0$.}
\label{phi-1fhs1}
\end{figure}

In Fig.~\ref{phi-1fhs1}, suitably normalized plots of
$\phi_{\bm Q}$ are shown for three points on the freeze-out
hypersurface, characterized by radial coordinate $r=0, 3$ and $6$ fm, 
for two choices of freezeout temperature $T_f$ and two values
of the diffusion parameter $D_0$. These plots of
$\phi_{\bm Q}$ should be compared to the equilibrium
$\bar\phi_{\bm Q}$ at three characteristic points: at $T=T_c$, at $T=T_f$
and at a point far away from critical, where $\xi=\xi_0$, shown by the dashed and dotted curves.  The left and right
plots differ by the choice of the freezeout temperature,
$T_f=156$ MeV and $140$ MeV, respectively.  As expected, $Q=0$ modes
are ``stuck'' at their initial values and are not affected by the
critical point.  (To see how different this would be in the absence of conservation,
Fig.~\ref{phi-1fhs1} can be compared with the results obtained in the case of Model A dynamics in Fig.~\ref{phi-1fhs1A} in Appendix~\ref{RRWYModelA}.)
At moderate $Q$ the ``memory'' effect weakens and at
large $Q$ the modes closely track their equilibrium values, which
rises and then falls as the critical point is approached and then passed.
By comparing the plots in Fig.~\ref{phi-1fhs1} for
different $T_f$, we can also see that at smaller, but not too small, $Q$, the
``memory'' causes the fluctuation measure $\phi_{\bm Q}$ to be larger
than its equilibrium value $\bar\phi_{\bm Q}$. This effect is more
pronounced for lower $T_f$, due to the fact that the equilibrium
fluctuations are smaller there.

\begin{figure}[t]
\begin{center}
\begin{subfigure}{0.49\textwidth}
\includegraphics[scale=0.8]{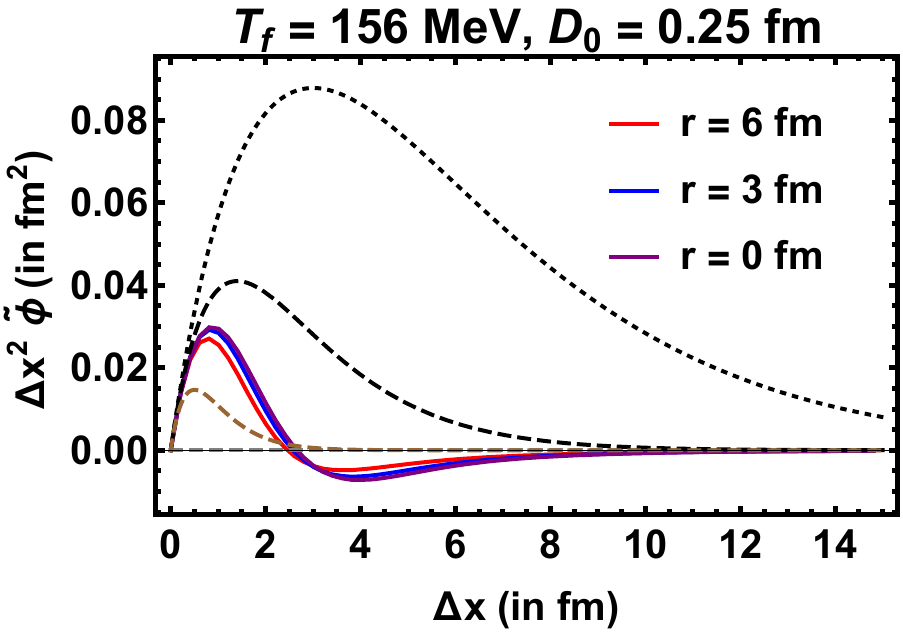}
\end{subfigure}
\begin{subfigure}{0.49\textwidth}
\includegraphics[scale=0.8]{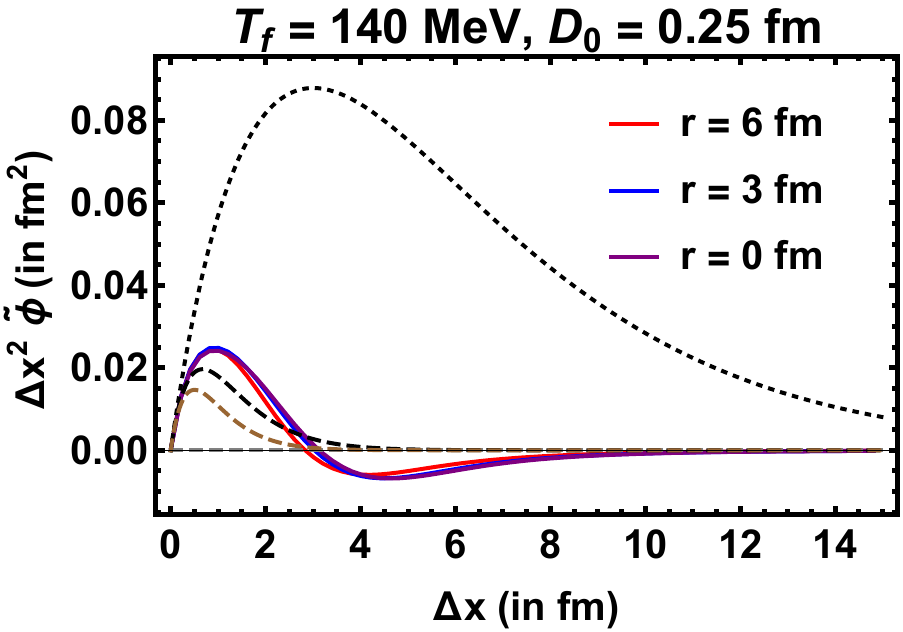}
\end{subfigure}
\begin{subfigure}{0.49\textwidth}
\includegraphics[scale=0.8]{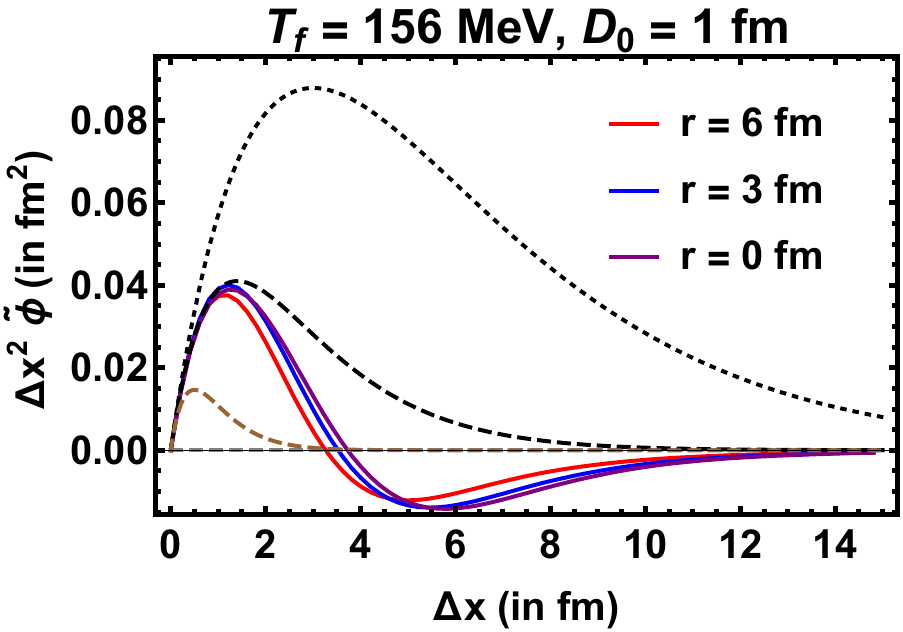}
\end{subfigure}
\begin{subfigure}{0.49\textwidth}
\includegraphics[scale=0.8]{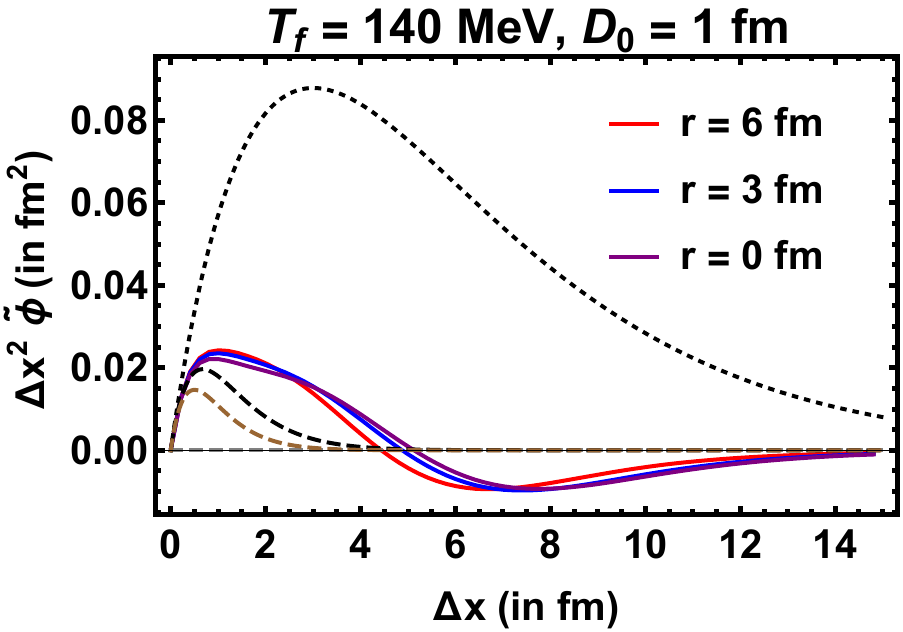}
\end{subfigure}
\end{center}
\caption{$\tilde{\phi}\times\Delta x^2$, the measure of fluctuations of $\hat{s}$ described by the correlator
$\langle \delta\hat s( x_+)  \delta\hat s(x_-)\rangle$,  at freezeout  as a function of the spatial separation between the points $\Delta x\equiv |\Delta {\bm x_\perp}|$.     In the calculations depicted in different panels, the $\phi_{\bm Q}$'s were evolved with two different $D_0$'s until freezeout at two different $T_f$'s, with the inverse Fourier transform to obtain $\tilde \phi ({\bm x_\perp)}$ performed at $T_f$.
In all panels, we have chosen a trajectory with $\xi_{\text{max}}=3\, \text{fm}$.    
The three $r$ values depicted via the colored curves 
correspond to three $r$ values on the freeze-out surface in the lab frame. The black dashed and dotted curves are the equilibrium curves at $T=T_f$ and $T=T_c$ respectively. The dashed brown curve is the (non-critical) equilibrium curve corresponding to $\xi=\xi_0$.}
\label{phiFT1}
\end{figure}

Having understood the effects of varying the parameter $D_0$ and the proximity to the critical point on the fluctuation measure $\phi_{\bm Q}$, as in Section~\ref{Sectbjk} the next step toward the calculation of observable particle multiplicity fluctuations is to compute 
$\tilde\phi(\Delta{\bm x_\perp})$, the inverse Fourier transform of $\phi_{\bm Q}$ defined in Eq.~(\ref{eq:phitilde}). In Fig.~\ref{phiFT1},
we plot $\Delta x^2 \tilde\phi(\Delta{\bm x_\perp})$ as a function of the spatial separation
$\Delta x\equiv |\Delta {\bm x_\perp}|$ between the two points in the correlator $\langle \delta \hat s(x_+)  \delta \hat s(x_-)\rangle$, see Eq.~(\ref{eq:<ss>=phiQ}).
By comparing to Fig.~\ref{phiQdelnaGH}(b), we see that the $D_0$-dependence is qualitatively
similar to that in the Bjorken scenario, discussed at length in Section~\ref{Sectbjk}.  The small 
$\Delta x$ (large $Q$) behavior of the fluctuations is not affected by changing $D_0$, while
at the same time the spatial correlator becomes longer ranged as $D_0$ is increased.
The correlator goes negative at larger values of $\Delta x$; this is a consequence of
conservation as can be seen by comparing Fig.~\ref{phiFT1} to Fig.~\ref{phiFT1A} in Appendix~\ref{RRWYModelA} and 
as explained in the context of the Bjorken scenario in Eq.~(\ref{eq:0>Q}). 
Finally, consistent with what we have already seen in Fig.~\ref{phi-1fhs1}, 
with either value of $D_0$ memory effects are strong enough that 
the magnitudes of the fluctuations are not much smaller if the freezeout temperature is $T_f=140$~MeV (well below
the critical point) as compared to their magnitudes if  $T_f=156$~MeV (very close to $T_c=160$~MeV) despite the fact that the equilibrium fluctuations at these two temperatures 
differ substantially.

\subsection{Variance of particle multiplicities}
\label{freeze2ptRRWY1}

Because of the greater symmetry in the simpler Bjorken scenario that we were employing there, in Section~\ref{Sectbjk} we were able to compute the rapidity correlator $C(\Delta y)$ 
of the multiplicity fluctuations, before integrating over a rapidity window to
obtain the variance of the particle multiplicity.
Here, with nontrivial radial dependence and radial flow we shall instead obtain
the variance of the multiplicity of particles of species $A$ 
	directly from $\tilde \phi$ by employing the more general expressions in Eqs.~(\ref{delNA2RRWYagain3}) and (\ref{eq:J-Bjorken}).  	
As we did in Section~\ref{Sectbjk}, we shall compute $\omega_A$, the ratio of the variance
of the multiplicity of species $A$ to its mean, see Eq.~(\ref{omegaAdef1}).
We can obtain the
        mean multiplicity of protons and pions for a rapidity
        acceptance window  $[-y_{\text{max}},y_{\text{max}}]$ and acceptance cuts in $p_T$ using the Cooper-Frye formula
\begin{eqnarray}\label{meanRRWY}
\<N_A\>=d_A \int dS_\mu\, \int_{-y_\text{max}}^{y_{\text{max}}} \frac{dy}{2\pi}\, \int_{p_\text{T,min}}^{p_{T,\text{max}}} \frac{p_T\,dp_T}{2\pi}\, \int_{0}^{2\pi}\frac{d\phi}{2\pi}\,e^{-\frac{p\cdot u}{T_f}}\, p^\mu \,,
\end{eqnarray}
employing
the flow velocity $u(x)$ profile and freeze-out hypersurface for
        the simulation from Ref.~\cite{Rajagopal:2019xwg} illustrated in Fig.~\ref{temperaturedensity}. 
The fluctuation measure $\omega_A$ is then obtained by taking the ratio of
Eq.~(\ref{delNA2RRWYagain3}) and Eq.~(\ref{meanRRWY}). 
In this
Subsection, we present the results for the normalized fluctuation measure $\tilde{\omega}_A$ as defined in
Eq.~(\ref{omegaAdef2}) as a function of $\xi_{\text{max}}$ (which is to say as a function of 
how close the trajectory in the phase diagram is to the critical point) for protons
and pions obtained with our two choices of $\dip$ and with our isothermal freezeout scenario with two different choices of
the freezeout temperature $T_f$, as discussed above. In all calculations, we choose
the acceptance cuts
$p_T\,\epsilon\,(0.4\, \text{GeV},\, 2\, \text{GeV})$ and
$y_{\text{max}}=1$. As already discussed above, $\tilde{\omega}_A$,
should not depend on the acceptance. This is explicitly the case in
equilibrium and we have verified that this remains approximately the case 
in all of our simulations.

\begin{figure}[t]
\begin{center}
\begin{subfigure}{0.49\textwidth}
\includegraphics[scale=0.6]{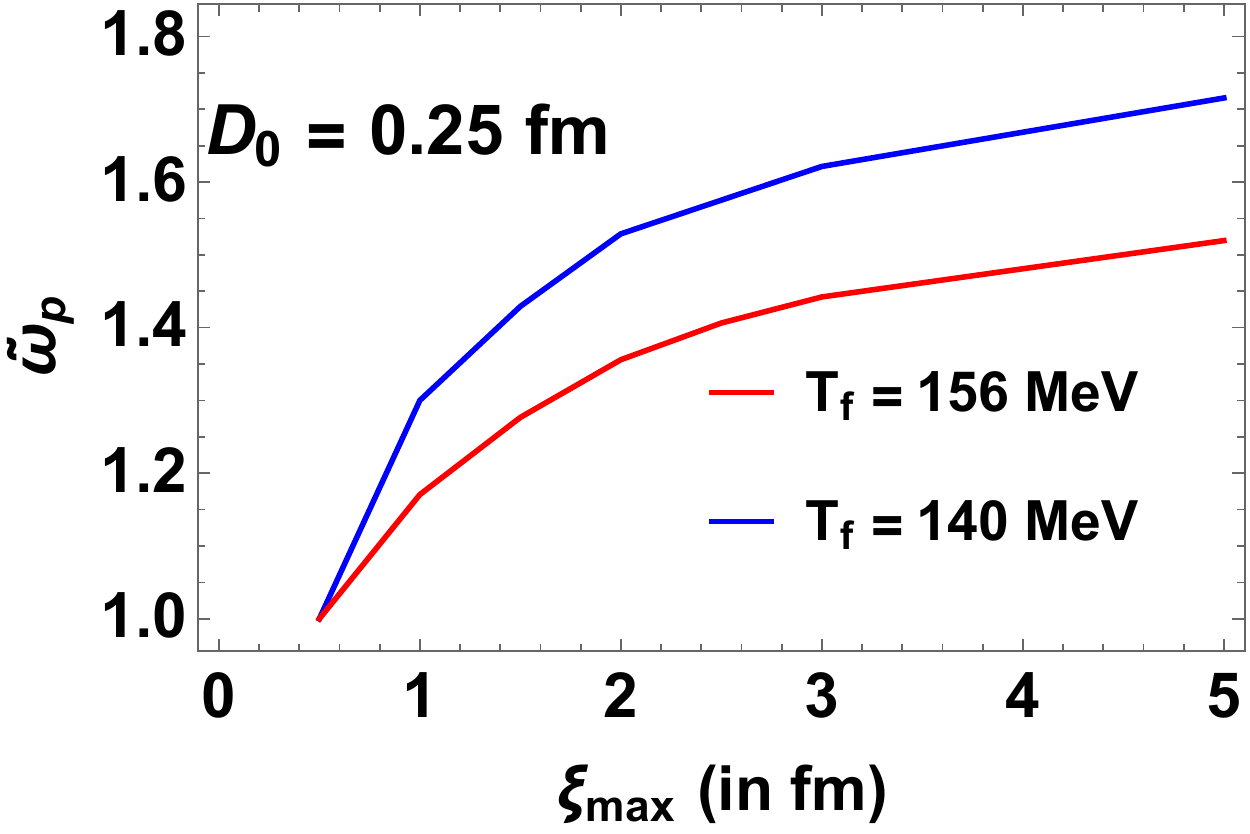}
\subcaption{}
\end{subfigure}
\begin{subfigure}{0.49\textwidth}
\includegraphics[scale=0.6]{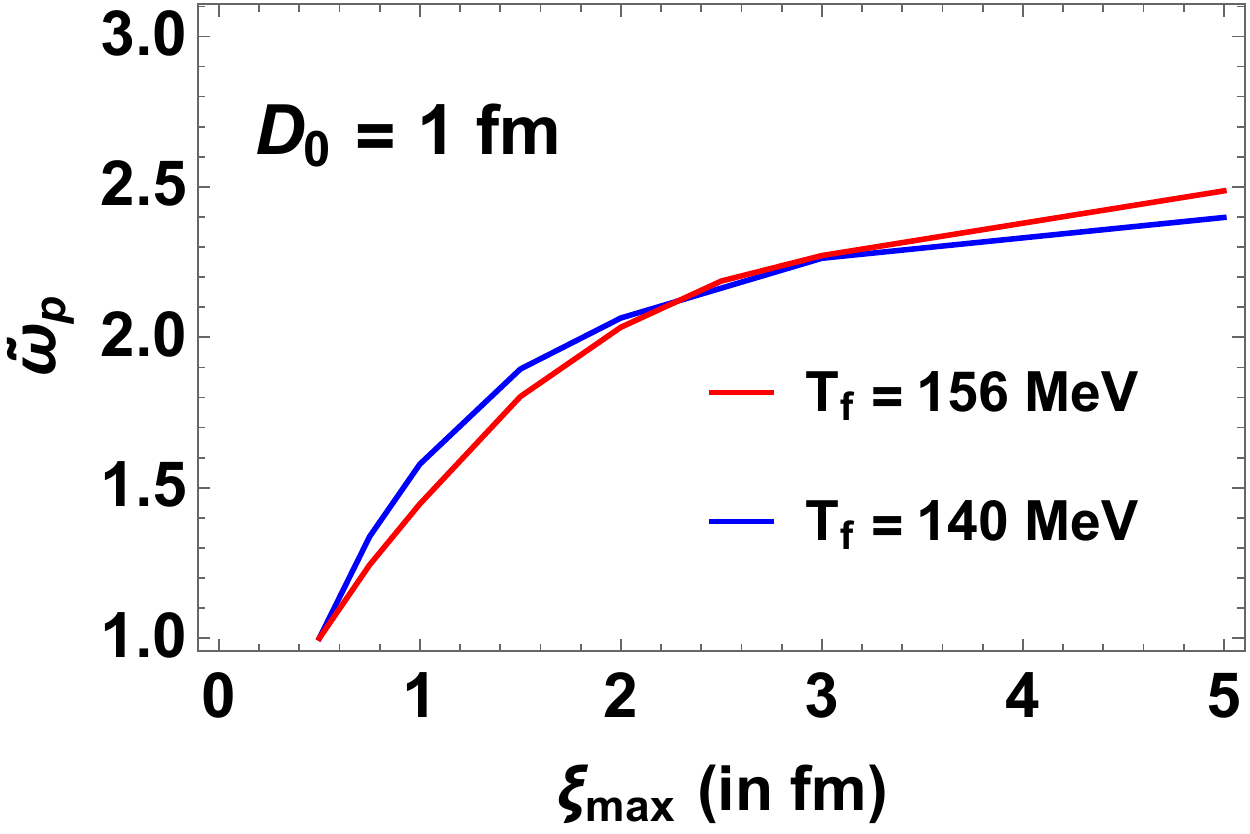}
\subcaption{}
\end{subfigure}
\begin{subfigure}{0.49\textwidth}
\includegraphics[scale=0.6]{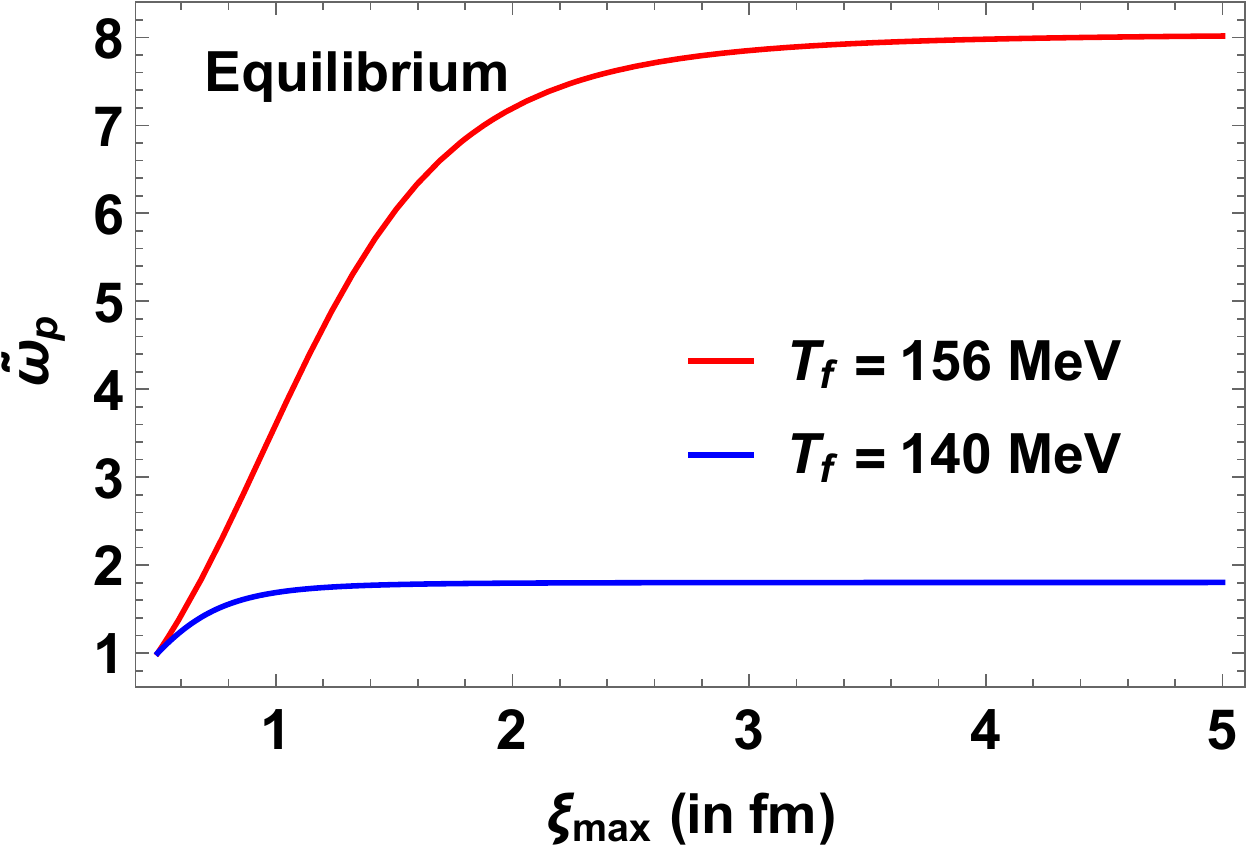}
\subcaption{}
\end{subfigure}
\end{center}
\caption{Normalized measure of the fluctuations in proton multiplicity, $\tilde{\omega}_p=\frac{\omega_p}{\omega^{\text{nc}}_p}$, as a function of the maximum equilibrium correlation length along the system trajectory, which is to say as a function of how closely the trajectory passes the critical point. As $\dip\rightarrow\infty$, the $\tilde{\omega}_p$'s approach their equilibrium values shown in panel (c).}
\label{fhs-protons}
\end{figure}

\begin{figure}[t]
\begin{center}
\begin{subfigure}{0.49\textwidth}
\includegraphics[scale=0.6]{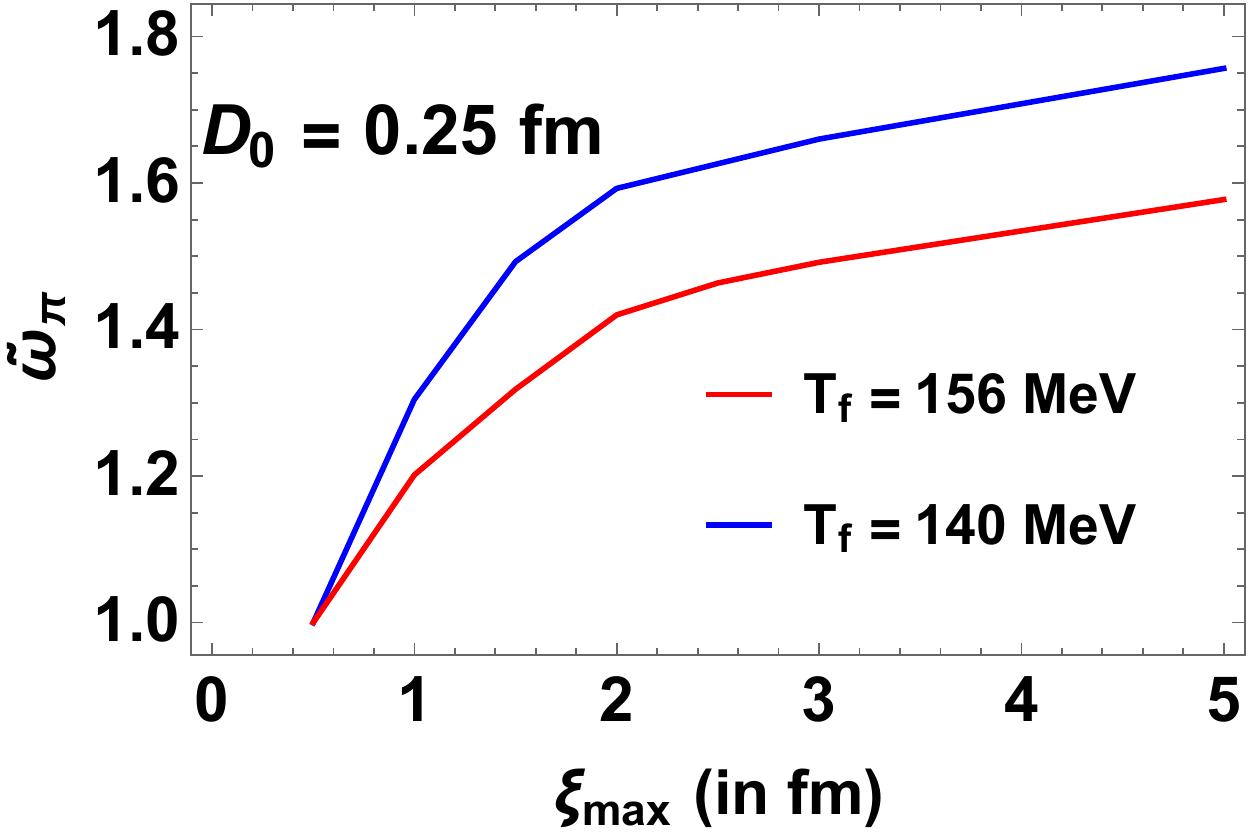}
\subcaption{}
\end{subfigure}
\begin{subfigure}{0.49\textwidth}
\includegraphics[scale=0.6]{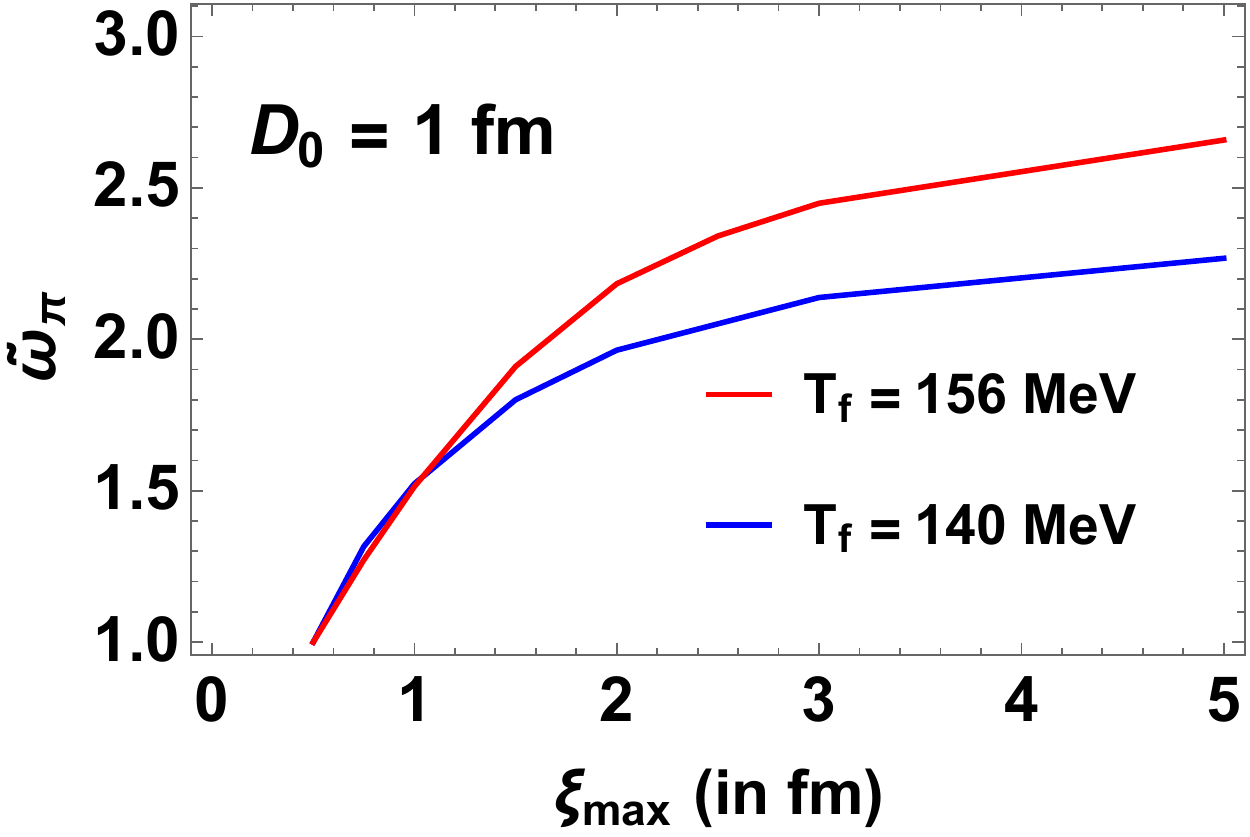}
\subcaption{}
\end{subfigure}
\begin{subfigure}{0.49\textwidth}
\includegraphics[scale=0.6]{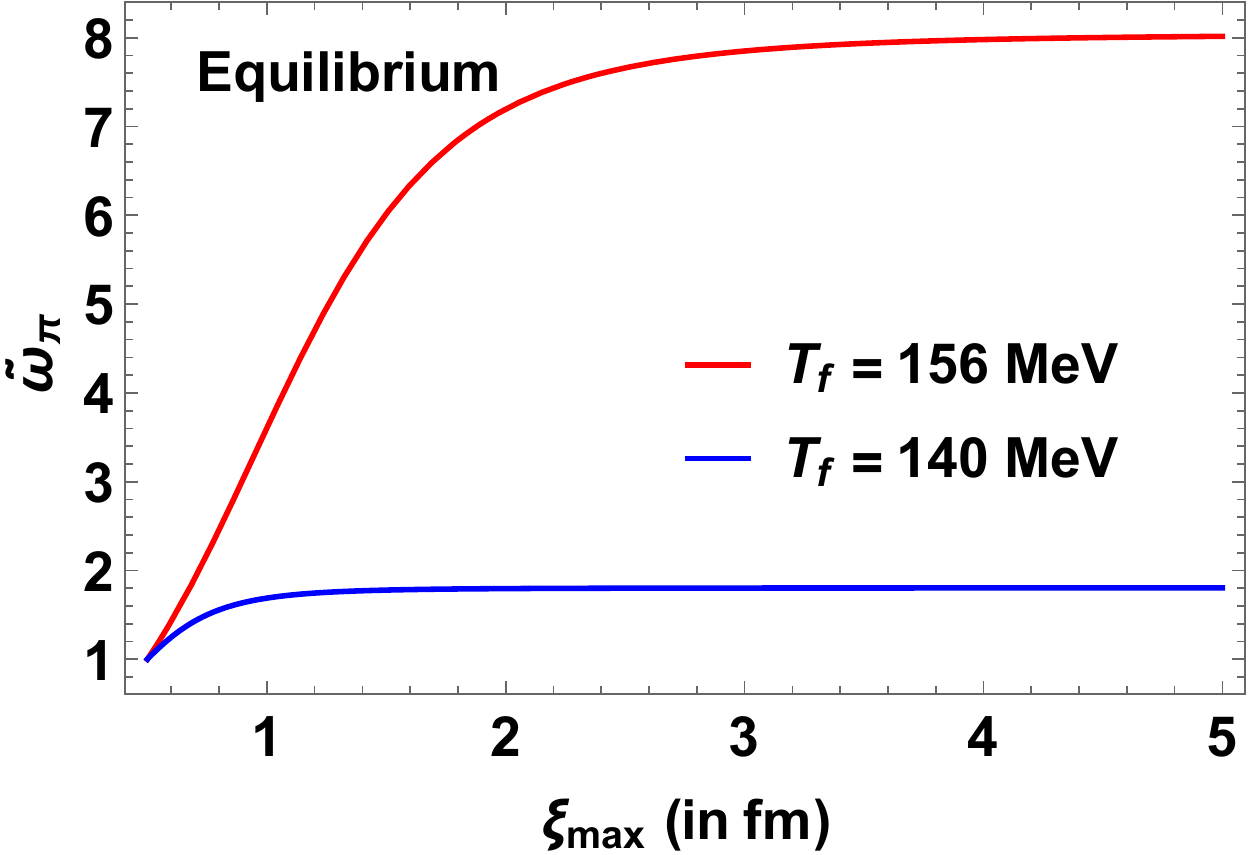}
\subcaption{}
\end{subfigure}
\end{center}
\caption{Normalized measure of the fluctuations in pion multiplicity, 
$\tilde{\omega}_\pi=\frac{\omega_\pi}{\omega^{\text{nc}}_\pi}$, as a function of the maximum equilibrium correlation length along the system trajectory, which is to say as a function of how closely the trajectory passes the critical point. As $\dip\rightarrow\infty$, the $\tilde{\omega}_\pi$'s approach their equilibrium values shown in panel (c). The definition of the normalized measure of fluctuations $\tilde \omega$ is such that it is species-independent in equilibrium, meaning that panel (c) here is identical to panel (c) in Fig.~\ref{fhs-protons}.}
\label{fhs-pions}
\end{figure}

We present our final results for the normalized fluctuation measure for protons and pions, 
$\tilde{\omega}_p$ and $\tilde{\omega}_\pi$, in Figs.~\ref{fhs-protons} and~\ref{fhs-pions}. These results demonstrate that for
trajectories passing closer to the critical point (i.e., for
trajectories with larger $\xi_{\rm max}$) the magnitude of
fluctuations is larger, as we have already seen in
Fig.~\ref{fhs-protonsHBjk} for the Bjorken expansion scenario. Again
as in the Bjorken scenario, the magnitude of the effect depends on the
rate of the diffusive relaxation of the fluctuations controlled by
parameter $D_0$. The conservation laws (i.e., ``memory'') lead to
significant suppression of the magnitude of fluctuations compared to
the prediction based on the assumption that fluctuations have enough
time to equilibrate. (The equilibrium predictions are shown in
Figs.~\ref{fhs-protons}(c) and~\ref{fhs-pions}(c).)  The smaller the value
of the diffusion parameter $D_0$, i.e., the slower is the diffusion,
the stronger is the suppression.

It is also interesting to compare the magnitude of fluctuations for
the same value $D_0$ and $\xi_{\rm max}$, but for two different
freezeout temperatures. Naturally, the equilibrium assumption leads to a
prediction of larger fluctuations at higher freezeout temperature,
since the higher temperature is closer to the critical point. However,
nonequilibrium effects not only suppress the magnitude of the
fluctuations relative to the equilibrium prediction but also
introduce memory effects that substantially reduce 
the decrease in the magnitude of the fluctuations
that occurs between $T=156$~MeV and the lower freezeout temperature $T_f=140$~MeV.
This makes the fluctuations obtained upon assuming a freezeout temperature of $T_f=140$~MeV 
much more similar in magnitude to those that would be obtained if $T_f=156$~MeV than is the
case in equilibrium.
So much so, in fact, that,
depending on the choice of the parameters such as $D_0$,
it is even possible to find larger fluctuations at $T_f=140$~MeV than would have been obtained at the higher
freezeout temperature $T_f=156$~MeV, see Figs.~\ref{fhs-protons}(a) and~\ref{fhs-pions}(a)!
This
effect arises because, as we have already discussed, the
fluctuations continue to grow even after the system has passed the
critical point, as long as their value is below the equilibrium, as
can be seen in Fig.~\ref{phi-ch2}. This effect is not seen in the
Bjorken scenario in Section~\ref{Sectbjk} because the drop of the temperature during the
one-dimensional longitudinal expansion there is much slower than in our calculations in this
Section which include radial expansion in addition.
As a result, in the Bjorken scenario the freezeout temperatures
are reached at somewhat later times, after the non-equilibrium
fluctuations have began to decrease. For the same reason, in our more realistic calculations in this Section this effect
disappears for larger values of $D_0$ which yields faster relaxation of
fluctuations toward equilibrium, see Figs.~\ref{fhs-protons}(b) and~\ref{fhs-pions}(b).

\section{Summary and Outlook}
\label{SummaryOutlook}


In this work, we have introduced a novel approach which connects
hydrodynamic fluctuations, which fall out of equilibrium during the
hydrodynamic stage of the expansion and cooling of the droplets of QGP produced
in heavy ion collisions, to subsequent fluctuations of
particle multiplicities, observable in experiments. In this approach,
we treat the hydrodynamic fluctuations in the Hydro+ formalism
\cite{Stephanov:2017ghc}, focusing on the critical, which is to say the most
singular as well as the slowest --- and thus the most out-of-equilibrium --- 
modes of fluctuations near the critical point. On the kinetic, or particle, side 
of the freezeout, such fluctuations are matched by
introducing a new critical scalar field $\sigma$ which couples to observable
particles. One can understand this field as an effective critical
field -- a collective phenomenon in the hadron gas -- a precursor of
the critical point.  It has the same quantum numbers as the $\sigma$ meson, meaning that its couplings to other hadrons, including protons, are such that fluctuations in $\sigma$ correspond to fluctuations in the masses of other hadrons, including protons.
In prior estimates of the observable consequences of critical
fluctuations, the $\sigma$ field and in particular its fluctuations were assumed to stay in equilibrium near the critical point for
lack, at the time, of an approach to describing its non-equilibrium evolution. In
this paper, we connect the fluctuations of this field, and consequent observable fluctuations in 
particle multiplicities, to the Hydro+
variable $\phi_{\bm Q}$, which describes the non-equiibrium evolution
of fluctuations in the earlier hydrodynamic stage of the collision.

Our approach generalizes the well-known and well-tested Cooper-Frye
freezeout~\cite{Cooper:1974mv}, which translates hydrodynamic degrees of freedom (but not their fluctuations) into
particle distributions. The Cooper-Frye procedure only specifies
event-averaged single-particle quantities (multiplicities and spectra of each hadron species)
and as such is not
sufficient to describe the freezeout of {\em fluctuations} or {\em
  correlations}. Our more general freezeout procedure allows us to perform such a translation, or
matching, of Hydro+ (hydrodynamic and fluctuation) degrees of freedom
to particle multiplicities and their fluctuations.

We demonstrate our generalized freezeout procedure in practice by freezing out a
simplified case (the Hydro+ description of boost-invariant Bjorken expansion with no transverse expansion) 
where the calculations can largely be pushed through analytically and that yields valuable intuition as
well as  a numerical Hydro+ simulation of a more realistic scenario with boost invariance and azimuthal symmetry that incorporates transverse radial
expansion, similar to the one considered in Ref.~\cite{Rajagopal:2019xwg}.
%
%
%
%
In both examples, we observed a significant suppression of the
out-of-equilibrium fluctuations relative to what their values would have been in equilibrium for
reasonable values of the parameters we considered. In addition, we also
noted that while the {\em equilibrium} fluctuations sensitively depend on
how soon, i.e., how far below the critical temperature, the freezeout
occurs, this sensitivity is almost
elliminated by non-equilibrium (``memory'')  effects, see
Figs.~\ref{fhs-protonsHBjk}, \ref{fhs-protons} and \ref{fhs-pions}.

The Hydro+ variable $\phi_{\bm Q}$ describes the magnitude of the fluctuations at
different wavevectors $\bm Q$. We observe (and, in the case of the Bjorken
scenario, can describe analytically as in Eq.~(\ref{eq:Qtauf}) )
that the characteristic value of the wavenumber $Q$ whose  $\phi_{\bm Q}$'s 
control
the magnitude of the multiplicities after freezeout is determined by
multiple factors including the scale of the inhomogeneity of the expanding fluid (the
Bjorken time $\tau$ in the Bjorken scenario), the radial flow, the typical thermal velocity spread of the
produced particles, as well as the acceptance window in rapidity 
if this
acceptance window is larger than the typical thermal spread in rapidity. 
This
characteristic $Q$ is 
small compared to microscopic scales as it is
typically of order $1/\tau_f$ (see Eq.~(\ref{eq:Qtauf})). Since the fluctuations at
small $Q$ are suppressed by conservation laws (another aspect of the out-of-equilibrium
dynamics that is in a sense also a  ``memory'' effect),
the smallness of the characteristic $Q$ relevant for the freezeout
contributes to the suppression of the fluctuations relative to the
equilibrium values.


Our study focused on Gaussian measures of fluctuations. The higher,
non-Gaussian, cumulants are more sensitive to the proximity of the
critical point~\cite{Stephanov:2008qz,Athanasiou:2010kw}. It is, therefore, important to
generalize our freezeout procedure to higher-order cumulants. This
can be done using Eq.~(\ref{kthcumulant}) and we leave implementation
and the analysis of the results to future work. 
While we demonstrated the application of our procedure to the
variance of particle multiplicities, it will also be straightforward to
generalize to the cross-correlation of different particle species, as
was done in Ref.~\cite{Athanasiou:2010kw} for equilibrium
fluctuations. We expect the conclusion from that work that cumulants involving protons are most sensitive to critical fluctuations to persist, but we leave
an investigation of how best to combine measurements of different (cross-)correlations
so as to optimize the sensitivity to critical fluctuations while reducing dilution of their effects
by backgrounds to future work.

We have focused on the {\em dependence} of the observable fluctuations on
the proximity of the critical point, either by varying
$\xi_{\rm max}$, which corresponds to varying freezeout $\mu_B$ via
changing collision energy $\sqrt s$ (see Fig.~\ref{fig:trajectory}),
or by varying the freezeout temperature (for the same trajectory). We
also studied the dependence on the (thus far unknown) value of the diffusion
parameter $D_0$. In order to illustrate these dependences, we chose to present our
results using normalized variables which did not depend on
the {\em absolute} magnitude of the effect. In order to predict the
absolute magnitude one would have to know the equation of state, i.e.,
the parameter $Z$ in Eq.~(\ref{eq:<s.s>=<sigma.sigma>}) which
determines the absolute strength of the singularity. Also, one would
have to determine the couplings $g_A$ of hadrons to the critical
collective field $\sigma$. These couplings are, of course, also
related to the equation of state and it would be interesting to make
this relation more explicit. We leave this to future work. For the purpose of
using our results to make crude estimates, one could take $Z\sim 1/T_c^2$ and follow Ref.~\cite{Athanasiou:2010kw}
and choose $g_p\sim 7$ and $g_\pi\sim 2$.

It would be also interesting to consider going beyond the leading
critical behavior of fluctuations to take into account less singular critical
contributions and modes which are not critical, including fluctuations
of pressure and flow velocity. Extending our work in this way 
would be necessary in order to attempt to develop a 
fluctuation freezeout procedure  
that could be applied away from the
critical point as well as near it, as in our paper.

It is also important to realize that we considered fluctuations and
freezeout on the
crossover side of the critical point (see
Fig.~\ref{fig:trajectory}). It would also be interesting and important to
understand what happens on the other side, where the first-order phase
transition occurs. The challenge in this domain begins already at the level of
hydrodynamics and is beyond the scope of this paper.
	
Although, as we have detailed in this Section, there is a scope for improvement and
generalization of the freezeout procedure that we have introduced and explored in this paper, 
we believe
that the procedure can already be
integrated into the full numerical simulation of heavy-ion collisions
relevant for the beam energy scan (BES) program aimed at the search for the
QCD critical point. With first results from high-statistics BES data taken at RHIC in 2019-2021
anticipated soon, this represents a high priority next step.


\begin{acknowledgments}
We gratefully acknowledge the contributions of Ryan Weller, who collaborated with us on this research project in its early stages.  We acknowledge helpful conversations with Travis Dore, Lipei Du, Jamie Karthein, Gregory Ridgway and Chun Shen. This work was supported by the U.S.~Department of Energy, Office of Science, Office of Nuclear Physics, within the framework of the Beam Energy Scan Theory (BEST) Topical Collaboration and grants DE-SC0011090 and DE-FG0201ER41195.  Y.Y.~acknowledges support from the Strategic Priority Research Program of the Chinese Academy of Sciences, Grant No.~XDB34000000.
\end{acknowledgments}

\appendix
\section{Equation of state used in the hydrodynamical evolution}
\label{appendeos}

The equation of state that we have used in the analytical calculations of Section~\ref{Sectbjk}, done within a Bjorken scenario, as well as in the numerical  Hydro+ simulations for a semi-realistic scenario done in Section~\ref{RRWY}, is taken from Ref.~\cite{Rajagopal:2019xwg}. We describe this equation of state briefly in this Appendix. An aspect of the physics that the Hydro+ formalism is well-suited to describe is the way in which the out-of-equilibrium fluctuations of the slow modes modify the equation of state~\cite{Stephanov:2017ghc}. However, it has been observed in Refs.~\cite{Rajagopal:2019xwg} 
and \cite{Du:2020bxp} that these backreaction effects are smaller than $1\,\%$ in most cases. For this reason, throughout the present work we neglect the feedback of Hydro+ modes on the equation of state. That is, in the notation of Refs.~\cite{Stephanov:2017ghc,Rajagopal:2019xwg,Du:2020bxp} we approximate the Hydro+ equation of state $p_{+}(\varepsilon)$ by the standard pressure $p(\varepsilon)$ given by 
\begin{eqnarray}
\label{press1}
p=\frac{s}{\beta}-\varepsilon
\end{eqnarray}
where $\varepsilon$ is the energy density, $\beta$ is the inverse temperature, and where the entropy density $s$ is given as a function of the local temperature by
	\begin{eqnarray}\label{SecRRWYdetails}
	s(T)=\int_0^{T}\, dT^{'}\, \frac{c_V(T^{'})}{T^{'}}
	\end{eqnarray}
with $c_V(T)$ being the specific heat capacity at fixed volume.  In Ref.~\cite{Rajagopal:2019xwg}, the equation of state is specified close to and away from a critical temperature $T_c$ by choosing
\begin{eqnarray}
c_V(T) =
    \begin{cases}\label{cvnoncrit}
      c_V^{\text{no C.P}}(T) & T\leq T_L\, \text{or}\, T\geq T_H\\ \label{cvcrit}
      c_V^{\text{crit}}(T)+\sum_{n=0}^{5} c_n \left(\frac{T-T_c}{\Delta T}\right)^n & T_L < T < T_H
    \end{cases} 
\end{eqnarray}
with $(T_L,T_H)=(T_c-\Delta T,T_c+\Delta T)$ and where $\Delta T$, which parametrizes the width of the critical region, is the same parameter that arises in Eq.~(\ref{xiparam1}). Here as there, we take $\Delta T = T_c/5$.  Following Ref.~\cite{Rajagopal:2019xwg}, we take $c_V^{\text{crit}}(T)$, the critical part of $c_V$ that shows the leading singular behavior near the critical point, to have the form
\begin{eqnarray}
c_V^{\text{crit}}(T)\equiv\frac{1}{2}\frac{1}{\xi_0^{3}}\frac{\xi(T)}{\xi_0}\,,
\label{CV_crit}
\end{eqnarray}
where the temperature dependence of the correlation length of critical fluctuations, $\xi(T)$, 
needs to be specified.  Following Ref.~\cite{Rajagopal:2019xwg}, we do so as in Eq.~({\ref{xiparam1}). 
The prefactor $1/2$ in Eq.~(\ref{CV_crit}) is a non-universal constant whose value depends on the mapping between the equation of state of the 3D Ising model and the equation of state of QCD, whose critical point is in the 3D Ising universality class. We have used a value that is reasonable for $\Delta T=T_c/5$; see the argument in Ref.~\cite{Rajagopal:2019xwg}. Continuing to follow Ref.~\cite{Rajagopal:2019xwg}, away from the critical point we choose the form of the specific heat capacity $c_V^{\text{no C.P}}(T)$ as follows:
\begin{eqnarray}
\frac{c_V^{\text{no C.P}}(T)}{T^{3}}\equiv \left[\frac{a_H+a_L}{2}+\frac{a_H-a_L}{2}\tanh{\frac{T-T_{\text{crossover}}}{\Delta T_{\text{crossover}}}}\right]
\end{eqnarray}
with
\begin{eqnarray}
a_L=0.1 \, a_{\text{QGP}} \, , \quad a_H=0.8 \, a_{\text{QGP}}\, , \quad \mathrm{and} \ a_{\text{QGP}}\equiv\frac{4\pi^2(N_c^2-1)+21\pi^2 N_f}{15}\,,
\end{eqnarray}
where $N_c=3$ and $N_f=3$ are the number of flavors and colors respectively, and with
\begin{eqnarray}
T_{\text{crossover}}=T_c\, , \quad \Delta T_{\text{crossover}}=0.6\, T_c\,.
\end{eqnarray}
 The specification of the equation of state is completed by choosing the six constant coefficients $c_n$ that appear in Eq.~(\ref{cvnoncrit}) so as to enforce that $c_V(T)/T^3$ and its first two derivatives are continuous at $T=T_L$ and at $T=T_H$.

\section{Analytical calculations in a Bjorken scenario}
\label{append1}

In this Appendix, we derive an explicit expression for the numerator of $C_A(\Delta y)$ defined in Eq.~(\ref{cdely0}), namely $ \big\langle\delta\frac{d N_A}{dy_+}\delta\frac{d N_A}{dy_-}\big\rangle_\sigma$, in the Bjorken symmetric background described in Section~\ref{Sectbjk}. We shall begin from the somewhat formal expression for $\<\delta N_A^2\>_{\sigma}$ that we obtained in Eq.~(\ref{delNA2RRWYagain3}), develop an explicit expression for this measure of fluctuations in the Bjorken background, and then see that we can obtain the explicit expression (\ref{NAdydyvariance2bjk2}) for the rapidity correlator $ \big\langle\delta\frac{d N_A}{dy_+}\delta\frac{d N_A}{dy_-}\big\rangle_\sigma$ essentially by inspection of the explicit form of $\<\delta N_A^2\>_{\sigma}$.

In the second half of the Appendix, as a bonus we shall use the analytic control that we have over
the calculation of $\<\delta N_A^2\>_{\sigma}$ to show that the low-$Q$ modes of $\phi_{\bm Q}$ make the dominant contribution to this observable.

We begin by noting that in the Bjorken scenario we have the following expressions for $x$ and $\Delta x_\perp$ at $\tau=\tau_f$:
\bes \label{xvarBjk}
\begin{eqnarray}
x&=&\tau_f\,\cosh\Delta \eta\, \hat{\tau}+\,x_T\\
\Delta x_\perp&=&\Delta x=2\tau_f\sinh \frac{\Delta \eta}{2}\,\hat{\eta}+\,\Delta x_T\,.
\end{eqnarray}
\ees
Eqs.~(\ref{xvarBjk}) are exact for two points $x_+$ and $x_-$ on the freeze-out hypersurface and can be obtained by substituting $u^{\tau}=1$, $u^{r}=0$, $\tau=\tau_f$ and $\Delta \tau=0$ in Eqs.~(\ref{xdelxsym1}). We have used $x_T$ and $\Delta x_T$ to denote the transverse parts of  $x$ and $\Delta x_\perp$, namely their projections onto the plane spanned by $\hat{r}$ and $\hat{\varphi}$. Note that because points on the freeze-out surface, including $x_+$ and $x_-$, all have the same $\tau$ in the Bjorken scenario,  $\Delta x_\perp=\Delta x$ in this setting and
Eq.~(\ref{delNA2RRWYagain3}) need not be corrected 
as described around Fig.~\ref{fig:delta-x}.

Next, we note that the $\tilde{\phi}$ that arises in the expression~(\ref{delNA2RRWYagain3}) that we wish to evaluate is the inverse Wigner transform of $\phi_\mathbf{Q}$, see Eq.~(\ref{sseq1uvFT}).  In the Bjorken scenario, this transform takes the form
  \begin{eqnarray}
		\label{sigbjk1}
		\tilde{\phi}(x,\mathbf{\Delta x})&=&\int \frac{d\mathbf{Q_\perp}dQ_\eta}{(2\pi)^3} e^{i 2Q_\eta \tau_f\sinh \frac{\Delta \eta}{2} +i\mathbf{Q_\perp}\cdot \mathbf{\Delta x_T}}\phi_\mathbf{Q}(\tau_f\cosh\Delta \eta)
		\end{eqnarray}
		where $Q_\eta\equiv Q\cdot \hat{\eta}$ and $\mathbf{Q_\perp}\equiv\mathbf{Q}-Q_\eta\hat{\eta}$.  Note that since points on the freeze-out hypersurface all have the same 
		$\tau$ in the Bjorken scenario, the expression for the two-point correlator of $\sigma$ between two points on the freeze-out hypersurface given by Eq.~(\ref{eq:<ss>=phiQ}) does not receive the correction described following that equation. Note that $\phi_{\bf Q}$ that enters Eq.~(\ref{sigbjk1})  is obtained by solving Eq.~(\ref{Bjkphiev}), as discussed in Section~\ref{Sectbjk}.
Later in this Appendix, we shall need the formal solution to Eq.~(\ref{Bjkphiev}) that satisfies the initial conditions~(\ref{phiqinit}); it is given by 
	\begin{eqnarray}\nonumber
	\phi_\mathbf{Q}(\tau)&=&\bar{\phi}_\mathbf{Q}(T_i)e^{-\int_{\tau_i}^{\tau}\, \Gamma(Q\xi(\tau^{'}))\, d\tau^{'}}
	+\int_{\tau_i}^{\tau}e^{-\int _{\tau^{''}}^{\tau}\, \Gamma(Q\xi(\tau^{'}))\, d\tau^{'}}\, \Gamma(Q\xi(\tau^{''}))\, \bar{\phi}_\mathbf{Q}(T(\tau^{''}))\, d\tau^{''}\\ 	\label{phiqchar_bjk}
	\end{eqnarray}
The functional form for the evolution of temperature $T(\tau)$, which also determines $\xi(\tau)$ through Eq.~(\ref{xiparam1}), can be obtained 
from the condition
   \begin{eqnarray} \label{sbjk}
    \tau s(T(\tau))=\tau_i s(T(\tau_i))\, 
    \end{eqnarray}
that follows from the isentropic nature and Bjorken symmetry of the flow and that must be satisfied for all $\tau_i<\tau<\tau_f$.
And, we employ the equation of state $s(T)$ from Ref.~\cite{Rajagopal:2019xwg} that we describe briefly in Appendix~\ref{appendeos}.

		With these preliminaries in place, we now substitute Eq.~(\ref{sigbjk1}) into Eq.~(\ref{delNA2RRWYagain3}) and integrate over $d\mathbf{x_T}$, $d\mathbf{\Delta x_T}$ and $d\mathbf{Q_\perp}$, obtaining
	\begin{eqnarray}\nonumber
	\<\delta N_A^2\>_\sigma&=&g_A^2\, A_\perp\,\tau_f^2 \,\int_{-\infty}^{\infty} d\eta\, \int_{-\infty}^{\infty} d\Delta \eta\, I_A\left(\eta_+, \eta_-\right)\,	\int \frac{dQ_\eta}{2\pi} e^{i 2Q_\eta \tau_f\sinh \frac{\Delta \eta}{2}}\, \phi_\mathbf{Q_\parallel}\left(\tau_f\cosh\Delta \eta\right) \,, \\ & &\label{NAvariance2bjk1}
	\end{eqnarray}
	where we have defined $\mathbf{Q_\parallel}\equiv Q_\eta \mathbf{\hat{\eta}}$, where $A_\perp$ is the transverse area in the plane spanned by $\hat{r}$ and $\hat{\varphi}$, and where we have defined 
	\begin{equation}\label{I_A_defn}
	I_A(\eta_+,\eta_-)\equiv n(\eta_+)\cdot J_A(\eta_+)\,n(\eta_-)\cdot J_A(\eta_-)\,.
	\end{equation}
	 Upon explicit evaluation, this function is given by
	\begin{eqnarray}\label{IAbjk1}
	I_A(\eta_+, \eta_-)=\int_{y_{\text{min}}}^{y_{\text{max}}}dy_+\, \int_{y_{\text{min}}}^{y_{\text{max}}} dy_- F_A(y_+-\eta_+)F_A(y_--\eta_-)
	\end{eqnarray}
	where
			\begin{eqnarray}
			\label{FAbj}
			F_A(y_\pm-\eta_\pm)\equiv \frac{d_A\,m_{A}}{T_f} \frac{dy}{2\pi}\, \int_{m_{T,\text{min}}}^{m_{T,\text{max}}} \frac{m_{T}dm_{T}}{2\pi} e^{-\frac{m_T\cosh\(y_\pm-\eta_\pm\)}{T_f}}
			\,.
			\end{eqnarray}
			Upon specifying $m_{T,\text{min}}=m_A$ and choosing $m_{T,\text{max}}=\infty$ and $y_{\text{min}}=-y_{\text{max}}$, $F_A$ is given by
	\bes
	\begin{eqnarray} \label{Fa1}
	F_A(x)&=&d_A \frac{m_{A}}{T_f} \int_{m_{A}}^{\infty} \frac{m_{T}dm_{T}}{(2\pi)^2} e^{-\frac{m_T\cosh x}{T_f}}\\ \label{Fa2}
	&=&d_A m_A (2\pi)^{-2} \text{sech}^2 x\,  \left(m_A \cosh x+T_f\right)\,e^{-\frac{m_A \cosh x}{T_f}}\,.
	\end{eqnarray}
\ees
			
		Using the expressions above, Eq.~(\ref{NAvariance2bjk1}) can be evaluated directly, numerically.  However, to elucidate its main features we ignore the subleading corrections due to the curvature of the freeze-out hypersurface and assume that $m_A\gg T$, in both cases as discussed in Section~\ref{azsbifo}. This allows us to make the following approximations:
\bes
\begin{eqnarray}
		\label{sigbjk2}
		\phi_\mathbf{Q_\parallel}\left(\tau_f\cosh\Delta \eta\right) &\approx&\phi_\mathbf{Q_\parallel}(\tau_f)
		\\ \label{Fa3}
	F_A(\eta_\pm)&\approx&d_A m_A^{2} (2\pi)^{-2} \text{sech} \,\eta\, e^{-\frac{m_A }{T_f}\left[\cosh \eta\pm\frac{ \sinh\eta}{2}\Delta \eta+\frac{ \cosh\eta}{8}\Delta \eta^2\right]}
	\end{eqnarray}
	\ees
		
The assumption $m_A\gg T$ has allowed us to simplify Eq.~(\ref{Fa3})
by expanding only the exponential term in $F_A$ as a function of $\Delta \eta$ and not its prefactor. We have verified by explicit calculation that this assumption is well justified for protons. 
With the above simplifications, after defining $\Delta y\equiv y_+-y_-$ and redefining the variables $\eta$ and $\Delta \eta$ according to $\eta\rightarrow\eta-(y_++y_-)/2$ and   $\Delta \eta\rightarrow\Delta \eta - \Delta y$,  Eq.~(\ref{NAvariance2bjk1}) becomes
 \begin{eqnarray}
	\nonumber
	& &\<\delta N_A^2\>_\sigma\\ \nonumber
	&\approx& g_A^2T_f A_\perp \tau_f^2\int_{-\infty}^{\infty} d\eta \int_{-\infty}^{\infty} d\Delta \eta F_A(\eta+)F_A(\eta_-)\int_{-y_\text{max}}^{y_{\text{max}}} dy_{+}\int_{-y_\text{max}}^{y_{\text{max}}} dy_{-}
	\int \frac{dQ_\eta}{2\pi}  e^{iQ_\eta\,\tau_f\left(\Delta\eta+\Delta y\right)}\phi_{\mathbf{Q_\parallel}}(\tau_f) \,. \\\label{NAvariance2bjk2a_app}
	\end{eqnarray}
This explicit expression for the observable measure of the fluctuations $\<\delta N_A^2\>_\sigma$ is the first main result of this Appendix.

Upon inspection of the result (\ref{NAvariance2bjk2a_app}), we see that the two point rapidity space correlator occurring in $C_A(\Delta y)$ is given by
\begin{eqnarray}\nonumber
& &\left<\delta\frac{d N_A}{dy_+}\delta\frac{d N_A}{dy_-}\right>_\sigma\approx g_A^2T_f A_\perp \tau_f^2\int_{-\infty}^{\infty} d\eta \int_{-\infty}^{\infty} d\Delta \eta F_A(\eta+)F_A(\eta_-)
	\int \frac{dQ_\eta}{2\pi}  e^{iQ_\eta\,\tau_f\left(\Delta\eta+\Delta y\right)}\phi_{\mathbf{Q_\parallel}}(\tau_f)\\   \label{delNadyplusminus_app_a}\\ \nonumber
&\approx& \frac{1}{8\pi^{7/2}}g_A^2\, d_A^2\, Z^{-1}m_A^{7/2} T_f^{1/2}\,A_\perp\,\tau_f^2\,\int d\eta\,\text{sech}^{5/2} \,\eta\,  e^{-\frac{2m_A \cosh \eta}{T_f}}  \,\int \frac{dQ_\eta}{2\pi} \, e^{iQ_\eta\tau_f\Delta y}\, e^{-\frac{Q_\eta^2\tau_f^2 T_f}{m_A\cosh \eta}}\,\phi_{\mathbf{Q_\parallel}}(\tau_f)\,. \\ \label{delNadyplusminus_app}
\end{eqnarray}
Eq.~(\ref{delNadyplusminus_app}), our second main result of this Appendix, is reproduced in Section~\ref{Sectbjk} as 
Eq.~(\ref{NAdydyvariance2bjk2})
and its implications are discussed there.

In the remainder of this Appendix, we shall demonstrate 
that the low $Q$ modes of $\phi_{\bf Q}$ contribute the most to the variance of particle multiplicities, $\left<\delta N_A^2\right>_\sigma$. 
We shall expand  $\phi_\mathbf{Q}$, 
given by Eq.~(\ref{phiqchar_bjk}),
in powers of $\mathbf Q$ to ${\cal O}(\mathbf{Q}^2)$ and 
compare the result for $\left<\delta N_A^2\right>_\sigma$ that we obtain starting from this expansion to the result that we obtain starting from the
full form of $\phi_\mathbf{Q}$. We denote the polynomial expansion for $\phi_\mathbf{Q}$ to quadratic order by
		\begin{eqnarray}\label{phiapp1}
		\phi_\mathbf{Q}\approx \phi^{(0)}+\phi^{(2)}Q^2
		\end{eqnarray}
	where
	\bes	
		\begin{eqnarray}\label{phiapp11}
\phi^{(0)}&=&Z\,T_f\,\xi^{2}(T_i)\\ \label{phiapp12}
\phi^{(2)}&=&-Z\,T_f\,\xi^4(T_i)+2\dip \xi_0\,Z\, T_f\, \int_{\tau_i}^{\tau_f}d\tau\,\left(\xi(T(\tau))-\frac{\xi^2(T_i)}{\xi(T(\tau))}\right)\,.
\end{eqnarray}
\ees
The expression (\ref{phiapp1}) is a good approximation to $\phi_{\bf Q}$ for its low-$Q$ modes,
as we illustrate in Fig.~\ref{phiQdelnaGHaappendix}.

\begin{figure}[t]
	    \centering
	    \begin{subfigure}{0.45\textwidth}
	    \includegraphics[width=\textwidth]{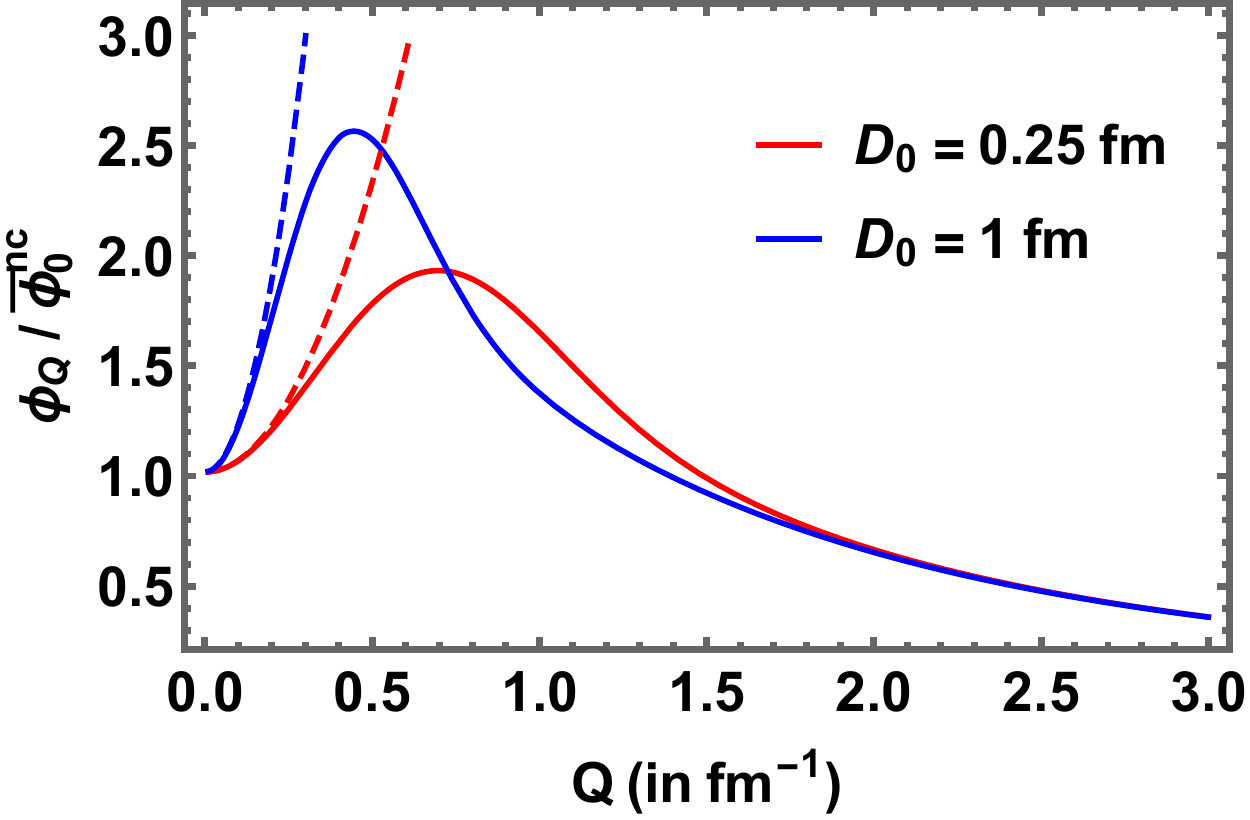}
	    \caption{$\phi$ at freeze-out}
	    \label{phiQdelnaGHaappendix}
	    \end{subfigure}
	    \begin{subfigure}{0.45\textwidth}
	    \includegraphics[width=\textwidth]{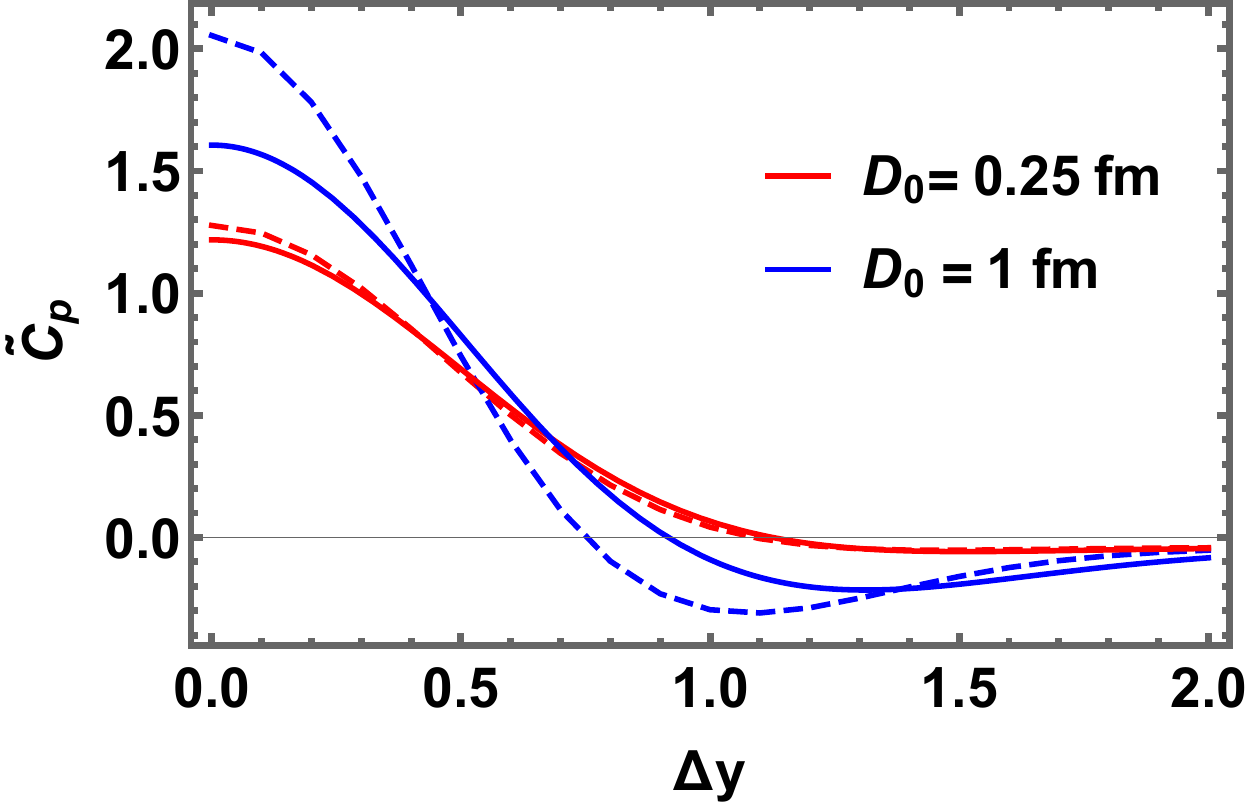}
	    \caption{$\tilde{C}(\Delta y)$ for protons}
	    \label{phiQdelnaGHbappendix}
	    \end{subfigure}
	    	\caption{Panel (a): Normalized $\phi$ as a function of $Q$ evolved according to Model H dynamics with two values of $\dip$, plotted at freeze-out $\tau=\tau_f$, corresponding to an equilibrium temperature of $T_f(\tau_f)=140\, \text{MeV}$.  The solid and dashed curves were obtained from the full solution~(\ref{phiqchar_bjk}) for $\phi_\mathbf{Q}$ and its truncated polynomial expansion~(\ref{phiapp1}) respectively. Panel (b): Normalized fluctuation measure observable (rapidity space correlator) for protons $\tilde{C}_p(\Delta y)$ obtained with the full form (solid) and truncated form (dashed) of $\phi_\mathbf{Q}$. The qualitative and even semi-quantitative agreement between the same colored curves in the right plot indicates that the low-$Q$ modes contribute significantly to the variance of particle multiplicities. In obtaining these plots, $\xi_{\text{max}}$ was set to $3\, \text{fm}$ and the fluctuations at $\tau_i=1\, \text{fm}$ were initialized to their equilibrium value at $\tau=\tau_i=1$~fm with $T_i(\tau_i)=235\,\text{MeV} $. }
	    \label{phiQdelnaGHappendix}
	\end{figure}

Upon making the low-$Q$ approximation and working to order $Q^2$ as in Eq.~(\ref{phiapp1}), we can perform the $Q_\eta$ integral in Eq.~(\ref{delNadyplusminus_app}), obtaining
	\begin{eqnarray}\nonumber
	& &\left<\delta\frac{d N_A}{dy_+}\delta \frac{d N_A}{dy_-}\right>_\sigma\\ \nonumber
	& & \approx \frac{g_A^2 d_A^2 A_\perp\,\tau_f Z^{-1}m_A^{4}}{(2\pi)^4}\int \frac{d\eta}{\cosh^2\eta}  e^{- \frac{2m_A\cosh\eta}{T_f} \left( 1 + \frac{\Delta y^2}{8}\right)}  
	\left[\phi^{(0)}+\frac{2 T_f m_A \cosh \eta- \Delta y^2 m_A^2 \cosh^2 \eta}{4\tau_f^2 T_f^2} \phi^{(2)}\right]\,. \\ \label{delNAdyplusminusq2app}
	\end{eqnarray}
	In Fig.~\ref{phiQdelnaGHbappendix}, we compare $\tilde C_p(\Delta y)$ (defined via Eqs.~(\ref{cdely0},\ref{Bjkmean},\ref{cdely1}))
	obtained from $\left\langle\delta\frac{d N_A}{dy_+}\delta \frac{d N_A}{dy_-}\right\rangle_\sigma$ computed without making a low-$Q$ approximation, namely Eq.~(\ref{delNadyplusminus_app}), which is plotted as the solid curves in Fig.~\ref{phiQdelnaGHbappendix}, to that computed upon working only to order  $Q^2$, namely Eq.~(\ref{delNAdyplusminusq2app}), which is plotted as the dashed curves in Fig.~\ref{phiQdelnaGHbappendix}. The qualitative, even semi-quantitative, agreement between them indicates that the low-$Q$ modes contribute significantly to the variance of particle multiplicities.

\section{Contrasting with Model A evolution}
\label{RRWYModelA}

In this Appendix we repeat the calculations of Section~\ref{RRWY} 
in a scenario in which the hydrodynamic background is the same (and 
hence also the same as in Ref.~\cite{Rajagopal:2019xwg}) but in which
the dynamical evolution of the fluctuations differs.  In Section~\ref{RRWY} 
we take conservation laws into account, following Model H dynamics.
Here, in contrast, we shall consider the case where the fluctuating slow mode
is not a conserved quantity, meaning that the appropriate dynamics for the relaxation
of the two-point function is that of Model A
in the classification of Halperin and Hohenberg \cite{Hohenberg:1977ym}, with the relaxation rate given by Eq.~(\ref{GammaA}), which we repeat here:
	\begin{eqnarray}
	\label{GammaAAppendixC}
          \Gamma(\bm Q)=\Gamma_0\frac{\xi_0^2}{\xi^{2}}\,
          \left(1+(Q\xi)^2\right)\,,
          \qquad \mbox {(model A).}
\end{eqnarray}
We have performed simulations with $\Gamma_0=1\, \text{fm}^{-1}$ 
and $8\, \text{fm}^{-1}$, which correspond to $\Gamma_0\xi_0^2=0.25\, \text{fm}$ 
and $2\, \text{fm}$, respectively. 
Comparing the ``shapes'' of all the results plotted in this Appendix to those in the analogous 
Figures in Section~\ref{RRWY} provides us with another way of seeing the impact of conservation
laws on the results from Section~\ref{RRWY}.  
The main difference between Model A and Model H evolution arises from the qualitatively different relaxation rate for the low $Q$ modes, $Q\xi\ll 1$, which goes as $\Gamma_0 \xi_0^{2}/\xi^{2}$ in Model A and as $(\dip \xi_0/\xi) Q^2$ in Model H, with the $Q^2$-suppression being a
manifestation of conservation. 
A second motivation for this Appendix is
that, because Model A dynamics is simpler to implement, in their pioneering
calculation the authors of Ref.~\cite{Rajagopal:2019xwg} used Model A dynamics, meaning that
in this Appendix we shall be freezing out the calculations of Ref.~\cite{Rajagopal:2019xwg}, turning
these Hydro+ simulations into particle multiplicity fluctuations.

 \begin{figure}[t]
\begin{center}
\begin{subfigure}{0.45\textwidth}
\includegraphics[scale=0.52]{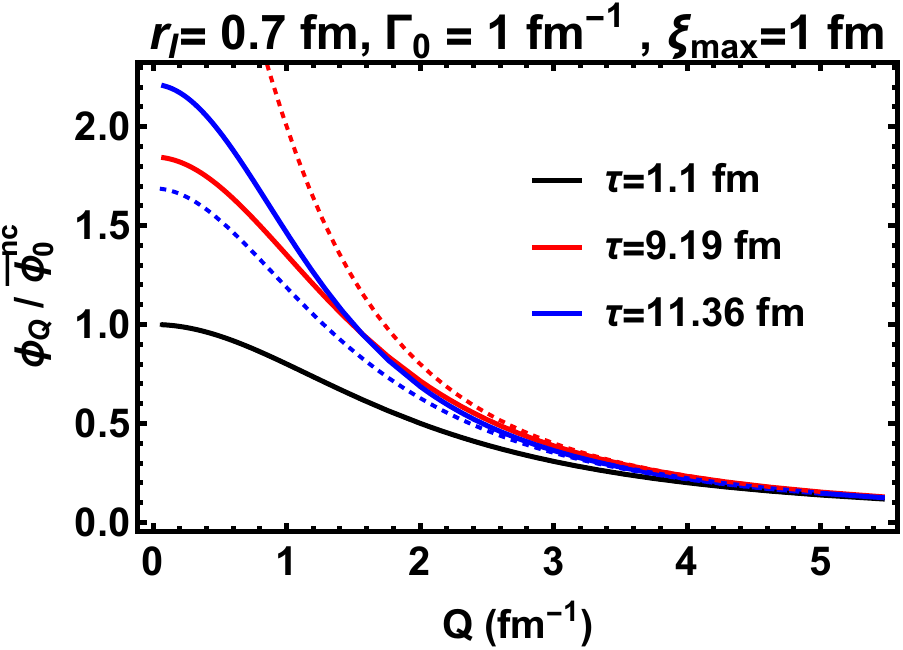}
\end{subfigure}
\begin{subfigure}{0.45\textwidth}
\includegraphics[scale=0.52]{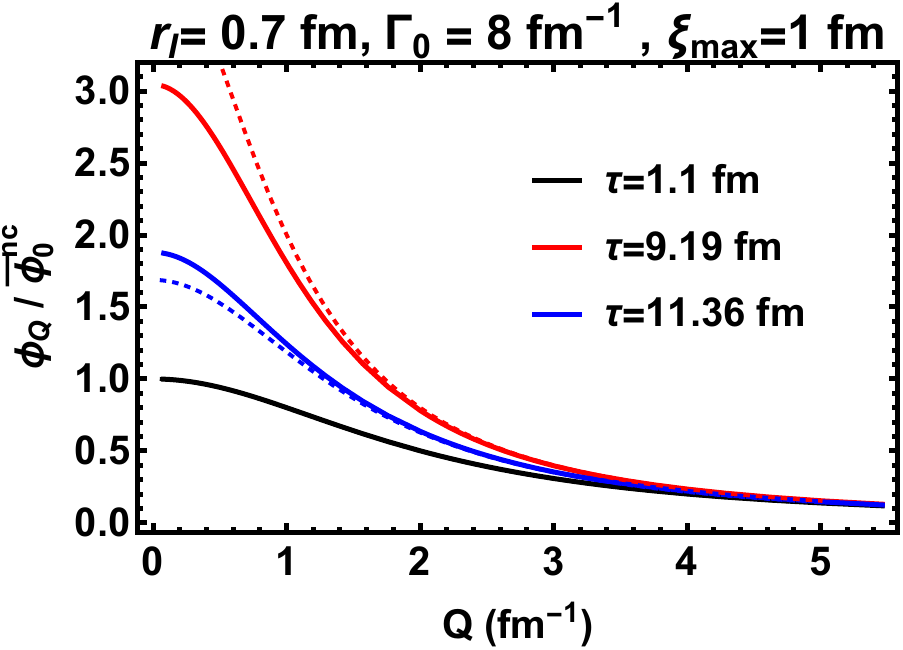}
\end{subfigure}
\begin{subfigure}{0.45\textwidth}
\includegraphics[scale=0.52]{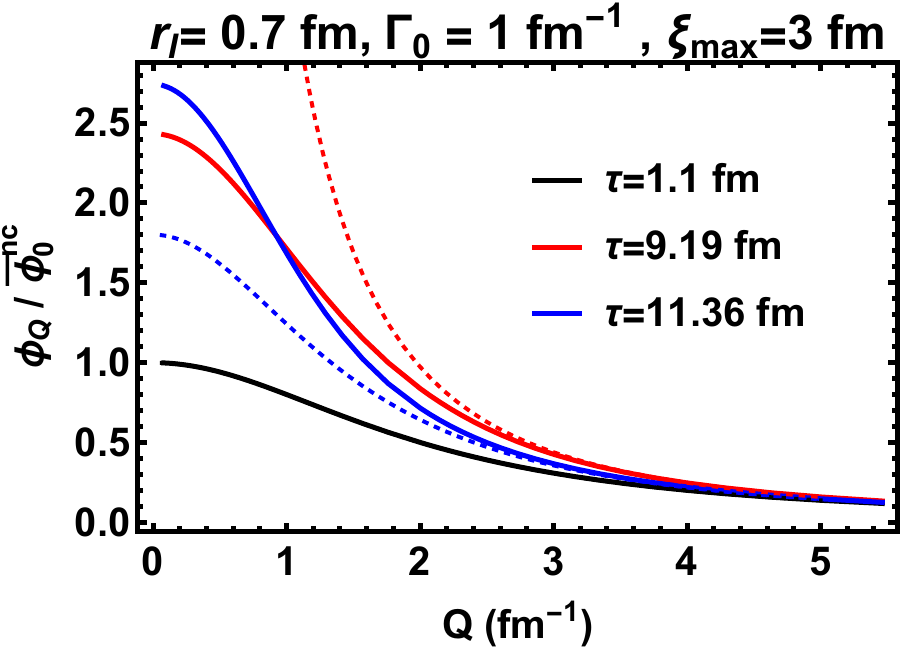}
\end{subfigure}
\begin{subfigure}{0.45\textwidth}
\includegraphics[scale=0.52]{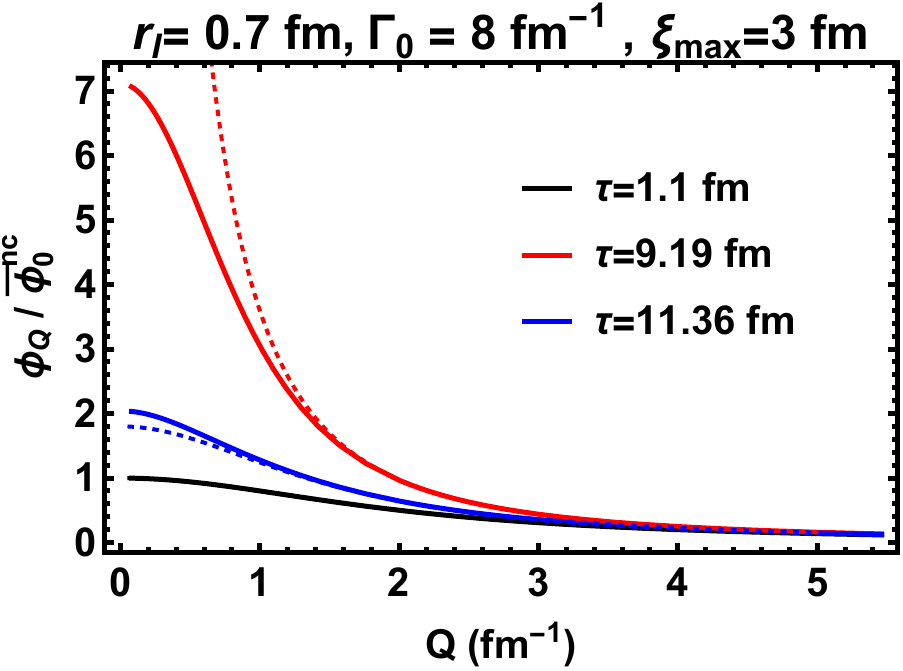}
\end{subfigure}
\begin{subfigure}{0.45\textwidth}
\includegraphics[scale=0.52]{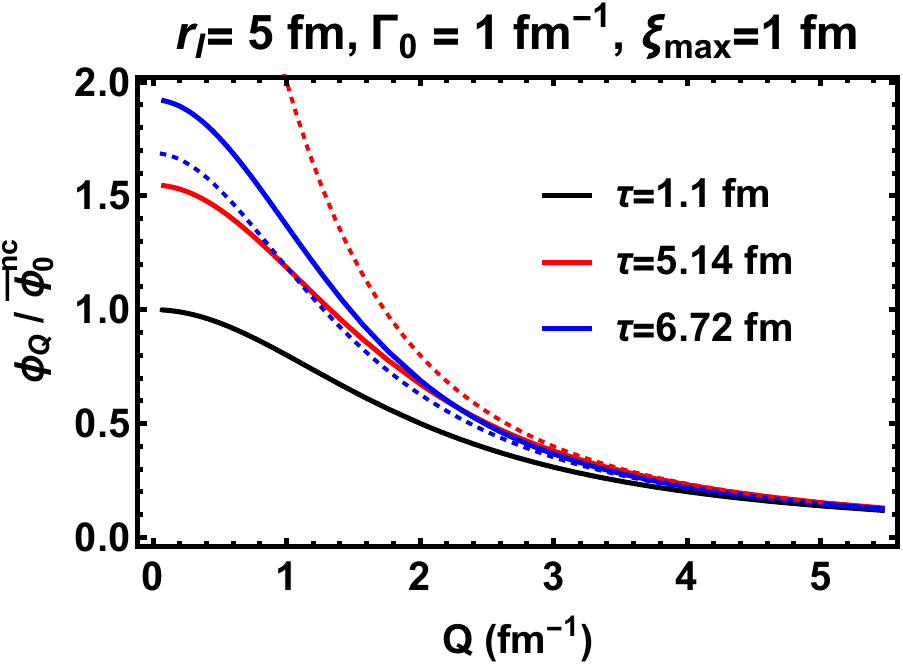}
\end{subfigure}
\begin{subfigure}{0.45\textwidth}
\includegraphics[scale=0.52]{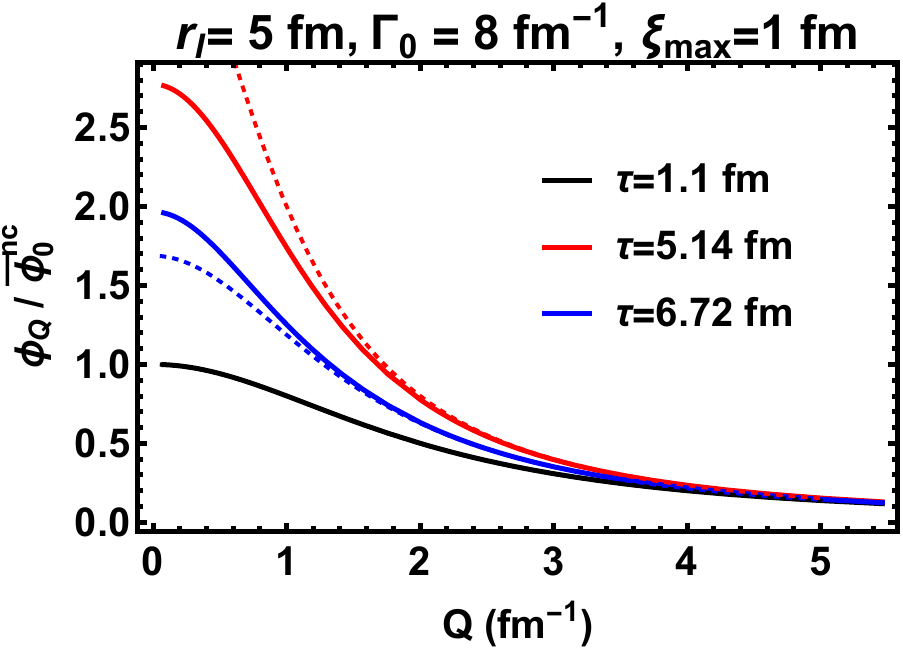}
\end{subfigure}
\begin{subfigure}{0.45\textwidth}
\includegraphics[scale=0.52]{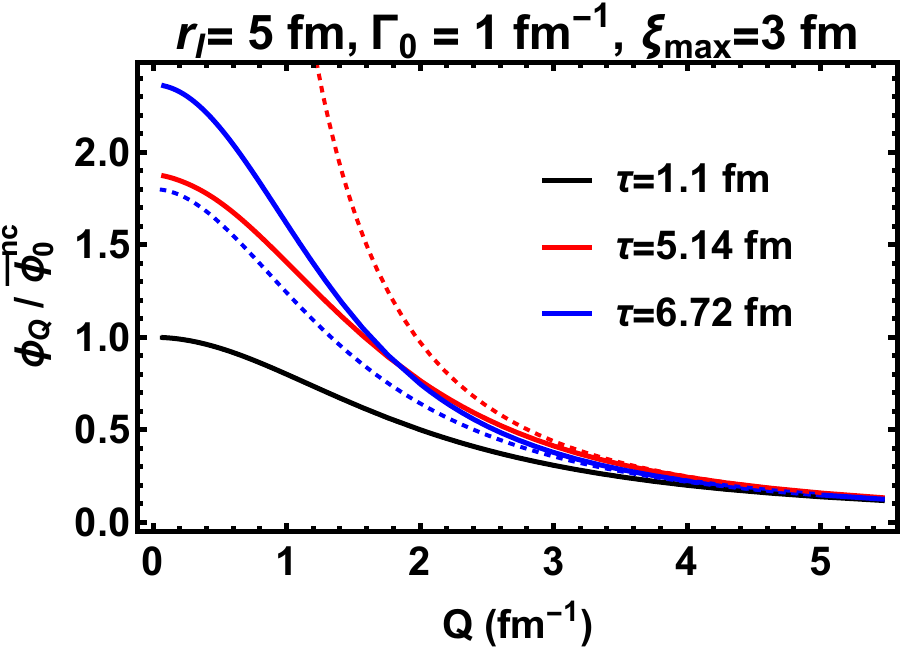}
\end{subfigure}
\begin{subfigure}{0.45\textwidth}
\includegraphics[scale=0.52]{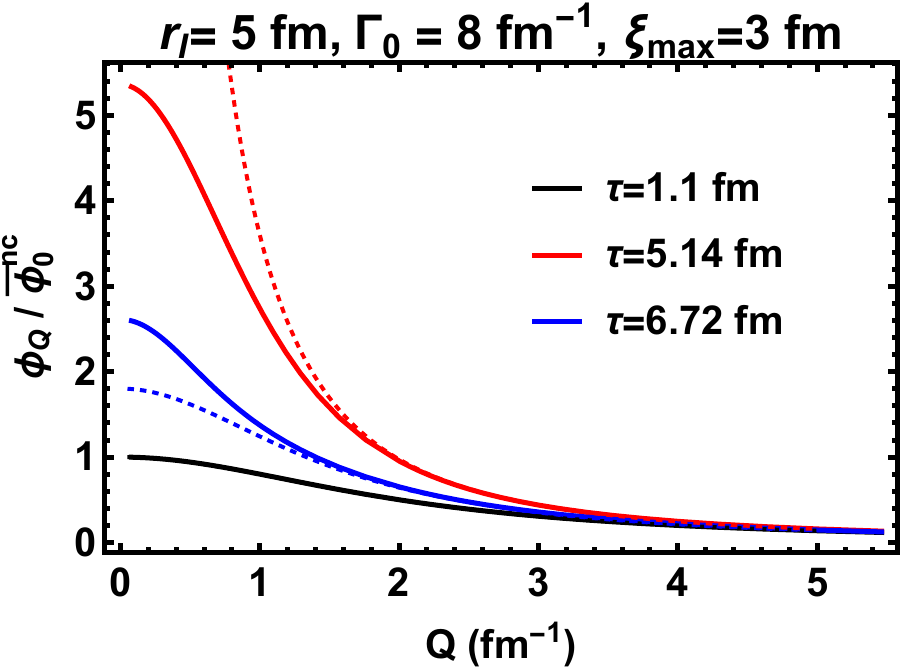}
\end{subfigure}
\end{center}
\caption{Hydro+ fluctuation measure $\phi_\mathbf{Q}$ evolved according to Model A dynamics along two hydrodynamic flow lines passing through $r=r_i$ at initial time $\tau=\tau_i$,
  with $r_i=0.7\, \text{fm}$ (top four panels) and $5\, \text{fm}$ (bottom
  four panels). Plots in the left (right) column are for
  $\Gamma_0 = 1\, \text{fm}^{-1}$ ($\Gamma_0 = 8\, \text{fm}^{-1}$), with
  $\xi_{\text{max}}=1\, \text{fm}$ and
  $\xi_{\text{max}}=3\, \text{fm}$ in alternating rows.  
  The solid (and dashed) curves
   are the $\phi_{\bm Q}$ (and $\bar{\phi}_{\bm Q}$),
normalized to the zero mode of the non-critical fluctuations. 
The black, red and blue curves correspond to $\phi_\mathbf{Q}$'s at the initial time $\tau_i$
and at the times 
when the equilibrium temperature reaches $160\, \text{MeV}$ and $140\, \text{MeV}$ respectively.}
\label{phi-1chA}
\end{figure}

\subsection{Evolution of $\phi_{\mathbf Q}$}

In Fig.~\ref{phi-1chA}, 
which can be compared to the analogous Model H results shown in Fig.~\ref{phi-1ch},
we illustrate the Model A dynamics of
$\phi_\mathbf{Q}$ 
for fluid cells following two different hydrodynamic flow lines, with two choices of $\Gamma_0$ 
and two choices of $\xi_{\text{max}}$. As in Section~\ref{RRWY}, varying $\xi_{\text{max}}$ corresponds to varying how close the cooling trajectory of the fluid cell comes to the critical point
on the phase diagram.
The central qualitative difference between the Model A results in Fig.~\ref{phi-1chA} and the 
Model H results in Fig.~\ref{phi-1ch} is that in Model H $\phi_\mathbf{Q}$ at $Q=0$ is unchanging in time, because of conservation, which means that as critical flucutations develop we see that
in Fig.~\ref{phi-1ch}  $\phi_\mathbf{Q}$ takes on a shape in which it first rises as a function of increasing $Q$ and then falls whereas in the Model A dynamics of this Appendix the maximum value of $\phi_\mathbf{Q}$ is found at $Q=0$, and this value is time dependent.
In Model A, here,  as in Model H in Fig.~\ref{phi-1ch}, the fluctuations 
$\phi_\mathbf{Q}$ fall out of equilibrium, lagging behind the 
equilibrium fluctuations $\bar{\phi}_\mathbf{Q}$ as the latter change with time.

\begin{figure}[t]
\begin{center}
\begin{subfigure}{0.32\textwidth}
\includegraphics[scale=0.55]{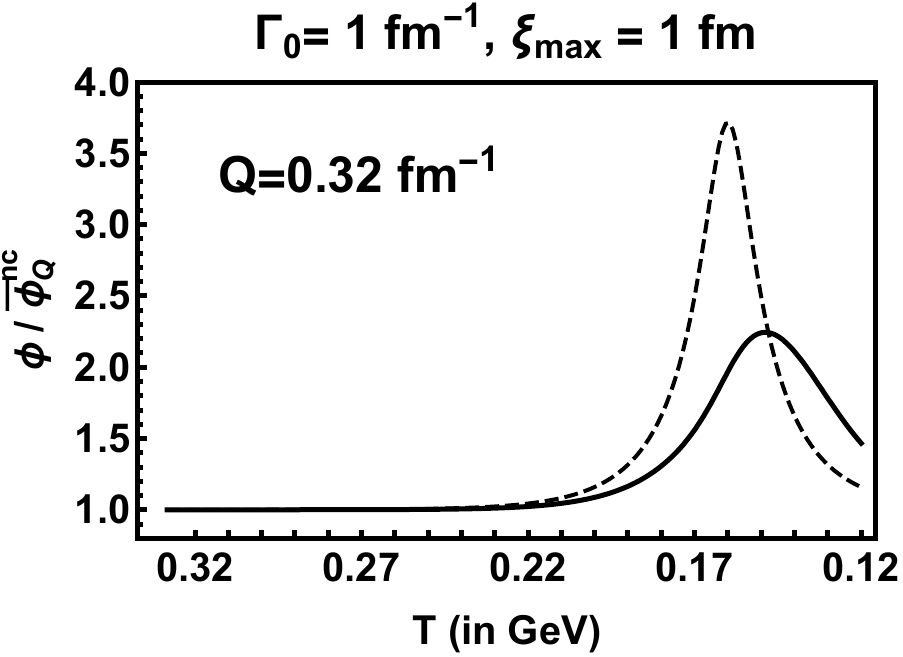}
\end{subfigure}
\begin{subfigure}{0.32\textwidth}
\includegraphics[scale=0.55]{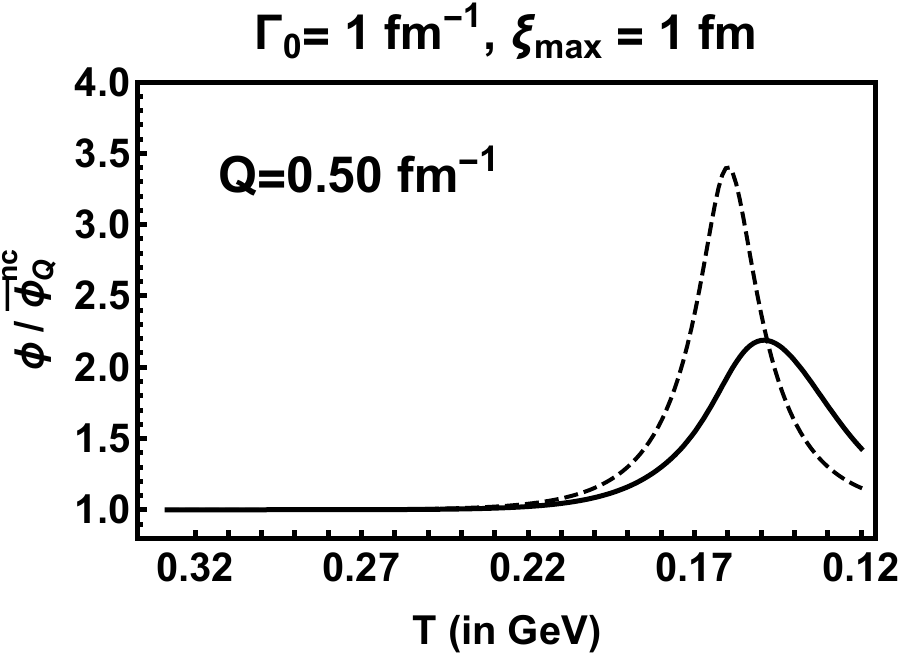}
\end{subfigure}
\begin{subfigure}{0.32\textwidth}
\includegraphics[scale=0.55]{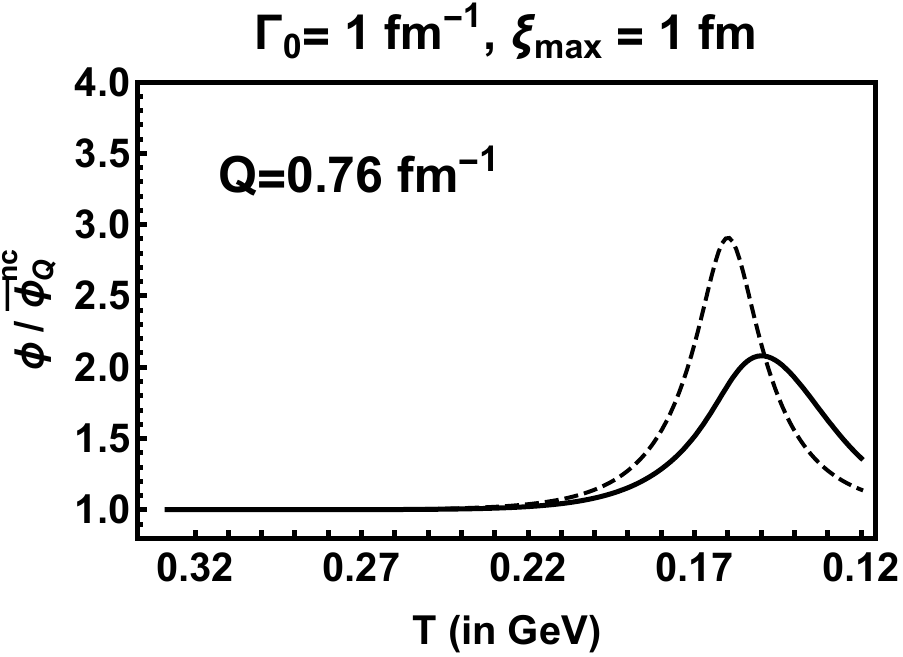}
\end{subfigure}
\begin{subfigure}{0.32\textwidth}
\includegraphics[scale=0.55]{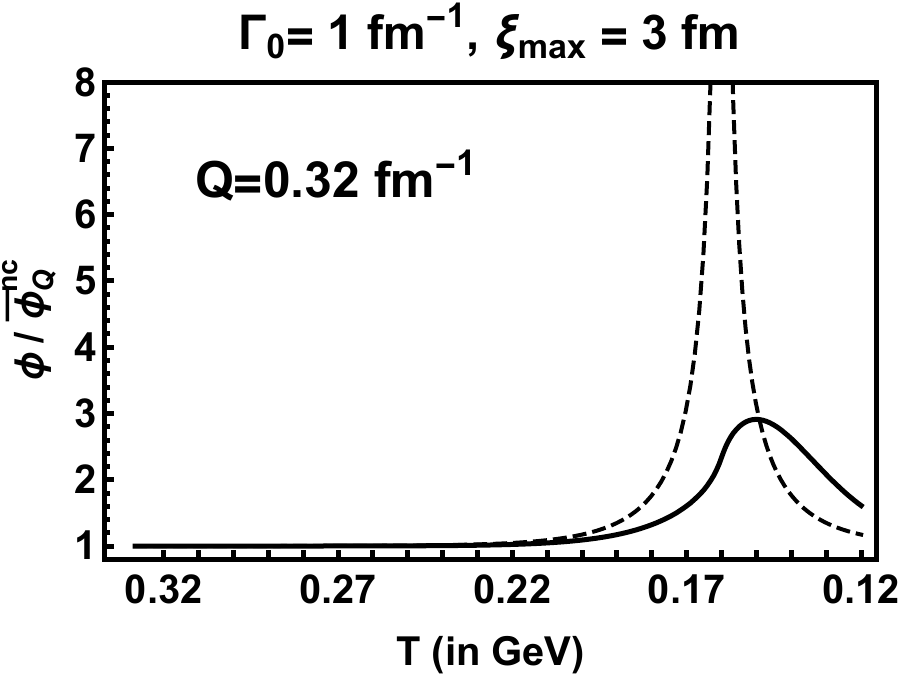}
\end{subfigure}
\begin{subfigure}{0.32\textwidth}
\includegraphics[scale=0.55]{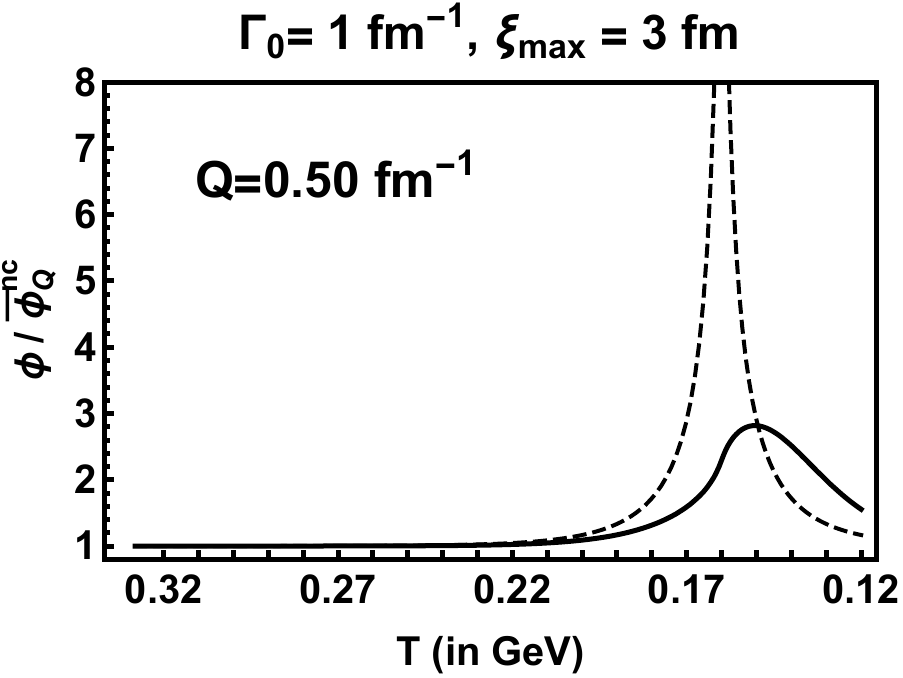}
\end{subfigure}
\begin{subfigure}{0.32\textwidth}
\includegraphics[scale=0.55]{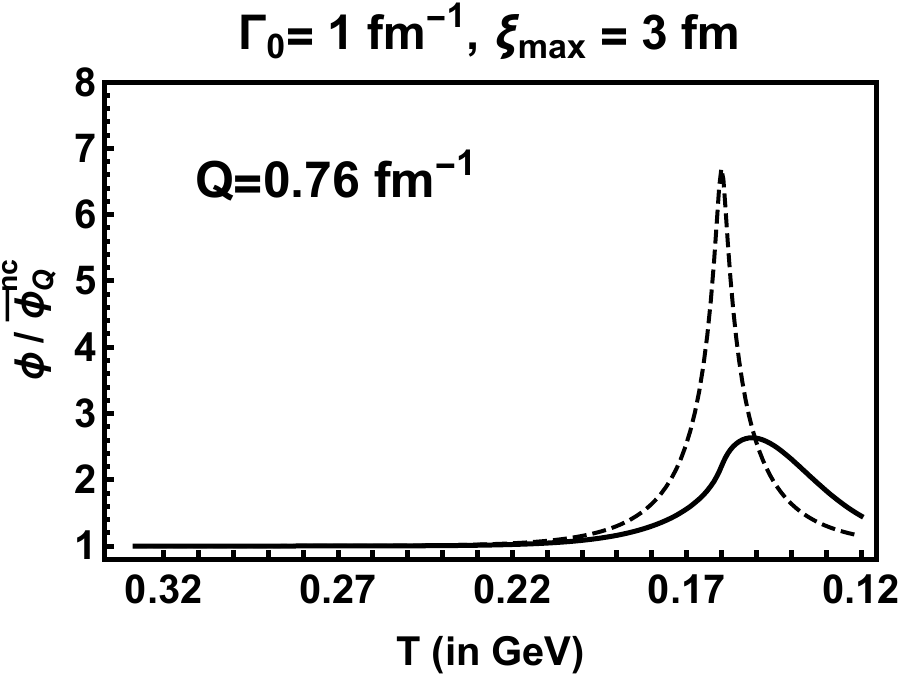}
\end{subfigure}
\begin{subfigure}{0.32\textwidth}
\includegraphics[scale=0.55]{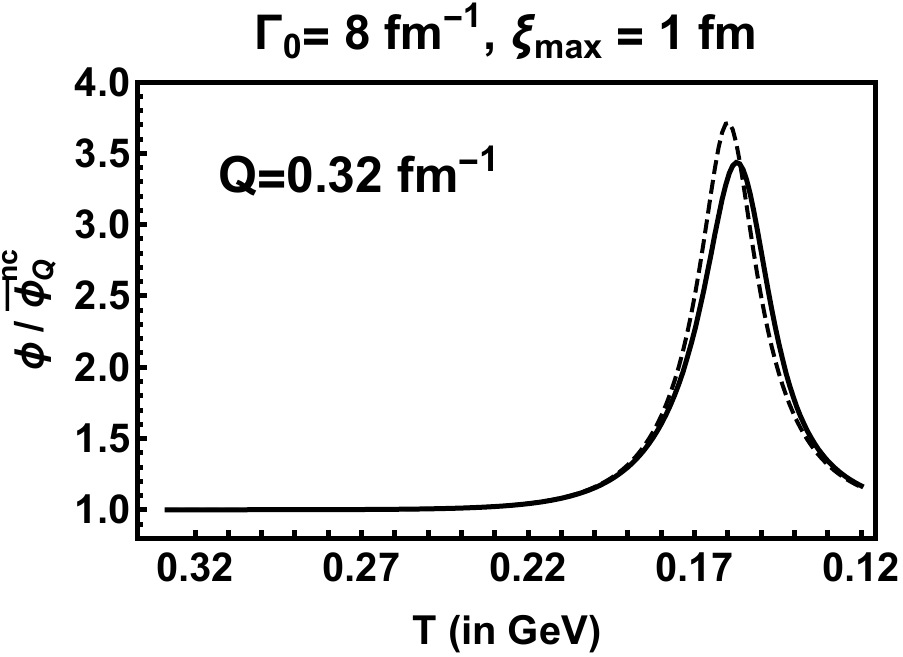}
\end{subfigure}
\begin{subfigure}{0.32\textwidth}
\includegraphics[scale=0.55]{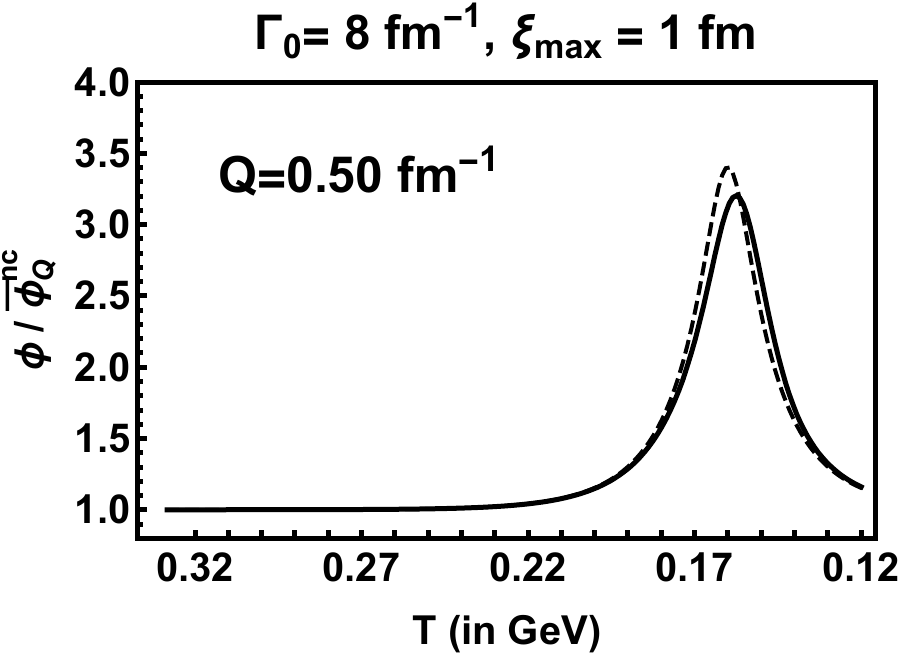}
\end{subfigure}
\begin{subfigure}{0.32\textwidth}
\includegraphics[scale=0.55]{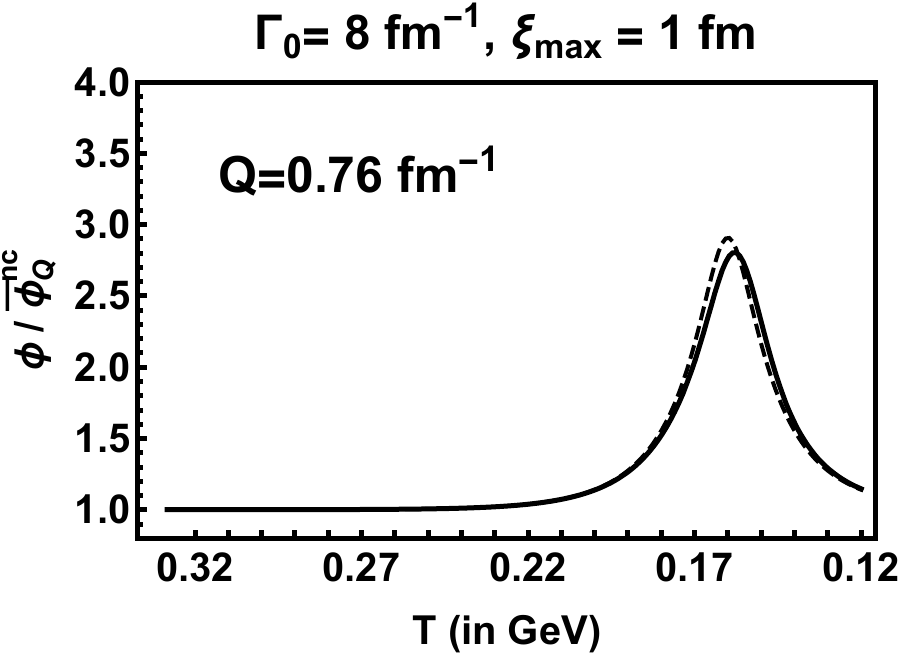}
\end{subfigure}
\begin{subfigure}{0.32\textwidth}
\includegraphics[scale=0.55]{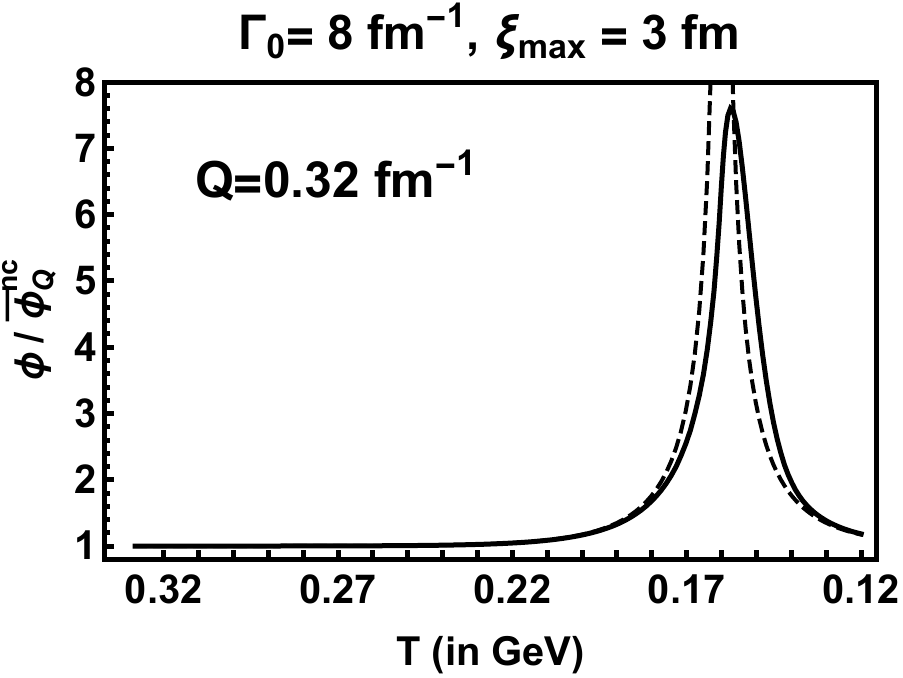}
\end{subfigure}
\begin{subfigure}{0.32\textwidth}
\includegraphics[scale=0.55]{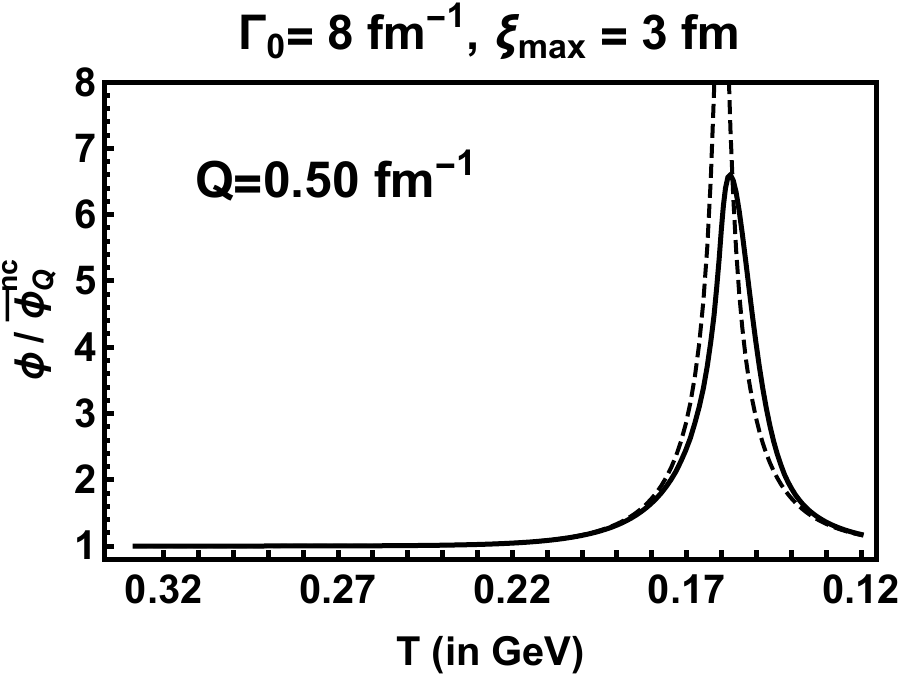}
\end{subfigure}
\begin{subfigure}{0.32\textwidth}
\includegraphics[scale=0.55]{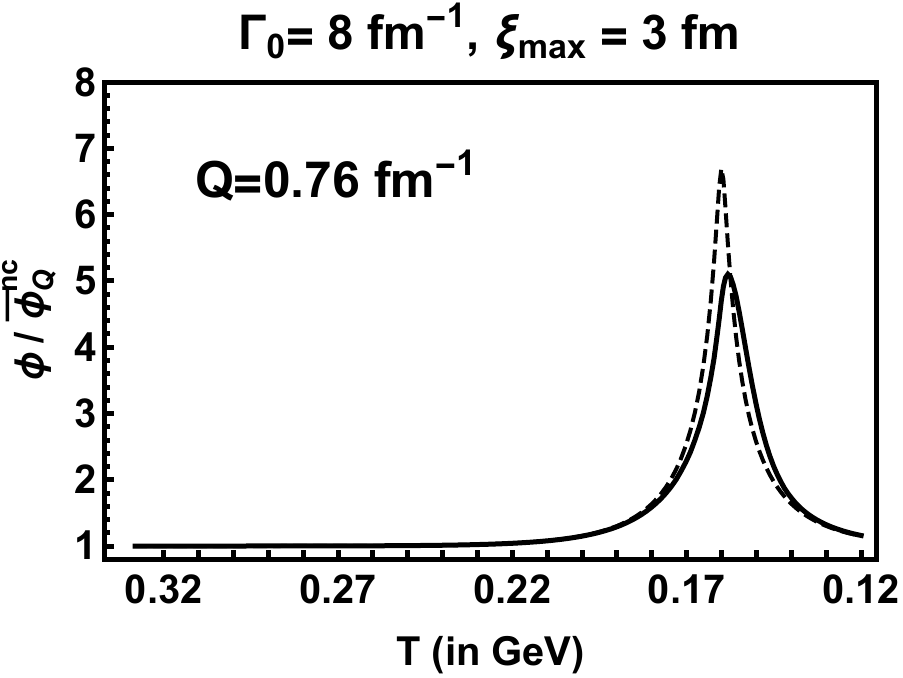}
\end{subfigure}
\end{center}
\caption{The values of $\phi_{\bm Q}$ (suitably normalized) for three
  representative values of $Q$ (same for each column), and for two values
  $\Gamma_0$ (same in top and bottom six panels)
  and $\xi_{\rm max}$ (same in alternating rows) as in Fig.~\ref{phi-1chA}. The values of
  $\phi_{\bm Q}$ are taken along a fluid cell trajectory and
  plotted as a function of temperature, which is a monotonous function
  of time $\tau$ along the trajectory. The trajectory chosen for these
  plots begins at $r_i=r(\tau_i)=1.8\, \text{fm}$. The dashed and solid curves
  represent the equilibrium $\bar{\phi}_{\bm Q}$ and non-equilibrium
  $\phi_{\bm Q}$, respectively.
}
\label{phi-ch2A}
\end{figure}

	 In Fig.~\ref{phi-ch2A}, which can be compared to the analogous 
	 Model H results shown in Fig.~\ref{phi-ch2}, for all three representative $Q$ modes that are plotted we notice the $\phi_\mathbf{Q}$'s lagging behind their respective $\bar{\phi}_\mathbf{Q}$s,
	 with the degree to which they fall out of equilibrium greater for smaller $\Gamma_0$, meaning slower relaxation toward equilibrium.  For the values of $\Gamma_0$ that we have 
	 considered in Fig.~\ref{phi-ch2A}, we can see that fluctuations do depend on whether we choose a freeze-out temperature of 156~MeV or 140~MeV. 
	 As we also observed in Fig.~\ref{phi-ch2}, $\phi_\mathbf{Q}$ has an inflection point at $T=T_c$ where the relaxation rate takes its minimum value and the growth of $\phi_\mathbf{Q}$ stops when $\phi_\mathbf{Q}$ equals the instantaneous $\bar{\phi}_\mathbf{Q}$.

\subsection{Fluctuations on the freezeout surface}

\begin{figure}[t]
\begin{center}
\begin{subfigure}{0.49\textwidth}
\includegraphics[scale=0.8]{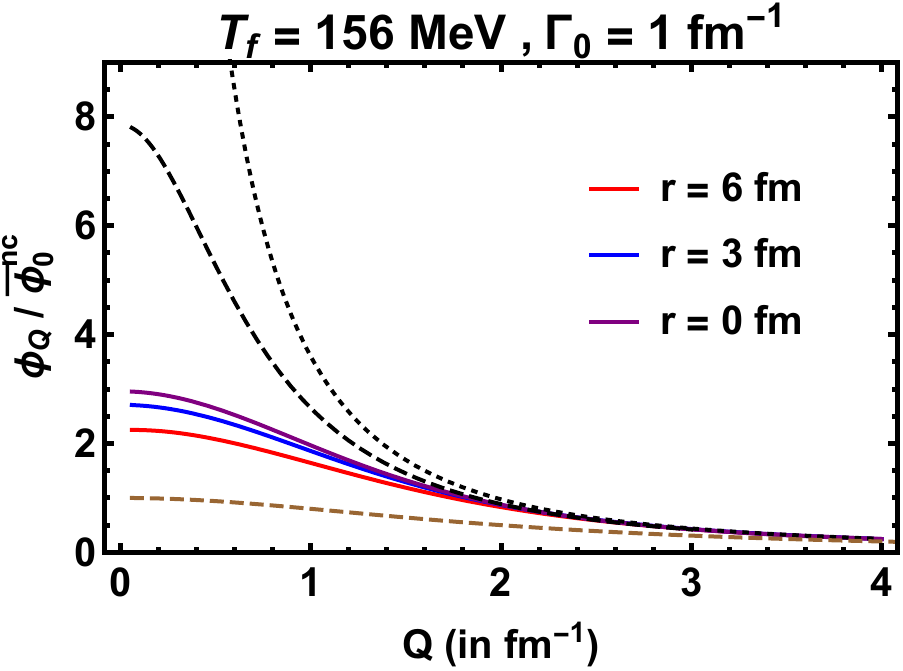}
\end{subfigure}
\begin{subfigure}{0.49\textwidth}
\includegraphics[scale=0.8]{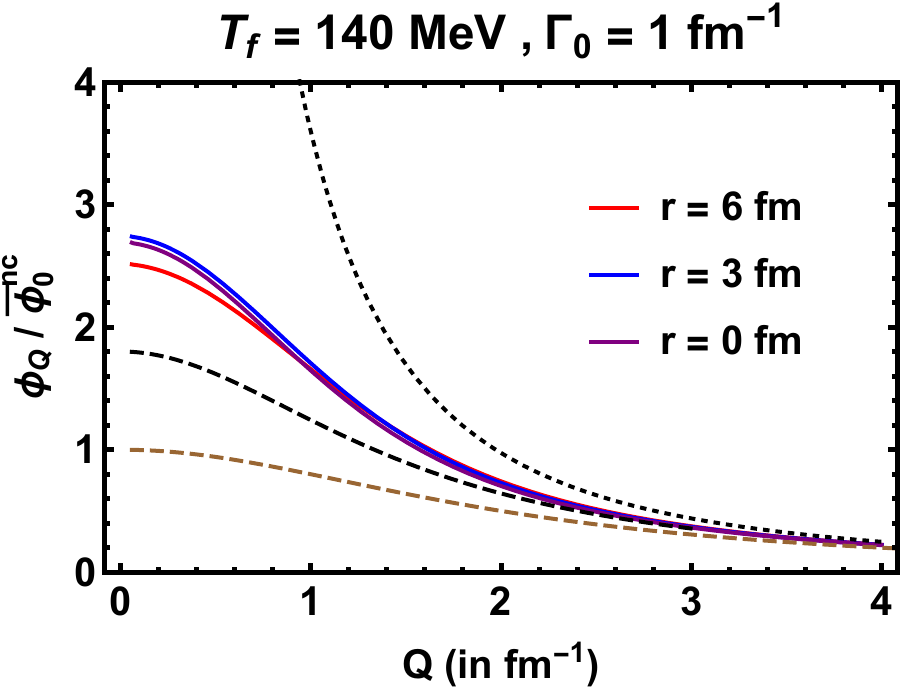}
\end{subfigure}
\begin{subfigure}{0.49\textwidth}
\includegraphics[scale=0.8]{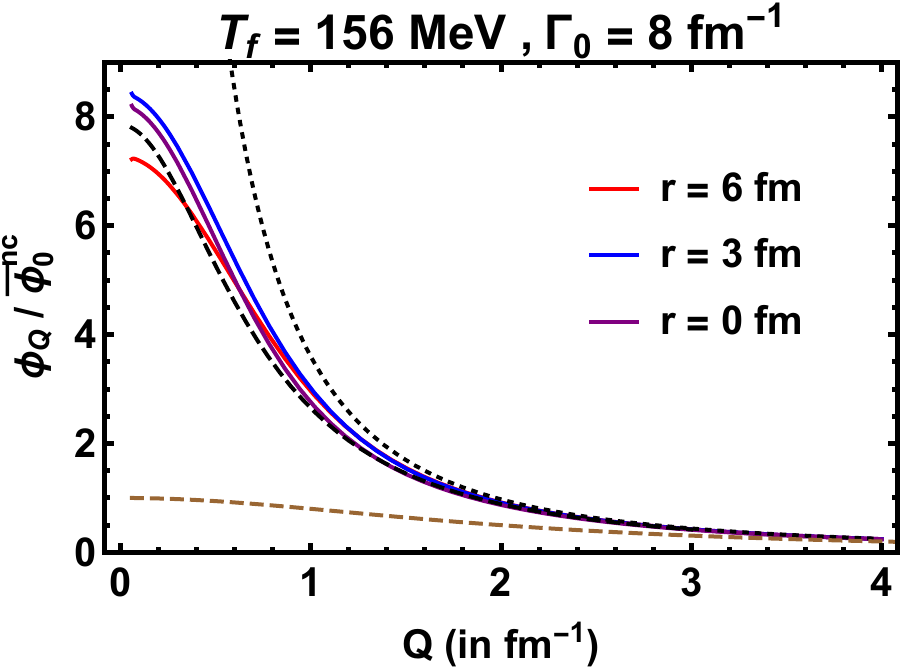}
\end{subfigure}
\begin{subfigure}{0.49\textwidth}
\includegraphics[scale=0.8]{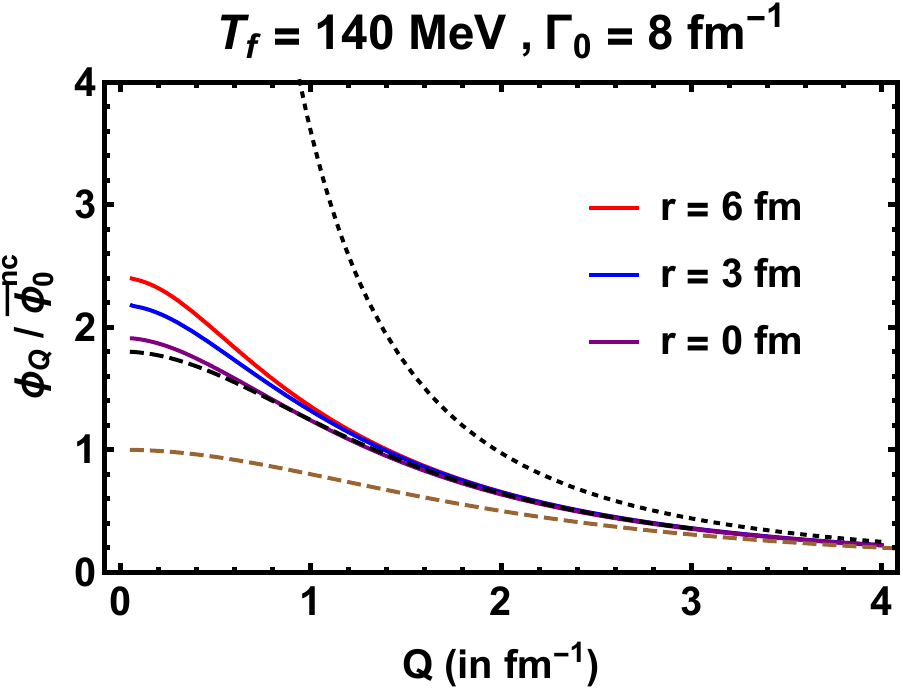}
\end{subfigure}
\end{center}
\caption{The Hydro+ variable $\phi_{\bm Q}$ (normalized to its value
  at $Q=0$ away from the critical point, where $\xi=\xi_0$) at
  freezeout evolved with $\Gamma_0=1$~fm$^{-1}$ (upper panels) and $8$~fm$^{-1}$
  (lower panels) and with $\xi_{\text{max}}=3\, \text{fm}$. 
 The left (right) panels show results for evolution until the decreasing temperature has reached 
  a higher (lower) freeze-out temperature.
    The
  blue, red and purple curves show the values of $\phi_{\bm Q}$ at
  different points on the freezeout hypersurface, characterized by the radial
  coordinate $r$. The black dashed and dotted
  curves are the equilibrium curves at $T=T_f$ and
  $T=T_c$ respectively. The dashed brown curve is
  the (non-critical) equilibrium curve corresponding to
  $\xi=\xi_0$.
}
\label{phi-1fhs1A}
\end{figure}

In Fig.~\ref{phi-1fhs1A}, which can be compared 
to Fig.~\ref{phi-1fhs1}, suitably normalized plots of
$\phi_{\bm Q}$ are shown for three points on the freeze-out
hypersurface, characterized by radial coordinate $r=0, 3$ and $6$ fm, 
for two choices of freezeout temperature $T_f$ and two values
of the parameter $\Gamma_0$.   Most of the discussion of Fig.~\ref{phi-1fhs1} in Section~\ref{RRWY} applies here also, with the one significant difference being that here 
the $Q=0$ modes are not ``stuck'' at their initial values.

\begin{figure}[t]
\begin{center}
\begin{subfigure}{0.49\textwidth}
\includegraphics[scale=0.8]{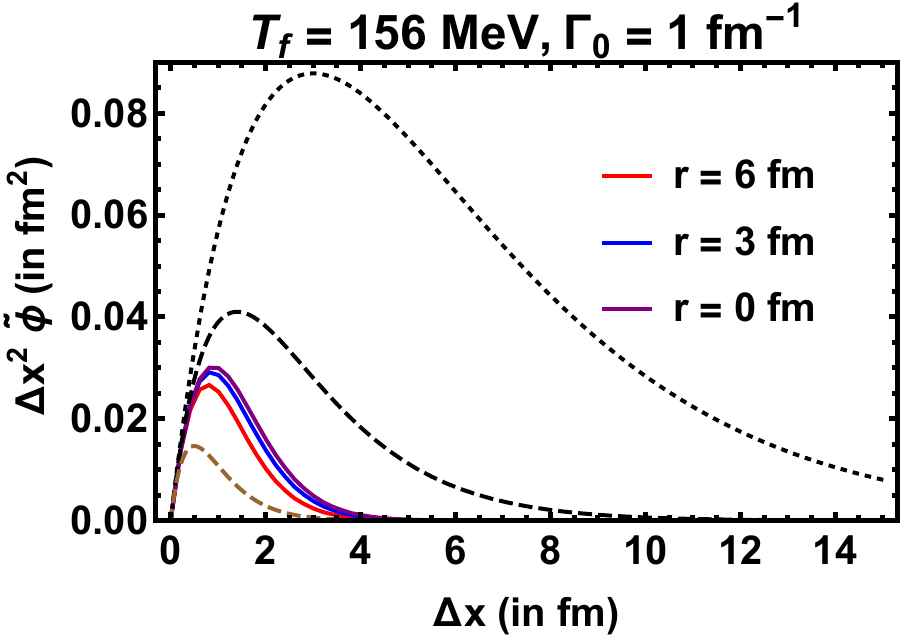}
\end{subfigure}
\begin{subfigure}{0.49\textwidth}
\includegraphics[scale=0.8]{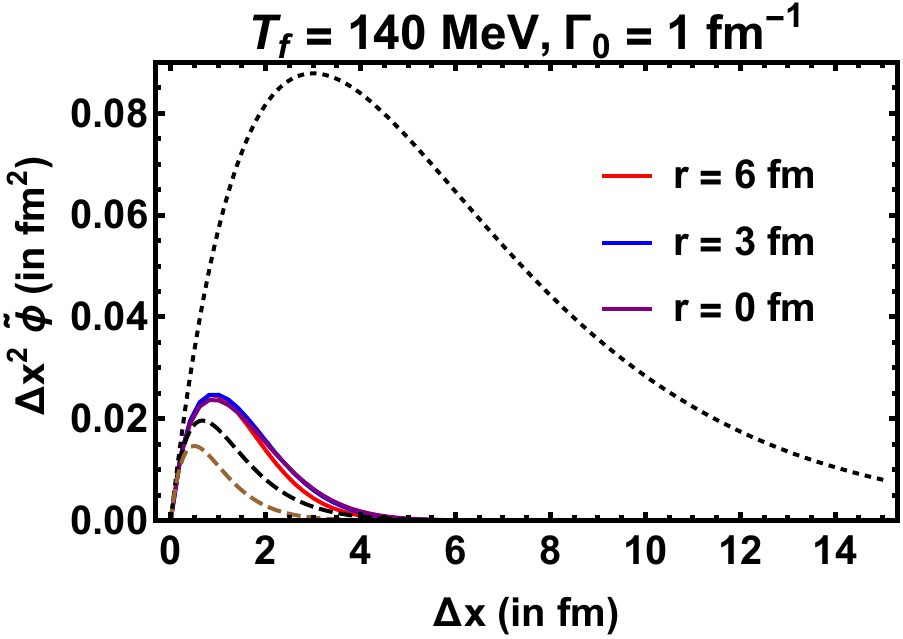}
\end{subfigure}
\begin{subfigure}{0.49\textwidth}
\includegraphics[scale=0.8]{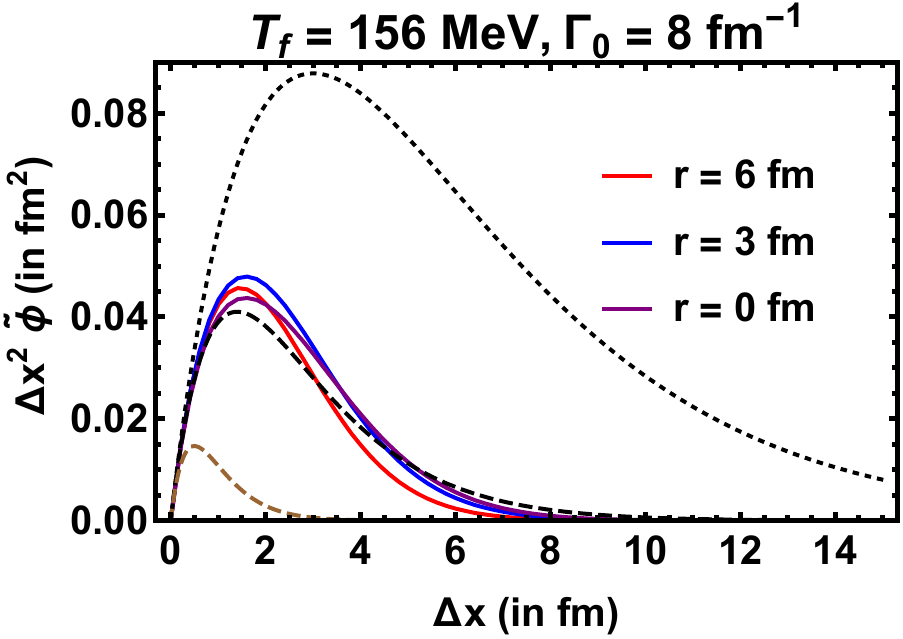}
\end{subfigure}
\begin{subfigure}{0.49\textwidth}
\includegraphics[scale=0.8]{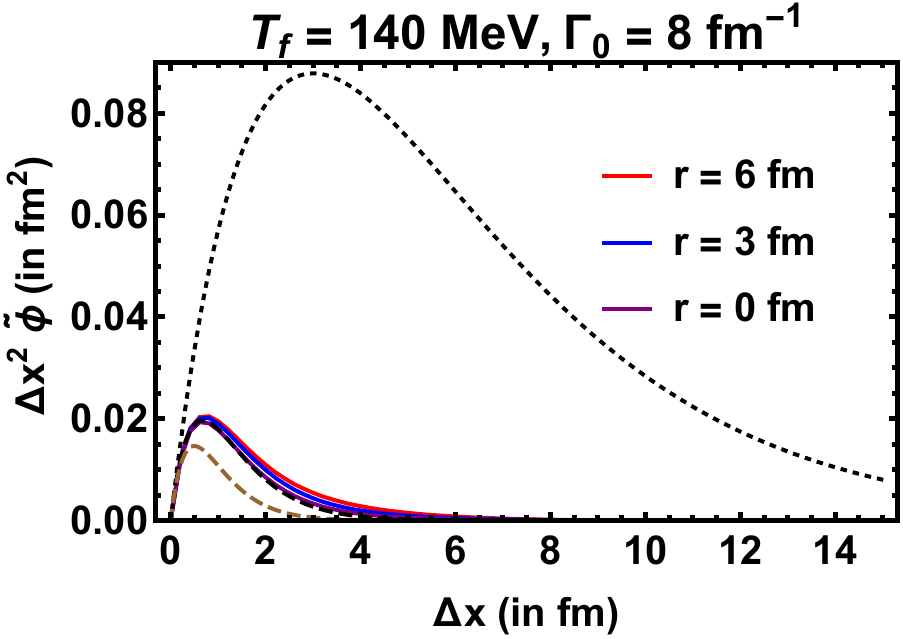}
\end{subfigure}
\end{center}
\caption{
$\tilde{\phi}\times\Delta x^2$, the measure of fluctuations of $\hat{s}$ described by the correlator
$\langle \delta\hat s(x_+)  \delta\hat s(x_-)\rangle$,  at freezeout  as a function of the spatial separation between the points $\Delta x\equiv |\Delta {\bm x_\perp}|$.     In the calculations depicted in different panels, the $\phi_{\bm Q}$'s were evolved with two different $\Gamma_0$'s until freezeout at two different $T_f$'s, with the inverse Fourier transform to obtain 
$\tilde \phi ({\bm x_\perp)}$ performed at $T_f$.
In all panels, we have chosen a trajectory with $\xi_{\text{max}}=3\, \text{fm}$. 
The three $r$ values depicted via the colored curves 
correspond to three $r$ values on the freeze-out surface in the lab frame. The black dashed and dotted curves are the equilibrium curves at $T=T_f$ and $T=T_c$ respectively. The dashed brown curve is the (non-critical) equilibrium curve corresponding to $\xi=\xi_0$.}
\label{phiFT1A}
\end{figure}

As in Fig.~\ref{phiFT1}, in Fig.~\ref{phiFT1A} 
we have computed 
$\tilde\phi(\Delta{\bm x_\perp})$, the inverse Fourier transform of $\phi_{\bm Q}$ defined in Eq.~(\ref{eq:phitilde}), and plotted
$\Delta x^2 \tilde\phi(\Delta{\bm x_\perp})$ as a function of the spatial separation
$\Delta x$ between the two points in the correlator $\langle \delta\hat s( x_+) \delta \hat s(x_-)\rangle$.
As in the Model H evolution of Fig.~\ref{phiFT1}, 
the small 
$\Delta x$ (large $Q$) behavior of the fluctuations in Fig.~\ref{phiFT1A}  
is not affected by changing $\Gamma_0$, while
at the same time the spatial correlator becomes longer ranged as $\Gamma_0$ is increased.
The central difference between the Model A dynamics here in Fig.~\ref{phiFT1A} 
and the Model H dynamics in 
Fig.~\ref{phiFT1} is that here $\tilde\phi(\Delta{\bm x_\perp})$ is positive at
large $\Delta x$: the fact that it becomes negative in the large $\Delta x$ region in Fig.~\ref{phiFT1}
is a direct consequence of conservation  in Model H.

\subsection{Variance of particle multiplicities}

\begin{figure}[t]
\begin{center}
\begin{subfigure}{0.49\textwidth}
\includegraphics[scale=0.6]{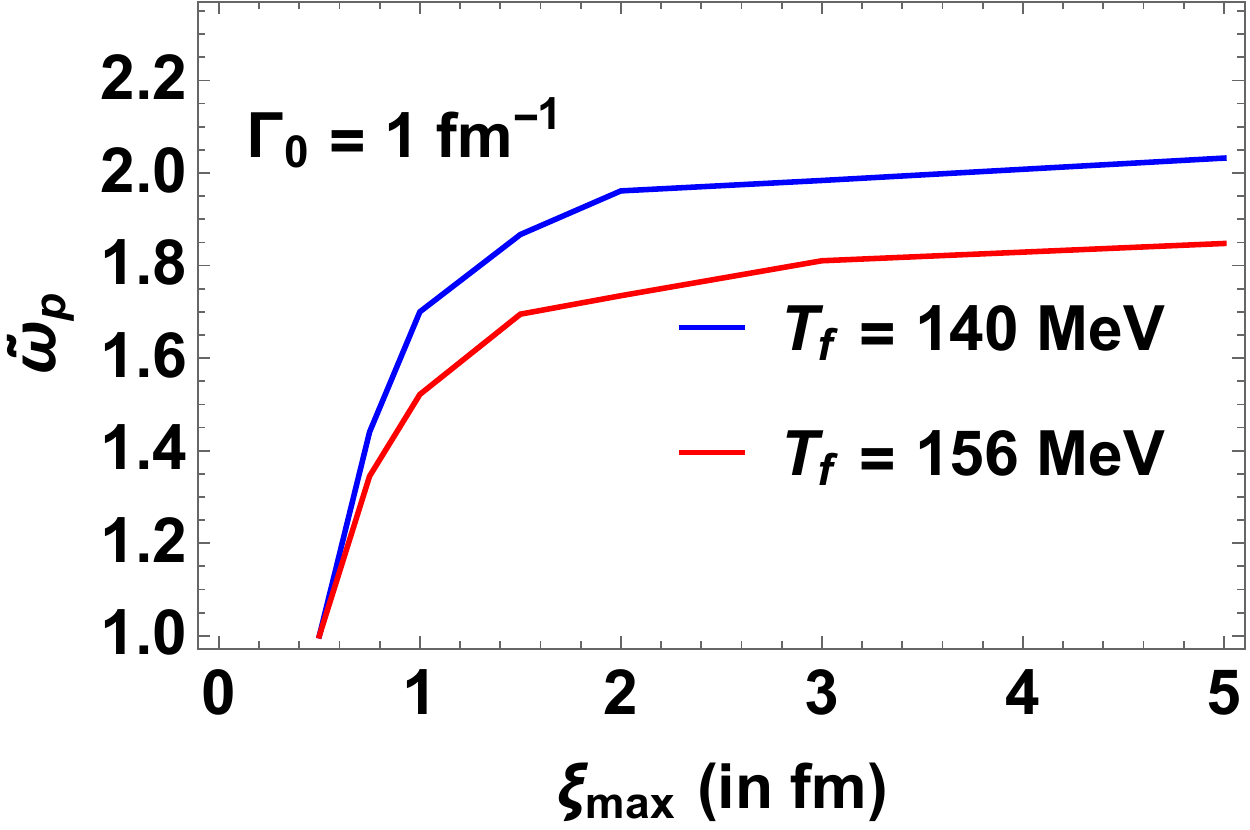}
\subcaption{}
\end{subfigure}
\begin{subfigure}{0.49\textwidth}
\includegraphics[scale=0.6]{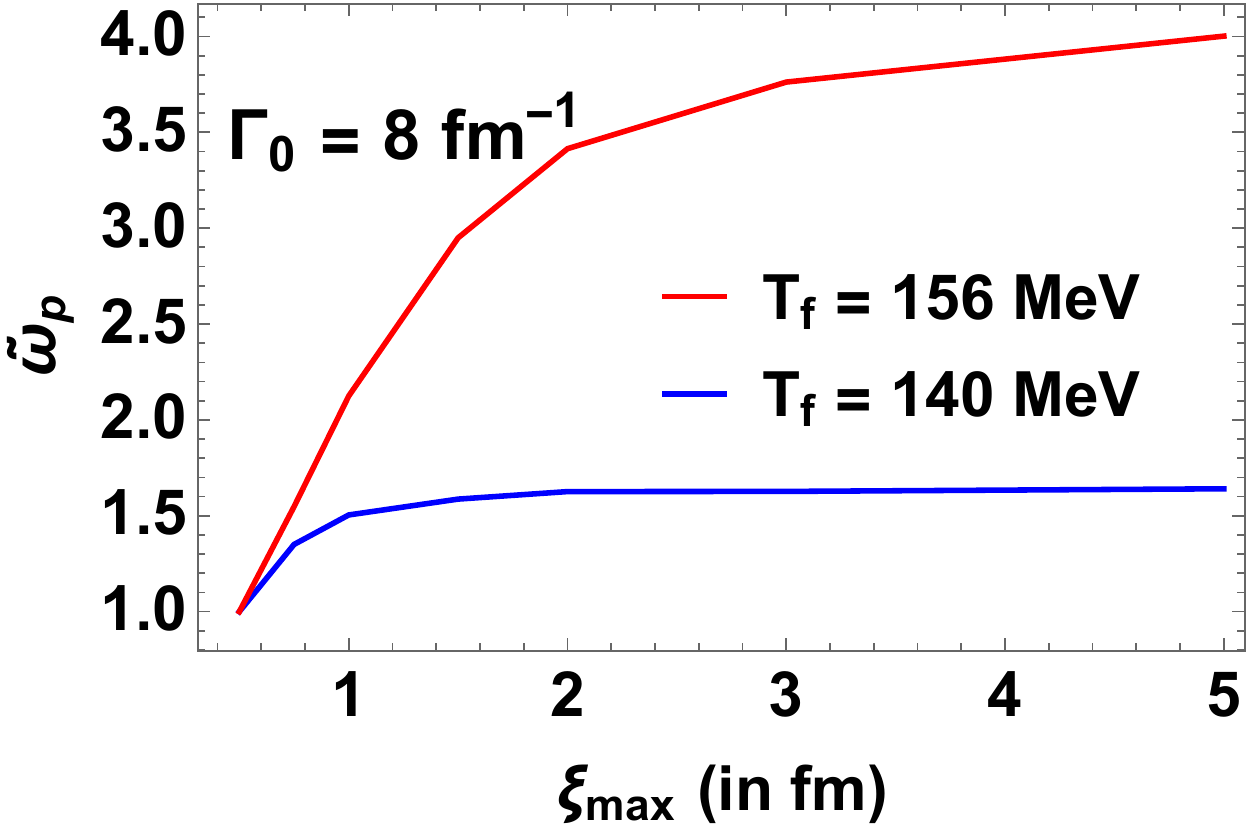}
\subcaption{}
\end{subfigure}
\begin{subfigure}{0.49\textwidth}
\includegraphics[scale=0.6]{images/plots_ModelH/omega_p_vsximax_equilibrium.pdf}
\subcaption{}
\end{subfigure}
\end{center}
\caption{Normalized measure of the fluctuations in proton multiplicity, $\tilde{\omega}_p=\frac{\omega_p}{\omega^{\text{nc}}_p}$, as a function of the maximum equilibrium correlation length along the system trajectory, which is to say as a function of how closely the trajectory passes the critical point. As $\Gamma_0\rightarrow\infty$, the $\tilde{\omega}_p$'s approach their equilibrium values shown in panel (c).}
\label{fhs-protonsA}
\end{figure}

\begin{figure}[t]
\begin{center}
\begin{subfigure}{0.49\textwidth}
\includegraphics[scale=0.6]{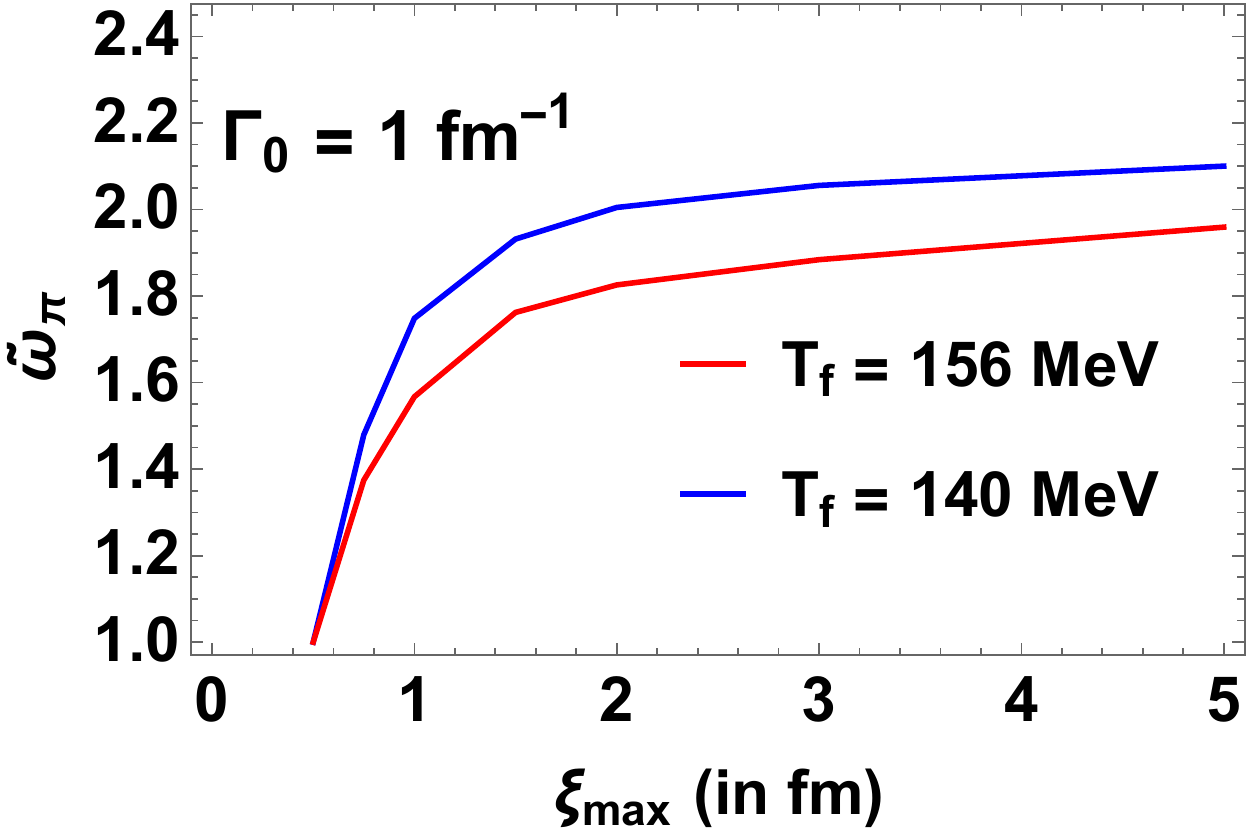}
\subcaption{}
\label{unusedlabel21a}
\end{subfigure}
\begin{subfigure}{0.49\textwidth}
\includegraphics[scale=0.6]{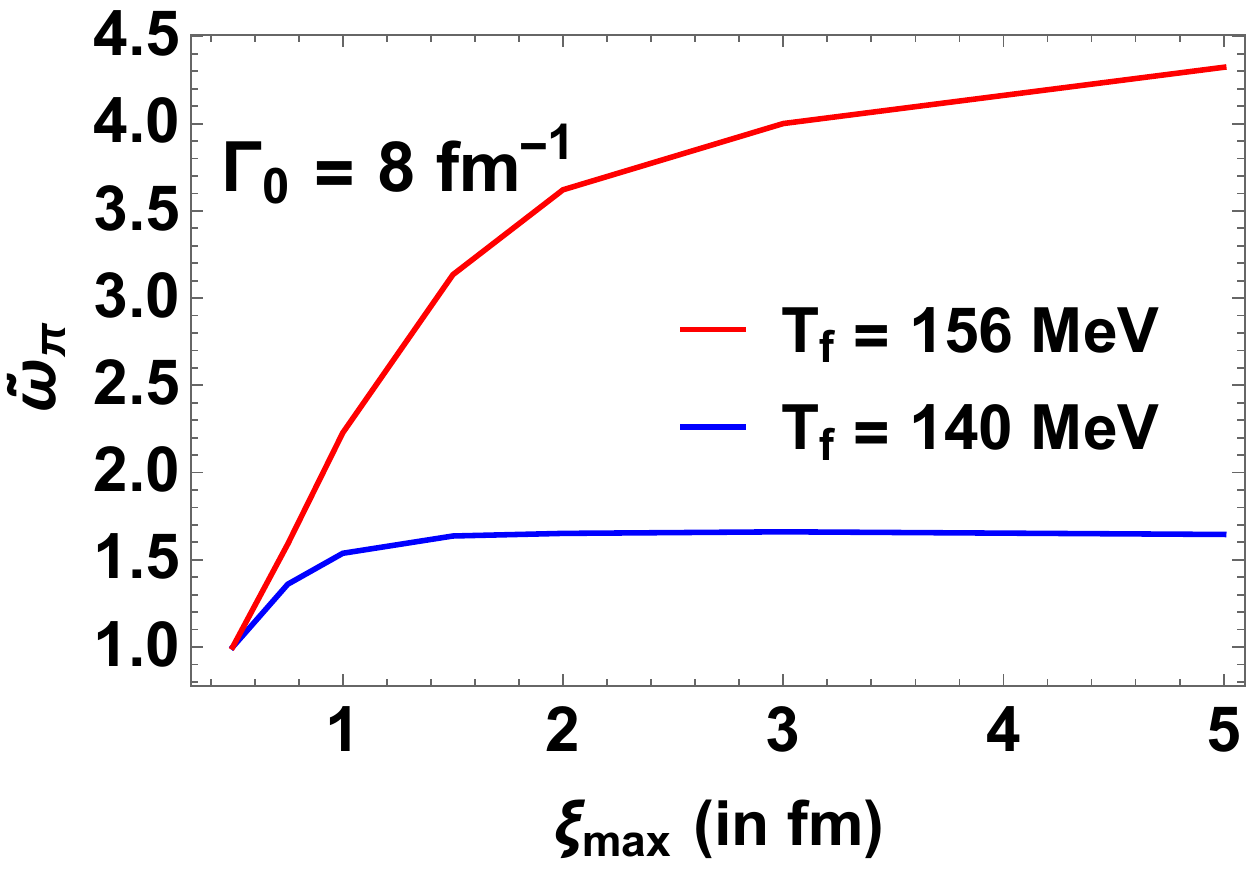}
\subcaption{}
\label{unusedlabel21b}
\end{subfigure}
\begin{subfigure}{0.49\textwidth}
\includegraphics[scale=0.6]{images/plots_ModelH/omega_pi_vsximax_equilibrium.pdf}
\subcaption{}
\label{unusedlabel21c}
\end{subfigure}
\end{center}
\caption{Normalized measure of the fluctuations in pion multiplicity, $\tilde{\omega}_\pi=\frac{\omega_\pi}{\omega^{\text{nc}}_\pi}$, as a function of the maximum equilibrium correlation length along the system trajectory, which is to say as a function of how closely the trajectory passes the critical point. As $\Gamma_0\rightarrow\infty$, the $\tilde{\omega}_\pi$'s approach their equilibrium values shown in panel (c).
}
\label{fhs-pionsA}
\end{figure}

As in Section~\ref{RRWY}, but here with Model A dynamics, we close by computing and plotting
the normalized fluctuation measure for protons and pions, $\tilde{\omega}_p$ and $\tilde{\omega}_\pi$, in Figs.~\ref{fhs-protonsA} and~\ref{fhs-pionsA}. As in Figs.~\ref{fhs-protons} and ~\ref{fhs-pions}, to which these figures can be compared, these results demonstrate that for
trajectories passing closer to the critical point (i.e., for
trajectories with larger $\xi_{\rm max}$) the magnitude of
fluctuations is larger. Again
as in Section~\ref{RRWY}, the magnitude of the effect depends on the
rate of the relaxation of the fluctuations, controlled here by
parameter $\Gamma_0$. 
We see that for large enough values of $\Gamma_0$, eg.~8~fm$^{-1}$,
the proton and pion fluctuations are able to come reasonably close to their
equilibrium values, which as an aside means that they depend quite sensitively on
the freeze-out temperature.  The differences that we discussed at length in Section~\ref{RRWY} originate from the effects of conservation.




\bibliography{refs}

\begin{thebibliography}{74}%
\makeatletter
\providecommand \@ifxundefined [1]{%
 \@ifx{#1\undefined}
}%
\providecommand \@ifnum [1]{%
 \ifnum #1\expandafter \@firstoftwo
 \else \expandafter \@secondoftwo
 \fi
}%
\providecommand \@ifx [1]{%
 \ifx #1\expandafter \@firstoftwo
 \else \expandafter \@secondoftwo
 \fi
}%
\providecommand \natexlab [1]{#1}%
\providecommand \enquote  [1]{``#1''}%
\providecommand \bibnamefont  [1]{#1}%
\providecommand \bibfnamefont [1]{#1}%
\providecommand \citenamefont [1]{#1}%
\providecommand \href@noop [0]{\@secondoftwo}%
\providecommand \href [0]{\begingroup \@sanitize@url \@href}%
\providecommand \@href[1]{\@@startlink{#1}\@@href}%
\providecommand \@@href[1]{\endgroup#1\@@endlink}%
\providecommand \@sanitize@url [0]{\catcode `\\12\catcode `\$12\catcode
  `\&12\catcode `\#12\catcode `\^12\catcode `\_12\catcode `\%12\relax}%
\providecommand \@@startlink[1]{}%
\providecommand \@@endlink[0]{}%
\providecommand \url  [0]{\begingroup\@sanitize@url \@url }%
\providecommand \@url [1]{\endgroup\@href {#1}{\urlprefix }}%
\providecommand \urlprefix  [0]{URL }%
\providecommand \Eprint [0]{\href }%
\providecommand \doibase [0]{https://doi.org/}%
\providecommand \selectlanguage [0]{\@gobble}%
\providecommand \bibinfo  [0]{\@secondoftwo}%
\providecommand \bibfield  [0]{\@secondoftwo}%
\providecommand \translation [1]{[#1]}%
\providecommand \BibitemOpen [0]{}%
\providecommand \bibitemStop [0]{}%
\providecommand \bibitemNoStop [0]{.\EOS\space}%
\providecommand \EOS [0]{\spacefactor3000\relax}%
\providecommand \BibitemShut  [1]{\csname bibitem#1\endcsname}%
\let\auto@bib@innerbib\@empty
\bibitem [{\citenamefont {Rajagopal}\ \emph {et~al.}(2020)\citenamefont
  {Rajagopal}, \citenamefont {Ridgway}, \citenamefont {Weller},\ and\
  \citenamefont {Yin}}]{Rajagopal:2019xwg}%
  \BibitemOpen
  \bibfield  {author} {\bibinfo {author} {\bibfnamefont {K.}~\bibnamefont
  {Rajagopal}}, \bibinfo {author} {\bibfnamefont {G.}~\bibnamefont {Ridgway}},
  \bibinfo {author} {\bibfnamefont {R.}~\bibnamefont {Weller}},\ and\ \bibinfo
  {author} {\bibfnamefont {Y.}~\bibnamefont {Yin}},\ }\bibfield  {title}
  {\bibinfo {title} {{Understanding the out-of-equilibrium dynamics near a
  critical point in the QCD phase diagram}},\ }\href
  {https://doi.org/10.1103/PhysRevD.102.094025} {\bibfield  {journal} {\bibinfo
   {journal} {Phys. Rev. D}\ }\textbf {\bibinfo {volume} {102}},\ \bibinfo
  {pages} {094025} (\bibinfo {year} {2020})},\ \Eprint
  {https://arxiv.org/abs/1908.08539} {arXiv:1908.08539 [hep-ph]} \BibitemShut
  {NoStop}%
\bibitem [{\citenamefont {Aprahamian}\ \emph {et~al.}(2015)\citenamefont
  {Aprahamian} \emph {et~al.}}]{Aprahamian:2015qub}%
  \BibitemOpen
  \bibfield  {author} {\bibinfo {author} {\bibfnamefont {A.}~\bibnamefont
  {Aprahamian}} \emph {et~al.},\ }\bibfield  {title} {\bibinfo {title}
  {{Reaching for the horizon: The 2015 long range plan for nuclear science}},\
  }\href@noop {} {\  (\bibinfo {year} {2015})}\BibitemShut {NoStop}%
\bibitem [{\citenamefont {Busza}\ \emph {et~al.}(2018)\citenamefont {Busza},
  \citenamefont {Rajagopal},\ and\ \citenamefont {van~der
  Schee}}]{Busza:2018rrf}%
  \BibitemOpen
  \bibfield  {author} {\bibinfo {author} {\bibfnamefont {W.}~\bibnamefont
  {Busza}}, \bibinfo {author} {\bibfnamefont {K.}~\bibnamefont {Rajagopal}},\
  and\ \bibinfo {author} {\bibfnamefont {W.}~\bibnamefont {van~der Schee}},\
  }\bibfield  {title} {\bibinfo {title} {{Heavy Ion Collisions: The Big
  Picture, and the Big Questions}},\ }\href
  {https://doi.org/10.1146/annurev-nucl-101917-020852} {\bibfield  {journal}
  {\bibinfo  {journal} {Ann. Rev. Nucl. Part. Sci.}\ }\textbf {\bibinfo
  {volume} {68}},\ \bibinfo {pages} {339} (\bibinfo {year} {2018})},\ \Eprint
  {https://arxiv.org/abs/1802.04801} {arXiv:1802.04801 [hep-ph]} \BibitemShut
  {NoStop}%
\bibitem [{\citenamefont {Bzdak}\ \emph {et~al.}(2020)\citenamefont {Bzdak},
  \citenamefont {Esumi}, \citenamefont {Koch}, \citenamefont {Liao},
  \citenamefont {Stephanov},\ and\ \citenamefont {Xu}}]{Bzdak:2019pkr}%
  \BibitemOpen
  \bibfield  {author} {\bibinfo {author} {\bibfnamefont {A.}~\bibnamefont
  {Bzdak}}, \bibinfo {author} {\bibfnamefont {S.}~\bibnamefont {Esumi}},
  \bibinfo {author} {\bibfnamefont {V.}~\bibnamefont {Koch}}, \bibinfo {author}
  {\bibfnamefont {J.}~\bibnamefont {Liao}}, \bibinfo {author} {\bibfnamefont
  {M.}~\bibnamefont {Stephanov}},\ and\ \bibinfo {author} {\bibfnamefont
  {N.}~\bibnamefont {Xu}},\ }\bibfield  {title} {\bibinfo {title} {{Mapping the
  Phases of Quantum Chromodynamics with Beam Energy Scan}},\ }\href
  {https://doi.org/10.1016/j.physrep.2020.01.005} {\bibfield  {journal}
  {\bibinfo  {journal} {Phys. Rept.}\ }\textbf {\bibinfo {volume} {853}},\
  \bibinfo {pages} {1} (\bibinfo {year} {2020})},\ \Eprint
  {https://arxiv.org/abs/1906.00936} {arXiv:1906.00936 [nucl-th]} \BibitemShut
  {NoStop}%
\bibitem [{\citenamefont {Abdallah}\ \emph {et~al.}(2021)\citenamefont
  {Abdallah} \emph {et~al.}}]{STAR:2021iop}%
  \BibitemOpen
  \bibfield  {author} {\bibinfo {author} {\bibfnamefont {M.}~\bibnamefont
  {Abdallah}} \emph {et~al.} (\bibinfo {collaboration} {STAR}),\ }\bibfield
  {title} {\bibinfo {title} {{Cumulants and Correlation Functions of
  Net-proton, Proton and Antiproton Multiplicity Distributions in Au+Au
  Collisions at RHIC}},\ }\href@noop {} {\  (\bibinfo {year} {2021})},\ \Eprint
  {https://arxiv.org/abs/2101.12413} {arXiv:2101.12413 [nucl-ex]} \BibitemShut
  {NoStop}%
\bibitem [{\citenamefont {Stephanov}\ \emph {et~al.}(1998)\citenamefont
  {Stephanov}, \citenamefont {Rajagopal},\ and\ \citenamefont
  {Shuryak}}]{Stephanov:1998dy}%
  \BibitemOpen
  \bibfield  {author} {\bibinfo {author} {\bibfnamefont {M.~A.}\ \bibnamefont
  {Stephanov}}, \bibinfo {author} {\bibfnamefont {K.}~\bibnamefont
  {Rajagopal}},\ and\ \bibinfo {author} {\bibfnamefont {E.~V.}\ \bibnamefont
  {Shuryak}},\ }\bibfield  {title} {\bibinfo {title} {{Signatures of the
  tricritical point in QCD}},\ }\href
  {https://doi.org/10.1103/PhysRevLett.81.4816} {\bibfield  {journal} {\bibinfo
   {journal} {Phys. Rev. Lett.}\ }\textbf {\bibinfo {volume} {81}},\ \bibinfo
  {pages} {4816} (\bibinfo {year} {1998})},\ \Eprint
  {https://arxiv.org/abs/hep-ph/9806219} {arXiv:hep-ph/9806219} \BibitemShut
  {NoStop}%
\bibitem [{\citenamefont {Stephanov}\ \emph {et~al.}(1999)\citenamefont
  {Stephanov}, \citenamefont {Rajagopal},\ and\ \citenamefont
  {Shuryak}}]{Stephanov:1999zu}%
  \BibitemOpen
  \bibfield  {author} {\bibinfo {author} {\bibfnamefont {M.~A.}\ \bibnamefont
  {Stephanov}}, \bibinfo {author} {\bibfnamefont {K.}~\bibnamefont
  {Rajagopal}},\ and\ \bibinfo {author} {\bibfnamefont {E.~V.}\ \bibnamefont
  {Shuryak}},\ }\bibfield  {title} {\bibinfo {title} {{Event-by-event
  fluctuations in heavy ion collisions and the QCD critical point}},\ }\href
  {https://doi.org/10.1103/PhysRevD.60.114028} {\bibfield  {journal} {\bibinfo
  {journal} {Phys. Rev. D}\ }\textbf {\bibinfo {volume} {60}},\ \bibinfo
  {pages} {114028} (\bibinfo {year} {1999})},\ \Eprint
  {https://arxiv.org/abs/hep-ph/9903292} {arXiv:hep-ph/9903292} \BibitemShut
  {NoStop}%
\bibitem [{\citenamefont {Hatta}\ and\ \citenamefont
  {Stephanov}(2003)}]{Hatta:2003wn}%
  \BibitemOpen
  \bibfield  {author} {\bibinfo {author} {\bibfnamefont {Y.}~\bibnamefont
  {Hatta}}\ and\ \bibinfo {author} {\bibfnamefont {M.~A.}\ \bibnamefont
  {Stephanov}},\ }\bibfield  {title} {\bibinfo {title} {{Proton number
  fluctuation as a signal of the QCD critical endpoint}},\ }\href
  {https://doi.org/10.1103/PhysRevLett.91.102003} {\bibfield  {journal}
  {\bibinfo  {journal} {Phys. Rev. Lett.}\ }\textbf {\bibinfo {volume} {91}},\
  \bibinfo {pages} {102003} (\bibinfo {year} {2003})},\ \bibinfo {note}
  {[Erratum: Phys.Rev.Lett. 91, 129901 (2003)]},\ \Eprint
  {https://arxiv.org/abs/hep-ph/0302002} {arXiv:hep-ph/0302002} \BibitemShut
  {NoStop}%
\bibitem [{\citenamefont {Stephanov}(2009)}]{Stephanov:2008qz}%
  \BibitemOpen
  \bibfield  {author} {\bibinfo {author} {\bibfnamefont {M.~A.}\ \bibnamefont
  {Stephanov}},\ }\bibfield  {title} {\bibinfo {title} {{Non-Gaussian
  fluctuations near the QCD critical point}},\ }\href
  {https://doi.org/10.1103/PhysRevLett.102.032301} {\bibfield  {journal}
  {\bibinfo  {journal} {Phys. Rev. Lett.}\ }\textbf {\bibinfo {volume} {102}},\
  \bibinfo {pages} {032301} (\bibinfo {year} {2009})},\ \Eprint
  {https://arxiv.org/abs/0809.3450} {arXiv:0809.3450 [hep-ph]} \BibitemShut
  {NoStop}%
\bibitem [{\citenamefont {Asakawa}\ \emph {et~al.}(2009)\citenamefont
  {Asakawa}, \citenamefont {Ejiri},\ and\ \citenamefont
  {Kitazawa}}]{Asakawa:2009aj}%
  \BibitemOpen
  \bibfield  {author} {\bibinfo {author} {\bibfnamefont {M.}~\bibnamefont
  {Asakawa}}, \bibinfo {author} {\bibfnamefont {S.}~\bibnamefont {Ejiri}},\
  and\ \bibinfo {author} {\bibfnamefont {M.}~\bibnamefont {Kitazawa}},\
  }\bibfield  {title} {\bibinfo {title} {{Third moments of conserved charges as
  probes of QCD phase structure}},\ }\href
  {https://doi.org/10.1103/PhysRevLett.103.262301} {\bibfield  {journal}
  {\bibinfo  {journal} {Phys. Rev. Lett.}\ }\textbf {\bibinfo {volume} {103}},\
  \bibinfo {pages} {262301} (\bibinfo {year} {2009})},\ \Eprint
  {https://arxiv.org/abs/0904.2089} {arXiv:0904.2089 [nucl-th]} \BibitemShut
  {NoStop}%
\bibitem [{\citenamefont {Athanasiou}\ \emph {et~al.}(2010)\citenamefont
  {Athanasiou}, \citenamefont {Rajagopal},\ and\ \citenamefont
  {Stephanov}}]{Athanasiou:2010kw}%
  \BibitemOpen
  \bibfield  {author} {\bibinfo {author} {\bibfnamefont {C.}~\bibnamefont
  {Athanasiou}}, \bibinfo {author} {\bibfnamefont {K.}~\bibnamefont
  {Rajagopal}},\ and\ \bibinfo {author} {\bibfnamefont {M.}~\bibnamefont
  {Stephanov}},\ }\bibfield  {title} {\bibinfo {title} {{Using Higher Moments
  of Fluctuations and their Ratios in the Search for the QCD Critical Point}},\
  }\href {https://doi.org/10.1103/PhysRevD.82.074008} {\bibfield  {journal}
  {\bibinfo  {journal} {Phys. Rev. D}\ }\textbf {\bibinfo {volume} {82}},\
  \bibinfo {pages} {074008} (\bibinfo {year} {2010})},\ \Eprint
  {https://arxiv.org/abs/1006.4636} {arXiv:1006.4636 [hep-ph]} \BibitemShut
  {NoStop}%
\bibitem [{\citenamefont {Stephanov}(2011)}]{Stephanov:2011pb}%
  \BibitemOpen
  \bibfield  {author} {\bibinfo {author} {\bibfnamefont {M.~A.}\ \bibnamefont
  {Stephanov}},\ }\bibfield  {title} {\bibinfo {title} {{On the sign of
  kurtosis near the QCD critical point}},\ }\href
  {https://doi.org/10.1103/PhysRevLett.107.052301} {\bibfield  {journal}
  {\bibinfo  {journal} {Phys. Rev. Lett.}\ }\textbf {\bibinfo {volume} {107}},\
  \bibinfo {pages} {052301} (\bibinfo {year} {2011})},\ \Eprint
  {https://arxiv.org/abs/1104.1627} {arXiv:1104.1627 [hep-ph]} \BibitemShut
  {NoStop}%
\bibitem [{\citenamefont {Ling}\ and\ \citenamefont
  {Stephanov}(2016)}]{Ling:2015yau}%
  \BibitemOpen
  \bibfield  {author} {\bibinfo {author} {\bibfnamefont {B.}~\bibnamefont
  {Ling}}\ and\ \bibinfo {author} {\bibfnamefont {M.~A.}\ \bibnamefont
  {Stephanov}},\ }\bibfield  {title} {\bibinfo {title} {{Acceptance dependence
  of fluctuation measures near the QCD critical point}},\ }\href
  {https://doi.org/10.1103/PhysRevC.93.034915} {\bibfield  {journal} {\bibinfo
  {journal} {Phys. Rev. C}\ }\textbf {\bibinfo {volume} {93}},\ \bibinfo
  {pages} {034915} (\bibinfo {year} {2016})},\ \Eprint
  {https://arxiv.org/abs/1512.09125} {arXiv:1512.09125 [nucl-th]} \BibitemShut
  {NoStop}%
\bibitem [{\citenamefont {Brewer}\ \emph {et~al.}(2018)\citenamefont {Brewer},
  \citenamefont {Mukherjee}, \citenamefont {Rajagopal},\ and\ \citenamefont
  {Yin}}]{Brewer:2018abr}%
  \BibitemOpen
  \bibfield  {author} {\bibinfo {author} {\bibfnamefont {J.}~\bibnamefont
  {Brewer}}, \bibinfo {author} {\bibfnamefont {S.}~\bibnamefont {Mukherjee}},
  \bibinfo {author} {\bibfnamefont {K.}~\bibnamefont {Rajagopal}},\ and\
  \bibinfo {author} {\bibfnamefont {Y.}~\bibnamefont {Yin}},\ }\bibfield
  {title} {\bibinfo {title} {{Searching for the QCD critical point via the
  rapidity dependence of cumulants}},\ }\href
  {https://doi.org/10.1103/PhysRevC.98.061901} {\bibfield  {journal} {\bibinfo
  {journal} {Phys. Rev. C}\ }\textbf {\bibinfo {volume} {98}},\ \bibinfo
  {pages} {061901} (\bibinfo {year} {2018})},\ \Eprint
  {https://arxiv.org/abs/1804.10215} {arXiv:1804.10215 [hep-ph]} \BibitemShut
  {NoStop}%
\bibitem [{\citenamefont {Berdnikov}\ and\ \citenamefont
  {Rajagopal}(2000)}]{Berdnikov:1999ph}%
  \BibitemOpen
  \bibfield  {author} {\bibinfo {author} {\bibfnamefont {B.}~\bibnamefont
  {Berdnikov}}\ and\ \bibinfo {author} {\bibfnamefont {K.}~\bibnamefont
  {Rajagopal}},\ }\bibfield  {title} {\bibinfo {title} {{Slowing
  out-of-equilibrium near the QCD critical point}},\ }\href
  {https://doi.org/10.1103/PhysRevD.61.105017} {\bibfield  {journal} {\bibinfo
  {journal} {Phys. Rev. D}\ }\textbf {\bibinfo {volume} {61}},\ \bibinfo
  {pages} {105017} (\bibinfo {year} {2000})},\ \Eprint
  {https://arxiv.org/abs/hep-ph/9912274} {arXiv:hep-ph/9912274} \BibitemShut
  {NoStop}%
\bibitem [{\citenamefont {Mukherjee}\ \emph {et~al.}(2015)\citenamefont
  {Mukherjee}, \citenamefont {Venugopalan},\ and\ \citenamefont
  {Yin}}]{Mukherjee:2015swa}%
  \BibitemOpen
  \bibfield  {author} {\bibinfo {author} {\bibfnamefont {S.}~\bibnamefont
  {Mukherjee}}, \bibinfo {author} {\bibfnamefont {R.}~\bibnamefont
  {Venugopalan}},\ and\ \bibinfo {author} {\bibfnamefont {Y.}~\bibnamefont
  {Yin}},\ }\bibfield  {title} {\bibinfo {title} {{Real time evolution of
  non-Gaussian cumulants in the QCD critical regime}},\ }\href
  {https://doi.org/10.1103/PhysRevC.92.034912} {\bibfield  {journal} {\bibinfo
  {journal} {Phys. Rev. C}\ }\textbf {\bibinfo {volume} {92}},\ \bibinfo
  {pages} {034912} (\bibinfo {year} {2015})},\ \Eprint
  {https://arxiv.org/abs/1506.00645} {arXiv:1506.00645 [hep-ph]} \BibitemShut
  {NoStop}%
\bibitem [{\citenamefont {Mukherjee}\ \emph {et~al.}(2016)\citenamefont
  {Mukherjee}, \citenamefont {Venugopalan},\ and\ \citenamefont
  {Yin}}]{Mukherjee:2016kyu}%
  \BibitemOpen
  \bibfield  {author} {\bibinfo {author} {\bibfnamefont {S.}~\bibnamefont
  {Mukherjee}}, \bibinfo {author} {\bibfnamefont {R.}~\bibnamefont
  {Venugopalan}},\ and\ \bibinfo {author} {\bibfnamefont {Y.}~\bibnamefont
  {Yin}},\ }\bibfield  {title} {\bibinfo {title} {{Universal off-equilibrium
  scaling of critical cumulants in the QCD phase diagram}},\ }\href
  {https://doi.org/10.1103/PhysRevLett.117.222301} {\bibfield  {journal}
  {\bibinfo  {journal} {Phys. Rev. Lett.}\ }\textbf {\bibinfo {volume} {117}},\
  \bibinfo {pages} {222301} (\bibinfo {year} {2016})},\ \Eprint
  {https://arxiv.org/abs/1605.09341} {arXiv:1605.09341 [hep-ph]} \BibitemShut
  {NoStop}%
\bibitem [{\citenamefont {Stephanov}\ and\ \citenamefont
  {Yin}(2018)}]{Stephanov:2017ghc}%
  \BibitemOpen
  \bibfield  {author} {\bibinfo {author} {\bibfnamefont {M.}~\bibnamefont
  {Stephanov}}\ and\ \bibinfo {author} {\bibfnamefont {Y.}~\bibnamefont
  {Yin}},\ }\bibfield  {title} {\bibinfo {title} {{Hydrodynamics with
  parametric slowing down and fluctuations near the critical point}},\ }\href
  {https://doi.org/10.1103/PhysRevD.98.036006} {\bibfield  {journal} {\bibinfo
  {journal} {Phys. Rev.}\ }\textbf {\bibinfo {volume} {D98}},\ \bibinfo {pages}
  {036006} (\bibinfo {year} {2018})},\ \Eprint
  {https://arxiv.org/abs/1712.10305} {arXiv:1712.10305 [nucl-th]} \BibitemShut
  {NoStop}%
\bibitem [{\citenamefont {Akamatsu}\ \emph {et~al.}(2017)\citenamefont
  {Akamatsu}, \citenamefont {Mazeliauskas},\ and\ \citenamefont
  {Teaney}}]{Akamatsu:2016llw}%
  \BibitemOpen
  \bibfield  {author} {\bibinfo {author} {\bibfnamefont {Y.}~\bibnamefont
  {Akamatsu}}, \bibinfo {author} {\bibfnamefont {A.}~\bibnamefont
  {Mazeliauskas}},\ and\ \bibinfo {author} {\bibfnamefont {D.}~\bibnamefont
  {Teaney}},\ }\bibfield  {title} {\bibinfo {title} {{A kinetic regime of
  hydrodynamic fluctuations and long time tails for a Bjorken expansion}},\
  }\href {https://doi.org/10.1103/PhysRevC.95.014909} {\bibfield  {journal}
  {\bibinfo  {journal} {Phys. Rev. C}\ }\textbf {\bibinfo {volume} {95}},\
  \bibinfo {pages} {014909} (\bibinfo {year} {2017})},\ \Eprint
  {https://arxiv.org/abs/1606.07742} {arXiv:1606.07742 [nucl-th]} \BibitemShut
  {NoStop}%
\bibitem [{\citenamefont {Akamatsu}\ \emph {et~al.}(2019)\citenamefont
  {Akamatsu}, \citenamefont {Teaney}, \citenamefont {Yan},\ and\ \citenamefont
  {Yin}}]{Akamatsu:2018vjr}%
  \BibitemOpen
  \bibfield  {author} {\bibinfo {author} {\bibfnamefont {Y.}~\bibnamefont
  {Akamatsu}}, \bibinfo {author} {\bibfnamefont {D.}~\bibnamefont {Teaney}},
  \bibinfo {author} {\bibfnamefont {F.}~\bibnamefont {Yan}},\ and\ \bibinfo
  {author} {\bibfnamefont {Y.}~\bibnamefont {Yin}},\ }\bibfield  {title}
  {\bibinfo {title} {{Transits of the QCD critical point}},\ }\href
  {https://doi.org/10.1103/PhysRevC.100.044901} {\bibfield  {journal} {\bibinfo
   {journal} {Phys. Rev. C}\ }\textbf {\bibinfo {volume} {100}},\ \bibinfo
  {pages} {044901} (\bibinfo {year} {2019})},\ \Eprint
  {https://arxiv.org/abs/1811.05081} {arXiv:1811.05081 [nucl-th]} \BibitemShut
  {NoStop}%
\bibitem [{\citenamefont {Bluhm}\ and\ \citenamefont
  {Nahrgang}(2020)}]{Bluhm:2019yfb}%
  \BibitemOpen
  \bibfield  {author} {\bibinfo {author} {\bibfnamefont {M.}~\bibnamefont
  {Bluhm}}\ and\ \bibinfo {author} {\bibfnamefont {M.}~\bibnamefont
  {Nahrgang}},\ }\bibfield  {title} {\bibinfo {title} {{Time-evolution of
  net-baryon density fluctuations across the QCD critical region}},\ }\href
  {https://doi.org/10.1007/978-3-030-53448-6_55} {\bibfield  {journal}
  {\bibinfo  {journal} {Springer Proc. Phys.}\ }\textbf {\bibinfo {volume}
  {250}},\ \bibinfo {pages} {351} (\bibinfo {year} {2020})},\ \Eprint
  {https://arxiv.org/abs/1911.08911} {arXiv:1911.08911 [nucl-th]} \BibitemShut
  {NoStop}%
\bibitem [{\citenamefont {Akamatsu}\ \emph {et~al.}(2018)\citenamefont
  {Akamatsu}, \citenamefont {Mazeliauskas},\ and\ \citenamefont
  {Teaney}}]{Akamatsu:2017rdu}%
  \BibitemOpen
  \bibfield  {author} {\bibinfo {author} {\bibfnamefont {Y.}~\bibnamefont
  {Akamatsu}}, \bibinfo {author} {\bibfnamefont {A.}~\bibnamefont
  {Mazeliauskas}},\ and\ \bibinfo {author} {\bibfnamefont {D.}~\bibnamefont
  {Teaney}},\ }\bibfield  {title} {\bibinfo {title} {{Bulk viscosity from
  hydrodynamic fluctuations with relativistic hydrokinetic theory}},\ }\href
  {https://doi.org/10.1103/PhysRevC.97.024902} {\bibfield  {journal} {\bibinfo
  {journal} {Phys. Rev. C}\ }\textbf {\bibinfo {volume} {97}},\ \bibinfo
  {pages} {024902} (\bibinfo {year} {2018})},\ \Eprint
  {https://arxiv.org/abs/1708.05657} {arXiv:1708.05657 [nucl-th]} \BibitemShut
  {NoStop}%
\bibitem [{\citenamefont {An}\ \emph {et~al.}(2019)\citenamefont {An},
  \citenamefont {Ba{\c s}ar}, \citenamefont {Stephanov},\ and\ \citenamefont
  {Yee}}]{An:2019osr}%
  \BibitemOpen
  \bibfield  {author} {\bibinfo {author} {\bibfnamefont {X.}~\bibnamefont
  {An}}, \bibinfo {author} {\bibfnamefont {G.}~\bibnamefont {Ba{\c s}ar}},
  \bibinfo {author} {\bibfnamefont {M.}~\bibnamefont {Stephanov}},\ and\
  \bibinfo {author} {\bibfnamefont {H.-U.}\ \bibnamefont {Yee}},\ }\bibfield
  {title} {\bibinfo {title} {{Relativistic Hydrodynamic Fluctuations}},\ }\href
  {https://doi.org/10.1103/PhysRevC.100.024910} {\bibfield  {journal} {\bibinfo
   {journal} {Phys. Rev.}\ }\textbf {\bibinfo {volume} {C100}},\ \bibinfo
  {pages} {024910} (\bibinfo {year} {2019})},\ \Eprint
  {https://arxiv.org/abs/1902.09517} {arXiv:1902.09517 [hep-th]} \BibitemShut
  {NoStop}%
\bibitem [{\citenamefont {Martinez}\ and\ \citenamefont
  {Sch\"afer}(2019)}]{Martinez:2018wia}%
  \BibitemOpen
  \bibfield  {author} {\bibinfo {author} {\bibfnamefont {M.}~\bibnamefont
  {Martinez}}\ and\ \bibinfo {author} {\bibfnamefont {T.}~\bibnamefont
  {Sch\"afer}},\ }\bibfield  {title} {\bibinfo {title} {{Stochastic
  hydrodynamics and long time tails of an expanding conformal charged fluid}},\
  }\href {https://doi.org/10.1103/PhysRevC.99.054902} {\bibfield  {journal}
  {\bibinfo  {journal} {Phys. Rev. C}\ }\textbf {\bibinfo {volume} {99}},\
  \bibinfo {pages} {054902} (\bibinfo {year} {2019})},\ \Eprint
  {https://arxiv.org/abs/1812.05279} {arXiv:1812.05279 [hep-th]} \BibitemShut
  {NoStop}%
\bibitem [{\citenamefont {Martinez}\ \emph {et~al.}(2019)\citenamefont
  {Martinez}, \citenamefont {Sch{\"a}fer},\ and\ \citenamefont
  {Skokov}}]{Martinez:2019bsn}%
  \BibitemOpen
  \bibfield  {author} {\bibinfo {author} {\bibfnamefont {M.}~\bibnamefont
  {Martinez}}, \bibinfo {author} {\bibfnamefont {T.}~\bibnamefont
  {Sch{\"a}fer}},\ and\ \bibinfo {author} {\bibfnamefont {V.}~\bibnamefont
  {Skokov}},\ }\bibfield  {title} {\bibinfo {title} {{Critical behavior of the
  bulk viscosity in QCD}},\ }\href
  {https://doi.org/10.1103/PhysRevD.100.074017} {\bibfield  {journal} {\bibinfo
   {journal} {Phys. Rev.}\ }\textbf {\bibinfo {volume} {D100}},\ \bibinfo
  {pages} {074017} (\bibinfo {year} {2019})},\ \Eprint
  {https://arxiv.org/abs/1906.11306} {arXiv:1906.11306 [hep-ph]} \BibitemShut
  {NoStop}%
\bibitem [{\citenamefont {An}\ \emph {et~al.}(2020)\citenamefont {An},
  \citenamefont {Ba\c{s}ar}, \citenamefont {Stephanov},\ and\ \citenamefont
  {Yee}}]{An:2019csj}%
  \BibitemOpen
  \bibfield  {author} {\bibinfo {author} {\bibfnamefont {X.}~\bibnamefont
  {An}}, \bibinfo {author} {\bibfnamefont {G.}~\bibnamefont {Ba\c{s}ar}},
  \bibinfo {author} {\bibfnamefont {M.}~\bibnamefont {Stephanov}},\ and\
  \bibinfo {author} {\bibfnamefont {H.-U.}\ \bibnamefont {Yee}},\ }\bibfield
  {title} {\bibinfo {title} {{Fluctuation dynamics in a relativistic fluid with
  a critical point}},\ }\href {https://doi.org/10.1103/PhysRevC.102.034901}
  {\bibfield  {journal} {\bibinfo  {journal} {Phys. Rev. C}\ }\textbf {\bibinfo
  {volume} {102}},\ \bibinfo {pages} {034901} (\bibinfo {year} {2020})},\
  \Eprint {https://arxiv.org/abs/1912.13456} {arXiv:1912.13456 [hep-th]}
  \BibitemShut {NoStop}%
\bibitem [{\citenamefont {An}\ \emph {et~al.}(2021{\natexlab{a}})\citenamefont
  {An}, \citenamefont {Ba\c{s}ar}, \citenamefont {Stephanov},\ and\
  \citenamefont {Yee}}]{An:2020vri}%
  \BibitemOpen
  \bibfield  {author} {\bibinfo {author} {\bibfnamefont {X.}~\bibnamefont
  {An}}, \bibinfo {author} {\bibfnamefont {G.}~\bibnamefont {Ba\c{s}ar}},
  \bibinfo {author} {\bibfnamefont {M.}~\bibnamefont {Stephanov}},\ and\
  \bibinfo {author} {\bibfnamefont {H.-U.}\ \bibnamefont {Yee}},\ }\bibfield
  {title} {\bibinfo {title} {{Evolution of Non-Gaussian Hydrodynamic
  Fluctuations}},\ }\href {https://doi.org/10.1103/PhysRevLett.127.072301}
  {\bibfield  {journal} {\bibinfo  {journal} {Phys. Rev. Lett.}\ }\textbf
  {\bibinfo {volume} {127}},\ \bibinfo {pages} {072301} (\bibinfo {year}
  {2021}{\natexlab{a}})},\ \Eprint {https://arxiv.org/abs/2009.10742}
  {arXiv:2009.10742 [hep-th]} \BibitemShut {NoStop}%
\bibitem [{\citenamefont {Bluhm}\ \emph {et~al.}(2020)\citenamefont {Bluhm}
  \emph {et~al.}}]{Bluhm:2020mpc}%
  \BibitemOpen
  \bibfield  {author} {\bibinfo {author} {\bibfnamefont {M.}~\bibnamefont
  {Bluhm}} \emph {et~al.},\ }\bibfield  {title} {\bibinfo {title} {{Dynamics of
  critical fluctuations: Theory \textendash{} phenomenology \textendash{}
  heavy-ion collisions}},\ }\href
  {https://doi.org/10.1016/j.nuclphysa.2020.122016} {\bibfield  {journal}
  {\bibinfo  {journal} {Nucl. Phys. A}\ }\textbf {\bibinfo {volume} {1003}},\
  \bibinfo {pages} {122016} (\bibinfo {year} {2020})},\ \Eprint
  {https://arxiv.org/abs/2001.08831} {arXiv:2001.08831 [nucl-th]} \BibitemShut
  {NoStop}%
\bibitem [{\citenamefont {Shen}\ and\ \citenamefont
  {Schenke}(2018)}]{Shen:2017bsr}%
  \BibitemOpen
  \bibfield  {author} {\bibinfo {author} {\bibfnamefont {C.}~\bibnamefont
  {Shen}}\ and\ \bibinfo {author} {\bibfnamefont {B.}~\bibnamefont {Schenke}},\
  }\bibfield  {title} {\bibinfo {title} {{Dynamical initial state model for
  relativistic heavy-ion collisions}},\ }\href
  {https://doi.org/10.1103/PhysRevC.97.024907} {\bibfield  {journal} {\bibinfo
  {journal} {Phys. Rev.}\ }\textbf {\bibinfo {volume} {C97}},\ \bibinfo {pages}
  {024907} (\bibinfo {year} {2018})},\ \Eprint
  {https://arxiv.org/abs/1710.00881} {arXiv:1710.00881 [nucl-th]} \BibitemShut
  {NoStop}%
\bibitem [{\citenamefont {Du}\ \emph {et~al.}(2020)\citenamefont {Du},
  \citenamefont {Heinz}, \citenamefont {Rajagopal},\ and\ \citenamefont
  {Yin}}]{Du:2020bxp}%
  \BibitemOpen
  \bibfield  {author} {\bibinfo {author} {\bibfnamefont {L.}~\bibnamefont
  {Du}}, \bibinfo {author} {\bibfnamefont {U.}~\bibnamefont {Heinz}}, \bibinfo
  {author} {\bibfnamefont {K.}~\bibnamefont {Rajagopal}},\ and\ \bibinfo
  {author} {\bibfnamefont {Y.}~\bibnamefont {Yin}},\ }\bibfield  {title}
  {\bibinfo {title} {{Fluctuation dynamics near the QCD critical point}},\
  }\href {https://doi.org/10.1103/PhysRevC.102.054911} {\bibfield  {journal}
  {\bibinfo  {journal} {Phys. Rev. C}\ }\textbf {\bibinfo {volume} {102}},\
  \bibinfo {pages} {054911} (\bibinfo {year} {2020})},\ \Eprint
  {https://arxiv.org/abs/2004.02719} {arXiv:2004.02719 [nucl-th]} \BibitemShut
  {NoStop}%
\bibitem [{\citenamefont {An}\ \emph {et~al.}(2021{\natexlab{b}})\citenamefont
  {An} \emph {et~al.}}]{An:2021wof}%
  \BibitemOpen
  \bibfield  {author} {\bibinfo {author} {\bibfnamefont {X.}~\bibnamefont {An}}
  \emph {et~al.},\ }\bibfield  {title} {\bibinfo {title} {{The BEST framework
  for the search for the QCD critical point and the chiral magnetic effect}},\
  }\href@noop {} {\  (\bibinfo {year} {2021}{\natexlab{b}})},\ \Eprint
  {https://arxiv.org/abs/2108.13867} {arXiv:2108.13867 [nucl-th]} \BibitemShut
  {NoStop}%
\bibitem [{\citenamefont {Kapusta}\ \emph {et~al.}(2012)\citenamefont
  {Kapusta}, \citenamefont {Muller},\ and\ \citenamefont
  {Stephanov}}]{Kapusta:2011gt}%
  \BibitemOpen
  \bibfield  {author} {\bibinfo {author} {\bibfnamefont {J.~I.}\ \bibnamefont
  {Kapusta}}, \bibinfo {author} {\bibfnamefont {B.}~\bibnamefont {Muller}},\
  and\ \bibinfo {author} {\bibfnamefont {M.}~\bibnamefont {Stephanov}},\
  }\bibfield  {title} {\bibinfo {title} {{Relativistic Theory of Hydrodynamic
  Fluctuations with Applications to Heavy Ion Collisions}},\ }\href
  {https://doi.org/10.1103/PhysRevC.85.054906} {\bibfield  {journal} {\bibinfo
  {journal} {Phys. Rev. C}\ }\textbf {\bibinfo {volume} {85}},\ \bibinfo
  {pages} {054906} (\bibinfo {year} {2012})},\ \Eprint
  {https://arxiv.org/abs/1112.6405} {arXiv:1112.6405 [nucl-th]} \BibitemShut
  {NoStop}%
\bibitem [{\citenamefont {Murase}\ and\ \citenamefont
  {Hirano}(2013)}]{Murase:2013tma}%
  \BibitemOpen
  \bibfield  {author} {\bibinfo {author} {\bibfnamefont {K.}~\bibnamefont
  {Murase}}\ and\ \bibinfo {author} {\bibfnamefont {T.}~\bibnamefont
  {Hirano}},\ }\bibfield  {title} {\bibinfo {title} {{Relativistic fluctuating
  hydrodynamics with memory functions and colored noises}},\ }\href@noop {} {\
  (\bibinfo {year} {2013})},\ \Eprint {https://arxiv.org/abs/1304.3243}
  {arXiv:1304.3243 [nucl-th]} \BibitemShut {NoStop}%
\bibitem [{\citenamefont {Young}\ \emph {et~al.}(2015)\citenamefont {Young},
  \citenamefont {Kapusta}, \citenamefont {Gale}, \citenamefont {Jeon},\ and\
  \citenamefont {Schenke}}]{Young:2014pka}%
  \BibitemOpen
  \bibfield  {author} {\bibinfo {author} {\bibfnamefont {C.}~\bibnamefont
  {Young}}, \bibinfo {author} {\bibfnamefont {J.~I.}\ \bibnamefont {Kapusta}},
  \bibinfo {author} {\bibfnamefont {C.}~\bibnamefont {Gale}}, \bibinfo {author}
  {\bibfnamefont {S.}~\bibnamefont {Jeon}},\ and\ \bibinfo {author}
  {\bibfnamefont {B.}~\bibnamefont {Schenke}},\ }\bibfield  {title} {\bibinfo
  {title} {{Thermally Fluctuating Second-Order Viscous Hydrodynamics and
  Heavy-Ion Collisions}},\ }\href {https://doi.org/10.1103/PhysRevC.91.044901}
  {\bibfield  {journal} {\bibinfo  {journal} {Phys. Rev. C}\ }\textbf {\bibinfo
  {volume} {91}},\ \bibinfo {pages} {044901} (\bibinfo {year} {2015})},\
  \Eprint {https://arxiv.org/abs/1407.1077} {arXiv:1407.1077 [nucl-th]}
  \BibitemShut {NoStop}%
\bibitem [{\citenamefont {Murase}\ and\ \citenamefont
  {Hirano}(2016)}]{Murase:2016rhl}%
  \BibitemOpen
  \bibfield  {author} {\bibinfo {author} {\bibfnamefont {K.}~\bibnamefont
  {Murase}}\ and\ \bibinfo {author} {\bibfnamefont {T.}~\bibnamefont
  {Hirano}},\ }\bibfield  {title} {\bibinfo {title} {{Hydrodynamic fluctuations
  and dissipation in an integrated dynamical model}},\ }\href
  {https://doi.org/10.1016/j.nuclphysa.2016.01.011} {\bibfield  {journal}
  {\bibinfo  {journal} {Nucl. Phys. A}\ }\textbf {\bibinfo {volume} {956}},\
  \bibinfo {pages} {276} (\bibinfo {year} {2016})},\ \Eprint
  {https://arxiv.org/abs/1601.02260} {arXiv:1601.02260 [nucl-th]} \BibitemShut
  {NoStop}%
\bibitem [{\citenamefont {Singh}\ \emph {et~al.}(2019)\citenamefont {Singh},
  \citenamefont {Shen}, \citenamefont {McDonald}, \citenamefont {Jeon},\ and\
  \citenamefont {Gale}}]{Singh:2018dpk}%
  \BibitemOpen
  \bibfield  {author} {\bibinfo {author} {\bibfnamefont {M.}~\bibnamefont
  {Singh}}, \bibinfo {author} {\bibfnamefont {C.}~\bibnamefont {Shen}},
  \bibinfo {author} {\bibfnamefont {S.}~\bibnamefont {McDonald}}, \bibinfo
  {author} {\bibfnamefont {S.}~\bibnamefont {Jeon}},\ and\ \bibinfo {author}
  {\bibfnamefont {C.}~\bibnamefont {Gale}},\ }\bibfield  {title} {\bibinfo
  {title} {{Hydrodynamic Fluctuations in Relativistic Heavy-Ion Collisions}},\
  }\href {https://doi.org/10.1016/j.nuclphysa.2018.10.061} {\bibfield
  {journal} {\bibinfo  {journal} {Nucl. Phys. A}\ }\textbf {\bibinfo {volume}
  {982}},\ \bibinfo {pages} {319} (\bibinfo {year} {2019})},\ \Eprint
  {https://arxiv.org/abs/1807.05451} {arXiv:1807.05451 [nucl-th]} \BibitemShut
  {NoStop}%
\bibitem [{\citenamefont {Nahrgang}\ \emph {et~al.}(2017)\citenamefont
  {Nahrgang}, \citenamefont {Bluhm}, \citenamefont {Sch\"afer},\ and\
  \citenamefont {Bass}}]{Nahrgang:2017oqp}%
  \BibitemOpen
  \bibfield  {author} {\bibinfo {author} {\bibfnamefont {M.}~\bibnamefont
  {Nahrgang}}, \bibinfo {author} {\bibfnamefont {M.}~\bibnamefont {Bluhm}},
  \bibinfo {author} {\bibfnamefont {T.}~\bibnamefont {Sch\"afer}},\ and\
  \bibinfo {author} {\bibfnamefont {S.}~\bibnamefont {Bass}},\ }\bibfield
  {title} {\bibinfo {title} {{Toward the description of fluid dynamical
  fluctuations in heavy-ion collisions}},\ }\href
  {https://doi.org/10.5506/APhysPolBSupp.10.687} {\bibfield  {journal}
  {\bibinfo  {journal} {Acta Phys. Polon. Supp.}\ }\textbf {\bibinfo {volume}
  {10}},\ \bibinfo {pages} {687} (\bibinfo {year} {2017})},\ \Eprint
  {https://arxiv.org/abs/1704.03553} {arXiv:1704.03553 [nucl-th]} \BibitemShut
  {NoStop}%
\bibitem [{\citenamefont {Nahrgang}\ \emph {et~al.}(2019)\citenamefont
  {Nahrgang}, \citenamefont {Bluhm}, \citenamefont {Schaefer},\ and\
  \citenamefont {Bass}}]{Nahrgang:2018afz}%
  \BibitemOpen
  \bibfield  {author} {\bibinfo {author} {\bibfnamefont {M.}~\bibnamefont
  {Nahrgang}}, \bibinfo {author} {\bibfnamefont {M.}~\bibnamefont {Bluhm}},
  \bibinfo {author} {\bibfnamefont {T.}~\bibnamefont {Schaefer}},\ and\
  \bibinfo {author} {\bibfnamefont {S.~A.}\ \bibnamefont {Bass}},\ }\bibfield
  {title} {\bibinfo {title} {{Diffusive dynamics of critical fluctuations near
  the QCD critical point}},\ }\href
  {https://doi.org/10.1103/PhysRevD.99.116015} {\bibfield  {journal} {\bibinfo
  {journal} {Phys. Rev. D}\ }\textbf {\bibinfo {volume} {99}},\ \bibinfo
  {pages} {116015} (\bibinfo {year} {2019})},\ \Eprint
  {https://arxiv.org/abs/1804.05728} {arXiv:1804.05728 [nucl-th]} \BibitemShut
  {NoStop}%
\bibitem [{\citenamefont {Nahrgang}\ and\ \citenamefont
  {Bluhm}(2020)}]{Nahrgang:2020yxm}%
  \BibitemOpen
  \bibfield  {author} {\bibinfo {author} {\bibfnamefont {M.}~\bibnamefont
  {Nahrgang}}\ and\ \bibinfo {author} {\bibfnamefont {M.}~\bibnamefont
  {Bluhm}},\ }\bibfield  {title} {\bibinfo {title} {{Modeling the diffusive
  dynamics of critical fluctuations near the QCD critical point}},\ }\href
  {https://doi.org/10.1103/PhysRevD.102.094017} {\bibfield  {journal} {\bibinfo
   {journal} {Phys. Rev. D}\ }\textbf {\bibinfo {volume} {102}},\ \bibinfo
  {pages} {094017} (\bibinfo {year} {2020})},\ \Eprint
  {https://arxiv.org/abs/2007.10371} {arXiv:2007.10371 [nucl-th]} \BibitemShut
  {NoStop}%
\bibitem [{\citenamefont {Sakai}\ \emph {et~al.}(2020)\citenamefont {Sakai},
  \citenamefont {Murase},\ and\ \citenamefont {Hirano}}]{Sakai:2020pjw}%
  \BibitemOpen
  \bibfield  {author} {\bibinfo {author} {\bibfnamefont {A.}~\bibnamefont
  {Sakai}}, \bibinfo {author} {\bibfnamefont {K.}~\bibnamefont {Murase}},\ and\
  \bibinfo {author} {\bibfnamefont {T.}~\bibnamefont {Hirano}},\ }\bibfield
  {title} {\bibinfo {title} {{Rapidity decorrelation of anisotropic flow caused
  by hydrodynamic fluctuations}},\ }\href
  {https://doi.org/10.1103/PhysRevC.102.064903} {\bibfield  {journal} {\bibinfo
   {journal} {Phys. Rev. C}\ }\textbf {\bibinfo {volume} {102}},\ \bibinfo
  {pages} {064903} (\bibinfo {year} {2020})},\ \Eprint
  {https://arxiv.org/abs/2003.13496} {arXiv:2003.13496 [nucl-th]} \BibitemShut
  {NoStop}%
\bibitem [{\citenamefont {Cooper}\ and\ \citenamefont
  {Frye}(1974)}]{Cooper:1974mv}%
  \BibitemOpen
  \bibfield  {author} {\bibinfo {author} {\bibfnamefont {F.}~\bibnamefont
  {Cooper}}\ and\ \bibinfo {author} {\bibfnamefont {G.}~\bibnamefont {Frye}},\
  }\bibfield  {title} {\bibinfo {title} {{Comment on the Single Particle
  Distribution in the Hydrodynamic and Statistical Thermodynamic Models of
  Multiparticle Production}},\ }\href {https://doi.org/10.1103/PhysRevD.10.186}
  {\bibfield  {journal} {\bibinfo  {journal} {Phys. Rev. D}\ }\textbf {\bibinfo
  {volume} {10}},\ \bibinfo {pages} {186} (\bibinfo {year} {1974})}\BibitemShut
  {NoStop}%
\bibitem [{\citenamefont {Berges}\ and\ \citenamefont
  {Rajagopal}(1999)}]{Berges:1998rc}%
  \BibitemOpen
  \bibfield  {author} {\bibinfo {author} {\bibfnamefont {J.}~\bibnamefont
  {Berges}}\ and\ \bibinfo {author} {\bibfnamefont {K.}~\bibnamefont
  {Rajagopal}},\ }\bibfield  {title} {\bibinfo {title} {{Color
  superconductivity and chiral symmetry restoration at nonzero baryon density
  and temperature}},\ }\href {https://doi.org/10.1016/S0550-3213(98)00620-8}
  {\bibfield  {journal} {\bibinfo  {journal} {Nucl. Phys. B}\ }\textbf
  {\bibinfo {volume} {538}},\ \bibinfo {pages} {215} (\bibinfo {year}
  {1999})},\ \Eprint {https://arxiv.org/abs/hep-ph/9804233}
  {arXiv:hep-ph/9804233} \BibitemShut {NoStop}%
\bibitem [{\citenamefont {Halasz}\ \emph {et~al.}(1998)\citenamefont {Halasz},
  \citenamefont {Jackson}, \citenamefont {Shrock}, \citenamefont {Stephanov},\
  and\ \citenamefont {Verbaarschot}}]{Halasz:1998qr}%
  \BibitemOpen
  \bibfield  {author} {\bibinfo {author} {\bibfnamefont {A.~M.}\ \bibnamefont
  {Halasz}}, \bibinfo {author} {\bibfnamefont {A.~D.}\ \bibnamefont {Jackson}},
  \bibinfo {author} {\bibfnamefont {R.~E.}\ \bibnamefont {Shrock}}, \bibinfo
  {author} {\bibfnamefont {M.~A.}\ \bibnamefont {Stephanov}},\ and\ \bibinfo
  {author} {\bibfnamefont {J.~J.~M.}\ \bibnamefont {Verbaarschot}},\ }\bibfield
   {title} {\bibinfo {title} {{On the phase diagram of QCD}},\ }\href
  {https://doi.org/10.1103/PhysRevD.58.096007} {\bibfield  {journal} {\bibinfo
  {journal} {Phys. Rev. D}\ }\textbf {\bibinfo {volume} {58}},\ \bibinfo
  {pages} {096007} (\bibinfo {year} {1998})},\ \Eprint
  {https://arxiv.org/abs/hep-ph/9804290} {arXiv:hep-ph/9804290} \BibitemShut
  {NoStop}%
\bibitem [{\citenamefont {Guida}\ and\ \citenamefont
  {Zinn-Justin}(1997)}]{Guida:1996ep}%
  \BibitemOpen
  \bibfield  {author} {\bibinfo {author} {\bibfnamefont {R.}~\bibnamefont
  {Guida}}\ and\ \bibinfo {author} {\bibfnamefont {J.}~\bibnamefont
  {Zinn-Justin}},\ }\bibfield  {title} {\bibinfo {title} {{3-D Ising model: The
  Scaling equation of state}},\ }\href
  {https://doi.org/10.1016/S0550-3213(96)00704-3} {\bibfield  {journal}
  {\bibinfo  {journal} {Nucl. Phys. B}\ }\textbf {\bibinfo {volume} {489}},\
  \bibinfo {pages} {626} (\bibinfo {year} {1997})},\ \Eprint
  {https://arxiv.org/abs/hep-th/9610223} {arXiv:hep-th/9610223} \BibitemShut
  {NoStop}%
\bibitem [{\citenamefont {Zinn-Justin}(2021)}]{Zinn-Justin:1989rgp}%
  \BibitemOpen
  \bibfield  {author} {\bibinfo {author} {\bibfnamefont {J.}~\bibnamefont
  {Zinn-Justin}},\ }\href@noop {} {\emph {\bibinfo {title} {{Quantum field
  theory and critical phenomena}}}},\ \bibinfo {series} {International Series
  of Monographs on Physics}, Vol.~\bibinfo {volume} {77}\ (\bibinfo
  {publisher} {Oxford University Press},\ \bibinfo {year} {2021})\BibitemShut
  {NoStop}%
\bibitem [{\citenamefont {Rajagopal}\ and\ \citenamefont
  {Wilczek}(1993)}]{Rajagopal:1992qz}%
  \BibitemOpen
  \bibfield  {author} {\bibinfo {author} {\bibfnamefont {K.}~\bibnamefont
  {Rajagopal}}\ and\ \bibinfo {author} {\bibfnamefont {F.}~\bibnamefont
  {Wilczek}},\ }\bibfield  {title} {\bibinfo {title} {{Static and dynamic
  critical phenomena at a second order QCD phase transition}},\ }\href
  {https://doi.org/10.1016/0550-3213(93)90502-G} {\bibfield  {journal}
  {\bibinfo  {journal} {Nucl. Phys. B}\ }\textbf {\bibinfo {volume} {399}},\
  \bibinfo {pages} {395} (\bibinfo {year} {1993})},\ \Eprint
  {https://arxiv.org/abs/hep-ph/9210253} {arXiv:hep-ph/9210253} \BibitemShut
  {NoStop}%
\bibitem [{\citenamefont {Rajagopal}\ and\ \citenamefont
  {Wilczek}(2000)}]{Rajagopal:2000wf}%
  \BibitemOpen
  \bibfield  {author} {\bibinfo {author} {\bibfnamefont {K.}~\bibnamefont
  {Rajagopal}}\ and\ \bibinfo {author} {\bibfnamefont {F.}~\bibnamefont
  {Wilczek}},\ }\bibinfo {title} {{The Condensed matter physics of QCD}},\ in\
  \href {https://doi.org/10.1142/9789812810458_0043} {\emph {\bibinfo
  {booktitle} {{At the frontier of particle physics. Handbook of QCD. Vol.
  1-3}}}},\ \bibinfo {editor} {edited by\ \bibinfo {editor} {\bibfnamefont
  {M.}~\bibnamefont {Shifman}}\ and\ \bibinfo {editor} {\bibfnamefont
  {B.}~\bibnamefont {Ioffe}}}\ (\bibinfo {year} {2000})\ pp.\ \bibinfo {pages}
  {2061--2151},\ \Eprint {https://arxiv.org/abs/hep-ph/0011333}
  {arXiv:hep-ph/0011333} \BibitemShut {NoStop}%
\bibitem [{\citenamefont {Parotto}\ \emph {et~al.}(2018)\citenamefont
  {Parotto}, \citenamefont {Bluhm}, \citenamefont {Mroczek}, \citenamefont
  {Nahrgang}, \citenamefont {Noronha-Hostler}, \citenamefont {Rajagopal},
  \citenamefont {Ratti}, \citenamefont {Schäfer},\ and\ \citenamefont
  {Stephanov}}]{Parotto:2018pwx}%
  \BibitemOpen
  \bibfield  {author} {\bibinfo {author} {\bibfnamefont {P.}~\bibnamefont
  {Parotto}}, \bibinfo {author} {\bibfnamefont {M.}~\bibnamefont {Bluhm}},
  \bibinfo {author} {\bibfnamefont {D.}~\bibnamefont {Mroczek}}, \bibinfo
  {author} {\bibfnamefont {M.}~\bibnamefont {Nahrgang}}, \bibinfo {author}
  {\bibfnamefont {J.}~\bibnamefont {Noronha-Hostler}}, \bibinfo {author}
  {\bibfnamefont {K.}~\bibnamefont {Rajagopal}}, \bibinfo {author}
  {\bibfnamefont {C.}~\bibnamefont {Ratti}}, \bibinfo {author} {\bibfnamefont
  {T.}~\bibnamefont {Schäfer}},\ and\ \bibinfo {author} {\bibfnamefont
  {M.}~\bibnamefont {Stephanov}},\ }\bibfield  {title} {\bibinfo {title}
  {{Lattice-QCD-based equation of state with a critical point}},\ }\href@noop
  {} {\  (\bibinfo {year} {2018})},\ \Eprint {https://arxiv.org/abs/1805.05249}
  {arXiv:1805.05249 [hep-ph]} \BibitemShut {NoStop}%
\bibitem [{\citenamefont {Pradeep}\ and\ \citenamefont
  {Stephanov}(2019)}]{Pradeep:2019ccv}%
  \BibitemOpen
  \bibfield  {author} {\bibinfo {author} {\bibfnamefont {M.~S.}\ \bibnamefont
  {Pradeep}}\ and\ \bibinfo {author} {\bibfnamefont {M.}~\bibnamefont
  {Stephanov}},\ }\bibfield  {title} {\bibinfo {title} {{Universality of the
  critical point mapping between Ising model and QCD at small quark mass}},\
  }\href {https://doi.org/10.1103/PhysRevD.100.056003} {\bibfield  {journal}
  {\bibinfo  {journal} {Phys. Rev.}\ }\textbf {\bibinfo {volume} {D100}},\
  \bibinfo {pages} {056003} (\bibinfo {year} {2019})},\ \Eprint
  {https://arxiv.org/abs/1905.13247} {arXiv:1905.13247 [hep-ph]} \BibitemShut
  {NoStop}%
\bibitem [{\citenamefont {Son}\ and\ \citenamefont
  {Stephanov}(2004)}]{Son:2004iv}%
  \BibitemOpen
  \bibfield  {author} {\bibinfo {author} {\bibfnamefont {D.~T.}\ \bibnamefont
  {Son}}\ and\ \bibinfo {author} {\bibfnamefont {M.~A.}\ \bibnamefont
  {Stephanov}},\ }\bibfield  {title} {\bibinfo {title} {{Dynamic universality
  class of the QCD critical point}},\ }\href
  {https://doi.org/10.1103/PhysRevD.70.056001} {\bibfield  {journal} {\bibinfo
  {journal} {Phys. Rev.}\ }\textbf {\bibinfo {volume} {D70}},\ \bibinfo {pages}
  {056001} (\bibinfo {year} {2004})},\ \Eprint
  {https://arxiv.org/abs/hep-ph/0401052} {arXiv:hep-ph/0401052 [hep-ph]}
  \BibitemShut {NoStop}%
\bibitem [{\citenamefont {Hohenberg}\ and\ \citenamefont
  {Halperin}(1977)}]{Hohenberg:1977ym}%
  \BibitemOpen
  \bibfield  {author} {\bibinfo {author} {\bibfnamefont {P.~C.}\ \bibnamefont
  {Hohenberg}}\ and\ \bibinfo {author} {\bibfnamefont {B.~I.}\ \bibnamefont
  {Halperin}},\ }\bibfield  {title} {\bibinfo {title} {{Theory of Dynamic
  Critical Phenomena}},\ }\href {https://doi.org/10.1103/RevModPhys.49.435}
  {\bibfield  {journal} {\bibinfo  {journal} {Rev. Mod. Phys.}\ }\textbf
  {\bibinfo {volume} {49}},\ \bibinfo {pages} {435} (\bibinfo {year}
  {1977})}\BibitemShut {NoStop}%
\bibitem [{\citenamefont {Kawasaki}(1970)}]{KAWASAKI19701}%
  \BibitemOpen
  \bibfield  {author} {\bibinfo {author} {\bibfnamefont {K.}~\bibnamefont
  {Kawasaki}},\ }\bibfield  {title} {\bibinfo {title} {Kinetic equations and
  time correlation functions of critical fluctuations},\ }\href
  {https://doi.org/https://doi.org/10.1016/0003-4916(70)90375-1} {\bibfield
  {journal} {\bibinfo  {journal} {Annals of Physics}\ }\textbf {\bibinfo
  {volume} {61}},\ \bibinfo {pages} {1} (\bibinfo {year} {1970})}\BibitemShut
  {NoStop}%
\bibitem [{\citenamefont {Borsanyi}\ \emph {et~al.}(2014)\citenamefont
  {Borsanyi}, \citenamefont {Fodor}, \citenamefont {Hoelbling}, \citenamefont
  {Katz}, \citenamefont {Krieg},\ and\ \citenamefont
  {Szabo}}]{Borsanyi:2013bia}%
  \BibitemOpen
  \bibfield  {author} {\bibinfo {author} {\bibfnamefont {S.}~\bibnamefont
  {Borsanyi}}, \bibinfo {author} {\bibfnamefont {Z.}~\bibnamefont {Fodor}},
  \bibinfo {author} {\bibfnamefont {C.}~\bibnamefont {Hoelbling}}, \bibinfo
  {author} {\bibfnamefont {S.~D.}\ \bibnamefont {Katz}}, \bibinfo {author}
  {\bibfnamefont {S.}~\bibnamefont {Krieg}},\ and\ \bibinfo {author}
  {\bibfnamefont {K.~K.}\ \bibnamefont {Szabo}},\ }\bibfield  {title} {\bibinfo
  {title} {{Full result for the QCD equation of state with 2+1 flavors}},\
  }\href {https://doi.org/10.1016/j.physletb.2014.01.007} {\bibfield  {journal}
  {\bibinfo  {journal} {Phys. Lett. B}\ }\textbf {\bibinfo {volume} {730}},\
  \bibinfo {pages} {99} (\bibinfo {year} {2014})},\ \Eprint
  {https://arxiv.org/abs/1309.5258} {arXiv:1309.5258 [hep-lat]} \BibitemShut
  {NoStop}%
\bibitem [{\citenamefont {Bazavov}\ \emph {et~al.}(2014)\citenamefont {Bazavov}
  \emph {et~al.}}]{HotQCD:2014kol}%
  \BibitemOpen
  \bibfield  {author} {\bibinfo {author} {\bibfnamefont {A.}~\bibnamefont
  {Bazavov}} \emph {et~al.} (\bibinfo {collaboration} {HotQCD}),\ }\bibfield
  {title} {\bibinfo {title} {{Equation of state in ( 2+1 )-flavor QCD}},\
  }\href {https://doi.org/10.1103/PhysRevD.90.094503} {\bibfield  {journal}
  {\bibinfo  {journal} {Phys. Rev. D}\ }\textbf {\bibinfo {volume} {90}},\
  \bibinfo {pages} {094503} (\bibinfo {year} {2014})},\ \Eprint
  {https://arxiv.org/abs/1407.6387} {arXiv:1407.6387 [hep-lat]} \BibitemShut
  {NoStop}%
\bibitem [{\citenamefont
  {Grad}(1949)}]{https://doi.org/10.1002/cpa.3160020403}%
  \BibitemOpen
  \bibfield  {author} {\bibinfo {author} {\bibfnamefont {H.}~\bibnamefont
  {Grad}},\ }\bibfield  {title} {\bibinfo {title} {On the kinetic theory of
  rarefied gases},\ }\href
  {https://doi.org/https://doi.org/10.1002/cpa.3160020403} {\bibfield
  {journal} {\bibinfo  {journal} {Communications on Pure and Applied
  Mathematics}\ }\textbf {\bibinfo {volume} {2}},\ \bibinfo {pages} {331}
  (\bibinfo {year} {1949})},\ \Eprint
  {https://arxiv.org/abs/https://onlinelibrary.wiley.com/doi/pdf/10.1002/cpa.3160020403}
  {https://onlinelibrary.wiley.com/doi/pdf/10.1002/cpa.3160020403} \BibitemShut
  {NoStop}%
\bibitem [{\citenamefont {Anderson}\ and\ \citenamefont
  {Witting}(1974)}]{ANDERSON1974466}%
  \BibitemOpen
  \bibfield  {author} {\bibinfo {author} {\bibfnamefont {J.}~\bibnamefont
  {Anderson}}\ and\ \bibinfo {author} {\bibfnamefont {H.}~\bibnamefont
  {Witting}},\ }\bibfield  {title} {\bibinfo {title} {A relativistic
  relaxation-time model for the boltzmann equation},\ }\href
  {https://doi.org/https://doi.org/10.1016/0031-8914(74)90355-3} {\bibfield
  {journal} {\bibinfo  {journal} {Physica}\ }\textbf {\bibinfo {volume} {74}},\
  \bibinfo {pages} {466} (\bibinfo {year} {1974})}\BibitemShut {NoStop}%
\bibitem [{\citenamefont {Pratt}\ and\ \citenamefont
  {Torrieri}(2010)}]{Pratt:2010jt}%
  \BibitemOpen
  \bibfield  {author} {\bibinfo {author} {\bibfnamefont {S.}~\bibnamefont
  {Pratt}}\ and\ \bibinfo {author} {\bibfnamefont {G.}~\bibnamefont
  {Torrieri}},\ }\bibfield  {title} {\bibinfo {title} {{Coupling Relativistic
  Viscous Hydrodynamics to Boltzmann Descriptions}},\ }\href
  {https://doi.org/10.1103/PhysRevC.82.044901} {\bibfield  {journal} {\bibinfo
  {journal} {Phys. Rev. C}\ }\textbf {\bibinfo {volume} {82}},\ \bibinfo
  {pages} {044901} (\bibinfo {year} {2010})},\ \Eprint
  {https://arxiv.org/abs/1003.0413} {arXiv:1003.0413 [nucl-th]} \BibitemShut
  {NoStop}%
\bibitem [{\citenamefont {McNelis}\ \emph {et~al.}(2021)\citenamefont
  {McNelis}, \citenamefont {Everett},\ and\ \citenamefont
  {Heinz}}]{McNelis:2019auj}%
  \BibitemOpen
  \bibfield  {author} {\bibinfo {author} {\bibfnamefont {M.}~\bibnamefont
  {McNelis}}, \bibinfo {author} {\bibfnamefont {D.}~\bibnamefont {Everett}},\
  and\ \bibinfo {author} {\bibfnamefont {U.}~\bibnamefont {Heinz}},\ }\bibfield
   {title} {\bibinfo {title} {{Particlization in fluid dynamical simulations of
  heavy-ion collisions: The iS3D module}},\ }\href
  {https://doi.org/10.1016/j.cpc.2020.107604} {\bibfield  {journal} {\bibinfo
  {journal} {Comput. Phys. Commun.}\ }\textbf {\bibinfo {volume} {258}},\
  \bibinfo {pages} {107604} (\bibinfo {year} {2021})},\ \Eprint
  {https://arxiv.org/abs/1912.08271} {arXiv:1912.08271 [nucl-th]} \BibitemShut
  {NoStop}%
\bibitem [{\citenamefont {McNelis}\ and\ \citenamefont
  {Heinz}(2021)}]{McNelis:2021acu}%
  \BibitemOpen
  \bibfield  {author} {\bibinfo {author} {\bibfnamefont {M.}~\bibnamefont
  {McNelis}}\ and\ \bibinfo {author} {\bibfnamefont {U.}~\bibnamefont
  {Heinz}},\ }\bibfield  {title} {\bibinfo {title} {{Modified equilibrium
  distributions for Cooper--Frye particlization}},\ }\href
  {https://doi.org/10.1103/PhysRevC.103.064903} {\bibfield  {journal} {\bibinfo
   {journal} {Phys. Rev. C}\ }\textbf {\bibinfo {volume} {103}},\ \bibinfo
  {pages} {064903} (\bibinfo {year} {2021})},\ \Eprint
  {https://arxiv.org/abs/2103.03401} {arXiv:2103.03401 [nucl-th]} \BibitemShut
  {NoStop}%
\bibitem [{\citenamefont {Denicol}\ \emph {et~al.}(2018)\citenamefont
  {Denicol}, \citenamefont {Gale}, \citenamefont {Jeon}, \citenamefont
  {Monnai}, \citenamefont {Schenke},\ and\ \citenamefont
  {Shen}}]{Denicol:2018wdp}%
  \BibitemOpen
  \bibfield  {author} {\bibinfo {author} {\bibfnamefont {G.~S.}\ \bibnamefont
  {Denicol}}, \bibinfo {author} {\bibfnamefont {C.}~\bibnamefont {Gale}},
  \bibinfo {author} {\bibfnamefont {S.}~\bibnamefont {Jeon}}, \bibinfo {author}
  {\bibfnamefont {A.}~\bibnamefont {Monnai}}, \bibinfo {author} {\bibfnamefont
  {B.}~\bibnamefont {Schenke}},\ and\ \bibinfo {author} {\bibfnamefont
  {C.}~\bibnamefont {Shen}},\ }\bibfield  {title} {\bibinfo {title} {{Net
  baryon diffusion in fluid dynamic simulations of relativistic heavy-ion
  collisions}},\ }\href {https://doi.org/10.1103/PhysRevC.98.034916} {\bibfield
   {journal} {\bibinfo  {journal} {Phys. Rev. C}\ }\textbf {\bibinfo {volume}
  {98}},\ \bibinfo {pages} {034916} (\bibinfo {year} {2018})},\ \Eprint
  {https://arxiv.org/abs/1804.10557} {arXiv:1804.10557 [nucl-th]} \BibitemShut
  {NoStop}%
\bibitem [{\citenamefont {Jiang}\ \emph {et~al.}(2016)\citenamefont {Jiang},
  \citenamefont {Li},\ and\ \citenamefont {Song}}]{Jiang:2015hri}%
  \BibitemOpen
  \bibfield  {author} {\bibinfo {author} {\bibfnamefont {L.}~\bibnamefont
  {Jiang}}, \bibinfo {author} {\bibfnamefont {P.}~\bibnamefont {Li}},\ and\
  \bibinfo {author} {\bibfnamefont {H.}~\bibnamefont {Song}},\ }\bibfield
  {title} {\bibinfo {title} {{Correlated fluctuations near the QCD critical
  point}},\ }\href {https://doi.org/10.1103/PhysRevC.94.024918} {\bibfield
  {journal} {\bibinfo  {journal} {Phys. Rev. C}\ }\textbf {\bibinfo {volume}
  {94}},\ \bibinfo {pages} {024918} (\bibinfo {year} {2016})},\ \Eprint
  {https://arxiv.org/abs/1512.06164} {arXiv:1512.06164 [nucl-th]} \BibitemShut
  {NoStop}%
\bibitem [{\citenamefont {Bluhm}\ \emph {et~al.}(2017)\citenamefont {Bluhm},
  \citenamefont {Nahrgang}, \citenamefont {Bass},\ and\ \citenamefont
  {Schaefer}}]{Bluhm:2016byc}%
  \BibitemOpen
  \bibfield  {author} {\bibinfo {author} {\bibfnamefont {M.}~\bibnamefont
  {Bluhm}}, \bibinfo {author} {\bibfnamefont {M.}~\bibnamefont {Nahrgang}},
  \bibinfo {author} {\bibfnamefont {S.~A.}\ \bibnamefont {Bass}},\ and\
  \bibinfo {author} {\bibfnamefont {T.}~\bibnamefont {Schaefer}},\ }\bibfield
  {title} {\bibinfo {title} {{Impact of resonance decays on critical point
  signals in net-proton fluctuations}},\ }\href
  {https://doi.org/10.1140/epjc/s10052-017-4771-3} {\bibfield  {journal}
  {\bibinfo  {journal} {Eur. Phys. J. C}\ }\textbf {\bibinfo {volume} {77}},\
  \bibinfo {pages} {210} (\bibinfo {year} {2017})},\ \Eprint
  {https://arxiv.org/abs/1612.03889} {arXiv:1612.03889 [nucl-th]} \BibitemShut
  {NoStop}%
\bibitem [{\citenamefont {Wu}\ \emph {et~al.}(2019)\citenamefont {Wu},
  \citenamefont {Wu},\ and\ \citenamefont {Song}}]{Wu:2018twy}%
  \BibitemOpen
  \bibfield  {author} {\bibinfo {author} {\bibfnamefont {S.}~\bibnamefont
  {Wu}}, \bibinfo {author} {\bibfnamefont {Z.}~\bibnamefont {Wu}},\ and\
  \bibinfo {author} {\bibfnamefont {H.}~\bibnamefont {Song}},\ }\bibfield
  {title} {\bibinfo {title} {{Universal scaling of the \ensuremath{\sigma}
  field and net-protons from Langevin dynamics of model A}},\ }\href
  {https://doi.org/10.1103/PhysRevC.99.064902} {\bibfield  {journal} {\bibinfo
  {journal} {Phys. Rev. C}\ }\textbf {\bibinfo {volume} {99}},\ \bibinfo
  {pages} {064902} (\bibinfo {year} {2019})},\ \Eprint
  {https://arxiv.org/abs/1811.09466} {arXiv:1811.09466 [nucl-th]} \BibitemShut
  {NoStop}%
\bibitem [{\citenamefont {Szyma\'nski}\ \emph {et~al.}(2020)\citenamefont
  {Szyma\'nski}, \citenamefont {Bluhm}, \citenamefont {Redlich},\ and\
  \citenamefont {Sasaki}}]{Szymanski:2019yho}%
  \BibitemOpen
  \bibfield  {author} {\bibinfo {author} {\bibfnamefont {M.}~\bibnamefont
  {Szyma\'nski}}, \bibinfo {author} {\bibfnamefont {M.}~\bibnamefont {Bluhm}},
  \bibinfo {author} {\bibfnamefont {K.}~\bibnamefont {Redlich}},\ and\ \bibinfo
  {author} {\bibfnamefont {C.}~\bibnamefont {Sasaki}},\ }\bibfield  {title}
  {\bibinfo {title} {{Net-proton number fluctuations in the presence of the QCD
  critical point}},\ }\href {https://doi.org/10.1088/1361-6471/ab614c}
  {\bibfield  {journal} {\bibinfo  {journal} {J. Phys. G}\ }\textbf {\bibinfo
  {volume} {47}},\ \bibinfo {pages} {045102} (\bibinfo {year} {2020})},\
  \Eprint {https://arxiv.org/abs/1905.00667} {arXiv:1905.00667 [nucl-th]}
  \BibitemShut {NoStop}%
\bibitem [{\citenamefont {Stephanov}(2010)}]{Stephanov:2009ra}%
  \BibitemOpen
  \bibfield  {author} {\bibinfo {author} {\bibfnamefont {M.~A.}\ \bibnamefont
  {Stephanov}},\ }\bibfield  {title} {\bibinfo {title} {{Evolution of
  fluctuations near QCD critical point}},\ }\href
  {https://doi.org/10.1103/PhysRevD.81.054012} {\bibfield  {journal} {\bibinfo
  {journal} {Phys. Rev. D}\ }\textbf {\bibinfo {volume} {81}},\ \bibinfo
  {pages} {054012} (\bibinfo {year} {2010})},\ \Eprint
  {https://arxiv.org/abs/0911.1772} {arXiv:0911.1772 [hep-ph]} \BibitemShut
  {NoStop}%
\bibitem [{\citenamefont {Bzdak}\ \emph {et~al.}(2013)\citenamefont {Bzdak},
  \citenamefont {Koch},\ and\ \citenamefont {Skokov}}]{Bzdak:2012an}%
  \BibitemOpen
  \bibfield  {author} {\bibinfo {author} {\bibfnamefont {A.}~\bibnamefont
  {Bzdak}}, \bibinfo {author} {\bibfnamefont {V.}~\bibnamefont {Koch}},\ and\
  \bibinfo {author} {\bibfnamefont {V.}~\bibnamefont {Skokov}},\ }\bibfield
  {title} {\bibinfo {title} {{Baryon number conservation and the cumulants of
  the net proton distribution}},\ }\href
  {https://doi.org/10.1103/PhysRevC.87.014901} {\bibfield  {journal} {\bibinfo
  {journal} {Phys. Rev. C}\ }\textbf {\bibinfo {volume} {87}},\ \bibinfo
  {pages} {014901} (\bibinfo {year} {2013})},\ \Eprint
  {https://arxiv.org/abs/1203.4529} {arXiv:1203.4529 [hep-ph]} \BibitemShut
  {NoStop}%
\bibitem [{\citenamefont {Vovchenko}\ \emph {et~al.}(2020)\citenamefont
  {Vovchenko}, \citenamefont {Poberezhnyuk},\ and\ \citenamefont
  {Koch}}]{Vovchenko:2020gne}%
  \BibitemOpen
  \bibfield  {author} {\bibinfo {author} {\bibfnamefont {V.}~\bibnamefont
  {Vovchenko}}, \bibinfo {author} {\bibfnamefont {R.~V.}\ \bibnamefont
  {Poberezhnyuk}},\ and\ \bibinfo {author} {\bibfnamefont {V.}~\bibnamefont
  {Koch}},\ }\bibfield  {title} {\bibinfo {title} {{Cumulants of multiple
  conserved charges and global conservation laws}},\ }\href
  {https://doi.org/10.1007/JHEP10(2020)089} {\bibfield  {journal} {\bibinfo
  {journal} {JHEP}\ }\textbf {\bibinfo {volume} {10}},\ \bibinfo {pages}
  {089}},\ \Eprint {https://arxiv.org/abs/2007.03850} {arXiv:2007.03850
  [hep-ph]} \BibitemShut {NoStop}%
\bibitem [{\citenamefont {Luzum}\ and\ \citenamefont
  {Petersen}(2014)}]{Luzum:2013yya}%
  \BibitemOpen
  \bibfield  {author} {\bibinfo {author} {\bibfnamefont {M.}~\bibnamefont
  {Luzum}}\ and\ \bibinfo {author} {\bibfnamefont {H.}~\bibnamefont
  {Petersen}},\ }\bibfield  {title} {\bibinfo {title} {{Initial State
  Fluctuations and Final State Correlations in Relativistic Heavy-Ion
  Collisions}},\ }\href {https://doi.org/10.1088/0954-3899/41/6/063102}
  {\bibfield  {journal} {\bibinfo  {journal} {J. Phys. G}\ }\textbf {\bibinfo
  {volume} {41}},\ \bibinfo {pages} {063102} (\bibinfo {year} {2014})},\
  \Eprint {https://arxiv.org/abs/1312.5503} {arXiv:1312.5503 [nucl-th]}
  \BibitemShut {NoStop}%
\bibitem [{\citenamefont {Floerchinger}\ and\ \citenamefont
  {Wiedemann}(2014)}]{Floerchinger:2013hza}%
  \BibitemOpen
  \bibfield  {author} {\bibinfo {author} {\bibfnamefont {S.}~\bibnamefont
  {Floerchinger}}\ and\ \bibinfo {author} {\bibfnamefont {U.~A.}\ \bibnamefont
  {Wiedemann}},\ }\bibfield  {title} {\bibinfo {title} {{Kinetic freeze-out,
  particle spectra and harmonic flow coefficients from mode-by-mode
  hydrodynamics}},\ }\href {https://doi.org/10.1103/PhysRevC.89.034914}
  {\bibfield  {journal} {\bibinfo  {journal} {Phys. Rev.}\ }\textbf {\bibinfo
  {volume} {C89}},\ \bibinfo {pages} {034914} (\bibinfo {year} {2014})},\
  \Eprint {https://arxiv.org/abs/1311.7613} {arXiv:1311.7613 [hep-ph]}
  \BibitemShut {NoStop}%
\bibitem [{\citenamefont {Bjorken}()}]{Bjorken:1982qr}%
  \BibitemOpen
  \bibfield  {author} {\bibinfo {author} {\bibfnamefont {J.~D.}\ \bibnamefont
  {Bjorken}},\ }\bibfield  {title} {\bibinfo {title} {{Highly Relativistic
  Nucleus-Nucleus Collisions: The Central Rapidity Region}},\ }\href
  {https://doi.org/10.1103/PhysRevD.27.140} {\bibfield  {journal} {\bibinfo
  {journal} {Phys. Rev. D}\ }\textbf {\bibinfo {volume} {27}},\ \bibinfo
  {pages} {140}}\BibitemShut {NoStop}%
\bibitem [{\citenamefont {Ohnishi}\ \emph {et~al.}(2016)\citenamefont
  {Ohnishi}, \citenamefont {Kitazawa},\ and\ \citenamefont
  {Asakawa}}]{Ohnishi:2016bdf}%
  \BibitemOpen
  \bibfield  {author} {\bibinfo {author} {\bibfnamefont {Y.}~\bibnamefont
  {Ohnishi}}, \bibinfo {author} {\bibfnamefont {M.}~\bibnamefont {Kitazawa}},\
  and\ \bibinfo {author} {\bibfnamefont {M.}~\bibnamefont {Asakawa}},\
  }\bibfield  {title} {\bibinfo {title} {{Thermal blurring of event-by-event
  fluctuations generated by rapidity conversion}},\ }\href
  {https://doi.org/10.1103/PhysRevC.94.044905} {\bibfield  {journal} {\bibinfo
  {journal} {Phys. Rev. C}\ }\textbf {\bibinfo {volume} {94}},\ \bibinfo
  {pages} {044905} (\bibinfo {year} {2016})},\ \Eprint
  {https://arxiv.org/abs/1606.03827} {arXiv:1606.03827 [nucl-th]} \BibitemShut
  {NoStop}%
\bibitem [{\citenamefont {Baier}\ and\ \citenamefont
  {Romatschke}(2007)}]{Baier:2006gy}%
  \BibitemOpen
  \bibfield  {author} {\bibinfo {author} {\bibfnamefont {R.}~\bibnamefont
  {Baier}}\ and\ \bibinfo {author} {\bibfnamefont {P.}~\bibnamefont
  {Romatschke}},\ }\bibfield  {title} {\bibinfo {title} {{Causal viscous
  hydrodynamics for central heavy-ion collisions}},\ }\href
  {https://doi.org/10.1140/epjc/s10052-007-0308-5} {\bibfield  {journal}
  {\bibinfo  {journal} {Eur. Phys. J. C}\ }\textbf {\bibinfo {volume} {51}},\
  \bibinfo {pages} {677} (\bibinfo {year} {2007})},\ \Eprint
  {https://arxiv.org/abs/nucl-th/0610108} {arXiv:nucl-th/0610108} \BibitemShut
  {NoStop}%
\bibitem [{\citenamefont {Baier}\ \emph {et~al.}(2006)\citenamefont {Baier},
  \citenamefont {Romatschke},\ and\ \citenamefont {Wiedemann}}]{Baier:2006um}%
  \BibitemOpen
  \bibfield  {author} {\bibinfo {author} {\bibfnamefont {R.}~\bibnamefont
  {Baier}}, \bibinfo {author} {\bibfnamefont {P.}~\bibnamefont {Romatschke}},\
  and\ \bibinfo {author} {\bibfnamefont {U.~A.}\ \bibnamefont {Wiedemann}},\
  }\bibfield  {title} {\bibinfo {title} {{Dissipative hydrodynamics and heavy
  ion collisions}},\ }\href {https://doi.org/10.1103/PhysRevC.73.064903}
  {\bibfield  {journal} {\bibinfo  {journal} {Phys. Rev. C}\ }\textbf {\bibinfo
  {volume} {73}},\ \bibinfo {pages} {064903} (\bibinfo {year} {2006})},\
  \Eprint {https://arxiv.org/abs/hep-ph/0602249} {arXiv:hep-ph/0602249}
  \BibitemShut {NoStop}%
\bibitem [{\citenamefont {Romatschke}(2007)}]{Romatschke:2007jx}%
  \BibitemOpen
  \bibfield  {author} {\bibinfo {author} {\bibfnamefont {P.}~\bibnamefont
  {Romatschke}},\ }\bibfield  {title} {\bibinfo {title} {{Causal viscous
  hydrodynamics for central heavy-ion collisions. II. Meson spectra and HBT
  radii}},\ }\href {https://doi.org/10.1140/epjc/s10052-007-0354-z} {\bibfield
  {journal} {\bibinfo  {journal} {Eur. Phys. J. C}\ }\textbf {\bibinfo {volume}
  {52}},\ \bibinfo {pages} {203} (\bibinfo {year} {2007})},\ \Eprint
  {https://arxiv.org/abs/nucl-th/0701032} {arXiv:nucl-th/0701032} \BibitemShut
  {NoStop}%
\end{thebibliography}%

\end{document}